\documentclass[twocolumn,tighten]{aastex63}
\usepackage{amsmath}
\usepackage{amstext}
\usepackage{colortbl}
\usepackage{enumitem}

\newcommand{\kms}{km~s$^{-1}$}
\newcommand{\Msolar}{M$_{\odot}$}
\newcommand{\numbin}{120}  
\newcommand{\numtrip}{six}
\newcommand{\Pcirc}{11.0\substack{+1.1 \\ -1.0} \text{ days}}

\shorttitle{The M67 Spectroscopic Binary Population}
\shortauthors{Geller et al.}

\begin{document}

\title{Stellar Radial Velocities in the Old Open Cluster M67 (NGC 2682). II.  The Spectroscopic Binary Population}

\author[0000-0002-3881-9332]{Aaron M.\ Geller}
\affiliation{Center for Interdisciplinary Exploration and Research in Astrophysics (CIERA) and Department of Physics and Astronomy, Northwestern University, 1800 Sherman Ave., Evanston, IL 60201, USA}
\affiliation{Adler Planetarium, Department of Astronomy, 1300 S. Lake Shore Drive, Chicago, IL 60605, USA}
\email{a-geller@northwestern.edu}

\author{Robert D.\ Mathieu}
\affiliation{Department of Astronomy, University of Wisconsin-Madison, 475 North Charter Street, Madison, WI 53706, USA}

\author{David W.\ Latham}
\affiliation{Center for Astrophysics $\vert$ Harvard \& Smithsonian, 60 Garden Street, 60 Garden Street, Cambridge, MA 02138, USA}

\author{Maxwell Pollack}
\affiliation{Department of Astronomy, University of Wisconsin-Madison, 475 North Charter Street, Madison, WI 53706, USA}

\author{Guillermo Torres}
\affiliation{Center for Astrophysics $\vert$ Harvard \& Smithsonian, 60 Garden Street, 60 Garden Street, Cambridge, MA 02138, USA}

\author{Emily M.\ Leiner}
\affiliation{Center for Interdisciplinary Exploration and Research in Astrophysics (CIERA) and Department of Physics and Astronomy, Northwestern University, 1800 Sherman Ave., Evanston, IL 60201, USA}
\affiliation{Department of Astronomy, University of Wisconsin-Madison, 475 North Charter Street, Madison, WI 53706, USA}

\begin{abstract}
We present and analyse \numbin\ spectroscopic binary and triple cluster members of the old (4~Gyr) open cluster M67 (NGC~2682).  As a cornerstone of stellar astrophysics, M67 is a key cluster in the WIYN Open Cluster Study (WOCS); radial-velocity (RV) observations of M67 are ongoing and extend back over 45 years, incorporating data from seven different telescopes, and allowing us to detect binaries with orbital periods $\lesssim10^4$ days. Our sample contains 1296 stars (604 cluster members) with magnitudes of $10 \le V \le 16.5$ (about 1.3 to 0.7 \Msolar), from the giants down to $\sim$4 mag below the main-sequence turnoff, and extends in radius to 30 arcminutes (7.4 pc at a distance of 850 pc, or $\sim$7 core radii). This paper focuses primarily on the main-sequence binaries, but orbital solutions are also presented for red giants, yellow giants and sub-subgiants.  Out to our period detection limit and within our magnitude and spatial domain, we find a global main-sequence incompleteness-corrected binary fraction of 34\%~$\pm$~3\%, which rises to 70\%~$\pm$~17\% in the cluster center.  We derive a tidal circularization period of $P_{\text{circ}} = \Pcirc$.  We also analyze the incompleteness-corrected distributions of binary orbital elements and masses.  The period distribution rises toward longer periods. The eccentricity distribution, beyond $P_{\text{circ}}$, is consistent with a uniform distribution.  The mass-ratio distribution is also consistent with a uniform distribution. Overall, these M67 binaries are closely consistent with similar binaries in the galactic field, as well as the old (7~Gyr) open cluster NGC 188. WIYN Open Cluster Study. 83.

\end{abstract}


\section{Introduction}

M67 (NGC 2682) is among the most well-studied open clusters.  Given its old age ($\sim$4 Gyr), proximity ($\sim$850 pc) and low extinction ($E(B-V)$ measurements between 0.015 and 0.056), M67 serves as a powerful probe into the nature of solar-age stars near solar metallicity ([Fe/H] between +0.05 and +0.10) and into the later stages of stellar and dynamical evolution, both orbital and cluster. M67 is located at $\alpha = 8^{\text{h}}51^{\text{m}}23.^{\text{s}}3$, $\delta = +11^\circ 49'02"$ (J2000), and is a core cluster of the WIYN Open Cluster Study \citep[WOCS,][]{mathieu2000}.  We refer the reader to \citet[][hereafter, Paper~1]{geller2015} for a thorough review of the cluster parameters and the associated literature.

M67 has been the target of radial-velocity (RV) surveys by our group, and others, for the past 45 years.  In Paper 1 we used the results from those surveys to derive improved cluster membership probabilities, to identify binaries and to explore the kinematics of the cluster.  However, we did not report the individual internal RVs in Paper 1, only the average RVs known at the time or the center-of-mass RVs for binaries with orbital solutions.  In this paper we report the full details of our orbital solutions for 83 single-lined binaries, 31 double-lined binaries,  three triple systems with simultaneous inner and outer orbits and an additional three candidate triple systems with only single orbit solutions. All are high-probability cluster members.  Furthermore, we include a table of all our individual RVs, for non-members as well as members.

Binary stars influence the dynamical evolution of star clusters.  Through close dynamical encounters in star clusters, energy can be exchanged between singles, binaries and higher-order systems, and later distributed throughout the cluster via two-body relaxation processes.  These close encounters, in turn, will also modify the binary population, for instance by changing their orbital periods, eccentricities, and/or mass ratios, by inducing stellar mergers or collisions, or by fully disrupting a binary.  These modifications to the birth binary population are imprinted on the binary frequency, and on distributions of orbital periods, eccentricities and mass ratios (among others).  Through long-term RV surveys, such as ours, we can begin to study these properties of the binary population in search of dynamical signatures. 

The binary population is not only a tracer of cluster dynamics, but also controls the production rate and characteristics of blue stragglers (and other binary evolution products), which are thought to arise from a combination of binary mass transfer, mergers and dynamical collisions \citep{leonard1989,mccrea1964,perets2009,geller2011,gosnell2015}.  Blue stragglers in particular have been used as probes of the accuracy of star cluster $N$-body models \citep[e.g.]{hurley2005,geller2013}, and may offer key insights into the details of the mass-transfer process.  

Furthermore, $N$-body star cluster models are sophisticated enough to attempt to match very detailed observed characteristics of real star clusters, including their binaries.  \citet{hurley2005} presented an $N$-body model specifically aimed at reproducing M67, with a focus on the blue stragglers. At the time, \citet{hurley2005} did not know the characteristics of the actual binaries in M67, and implemented common theoretical choices for initial conditions.  \citet[][hereafter GM12]{geller2012} question the accuracy of the binary population in the \citet{hurley2005} M67 model, showing that it is inconsistent with observations of another old open cluster, NGC 188.   \citet{geller2013} created an $N$-body model specifically for NGC 188, with an updated binary population to match observations, and found stark differences in the blue straggler populations between the M67 and NGC 188 models, due to the different initial binary populations.  Here we compare the observed M67 and NGC 188 binaries, finding close similarities between the two.  Empirical studies of binary populations offer critical tests of star cluster models.  In this paper, we provide a cornerstone binary population in an old open cluster to help guide such models. 

M67 also contains a rich population of blue straggler stars that lie blueward of the main-sequence turnoff (Paper~1). We reserve a discussion of this population of stars, including their binary orbits, for a subsequent paper.     

In Section~\ref{sec:samp} we summarize our stellar sample and observations, and we provide the RV measurements in Table~\ref{tab:rv}.  In Section~\ref{sec:orbfit}, we discuss our orbit fitting procedures for both SB1s and SB2s. We present three triple stars with simultaneous solutions for the inner and outer orbits plus three additional candidate triple stars with single orbits in  Section~\ref{sec:triples}. Orbital parameters are presented in  Tables~\ref{tab:sb1pars},~\ref{tab:sb2pars}~and~\ref{tab:triplepars}, and the orbit plots are shown in Figures~\ref{fig:sb1orbs},~\ref{fig:sb2orbs}~and~\ref{fig:tripleorbs}. In Section~\ref{sec:incomp}, we discuss our completeness in binary detection and orbital elements. We use a resulting completeness correction to investigate the binary frequency (Section~\ref{sec:bfreq}) and distributions of binary orbital parameters and secondary masses (Section~\ref{sec:orbdist}).  Throughout the paper, we compare the M67 binaries to similar binaries in the galactic field \citep[][hereafter R10]{raghavan2010} and in the old (7 Gyr) open cluster NGC 188 (GM12); we synthesize this comparison in Section~\ref{sec:discuss}.  Finally in Section~\ref{sec:summary} we provide a summary.

\section{Stellar Sample and Observations} \label{sec:samp}
Our database of RV measurements for M67 stars incorporates observations from seven instrument/telescope pairs (see Table~1 from Paper~1), and spans a baseline of over 45 years, allowing detection of binaries out to periods of $\sim$10$^4$ days. In this section, we present a brief overview of our RV sample;  a full discussion of its chronology and evolution is given in Paper~1.  In total, our stellar sample in M67 contains 1296 stars.

Of the 13,775 RVs used for Paper 1, 42\% came from the WIYN 3.5m telescope using the Hydra multi-object spectrograph (MOS) and 55\% from CfA facilities. Over the five years since Paper 1 was finalized, we have continued to monitor detected velocity variables that lacked orbital solutions and also stars with uncertain membership probabilities, adding 1339 WIYN/Hydra RVs and 192 from the Tillinghast Reflector Echelle Spectrograph (TRES) to the orbital solutions and analysis presented here.

\subsection{CfA Sample and Data}
Our RV survey of M67 began with the dissertation work of \citet{mathieu1983}, using the CfA Digital Speedometers (DS) on the MMT and the 1.5m Tillinghast Reflector of the Fred Lawrence Whipple Observatory, both on Mount Hopkins, Arizona. A third DS was later commissioned at the 1.5m Wyeth Reflector at Oak Ridge Observatory in the Town of Harvard, Massachusetts, where M67 data were also collected. Earlier RV data from the Palomar Hale 5m observatory and the CORAVEL instrument at Haute-Provence were also added to the sample \citep{mathieu1986b}. Subsequently, observational facilities were expanded to include the TRES. Details about the observations and reduction procedures for data taken with TRES and the CfA DSs may be found in \citet{latham1992} and \citet{latham2002}. 

\pagebreak
\subsection{WIYN Sample}
Observations of M67 with the Hydra MOS on the WIYN 3.5m began on January 15, 2005 and continue to the present day. The WIYN sample includes all stars within 30 arcmin of the cluster center, with V-band magnitudes in the range $10\le V \le 16.5$ and with colors $(B - V)_0 > 0.4$.  We note that many blue stragglers lie blueward of the color limit, but are nonetheless included in the sample due to their scientific interest. Information about the spectra and reduction procedure for WIYN data can be found in \citet{geller2008,geller2010} and \citet{hole2009}.

Additional chronological information about the sample, including date ranges, number of stars observed, and number of observations from each telescope/instrument pair, can be found in Paper~1. 

\section{Orbit Fitting \label{sec:orbfit}}

Using our RV measurements of single- and double-lined binary systems, we fit orbital solutions to M67 RV members following the procedures outlined in \citet{geller2009}, \citet{latham2002} and \citet{goldberg2002}.  Orbital solutions were checked independently by both the WOCS and CfA groups using similar, though not identical, codes.  
In Table~\ref{tab:rv}, we provide the heliocentric RV measurements for all stars in our database, excluding the blue straggler stars (which we will publish separately in a forthcoming paper).  This table includes all the binaries with orbital solutions presented here, and also single stars, detected binaries without orbital solutions and non-members.  Within Table~\ref{tab:rv}, we provide the WOCS ID, a cross-match ID (XID)\footnote{If available the cross-reference ID is taken from  \citet{sanders1977} and denoted by the prefix ``S''.  If there is no \citet{sanders1977} match, we provide the \citet{montgomery1993} ID, if available, denoted by the prefix ``M''.  If both of those studies lack a match, we provide the \citet{fan1996} IDs, if available, denoted by the prefix ``F''. If all of these studies lack this star, we provide the \citet{yadav:08} ID, if available, denoted by the prefix ``Y". For all sources, we also provide their RA and Dec.\ positions for matching to other catalogs.}, the right ascension (RA), declination (Dec), the Julian Date, the telescope and instrument pair of the observations (Tel.; ``P" for Palomar Hale 5~m, ``TD" for Tillinghast 1.5~m + DS, ``C" for CORAVEL, ``MD" for MMT + DS, ``WD" for Wyeth 1.5~m + DS, ``WH" for WIYN + Hydra, and ``TT" for Tillinghast 1.5~m + TRES; see Paper~1, and references therein, for more information on each telescope and instrument), and the heliocentric RV measurement for the primary star (RV$_1$), secondary star (RV$_2$) and tertiary star (RV$_3$) if relevant. For CfA RVs of single-lined stars, we provide a direct estimate of the uncertainty on the primary RV (RV$_{1,e}$; \citealt{kurtz1998}).  For WIYN RVs of single-lined stars, we provide the cross-correlation peak height for the primary star, as an indicator of the uncertainty on the RV measurement \citep[see][]{geller2008,geller2010}.  For binaries with orbital solutions, we provide the residuals for the primary star from the orbital solution fit ($(O-C)_1$).  If the orbital solution is double lined, we also provide the secondary residuals from the orbital solution ($(O-C)_2$).  For triple systems that have a solution to the outer orbit, we provide the residuals from this outer orbit as $(O-C)_{1,out}$ and  $(O-C)_{2,out}$ (and see Section~\ref{sec:triples} for further details).  Finally, for binaries with orbits we provide the phase of the observation (``Phase"), and for triples with orbits we also provide the phase related to the outer orbit (``Phase$_{out}$"). Only a portion of the table is shown in this paper; the entire table is available online in machine-readable format.   Summary information for these stars (including, their  photometry, membership and binarity status, etc.) is available in Table 2 of Paper~1.  

Membership is assessed using the center-of-mass $\gamma$-velocity of the system within the formalism, and using the same membership criteria, as in Paper~1.  Only orbits of M67 binaries that are both RV and proper-motion members are presented. A subset of these binaries were previously presented in \citet{mathieu1990}.  We confirm all those orbital solutions with our expanded data set.  For two of the \citet{mathieu1990} SB1s, we now detect the secondary spectra and derive SB2 orbits (WOCS IDs 3015 and 6010).

For an orbital solution to be secure, the phased RV measurements clearly follow the visual trend of the orbit curve, relative errors on orbital parameters are low, and root mean square (RMS) residuals ($\sigma$) are near the RV measurement precision (0.5 \kms\ overall for WIYN; see Paper~1 for more details and for the precisions of other data sources). Candidate orbit solutions not meeting these criteria have either small orbital amplitudes, insufficient phase coverage, or simply too few observations. Most of the solutions presented here are very clearly secure; however those with high eccentricities should be treated with caution if phase coverage at periastron is sparse.

\subsection{Single-Lined Binaries}

Single-lined spectroscopic binaries (SB1s) are RV-variable stars with only one distinguishable peak in their cross-correlation functions. The orbital parameters for SB1s are presented in Table~\ref{tab:sb1pars}, and the corresponding plots of the orbital solutions are presented in Figure~\ref{fig:sb1orbs}.  For each SB1, in Table~\ref{tab:sb1pars} we list its WOCS ID (``ID"), a cross reference ID (``XID", see above), orbital period ($P$), number of orbital cycles covered in our data, center-of-mass RV of the binary ($\gamma$), RV semi-amplitude of the orbital solution ($K$), eccentricity ($e$), longitude of periastron ($\omega$), the Julian date of periastron (or maximum primary velocity, for circular orbits) nearest to the average date of observations ($T_0$), projected semi-major axis ($a \sin i$), mass function ($f(m)$), RMS of the velocity residuals from the orbital solution ($\sigma$), and number of RV measurements ($N$).

\subsection{Double-Lined Binaries}
Double-lined spectroscopic binaries (SB2s) are RV-variable stars with two distinguishable peaks in their cross-correlation functions. We extract primary and secondary RVs from SB2 spectra using the TwO-Dimensional CORrelation technique TODCOR \citep{zucker1994}.  We have two independent implementations, one a code by G.T.\ used for the CfA data, the other by A.G.\ used by the WOCS team.  An advantage of TODCOR is that the templates for the primary and secondary stars can be optimized to match the observed spectra.  The CfA code uses a library of synthetic templates \citep{latham2002,goldberg2002}.  The WOCS code can also use a similar library of synthetic templates, but for this work we choose to simply use the same high signal-to-noise sky spectrum for the SB1 and SB2 (TODCOR) analyses of WIYN data.  (The WOCS team has not found a significant improvement in the precision of RVs for these, mostly, solar-type and slowly rotating stars to justify using the more computationally intensive approach of the template grid for this project.) 
TODCOR simultaneously solves for the RVs of both the primary and secondary stars. This allows for precise measurement of RVs even in some cases where cross-correlation function peaks are highly blended. Our orbit-fitting procedure for SB2s involves first fitting an SB1 orbit to the primary RVs, then using the orbital parameters of that solution as initial guesses for an SB2 orbit fit \citep[e.g.][]{goldberg2002,geller2009}. In Figure \ref{fig:sb2orbs}, we present orbital solutions for RV-member SB2s for which we were able to obtain reliable RV measurements of both the primary and secondary peaks. The parameters of these orbits are listed in Table \ref{tab:sb2pars}, and follow the same format as Table \ref{tab:sb1pars}, except that instead of the mass function $f(m)$, we provide the $m \sin^3 i$ values for both the primary and secondary, along with the mass ratio $q=m_2/m_1$.

Finding a reliable orbital solution for double-lined binaries is complicated by increased RV measurement errors due to blending and the lower signal-to-noise of the secondary spectrum. For these reasons, SB2 $\sigma$ values are typically somewhat larger than SB1s, particularly for the orbital parameters for the secondaries.

\subsection{Triple Candidates}
\label{sec:triples}

R10 found that about 10\% of the targets in a volume-limited sample of 454 nearby solar-type stars are actually triple systems, composed of three stars in bound orbits.  Indeed triples are observed to be abundant within short-period binary samples \citep{tokovinin2006}.  We have identified \numtrip\ members of M67 that appear to be hierarchical triples.  (The lower observed triple fraction in M67 as compared to that of the field may simply be due to our study only including RV triples; nonetheless this would be an interesting topic for follow-up investigations.)  For three of the M67 candidate triple systems we have solved for both the inner and the outer orbits, thus demonstrating that those system are bound.  For the other three system we have solutions for the inner orbits, and evidence for an outer orbit with a period much longer than the span of our observations. The orbital parameters and plots for these triple systems are provided in Table~\ref{tab:triplepars} and Figure~\ref{fig:tripleorbs}.  These follow a similar format as for the SB1 and SB2 binaries.  We also describe each triple in greater detail below.  

\paragraph{WOCS 3012 (S1077)}
This is a hierarchical triple-lined system with a simultaneous solution for the inner and outer orbits.  There are 57 observations from the CfA DSs, 29 from WIYN/Hydra, and 16 from TRES.  The primary is the brighter star in the inner binary, the secondary is the distant third star, and the tertiary is the faint companion in the inner binary.  Velocities for all three stars were extracted from the TRES spectra using TRICOR \citep{zucker1995}. There are hints of the tertiary in the WIYN/Hydra correlation plots, but no attempt was made to extract velocities beyond using TODCOR for the primary and secondary. Also for the DS observations, only primary and
secondary velocities using TODCOR were extracted.  We derived the orbital solutions of this system in two steps. Initially we used the TODCOR velocities for the primary and secondary from all three sets of observations to solve for the inner and outer orbits simultaneously, using the code \texttt{dst} developed by G.T. 
This provided a good starting point for a TRICOR analysis of the TRES observations.  We then combined all the velocities using a second special code, \texttt{tst1} also developed by G.T., and 
including the TRES TRICOR velocities for the faint tertiary in order to derive a mass ratio for the inner binary.

\paragraph{WOCS 4008 (S2206)}
This system has 91 observations with the CfA DSs, but no observations with WIYN/Hydra or TRES.  The spectrum is clearly double-lined in a few of the DS observations, but most of the time the primary and secondary peaks of the correlation function are blended.  The best template for the primary is slightly evolved and cooler than a slightly hotter main-sequence template for the secondary.  There is no evidence for variation in the velocity of the secondary, which is constant at 37.0 \kms\ with an RMS of 1.6 \kms\ (see Figure~\ref{fig:tripleorbs}).  The RVs of the primary yield an orbital solution with a period of $P = 18.377$~days and semi-amplitude of $K = 12.28 \pm 0.63$~\kms, implying a minimum mass for the companion of about 0.2 $M_\odot$. The scatter of the velocity residuals is large, about 4 \kms, but there is no obvious trend in them.  The center-of-mass velocity for the primary is $\gamma = 31.3 \pm 0.4$~\kms, 2.3~\kms\ lower than the cluster mean RV of 33.64~\kms\ reported in Paper 1.  The secondary is 3.4~\kms\ higher.  One possible explanation might be that the two stars are not a bound triple system, but rather that they are two low-probability cluster members that lie by chance on the same line of sight.  A more appealing explanation might be that they are a bound triple with a very long period for the outer orbit.  Assuming that the $\gamma$ velocity of the triple system is the same as the cluster mean, the mass ratio would be about 1.5, consistent with an inner orbit composed of a slightly evolved star and late M dwarf companion, and an outer orbit around a main-sequence star near the cluster turnoff.  We also note that \citet{mathieu1990} performed a careful photometry analysis of this system and determined a similar physical description. 

\paragraph{WOCS 4030 (S1416)}
This is a hierachical triple system with 21 observations from the CfA DSs.  All of the spectra are single-lined. These 21 RVs yield solutions for the inner and outer orbits when solved simultaneously.  The secondary in the inner orbit and the distant third star in the outer orbit are both too faint to yield velocities.   The code \texttt{sst} developed by G.T.\ was used to derive solutions for the inner and outer orbits.

\paragraph{WOCS 7008 (S1234)}
This is a hierarchical triple system with a simultaneous solution for the inner and outer orbits.  Its architecture is the same as WOCS 3012 (S1077), except the primary is the distant third star in the outer orbit and the secondary is the brighter star in the inner orbit.  The companion in the inner orbit is too faint to yield velocities, so only TODCOR was used to extract velocities.  There are 51 observations from the CfA DSs, 8 from WIYN/Hydra, and 18 from TRES. Seven of the DS exposures were too weak to yield TODCOR velocities for the secondary.  We note that \citet{mathieu1990} also analyzed this system and presented a similar physical description.

\paragraph{WOCS 10012 (S796)}
This single-lined system has 28 WIYN/Hydra RVs.  The impression from a visual inspection of the velocity history is a periodic variation of a couple thousand days on top of a slow drift upwards. An orbit with period of 1884 days fits the data reasonably well, with orbital semi-amplitude $K=1.3$~\kms\ and velocity residuals of $\sigma=0.6$~\kms. The residuals show a slow drift of 2 \kms\ over a period of 4826 days, with no obvious curvature.  This is a good candidate for a bound triple with a long-period inner binary and a distant third star in a much longer period outer orbit.  In order to fit an orbit to the inner system, we fit the following linear equation to these data, $RV = 3.5027\times10^4 JD - 19.4531$.  We then subtract this linear fit from our observed RVs, and fit an orbital solution.  In Figure~\ref{fig:tripleorbs}, we show this orbital solution, and also the observed RVs, the fitted trend line and the RVs resulting from subtracting this trend from the observations.  The resulting orbital parameters are provided in Table~\ref{tab:triplepars}, and the observed RVs are provided in Table~\ref{tab:rv}.  Accounting for this slow drift in the residuals reduces the velocity residuals on the orbital solution to $\sigma=0.33$~\kms.

\paragraph{WOCS 21005 (S1278)}
This is a double-lined system with 12 recent TRES observations, 16 early CfA DS observations, and 5 WIYN/Hydra observations that sit in the gap between the CfA data sets.  The orbital solution using all of the data (and included in Table~\ref{tab:triplepars} and Figure~\ref{fig:tripleorbs}) has residuals of 1.1 and 2.2 \kms\ for the primary and secondary stars, respectively.  Using only the CfA TODCOR velocities yields a similar orbit with velocity residuals of only 0.82 and 1.75 \kms.  Inspection of the phased velocity plot for the orbital solution suggests that the WIYN/Hydra velocities are systematically shifted to higher values, and this is confirmed by plots of the residuals versus time, which show a shift on the order of 2 \kms\ (see Figure~\ref{fig:tripleorbs}).  This shift is much too large to be a systematic error in the WIYN/Hydra velocities, and a straightforward explanation is that we are seeing curvature in the velocities due to the pull of a distant third star in a bound outer orbit with a period several times longer than the time span of the observations.

\subsection{Notes for Specific Binaries}

\paragraph{WOCS 4021 (S815)} This SB1 binary is highly eccentric ($e = 0.83 \pm 0.07$).  We do not have observations precisely at periastron; the uncertainties on the orbital semi-amplitude derived from the orbit fit (and provided in Table~\ref{tab:sb1pars}) likely can be substantially reduced with future timely observations.

\paragraph{WOCS 10025 (S628)} This system is an SB2, where at least one star appears to be rapidly rotating.  Determining which star is the primary and which is the secondary in each observation has proven challenging.  Our best attempt (included in Table~\ref{tab:rv}) results in primary RVs that span a range of $\sim$30~\kms, and secondary RVs that span a range approaching 100 \kms.  We have not been able to derive a reliable orbit solution.

\paragraph{WOCS 13004 (S1011)} This SB1 binary appears to have an orbital period of $\sim$14620 days, and is included in Table~\ref{tab:sb1pars} and Figure~\ref{fig:sb1orbs}.  However, we caution the reader that we have only covered 80\% of the orbital phase, and so the orbital elements may be uncertain and subject to change once observations covering the full orbit are obtained and included in the fit. 
\begin{deluxetable}{l l l}
    \tabletypesize{\footnotesize}
    \tablecaption{Photometric Variables and X-ray Sources with RV Orbital Solutions \label{tab:binPVXray}}
    \tablehead{\colhead{ID} & \colhead{XID} & \colhead{Notes}}
     
    \startdata
    1001  & S1024 & SB2, PV, CX111 \\
    1015  & S1237 & SB1, X52, CX47, YG \\
    2002  & S1040 & SB1, PV, X10, CX6, YG \\
    2003  & S1045 & SB2, PV, X41, CX88  \\
    2008  & S1072 & SB1, X37, CX24, YG \\
    2016  & M6176 & SB1, PV? \\
    3004  & S1264 & SB1, PV? \\
    3012  & S1077 & SB2, triple, PV, HT Cnc, X7, CX10  \\
    4001  & S1029 & SB1, PV \\
    4003  & S1036 & SB1, RR, PV, W UMa, EV Cnc, X45, CX19  \\
    4007  & S986  & SB1, PV, HV Cnc, CX157  \\
    4016  & S1224 & SB2, PV? \\
    5037  & S1508 & SB1, PV \\
    6005  & S999  & SB2, PV, HW Cnc, X13, CX9  \\
    6008  & S1242 & SB1, PV, X50, CX49  \\
    6010  & S1272 & SB2, CX155 \\
    6016  & S760  & SB1, X49 \\
    7002  & S1019 & SB2, PV, X11, CX5  \\
    7008  & S1234 & SB2, triple, X53, CX36 \\
    7022  & S745  & SB1, PV \\
    9011  & S773  & SB1, CX72 \\
    11022 & S1109 & SB1, PV \\
    12004 & S1009 & SB1, CX78 \\
    13008 & S1063 & SB1, PV, HU Cnc, X8, CX1, SSG \\
    14007 & S1070 & SB2, PV, HX Cnc, X38, CX48  \\
    15028 & S1113 & SB2, PV, X26, SSG \\
    17004 & S1050 & SB1, CX104 \\
    21006 & S996 & SB2, CX81 \\
    24021 & S819 & SB2, PV, HR Cnc, CX58 \\
    \enddata
\end{deluxetable}

\begin{deluxetable}{l l l c c}
 \tabletypesize{\footnotesize}
 \tablecaption{Candidate Binaries without Orbital Solutions\label{tab:BLMBU}}
 \tablehead{\colhead{ID} & \colhead{XID} & \colhead{Class\tablenotemark{a}} & \colhead{$e/i$} & \colhead{$\Delta$JD}}
 
\startdata
4017 & S1463 & BU & 3.33 & 17242.8 \\
5032 & S1406 & BLM & 4.15 & 5177.4 \\
6025 & S1431 & BLM & 10.78 & 5178.4 \\
8004 & S2219 & BLM & 4.65 & 11756.0 \\
9029 & S1604 & BLM & 4.62 & 5179.1 \\
10018 & S1462 & BLM & 4.22 & 5178.4 \\
10024 & S1432 & BU & 12.30 & 5179.2 \\
10025\tablenotemark{b} & S628 & BU & 14.42 & 5468.3 \\
11043 & S1579 & BLM & 14.74 & 2274.8 \\
12032 & S1405 & BLM & 7.37 & 4845.1 \\
13013 & S1251 & BLM & 4.87 & 11350.3 \\
14012 & S965 & BLM & 8.21 & 11020.1 \\
14025 & S1112 & BLM & 31.97 & 4822.0 \\
16024 & Y589 & BLM & 3.29 & 4819.8 \\
16029 & S636 & BLM & 5.07 & 4336.1 \\
16036 & S1179 & BLM & 5.75 & 5030.2 \\
19020 & S800 & BU & 3.87 & 11334.4 \\
19032 & M6452 & BLM & 8.58 & 1896.1 \\
20059 & S243 & BU & 25.97 & 2281.7 \\
21036 & S637 & BU & 19.46 & 2273.7 \\
21037 & F4146 & BLM & 3.94 & 1536.8 \\
21044 & F2832 & BLM & 26.30 & 1896.1 \\
23015 & M5821 & BLM & 62.98 & 1892.1 \\
23026 & M6441 & BLM & 74.87 & 1898.1 \\
24007 & S1286 & BLM & 6.03 & 11350.3 \\
24012 & S797 & BU & 21.54 & 5135.2 \\
25008 & S1042 & BU & 12.78 & 5135.2 \\
25013 & S764 & BU & 6.63 & 5179.1 \\
25039 & S1709 & BLM & 41.52 & 5179.1 \\
27011 & M5380 & BLM & 7.96 & 5179.1 \\
28008 & F3467 & BLM & 6.89 & 5178.1 \\
28028 & M5285 & BU & 26.92 & 2153.1 \\
28059 & S250 & BLM & 5.00 & 4827.8 \\
34046 & F5460 & BLM & 15.64 & 1538.8 \\
\enddata
\tablenotetext{a}{Specifically, the classification ``BLM" is for ``Binary Likely Members"; these binaries have both RV and PM membership probabilities $\geq$50\% (see also Paper 1 for how we incorporate the multiple PM surveys; here we have also accounted for Gaia PMs).  The classification ``BU" is for ``Binary with Unknown Membership"; these binaries are PM members, but the RV membership calculated from the mean value of the current RVs is $<$50\%.  We include ``BU" binaries in our member sample because M67 has excellent PM information.  These binaries are included in our binary frequency analysis, and are accounted for by our incompleteness analysis in the orbital parameter distributions below.}
\tablenotetext{b}{WOCS 10025 has a double-lined spectrum; we show the $e/i$ value and class derived from the primary velocities.}

\end{deluxetable}

\begin{deluxetable}{l l c c c}
    \tabletypesize{\footnotesize}
    \tablecaption{Candidate Low-Amplitude Binaries\label{tab:binSM}}
    \tablehead{\colhead{ID} & \colhead{XID} & \colhead{$e/i$} & \colhead{$\Delta$JD}}
    
    \startdata
    2001 & S1027 & 1.34 &  10639.0 \\
    2004 & S1016 & 1.27 & 13424.3 \\
    3107 & S80   & 1.39 &  12785.1 \\
    4015 & S1456 & 0.81 &  11663.1 \\
    5004 & S2212 & 1.88 &  16870.9 \\
    5020 & S1458 & 1.77 &  8704.2 \\
    5025 & S827  & 1.50 &  10252.3 \\
    5026 & S1337 & 1.20 &  10158.2 \\
    5051 & S368  & 0.39 &  11667.2 \\
    5058 & S341  & 0.90 &  9105.1 \\
    5066 & S671  & 0.69 &  10602.0 \\
    6011 & S775  & 1.14 &  14730.9 \\
    6023 & S618  & 1.13 &  10189.1 \\
    7005 & S1274 & 0.81 &  11791.9 \\
    7014 & S1092 & 2.31 &  8761.1 \\
    8005 & M5951 & 1.40 &  10610.9 \\
    10005 & S2208 & 1.24 &  12064.9 \\
    10026 & S1470 & 2.54 &  1747.3 \\
    13012 & S1218 & 1.81 &  10546.0 \\
    15009 & S991  & 1.60 &  11293.1 \\
    15010 & S1252 & 0.88 &  10229.1 \\
    27035 & F4898 & 2.18 &  1393.2 \\
    \enddata

\end{deluxetable}

\paragraph{Photometric Variables and X-ray Sources}
In Table~\ref{tab:binPVXray} we provide a list of the known photometric variables and X-ray sources for which we have RV orbital solutions. The column labeled ``ID" provides the WOCS ID, and the column ``XID" provides a cross-match ID (in the same format as Table~\ref{tab:rv}, and here mostly from \citealt{sanders1977}). Within the ``Notes" columns, X-ray sources identified by \citet{belloni1998} with \textit{ROSAT} are labeled with ``X", followed by the source number given in their paper. Likewise X-ray sources identified by \citet{vandenberg2004} with \textit{Chandra} are labeled with ``CX" followed by the source number given in their paper. We draw our sample of photometric variables from a number of references, provided in Paper~1, and label them as ``PV", or ``PV?" if the authors identify the photometric variability as uncertain.  We also provide the GCVS names for photometric variables, where available (e.g., HT Cnc, EV Cnc, etc.).  

\paragraph{Candidate Binaries without Orbital Solutions}
Table~\ref{tab:BLMBU} contains velocity variables that we identified as binaries (having $e/i \geq 3$), but do not yet have orbital solutions.  (The $e/i$ value measures the spread in RVs relative to the expected precision of the RVs; see Paper~1 and \citealt{geller2008} for more details.)  In this table, we include the WOCS ID (column ``ID"), a cross-match ID (``XID"), the ``class" (see Paper 1), the  $e/i$ value, and the time span of our observations ($\Delta$JD, in days), as a possible lower limit to the true orbital period. (Where necessary, all values have been updated from Paper~1 to include the current set of observations in Table~\ref{tab:rv}.) The RV memberships for all of these binaries are somewhat uncertain, because we do not yet know the center-of-mass RV for each system.  However, all of these are PM members.  

\begin{figure*}[!ht]
\plotone{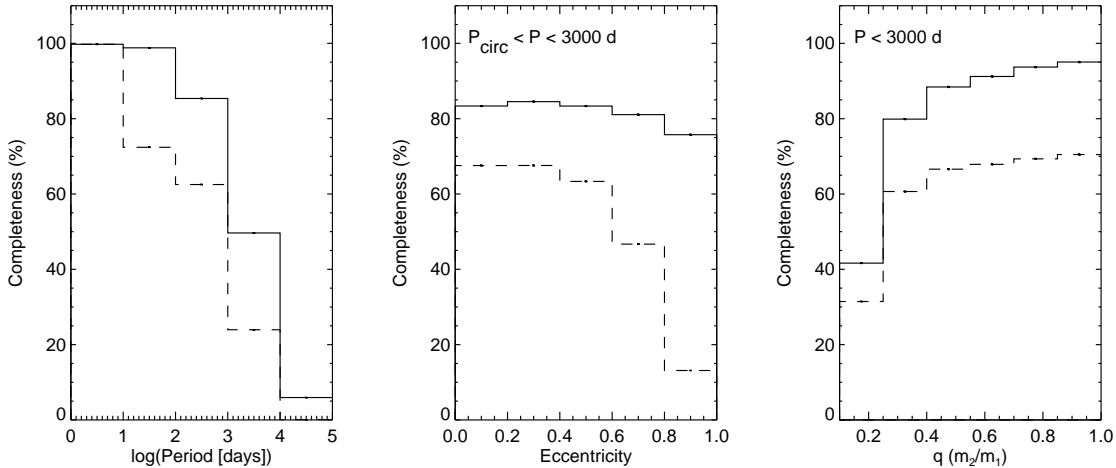}
\caption{Completeness in binary detection (solid lines) and binary orbits (dashed lines) as a function of period (left), eccentricity (middle) and mass-ratio (right).}  
\label{fig:incomp}
\end{figure*}

\paragraph{Candidate Low Amplitude and/or Long Period Binaries}
Table~\ref{tab:binSM} and Figure~\ref{fig:trends} contain stars that do not pass our criteria for RV variability of having $e/i \geq 3$, but do show an obvious trend of RV vs.\ time when visually inspected.  The columns in the table are the same as in Table~\ref{tab:BLMBU}, but we exclude the ``class" column (since all are classified as ``Single Members", SM).  For statistical purposes, we include these stars as RV non-variables (i.e., ``singles") in our binary frequency analysis in Section~\ref{sec:bfreq}, as they should be accounted for in our incompleteness correction.  Additional precise RV observations will be important to derive orbital solutions for these binaries.

\section{Completeness in Binary Detection and Orbital Elements \label{sec:incomp}}

In Section~\ref{sec:samp} we discuss our sample selection, and further details are available in Paper~1 along with information on the completeness of our survey. In short, we have derived RV membership probabilities and characterized the RV variability of $\sim$91\% of our complete stellar sample (down to $V=16.5$) and $>99.5$\% of the stars in our sample with $V<15.5$. We find a total of 604 cluster members, 169 of which show significant RV variability ($e/i > 3$).  However, our RV survey is likely not sensitive to all binaries in the cluster.

For example, the maximum baseline of RV observations for a given star in our survey is $\sim$10$^4$ days.  Binaries in the galactic field have periods extending well beyond this timespan (R10).  In a star cluster, dynamical encounters will break up binaries at periods beyond the ``hard-soft boundary" \citep{Heggie1975}.  Using the virial theorem and assuming the velocity dispersion is isotropic with a one-dimensional velocity dispersion for solar-type main-sequence stars in M67 of $0.59\substack{+0.07 \\ -0.06}$ \kms\ (Paper~1), we estimate that the orbital period at the hard-soft boundary in M67 for solar-mass stars is approximately $10^6$ days.  Therefore all of the binaries in our sample are hard.  Furthermore, there are assuredly binaries in M67 that we have not detected, due to having long periods, high eccentricities and/or small mass ratios, that are beyond the detection limits of our observational survey.  

To estimate the completeness of our survey, we follow a similar method as in GM12 and updated in Paper~1. Briefly, we run a Monte Carlo analysis to generate a population of synthetic binary stars that follow the M67 binary primary-mass distribution (as estimated relative to a 4 Gyr isochrone) and the orbital-parameter and mass-ratio distributions of the solar-type galactic field binaries from R10.  We produce synthetic RVs for these binaries on the true observing dates of our M67 survey, using realistic RV uncertainties for observations from each telescope (see Paper~1).  We then analyze these RVs in the same manner as our real observations. 

A synthetic binary is considered detected if the $e/i$ value is $\geq3$, as we also defined for our observationally detected binaries in Paper~1.  Using this method we conclude that within our magnitude and spatial limits we have detected 74\% of the binaries out to a period of $10^4$ days in M67.  For orbital periods beyond $10^4$ days, our detection completeness drops precipitously (Figure~\ref{fig:incomp}).

To estimate our completeness in binary orbital solutions, we use the same software used in this paper to derive the orbital solutions from our true observations, but here with RVs from our synthetic binaries.  As in GM12, we consider an orbital solution acceptable in these simulations if it has RVs covering at least one period, errors on $P$, $e$, and the orbital semi-amplitude $K$ of less than 30\% of the derived values, an RMS residual velocity of less than 1 \kms, and a range in RVs covering at least 75\% of the orbital amplitude (mainly applicable for highly eccentric binaries). 

The results of this completeness analysis are shown in Figure~\ref{fig:incomp}. Given the similar survey duration and methods as that of GM12 for NGC 188, our completeness curves look very similar to Figure 1 in GM12.  We estimate that for orbital periods of $<$3000 days, we have derived orbital solutions for $\gtrsim$50\% of our sample.  Beyond 3000 days, our completeness drops quickly to zero (reaching nearly 0\% at $>$10$^4$ days).  Therefore for much of the binary orbital parameter analysis presented below, we will limit our investigation to binaries with periods $<$3000 days.

\section{The Main-Sequence Hard-Binary Frequency\label{sec:bfreq}}

We define a main-sequence sample as those stars with $V<15.5$, $(B-V)>0.5$ and $V>8.4\times(B-V)+7.25$ (using the photometry in Paper~1, which is mostly drawn from \citealt{montgomery1993}).  The faint limit of $V<15.5$ is set by our observational completeness (see Paper~1).  The blue limit excludes potential blue stragglers, not already identified in Paper~1.  The last condition was determined by eye to include the main-sequence locus up through the turnoff.  (For reference, see the color-magnitude diagram in Figure~8 of Paper~1.)  Following a similar method to GM12 we also exclude the few binaries from this frequency analysis that, after correcting for secondary light, would have primaries with $V\geq15.5$.  

Of the 400 main-sequence stars in our sample of M67, we identify 101 binaries. When corrected for incompleteness out to an orbital period limit of $10^4$ days, this results in a main-sequence hard-binary frequency of 34\% $\pm$ 3\%.  

For comparison, if we limit the NGC 188 sample to the same (physical) radial extent as our M67 sample, we find a solar-type main-sequence binary fraction of 32\% $\pm$ 4\% out to orbital periods of $10^4$ days, nearly identical to that of M67.

\begin{figure}
\plotone{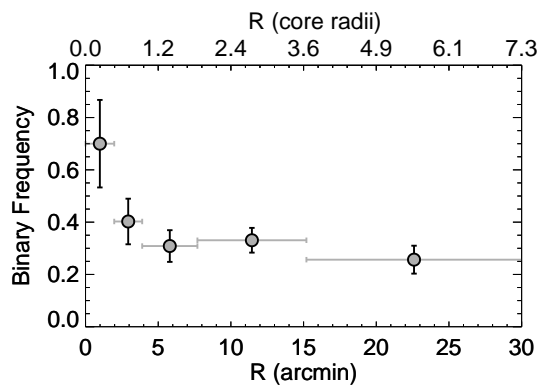}
\caption{Binary frequency as a function of radius from the cluster center.  We take equal bins in the log over our survey extent; horizontal bars extend to the respective bin edges.  The points are corrected for observational incompleteness out to periods of $10^4$ days; vertical bars show the uncertainties on our binary frequency measurements.  We provide distances in arcmin and also converted to core radii using a core radius of 4.12 arcmin \citep[]{davenport2010}, which at a distance of 850 pc is equivalent to 1.02 pc. }  
\label{fig:bfreq}
\end{figure}

In Figure~\ref{fig:bfreq} we divide the main-sequence sample in bins of radius to compare the incompleteness-corrected binary frequency in the core to that in the halo.  In each bin, we run our Monte Carlo completeness analysis as described above, and find a similar detection completeness of $\sim$74\% (out to orbital periods of 10$^4$ days) in each bin.  Inside about 0.4 core radii, the binary frequency reaches 70\% $\pm$ 17\%.  Outside of 0.4 core radii, the binary frequency is nearly constant at about 30\% to 40\% to the limit of our survey.

The binaries in M67 are known to be centrally concentrated (\citealt{mathieu1986}, Paper~1).
Indeed, a rise in binary frequency toward the cluster center is expected as a well-known effect of mass segregation.  Mass segregation in M67 has also been noted previously \citep[e.g.][without distinction to binaries]{sarajedini1999}, and is also seen in other clusters \citep[e.g.][GM12]{mathieu1985,milone2012}, and in $N$-body simulations of old open clusters \citep[e.g.][]{hurley2005, geller2013}.

\begin{figure}
\plotone{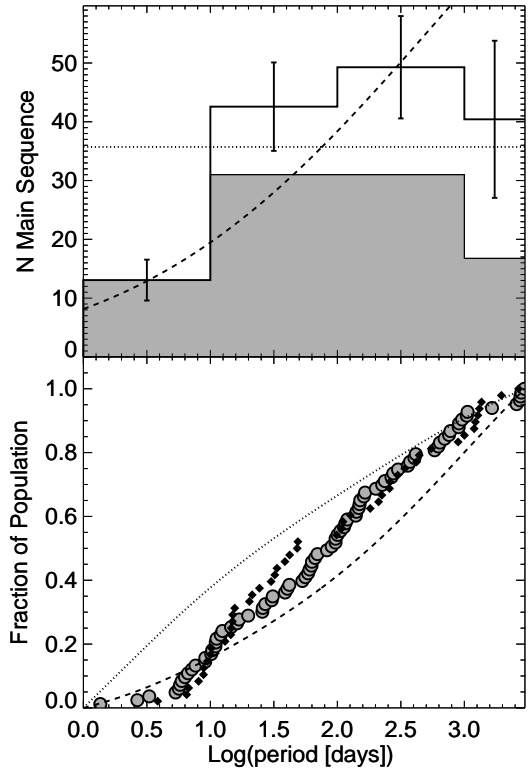}
\caption{Period distribution for M67 main-sequence binaries in both histogram (top) and cumulative distribution (bottom) form.  In gray we show the observed binaries in M67.  In the top panel, we apply our incompleteness correction in the solid-lined histogram.  In both panels, we plot (for reference) the log-normal period distribution from the solar-type binaries in the field (R10) as the dashed line, and also a log-uniform distribution commonly used in theoretical models as the dotted line.  In the top panel, these lines are normalized such that the integral under the curves produce the incompleteness-corrected number of M67 binaries in this sample.  In the bottom panel we apply our observational incompleteness to these curves in order to compare directly to our measurements.  Also in the bottom panel, we show the cumulative period distribution from the old open cluster NGC 188 in black squares (GM12). }
\label{fig:pfreq}
\end{figure}

\begin{figure}
\plotone{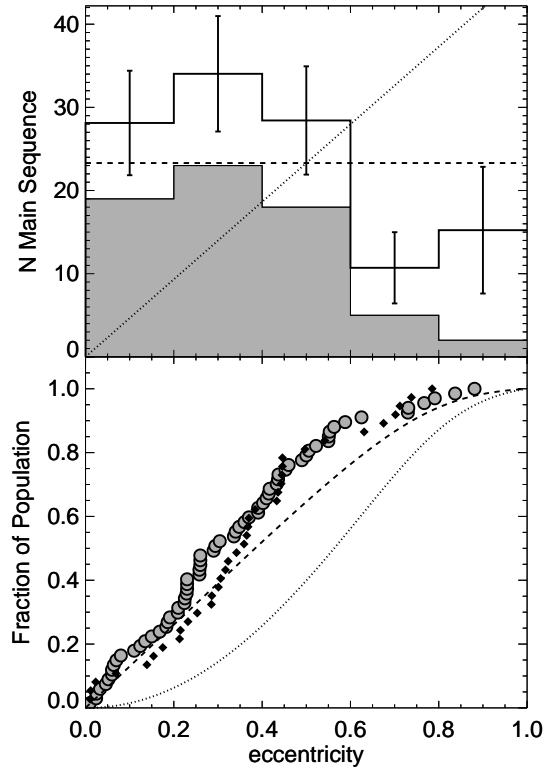}
\caption{Eccentricity distribution for the M67 main-sequence binaries in the same format as in Figure~\ref{fig:pfreq}.  Here we only show binaries with periods between the tidal circularization period (see Section~\ref{sec:tcirc}) and 3000 days.  We compare with a uniform distribution (dashed line), consistent with solar-type binaries in the field (R10), and a thermal distribution (dotted line), as is often used in theoretical models.  In the bottom panel, both of these lines have our observational incompleteness applied so that they can be compared directly with our observations.   As in Figure~\ref{fig:pfreq}, we show the cumulative eccentricity distribution from the old open cluster NGC 188 in the bottom panel in black squares (GM12). }
\label{fig:efreq}
\end{figure}

\section{Binary Orbital Parameter and Mass Distributions\label{sec:orbdist}}

In the following subsections, we analyze the distributions of binary orbital parameters and secondary masses for the main-sequence binaries in our sample in M67.  We exclude triples here, as their orbital elements (particularly for the inner binaries) may have been modified through internal dynamics during the cluster lifetime.  We utilize our completeness analysis and also compare against theoretical and observed distributions, where appropriate.  In particular, we will compare against the main-sequence binaries in the old open cluster NGC 188, which also has a very thorough (and similar) analysis of the hard-binary population (GM12).  We present the figures here with a statistical analysis, and save most of the discussion for Section~\ref{sec:discuss}.

\subsection{Periods and Eccentricities\label{sec:epdist}}

\begin{figure}
\plotone{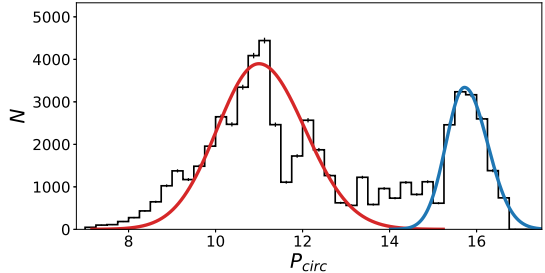}
\caption{Distribution of circularization periods ($P_{\text{circ}}$) resulting from our bootstrap analysis (black histogram), fit with two asymmetric Gaussian functions.  The red curve shows the asymmetric Gaussian fit to the bulk of the distribution and results in $P_{\text{circ}} = \Pcirc$.  (The fit represented by the blue curve results in a value of $15.7\substack{+0.5 \\ -0.4}$  days.)}
\label{fig:pcircDist}
\end{figure}

In the top panels of Figures~\ref{fig:pfreq}~and~\ref{fig:efreq} we show the distributions of orbital periods and eccentricities, respectively, for the M67 main-sequence binaries in our sample. The observations are shown in gray, with the incompleteness correction applied to the histogram (top panel) shown in solid lines with error bars.  

Beginning with the period distribution (Figure~\ref{fig:pfreq}), with the dashed lines we compare against a log-normal fit to the period distribution of the solar-type binaries in the galactic field, from R10; specifically this log-normal distribution has a mean value of $\log_{10}\left(P \mathrm{[days]}\right) = 5.03$ and $\sigma_{\log_{10}\left(P\right)} = 2.28$.  In the top panel, we show this distribution in comparison to our incompleteness-corrected observations.  In the bottom panel, we apply our observational incompleteness to the R10 log-normal distribution in order to compare directly with our observations. 

A one-sample Kolmogorov-Smirnov (K-S) test comparing the M67 observations against this field log-normal distribution (with our incompleteness applied) returns a marginal distinction with a $p$-value of 5.0$\times10^{-3}$, or 2.81$\sigma$.  It appears that the M67 binaries may be marginally depleted at the longer periods as compared to the field; we return to this in Section~\ref{sec:discuss}. With the dotted line, we compare against a period distribution that is uniform in the log, as is often used in theoretical models \citep[including the M67 $N$-body model of ][]{hurley2005}.  A one-sample K-S test comparing the observations against the incompleteness-applied log-uniform period distribution returns a $p$-value of 7.6$\times10^{-5}$, distinct at 3.96$\sigma$.  

In the bottom panel of Figure~\ref{fig:pfreq}, we also compare against the observed binaries in NGC 188, from GM12.  The NGC 188 survey has very similar completeness as our M67 survey, and therefore we do not apply any correction to the NGC 188 data.  A two-sample K-S test comparing the NGC 188 and M67 observations returns no significant distinction  (with a $p$-value of 0.60).  

Turning to orbital eccentricity, we plot in Figure~\ref{fig:efreq} the eccentricity distribution for the subset of M67 (and NGC 188) binaries with periods between the circularization period and 3000 days.  In this figure, we compare against both a uniform distribution (dashed line, as suggested for the galactic field solar-type binaries with periods beyond the circularization limit; R10), and a thermal distribution \citep[dotted line;][]{Jeans1919, Ambartsumian1937, Heggie1975}. The thermal distribution is often used in theoretical models due to its elegance and simplicity, but there is little support for this distribution from observations \citep[GM12,][]{Duchene2013,Moe2017,geller2019}. 
A one-sample K-S test comparing the M67 observations to the incompleteness-applied thermal distribution returns a $p$-value of 2.9$\times10^{-11}$, distinct at 6.65$\sigma$. 
On the other hand, a one-sample K-S test comparing the M67 observations to the incompleteness-applied uniform distribution returns no significant distinction (with a $p$-value of 0.046).  

We also compare the M67 eccentricity distribution to that of NGC 188, in the bottom panel of Figure~\ref{fig:efreq}.  A two-sample K-S test comparing these two observed samples returns no significant distinction (with a $p$-value of 0.38).  In summary, the M67 main-sequence binary eccentricity distribution is consistent with that of similar binaries in NGC 188 and the field, and can be described by a uniform distribution.

\subsection{Tidal Circularization}\label{sec:tcirc}

\begin{figure}
\plotone{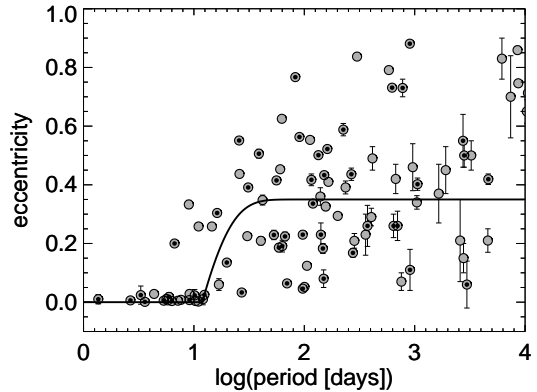}
\caption{The period-eccentricity ($e-\log P$) distribution of M67 main-sequence binaries.  Error bars are drawn for all orbits, though in many cases these are smaller than the points themselves. The points with filled black circles in the interior mark the unevolved main-sequence binaries used in fitting for the circularization period ($P_{\text{circ}}$). The black line shows our fit of the circularization function from MM05, resulting in $P_{\text{circ}} = \Pcirc$.}
\label{fig:elogp}
\end{figure}

The rate of tidal circularization depends on the strength of the tidal force, which scales as the inverse cube of distance. Consequently, within coeval populations such as open clusters nearly all main-sequence binaries of similar primary mass below a characteristic orbital period are found to be circular. As the population ages and wider binaries finish circularizing, this ``circularization period" ($P_{\text{circ}}$) increases to longer orbital periods \citep[][hereafter MM05]{mayor1984,meibom2005}.

\citet{mathieu1988}, \citet{lathammathieu1992} and MM05 studied the circularization period in M67 using a subset of the binaries included in our study.  We revisit the circularization period here, using our larger binary sample.  Notably, our study extends further down the main sequence, allowing us to select a sample of stellar binaries that is less affected by the rapidly increasing radius near the cluster turnoff.  Specifically, we use the subset of M67 main-sequence binaries with $V > 13.5$.  This bright magnitude limit is approximately where the main sequence detaches from the zero-age main sequence (ZAMS; see Figure 8 in Paper~1).  We also exclude the candidate triples discussed Section~\ref{sec:triples} and the blue lurkers introduced by \citet{leiner2019}.  Our aim is to include only unevolved main-sequence binaries that have not undergone internal eccentricity evolution due to triple dynamics or mass-transfer processes. This sample includes 52 binaries, which are identified in the $e - \log P$ diagram of Figure~\ref{fig:elogp}.

\begin{figure}
\epsscale{1.0}
\plotone{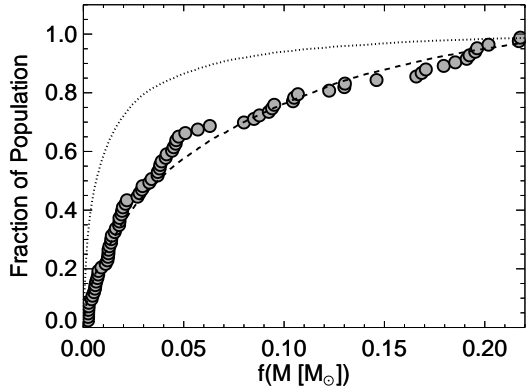}
\caption{Distribution of mass functions for the M67 main-sequence binaries (gray dots).  For comparison, we also show the mass-function distributions resulting from drawing primary masses from our observed distribution and  mass ratios from a uniform distribution (dashed line) as well as  secondary masses drawn from a \citet{kroupa2001} IMF (dotted line). Our incompleteness corrections have been applied to both model distributions.}
\label{fig:fmdist}
\end{figure}

We fit these data with the ``circularization function" from MM05 (Equation 1 therein) in order to derive the circularization period. Our method differs slightly from that of MM05, as follows.  We use an orthogonal distance regression method to fit the circularization function, which accounts for the uncertainties on both the period and eccentricity.  We also employ a bootstrapping technique to derive a distribution of measured circularization periods.  Specifically, in each of 10$^5$ iterations we select 52 binaries randomly from this M67 binary sample, with replacement, and  derive the circularization period from this random sample.   The resulting distribution of circularization periods is shown in the black-lined histogram in Figure~\ref{fig:pcircDist}.  

The distribution shows two peaks, one at approximately 11 days and another at approximately 16 days.  We  choose to fit an asymmetric Gaussian function to each of these peaks in the distribution (separately).  Each asymmetric Gaussian is defined by a single mean and a single amplitude but with two $\sigma$ values, one on either side of the mean.   We show the Gaussian fits to these two peaks in the red and blue curves (respectively).  The portion of the distribution fit by the red curve has a far larger integrated area, and we therefore consider this the cluster's circularization period, with a value of $P_{\text{circ}} = \Pcirc$.  (The uncertainties show the high and low 1$\sigma$ values derived directly from the fitting function.) This circularization period is in agreement  (to within the uncertainties) with that found by MM05 for M67 of $12.1\substack{+1.0 \\ -1.5}$ days, who had a smaller sample and also included brighter binaries closer to the turnoff.  

Statistically, the secondary peak at $\sim$16 days arises from the few somewhat longer period, nearly circular binaries in the sample, in iterations of the bootstrap technique that by chance also exclude some shorter-period binaries.  Physically, it is not immediately clear what meaning should be given to this peak, though we note that a similar double-peaked distribution was recently seen (using a similar technique) in the spectroscopic binaries in NGC 7789 \citep{nine2020}.

\subsection{Secondary-Mass and Mass-Ratio Distributions\label{sec:mrd}}

For all binary orbits, we can derive the mass function:
\begin{equation}\label{eq:fm}
f(m) = \frac{m_2^3 sin^3 i}{\left(m_1 + m_2\right)^2} = \frac{P K^3}{2\pi G},
\end{equation}
where $m_1$ and $m_2$ are the primary and secondary masses, respectively, $i$ is the orbital inclination to our line of sight, $P$ is the orbital period, and $K$ is the RV semi-amplitude of the orbit.  $P$ and $K$ are known from our orbital solutions.

In Figure~\ref{fig:fmdist} we show the cumulative distribution of mass functions for the main-sequence binaries in our M67 sample (gray points).  For comparison, we also show mass functions corresponding to binaries with secondaries drawn from a uniform mass-ratio distribution (dashed line) and from a \citet{kroupa2001} IMF (dotted line).  

To generate these theoretical curves, we follow a similar procedure as in GM12, which requires an estimate of the distribution of primary masses in our binary sample.  We estimate primary masses for the binaries using the $BV$ photometry and a 4 Gyr Padova isochrone.     

A one-sample K-S test comparing the observations against the uniform mass-ratio curve shows no significant distinction (with a $p$-value of 0.46).  On the other hand, a similar K-S test comparing the observations against the IMF results in a $p$-value of 1.3$\times10^{-12}$, a 7.10$\sigma$ distinction.  

\begin{figure}
\epsscale{1.0}
\plotone{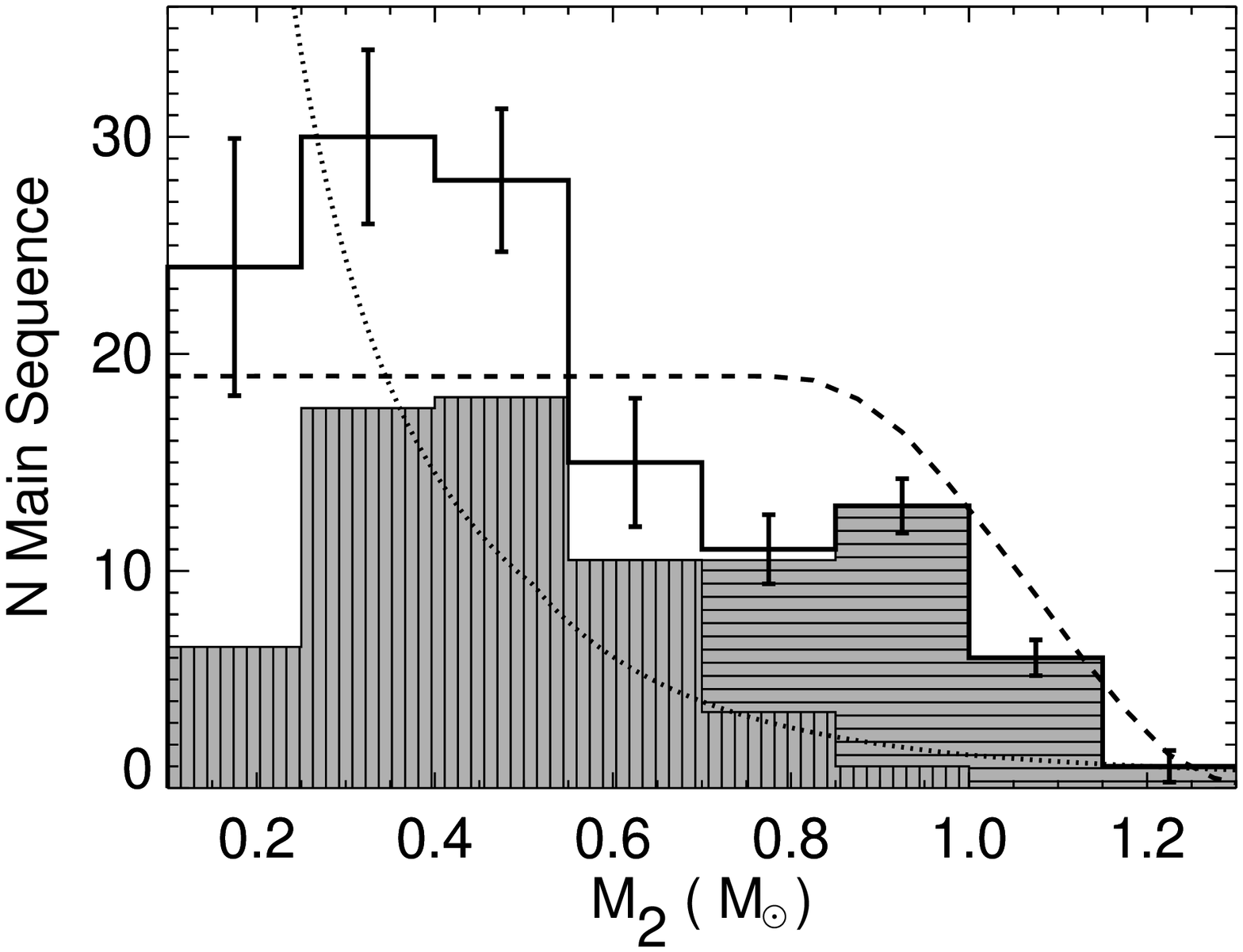}
\plotone{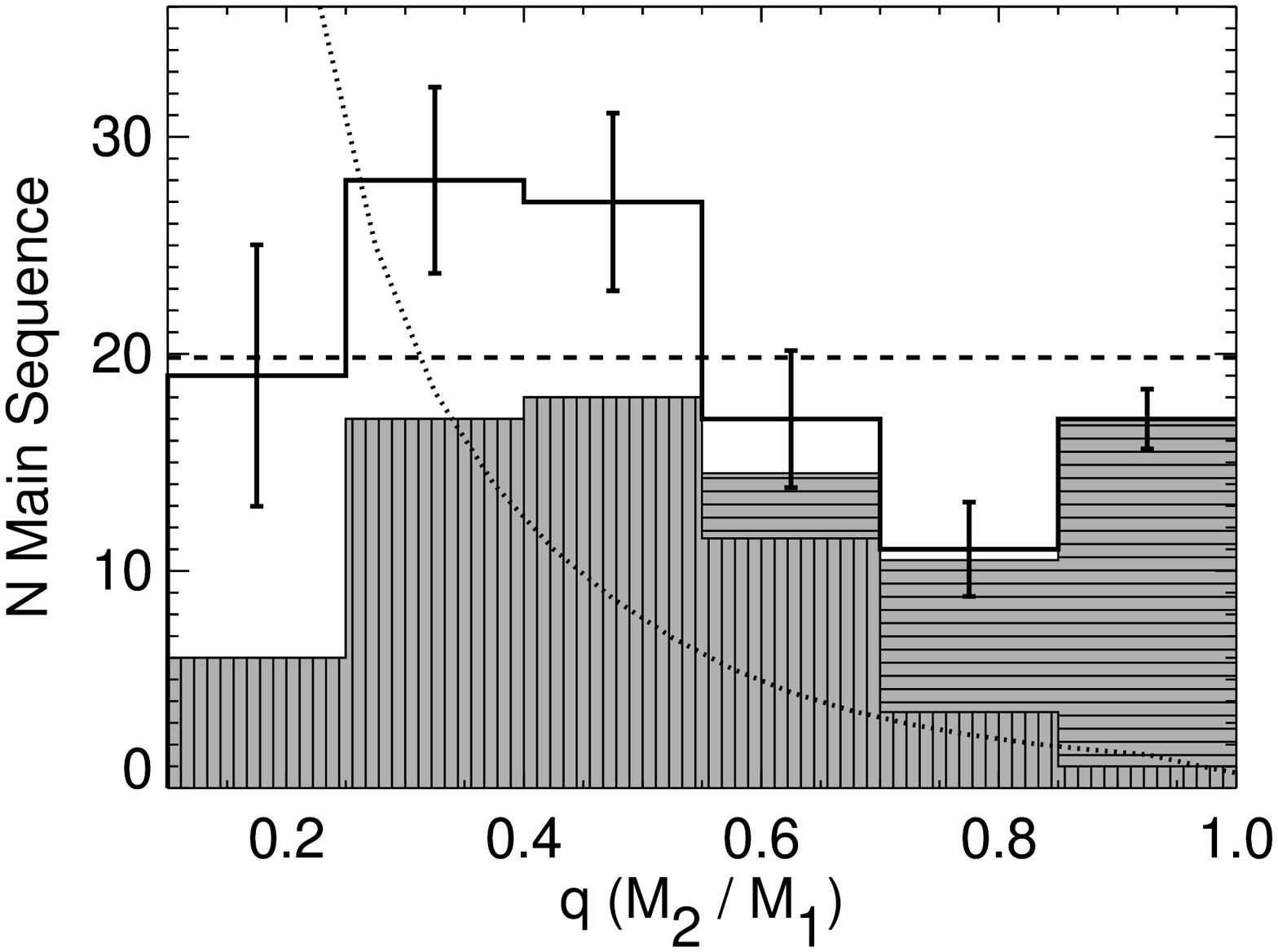}
\caption{Distributions of secondary masses (top) and mass ratio (bottom) of M67 main-sequence binaries, estimated by the statistical algorithm of \citet{mazeh1992}. The observations are show in gray, with SB1s in vertically hatched regions and SB2s in horizontally hatched regions.  The incompleteness-corrected distributions are shown in solid lines.  For comparison, in the top panel we also plot a \citet{kroupa2001} IMF (dotted line) and the secondary mass distribution that results from a uniform mass-ratio distribution (dashed line, see main text for details). With the dotted line in the bottom panel, we show the mass-ratio distribution resulting from choosing secondaries from the IMF and primaries from our observed M67 primary mass distribution.  Finally with the dashed line in the bottom panel, we show the uniform mass-ratio distribution. 
\label{fig:mqdist}}
\end{figure}

In order to derive the distribution of secondary masses, we utilize the statistical algorithm of \citet{mazeh1992} to convert our distributions of mass functions and primary masses to the distributions of secondary masses and mass ratios for the M67 main-sequence binaries. (See GM12 for further details.)  The results of this analysis are shown in Figure~\ref{fig:mqdist}.

\begin{figure}
\epsscale{1.0}
\plotone{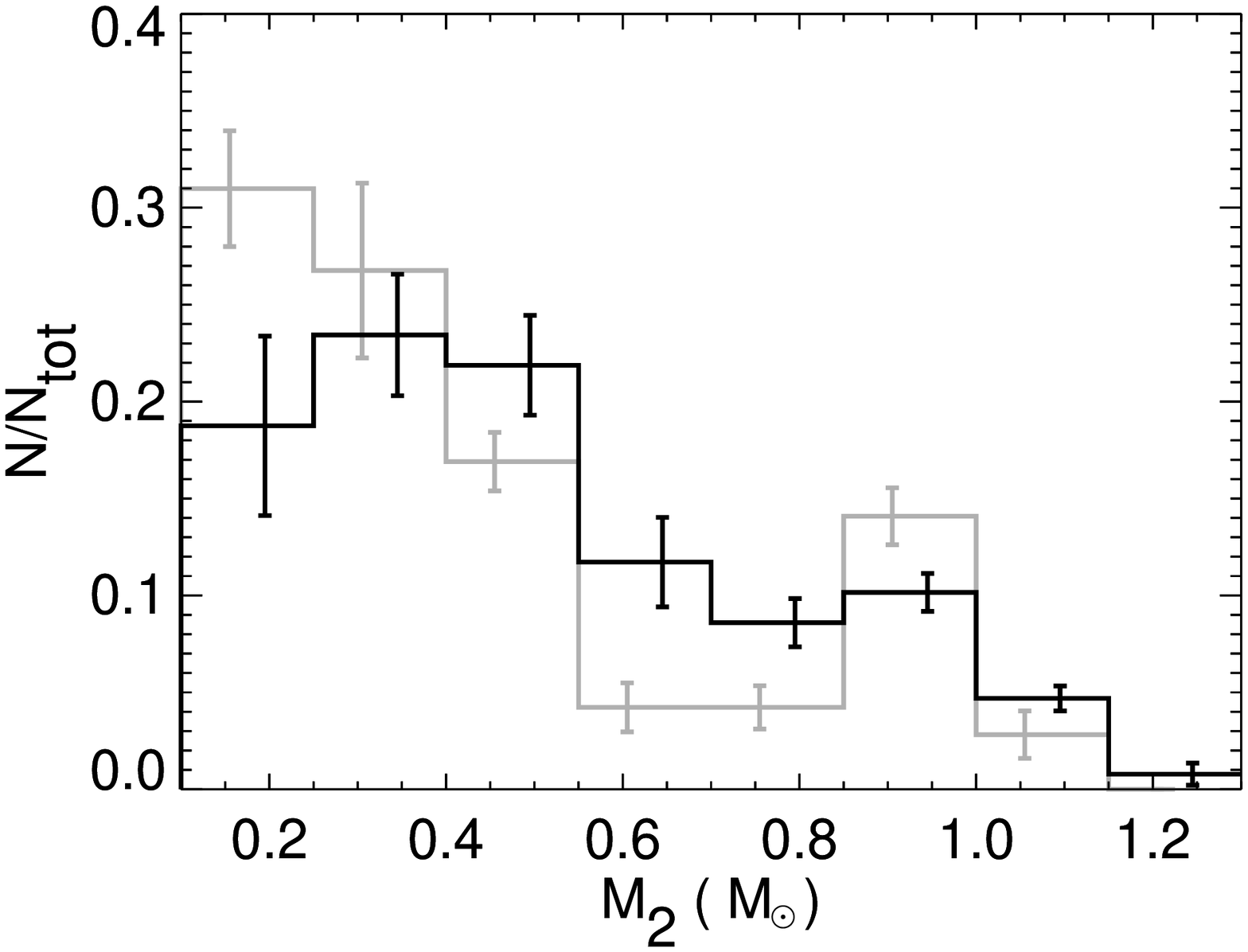}
\plotone{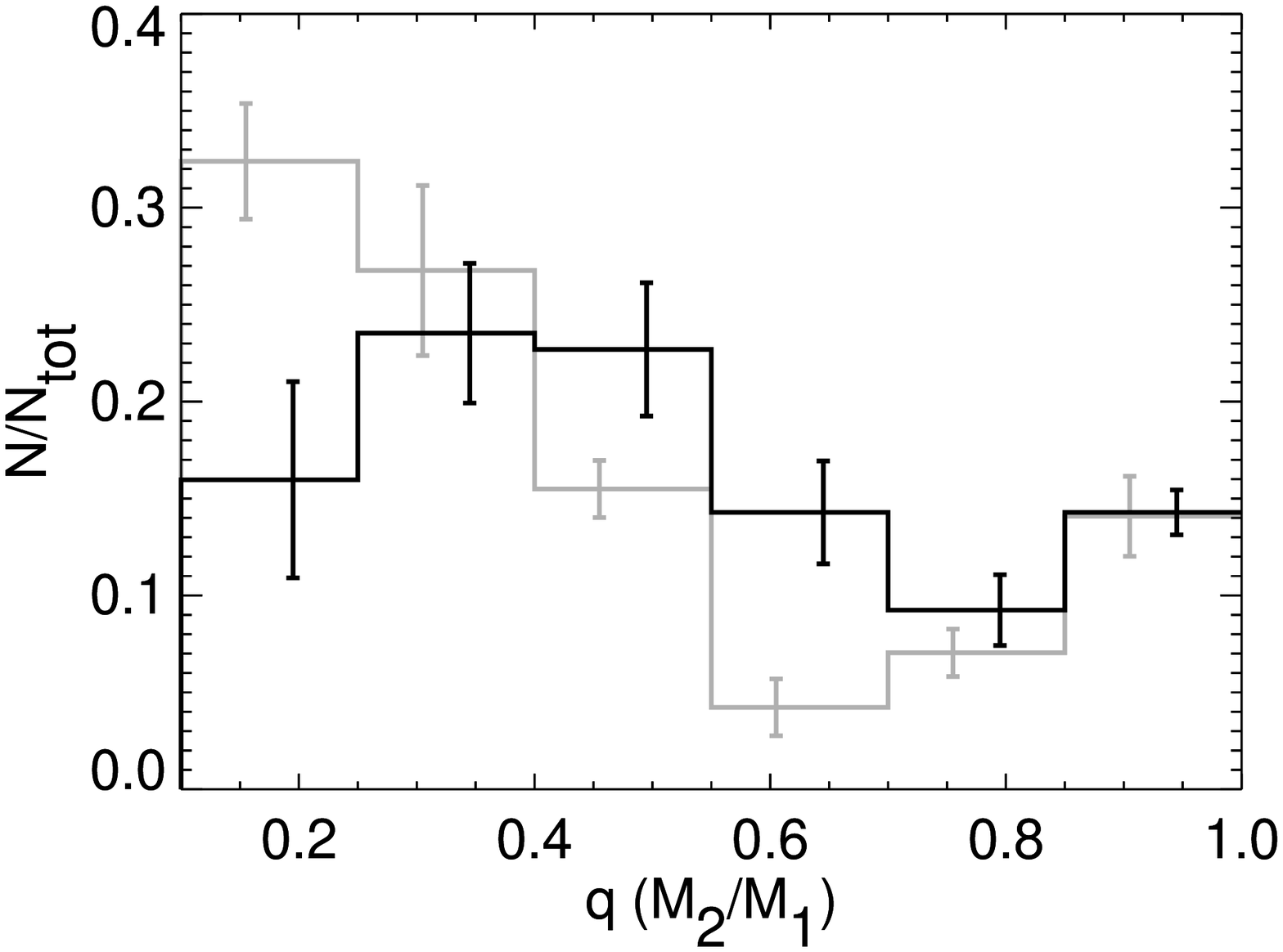}
\caption{Distribution of secondary masses (top) and mass ratios (bottom) of the main-sequence binaries in M67 (black) in comparison to the main-sequence binaries in the open cluster NGC 188 (gray).  We show only the incompleteness corrected distributions here. Error bars are shifted slightly off center from each bin for clarity. }
\label{fig:mdist188}
\end{figure}

In Figure \ref{fig:mqdist}, the observations are shown in gray (with SB1s in vertically hatched regions and SB2s in horizontally hatched regions), and the incompleteness-corrected distributions are show in the solid lines with error bars.  For comparison, we also plot curves showing a uniform mass-ratio distribution (dotted lines) and the \citet{kroupa2001} IMF (dotted lines), normalized to our observed M67 sample size.  To construct the dotted line in the top panel, we  draw primaries from our observed primary-mass distribution and mass ratios from a uniform distribution, to obtain a distribution of secondary masses defined by a uniform mass-ratio distribution.  The downturn in this distribution toward higher secondary masses is due to a similar downturn in the primary-mass distribution.  In the bottom panel, to derive the dotted line, we first draw primary masses from our observed primary-mass distribution, then secondary masses from the IMF up to the mass of the respective primary mass (thereby forcing the mass ratio to be less than unity), and we then plot the resulting mass-ratio distribution.  

Though we do not use these plots for a quantitative comparison (as we prefer a comparison to the observed mass functions instead), we can see a similar result by eye as is shown in Figure~\ref{fig:fmdist}.  Qualitatively, (dashed) lines resulting from the uniform mass-ratio distribution follow the observed distributions more closely than those resulting from the IMF. 

We do not observe a strong peak at mass ratios of unity in M67 (or NGC 188, see Figure~\ref{fig:mdist188}).  Interestingly, such ``twin" binaries are abundant in solar-type stars in the field.  For example, the field binary mass-ratio distribution in R10 (their Figure~16) shows a peak at $q\sim1$ that is approximately twice as high as the bins at lower mass ratio (though we note that their bin sizes are much smaller than those we use here in Figure~\ref{fig:mqdist}).

Finally, in Figure~\ref{fig:mdist188} we compare the M67 incompleteness corrected secondary-mass and mass-ratio distributions (in black) directly to those of NGC 188 (in gray, from GM12).  Here we do not compare using the mass functions because our observations in each cluster sample a slightly different primary-mass regime.  The shapes of the NGC 188 and M67 secondary-mass distributions are qualitatively similar (both rising toward lower masses), as is also true for the mass-ratio distributions (both showing some evidence of a rise toward lower mass ratios). However, formally a $\chi^2$ test comparing the M67 and NGC 188 secondary-mass distributions or mass-ratio distributions results in $>3\sigma$ distinctions, respectively.  We return to this in the following section.

\section{Discussion \label{sec:discuss}}

In the previous sections we compared the  properties of solar-type main-sequence binaries in M67 to similar binaries in the older (7 Gyr) open cluster NGC 188 and the field.  In this section we synthesize these results in the context of star-cluster dynamical evolution.  

If we investigate the binary fraction of the solar-type main-sequence binaries in the field (R10) out to the same limiting orbital period as for M67 (10$^4$ days), we find a binary fraction of 19\%~$\pm$~2\%.  The M67 solar-type main-sequence binary fraction out to the same period cutoff is significantly (4.27$\sigma$) higher than this at 34\%~$\pm$~3\%.  The NGC 188 solar-type main-sequence binary fraction within the same period and physical radius of 7 core radii, is nearly identical to that of M67 at 32\%~$\pm$~4\%.

Solar-type binaries in two additional WOCS clusters, M35 (NGC 2168; 150 Myr) and NGC 6819 (2.5 Gyr) have also been surveyed at a similar level of completeness as M67 and NGC 188.  \citet{leiner2015} find M35 to have a main-sequence binary fraction out to the same period limit and within $\sim$4 core radii of 24\%~$\pm$~3\%.  \citet{milliman2014} find NGC 6819 to have a main-sequence binary fraction out to the same period cutoff and within $\sim$12 core radii of 22\%~$\pm$~3\%.  These open cluster binary fractions are all statistically consistent.  A more careful comparison that accounts for the different spatial extents of each survey is of interest, in particular to investigate for any trends in binary fraction with cluster age or other parameters; however, that level of analysis is beyond the scope of this paper.

Turning to the period distribution, M67 shows a marginal (2.81$\sigma$) distinction from that of the field, driven primarily by the lack of long-period binaries in our sample.  On the other hand, the distributions of orbital periods in M67 are consistent with those in NGC 188.  

The eccentricity distributions, for main-sequence binaries with periods between the circularization period and 3000 days, in M67, NGC 188 and the field are all consistent with a uniform distribution (and inconsistent with a thermal distribution).

Turning to the observed distribution of mass functions, our M67 sample is consistent with mass ratios drawn from a uniform distribution, as is also observed for solar-type main-sequence binaries in NGC 188 and the field. We do not observe an abundance of twin ($q=1$) binaries in M67 or NGC 188, in contrast with the results of R10 for the field.  However, comparing the incompleteness-corrected mass-ratio and secondary-mass distributions between NGC 188 and M67 suggests that M67 lacks binaries with low-mass companions as compared to NGC 188. 

One explanation for these subtle differences between the M67 binaries and these two comparison samples could simply be differences in the progenitor binary populations, perhaps due to formation processes.  Another explanation could be that they were born with similar binary properties and dynamics subsequently modified them.  

Indeed, the results of these comparisons are consistent with expectations of cluster dynamics.  The larger binary fraction in our M67 (and NGC 188) sample as compared to the field may be due to single stars, which have lower masses on average as compared to the binaries, being preferentially lost from the cluster, or preferentially located at larger radial extent, due to two-body relaxation and mass-segregation processes (as was also discussed in GM12 for NGC 188). The lack of long-period binaries as compared to the field may be due to the long-period binaries being ``hardened" toward shorter periods.  The lack of low-mass companions among the M67 binaries as compared to NGC 188 may be tied to the difference in radial extent of our samples, and the processes of mass segregation and dynamical encounters.  The NGC 188 sample extends to $\sim$13 core radii, while the M67 sample is limited to $\sim$7 core radii.  Mass segregation leads to binaries with lower mass companions likely having larger radial distributions.  Also, cluster cores are generally the densest regions of the cluster, where encounters are most frequent; exchange encounters tend to drive binaries closer to mass ratios of unity.

\section{Summary\label{sec:summary}}
In this paper we expand upon the M67 RV study of Paper~1 by presenting orbital solutions for \numbin\ solar-type binary- and triple-star cluster members in M67 (4 Gyr).  Our full stellar sample spans a primary mass range of $\sim0.7 - 1.3$ \Msolar, and a spatial extent of about 7.4 pc ($\sim$7 core radii) in projection, and contains observations for some stars that extend back over 45 years.  Our orbital solutions span in orbital period from a few days out to a maximum orbital period of $\sim$14,000 days. Among this sample are \numtrip\ triples.  We present all orbit plots for SB1 and SB2 binaries and triples in Figures~\ref{fig:sb1orbs},~\ref{fig:sb2orbs}~and~\ref{fig:tripleorbs} with tabulated orbital parameters in Tables~\ref{tab:sb1pars},~\ref{tab:sb2pars}~and~\ref{tab:triplepars}, respectively.  All RVs for binaries, as well as single stars in our sample, are provided in Table~\ref{tab:rv}.  Though we focus our analysis in this paper on the main sequence binaries, we also provide orbital solutions for red giants, yellow giants and sub-subgiants (see the color-magnitude diagram in Figure~8 of Paper~1, for reference).

We analyze the main-sequence binary frequency and distributions of orbital parameters, correcting for observational incompleteness, and find the following results:
\begin{itemize}[topsep=0.1em, partopsep=0.1em, itemsep=0.1em, parsep=0.1em]
  \item Within our mass and spatial domains, the M67 main-sequence stars have a binary frequency of 34\% $\pm$ 3\% out to a period limit of 10$^4$ days.  This binary frequency increases to 70\% $\pm$ 17\% when only considering the inner 0.4 core radii.
  \item We find a circularization period for M67 of $\Pcirc$ by fitting a circularization function to the $e$-$\log P$ distribution of unevolved ($V>13.5$) main-sequence binaries, in agreement with previous determinations.
  \item The orbital period distribution rises toward longer periods, and is formally consistent with solar-type binaries in the field and in NGC 188.  There is, however, a hint (at 2.81$\sigma$) of missing long-period binaries as compared to the field log-normal distribution, possibly due to dynamical hardening of the wider binaries by stellar encounters that drive them toward shorter periods.  The period distribution is statistically inconsistent with a log-uniform distribution (at 3.96$\sigma$), a popular initial condition for binary population synthesis and star cluster models. 
  \item The eccentricity distribution for main-sequence binaries in M67 with periods between the circularization period and 3000 days is consistent with similar binaries in NGC 188 and the field, which can be described by a uniform distribution.  A thermal distribution, which is popular in theoretical models, is ruled out at very high confidence (6.65$\sigma$).  
  \item The distribution of mass functions for the M67 binaries is consistent with a uniform mass-ratio distribution, but inconsistent with a distribution derived by drawing companions from an IMF (at 7.10$\sigma$).  This result is consistent with solar-type binaries in NGC 188 and the galactic field.  
\end{itemize}

These results focus on the main-sequence binary population, which dominates the mass of the cluster. They provide an important touchstone for star cluster dynamical models and studies of solar-type stars more broadly.

\acknowledgments
The authors would like to thank the many individuals who
helped obtain these spectra, determine the stellar radial
velocities and solve for orbital solutions, 
both at the CfA: 
Jim Peters, Bob Davis, Ed Horine, Perry Berlind, Ale Milone, Robert Stefanik, Mike Calkins, Gil Esquerdo, Jessica Mink and Christian Latham,
and at the University of Wisconsin-Madison: 
Natalie Gosnell, Katelyn Milliman, Ben Tofflemire and David Szymulewski.   
This research was made possible by the National Science Foundation grant AST-1714506 and the Wisconsin Alumni Research Foundation at the University of Wisconsin~-~Madison.

\begin{deluxetable*}{llllccccccccccccc}
 \tabletypesize{\tiny}
 \tablecaption{Radial-Velocity Data Table\label{tab:rv}}
 \tablehead{\colhead{ID} & \colhead{XID} & \colhead{RA} & \colhead{Dec} & \colhead{HJD-2,400,000} & \colhead{Tel.} & \colhead{RV$_1$} & \colhead{RV$_2$} & \colhead{RV$_3$} & \colhead{RV$_1,e$} & \colhead{Correlation$_1$}& \colhead{$(O-C)_1$} & \colhead{$(O-C)_2$} & \colhead{$(O-C)_{1,out}$}  & \colhead{$(O-C)_{2,out}$} & \colhead{Phase} & \colhead{Phase$_{out}$} \\
 \colhead{} & \colhead{} & \colhead{} & \colhead{} & \colhead{(days)} & \colhead{} & \colhead{(\kms)} & \colhead{(\kms)} & \colhead{(\kms)} & \colhead{(\kms)} & \colhead{Height} & \colhead{(\kms)} & \colhead{(\kms)} & \colhead{(\kms)} & \colhead{(\kms)} & \colhead{} & \colhead{}}
\rotate
\startdata
1001 & S1024 & 08:51:22.91 & +11:48:49.40 & \nodata & \nodata & \nodata & \nodata &\nodata & \nodata & \nodata & \nodata & \nodata &\nodata & \nodata & \nodata & \nodata \\
\nodata & \nodata & \nodata & \nodata & 45784.8376 & TD & 10.60 & 56.86 & \nodata & \nodata & \nodata & 0.28 & 0.32 & \nodata & \nodata & 0.0645 & \nodata \\
\nodata & \nodata & \nodata & \nodata & 45807.6793 & TD & -31.66 & 101.01 & \nodata & \nodata & \nodata & 2.97 & -0.88 & \nodata & \nodata & 0.2549 & \nodata \\
\nodata & \nodata & \nodata & \nodata & 46065.0463 & MD & -30.25 & 99.10 & \nodata & \nodata & \nodata & 0.16 & 1.46 & \nodata & \nodata & 0.2022 & \nodata \\
\nodata & \nodata & \nodata & \nodata & 46072.0582 & TD & -27.84 & 94.82 & \nodata & \nodata & \nodata & -1.03 & 0.82 & \nodata & \nodata & 0.1815 & \nodata \\
\nodata & \nodata & \nodata & \nodata & 46125.8883 & TD & 93.85 & -32.79 & \nodata & \nodata & \nodata & -2.49 & -2.56 & \nodata & \nodata & 0.7002 & \nodata \\
\nodata & \nodata & \nodata & \nodata & 46162.6236 & TD & 97.02 & -28.00 & \nodata & \nodata & \nodata & 2.06 & 0.85 & \nodata & \nodata & 0.8311 & \nodata \\
\nodata & \nodata & \nodata & \nodata & 46393.0446 & MD & 29.17 & 35.04 & \nodata & \nodata & \nodata & -2.16 & -0.30 & \nodata & \nodata & 0.0147 & \nodata \\
\nodata & \nodata & \nodata & \nodata & 46423.0430 & TD & -30.90 & 97.89 & \nodata & \nodata & \nodata & -0.12 & -0.12 & \nodata & \nodata & 0.2047 & \nodata \\
\nodata & \nodata & \nodata & \nodata & 46423.9877 & TD & -27.12 & 94.08 & \nodata & \nodata & \nodata & -0.45 & 0.22 & \nodata & \nodata & 0.3367 & \nodata \\
\enddata
\tablecomments{Table 4 is published in its entirety in the electronic 
edition of the {\it Astrophysical Journal}.  A portion is shown here 
for guidance regarding its form and content.}
\end{deluxetable*}

\begin{figure*}
\gridline{\fig{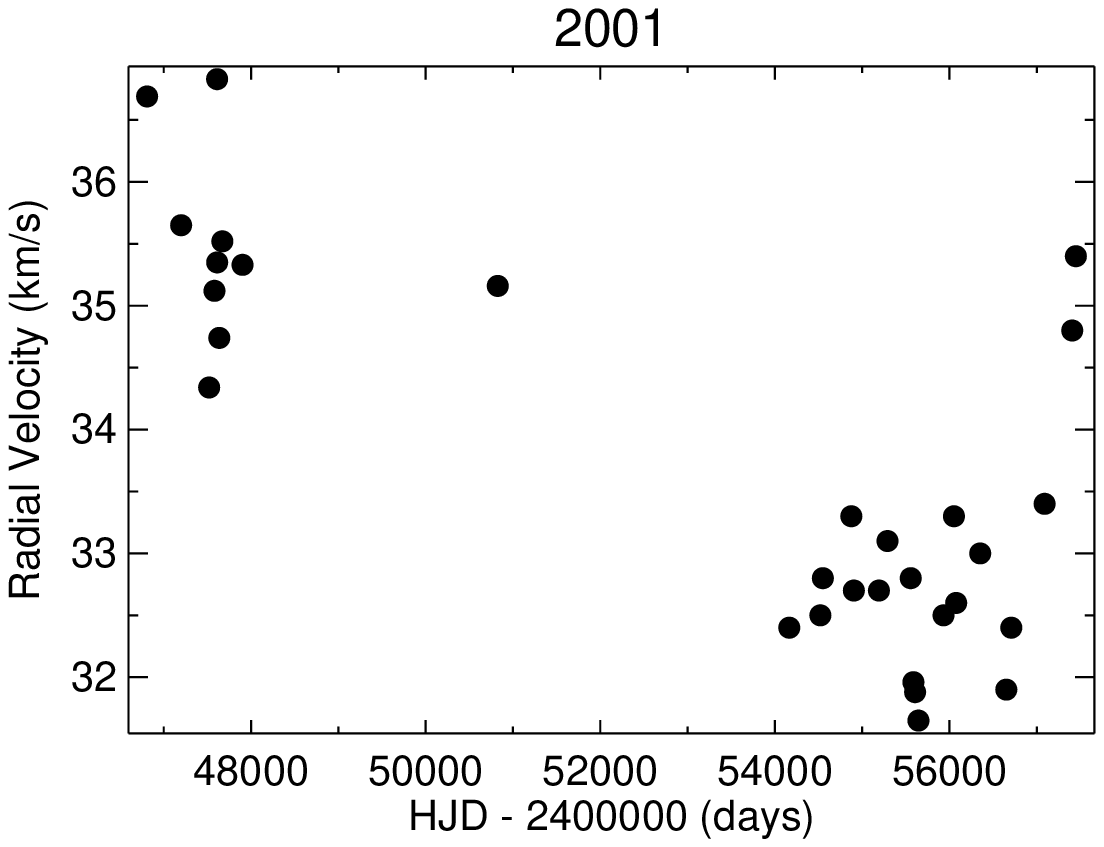}{0.3\textwidth}{}
    \fig{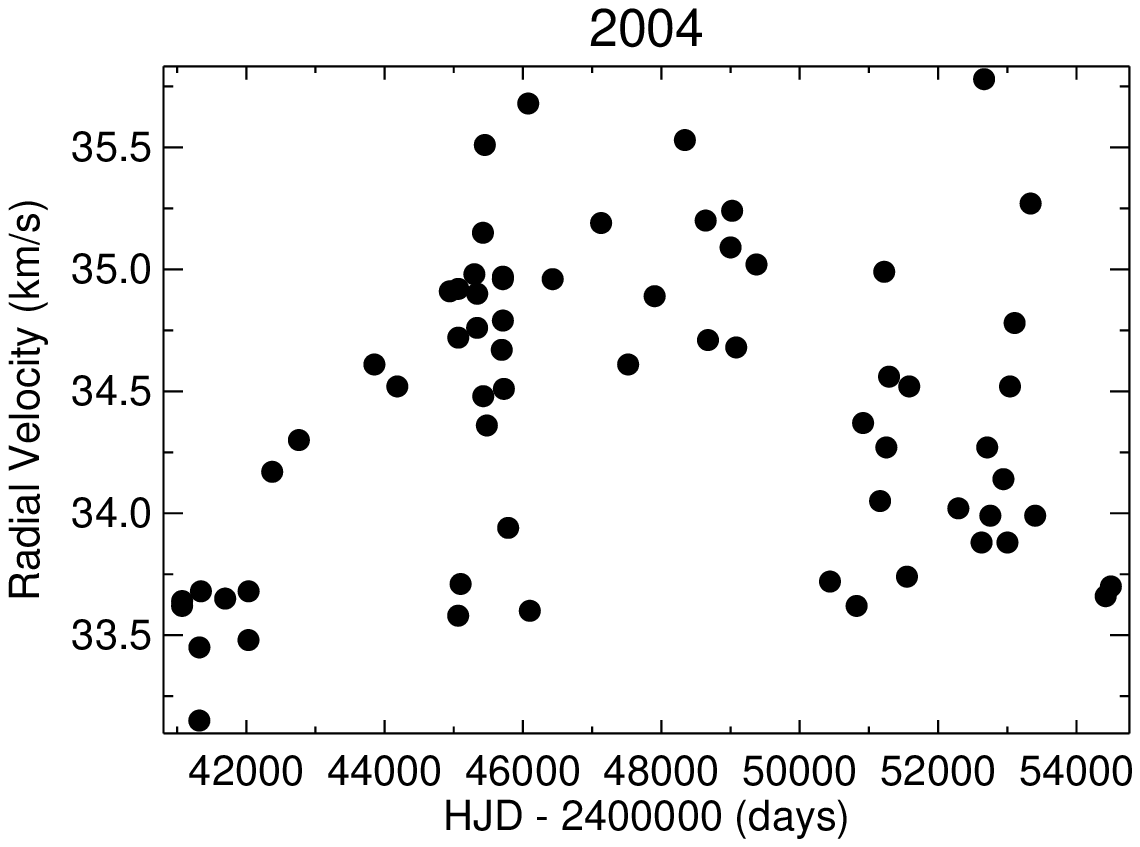}{0.3\textwidth}{}
    \fig{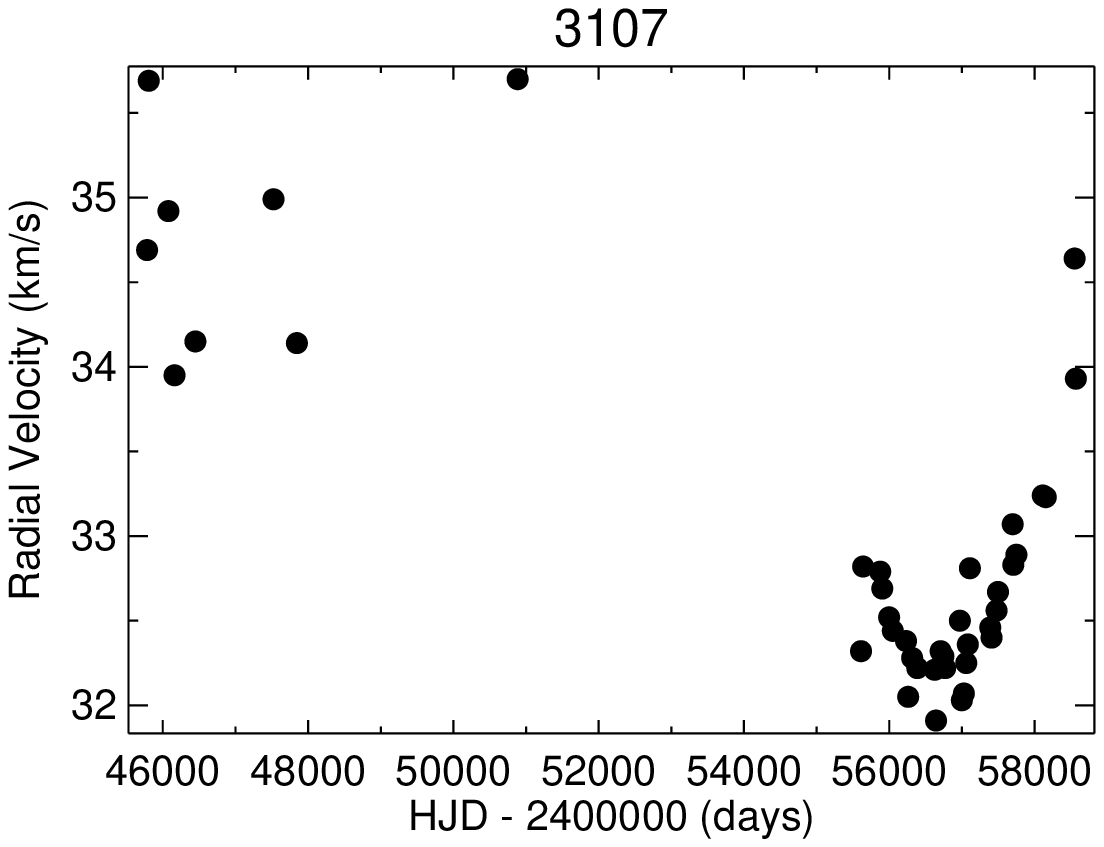}{0.3\textwidth}{}}
\gridline{\fig{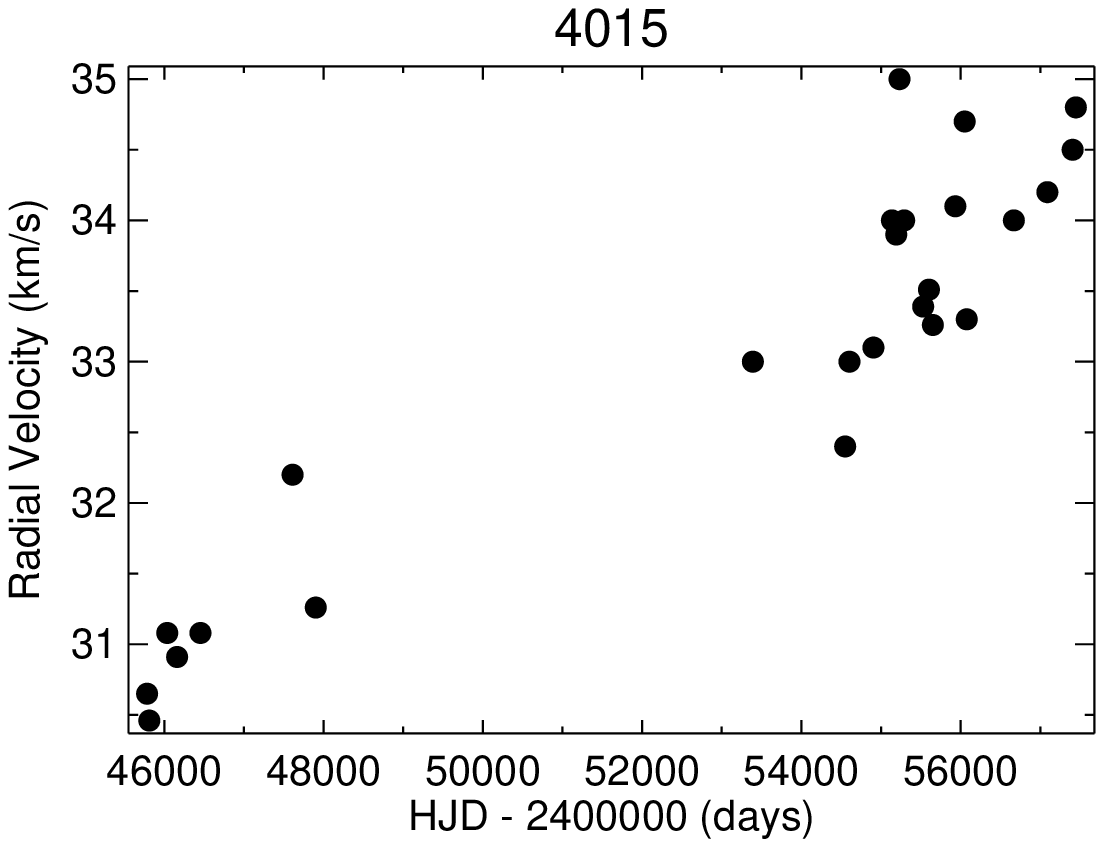}{0.3\textwidth}{}
    \fig{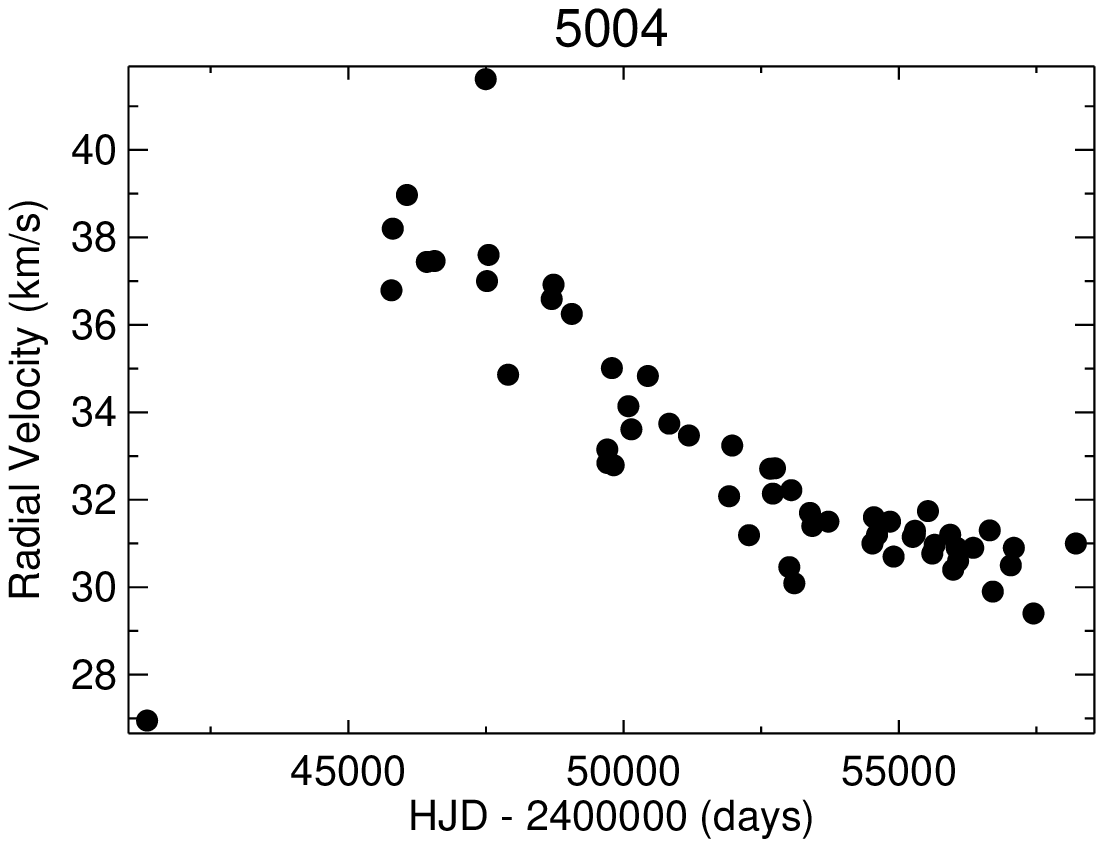}{0.3\textwidth}{}
    \fig{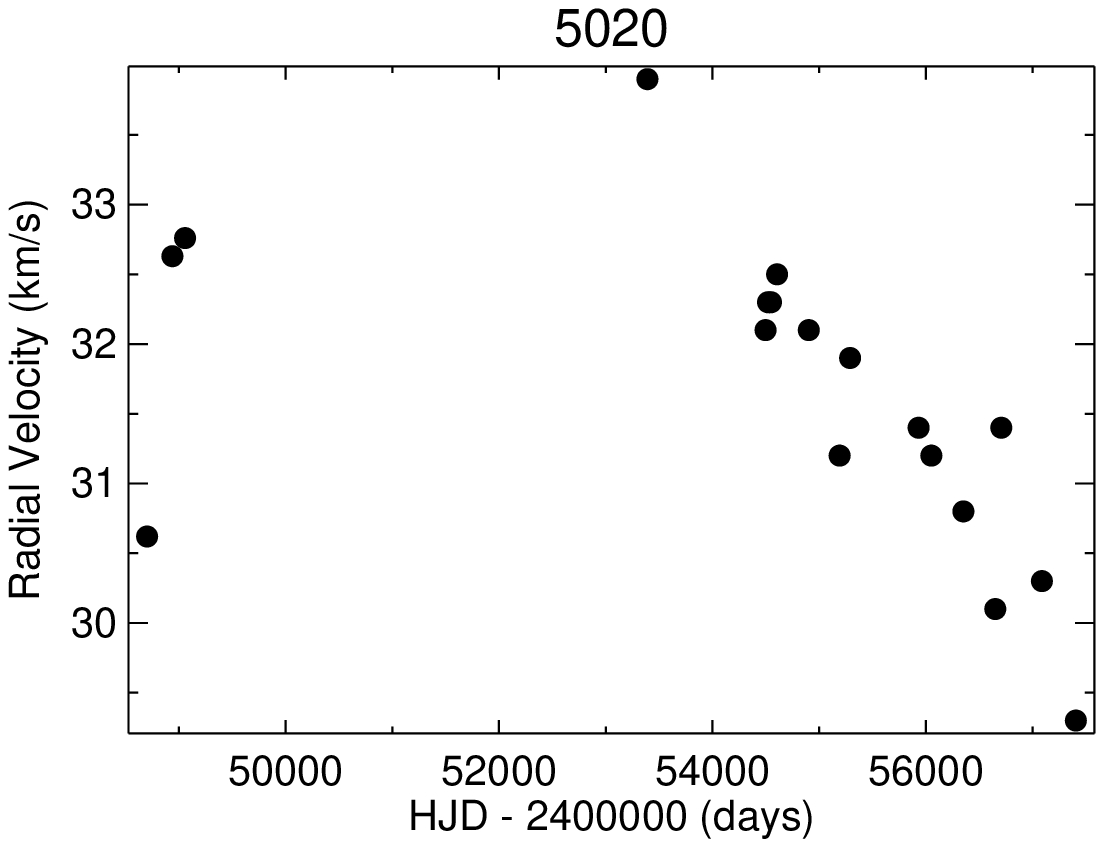}{0.3\textwidth}{}}
\gridline{\fig{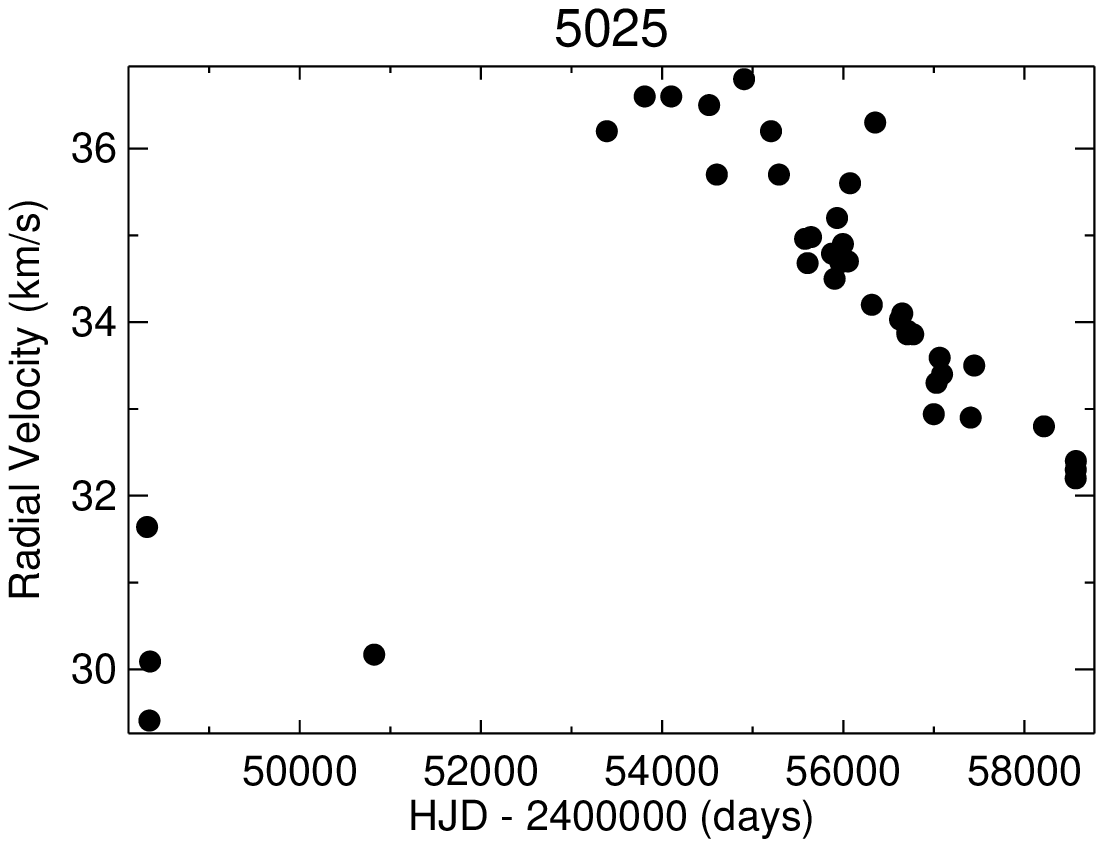}{0.3\textwidth}{}
    \fig{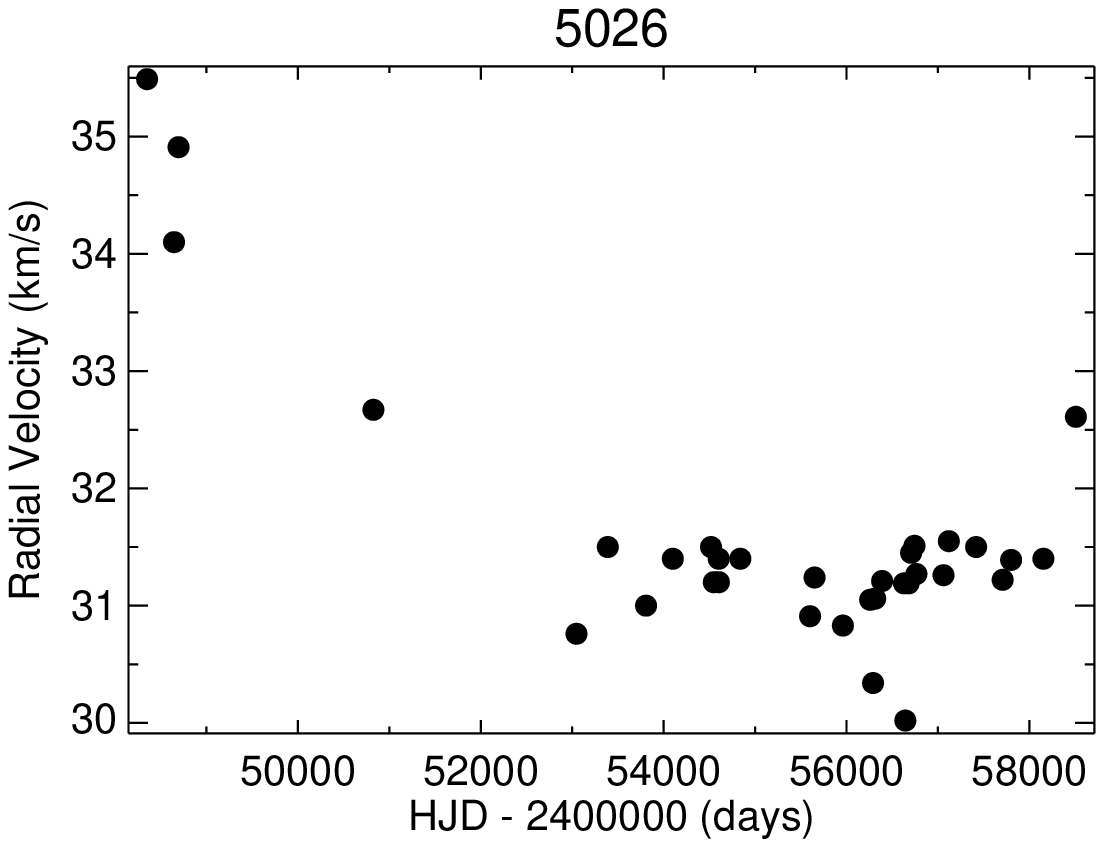}{0.3\textwidth}{}
    \fig{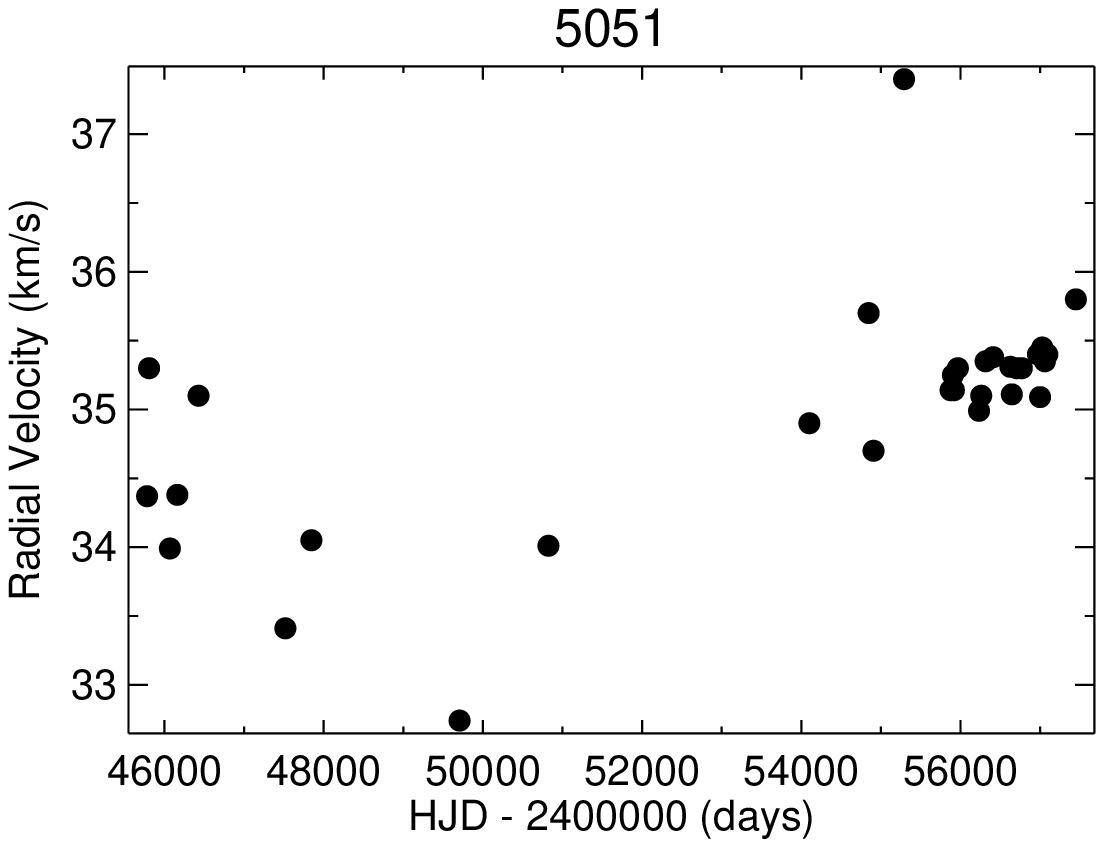}{0.3\textwidth}{}}
\gridline{\fig{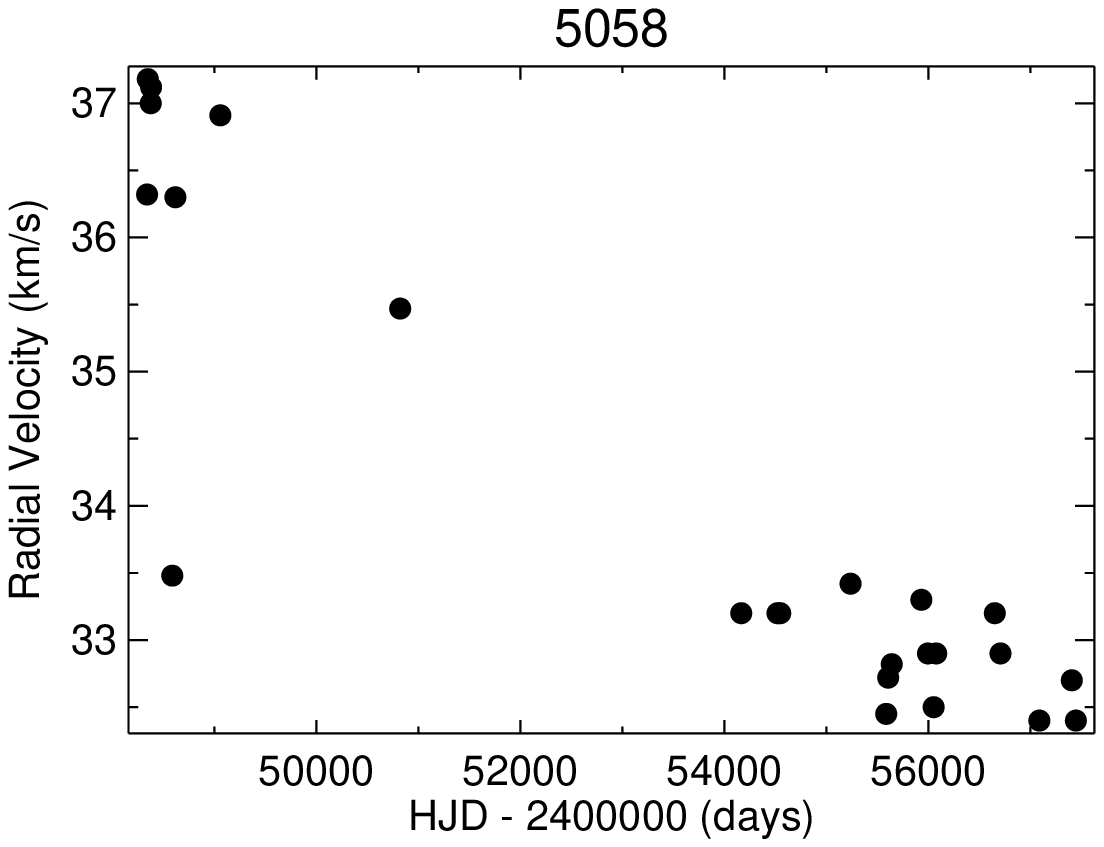}{0.3\textwidth}{}
    \fig{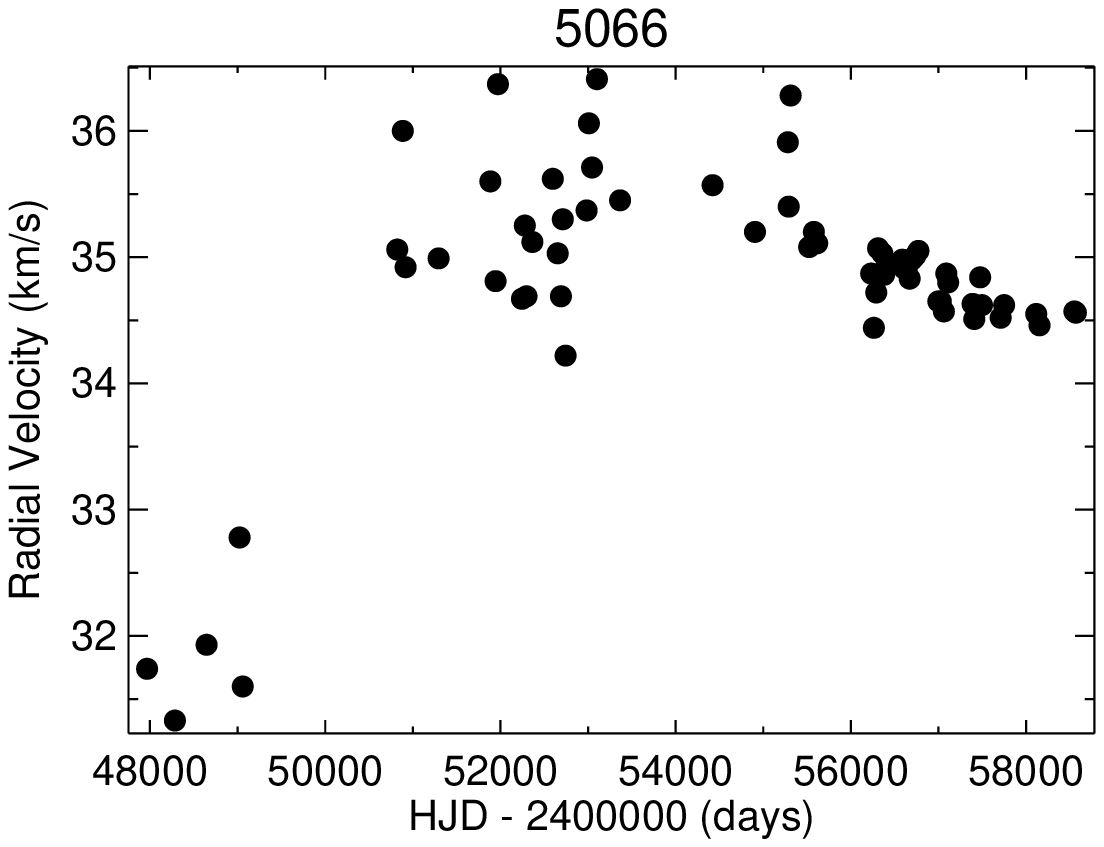}{0.3\textwidth}{}
    \fig{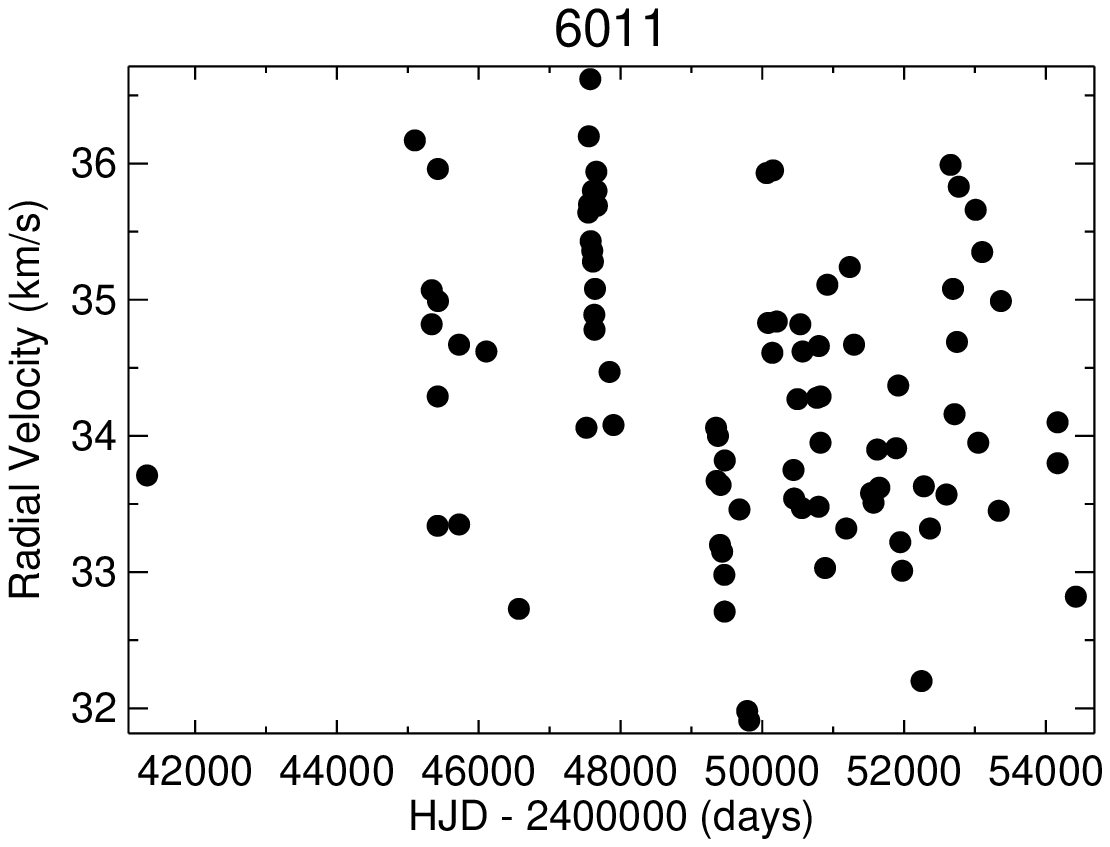}{0.3\textwidth}{}}
\caption{Candidate M67 binaries that visibly show low-amplitude trends in radial velocity (RV) versus time, but are not identified  as binaries in our statistical analysis (see Table~\ref{tab:binSM}).  For each panel we show the observed RV versus time in black filled circles, and provide the WOCS ID above.\label{fig:trends}}
\end{figure*}

\begin{figure*}
\figurenum{10}
\gridline{\fig{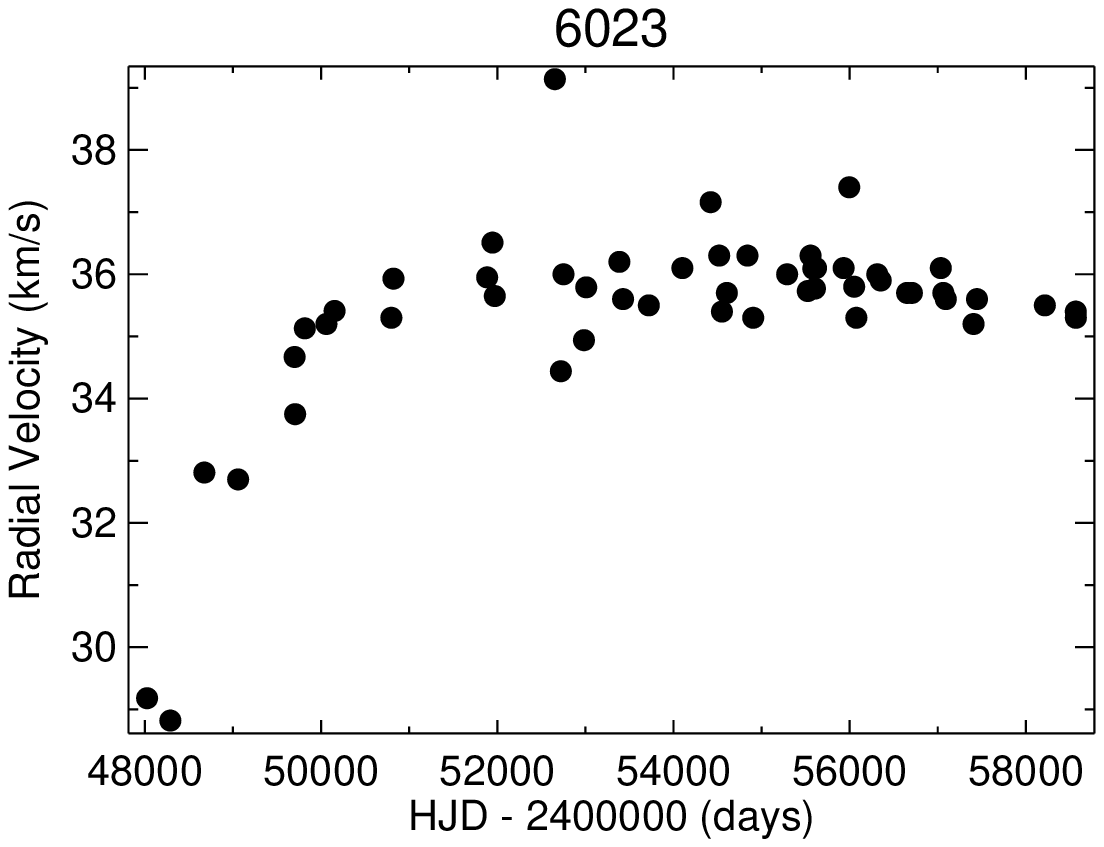}{0.3\textwidth}{}
    \fig{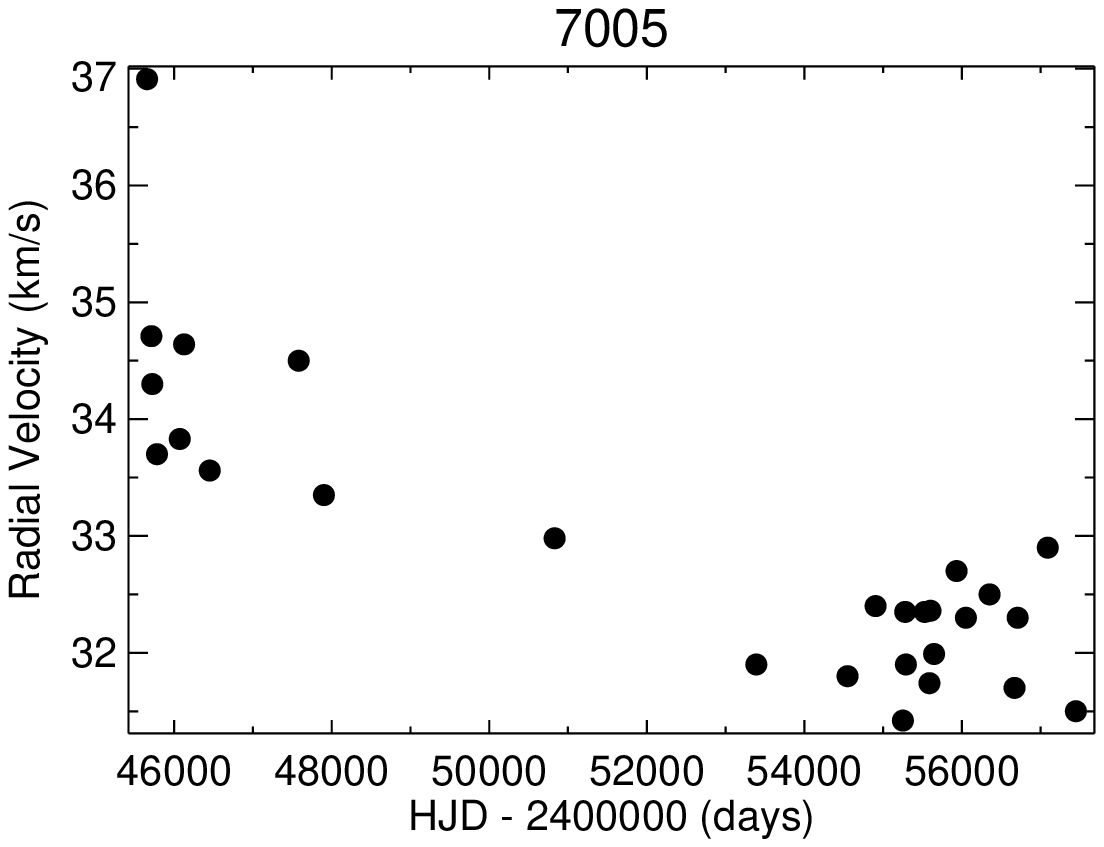}{0.3\textwidth}{}
    \fig{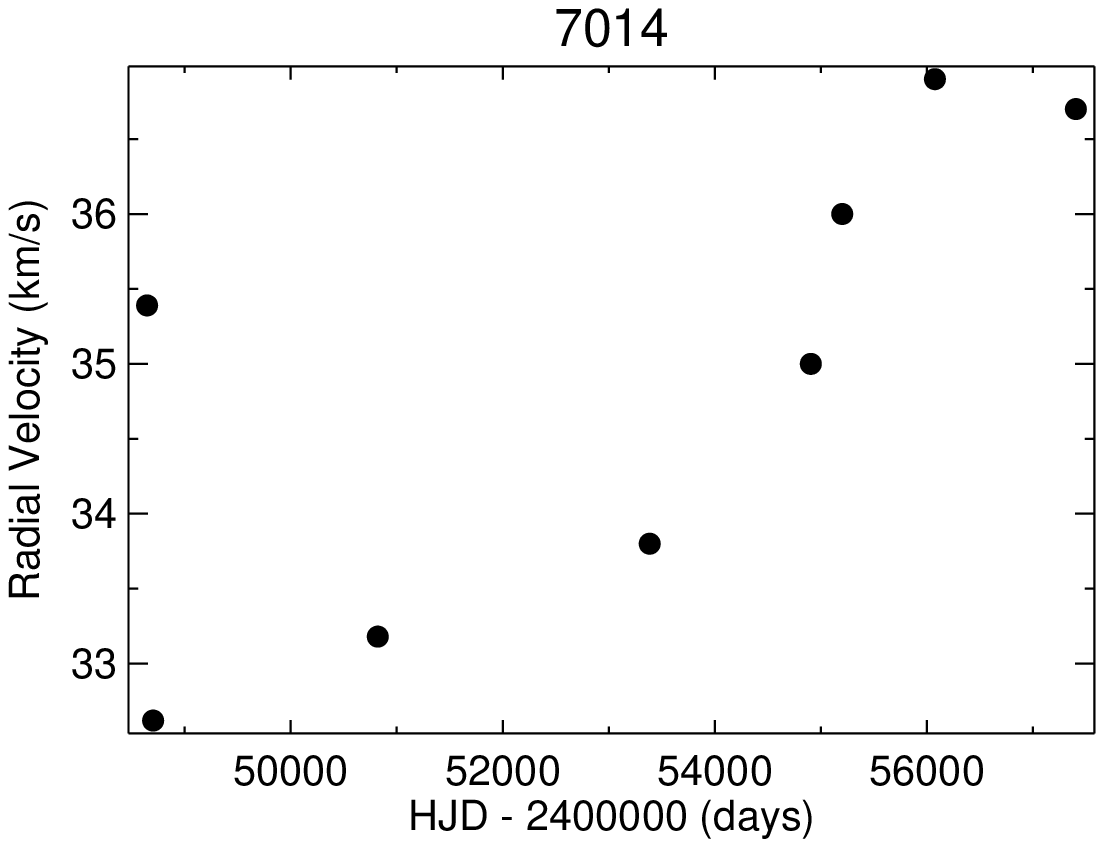}{0.3\textwidth}{}}
\gridline{\fig{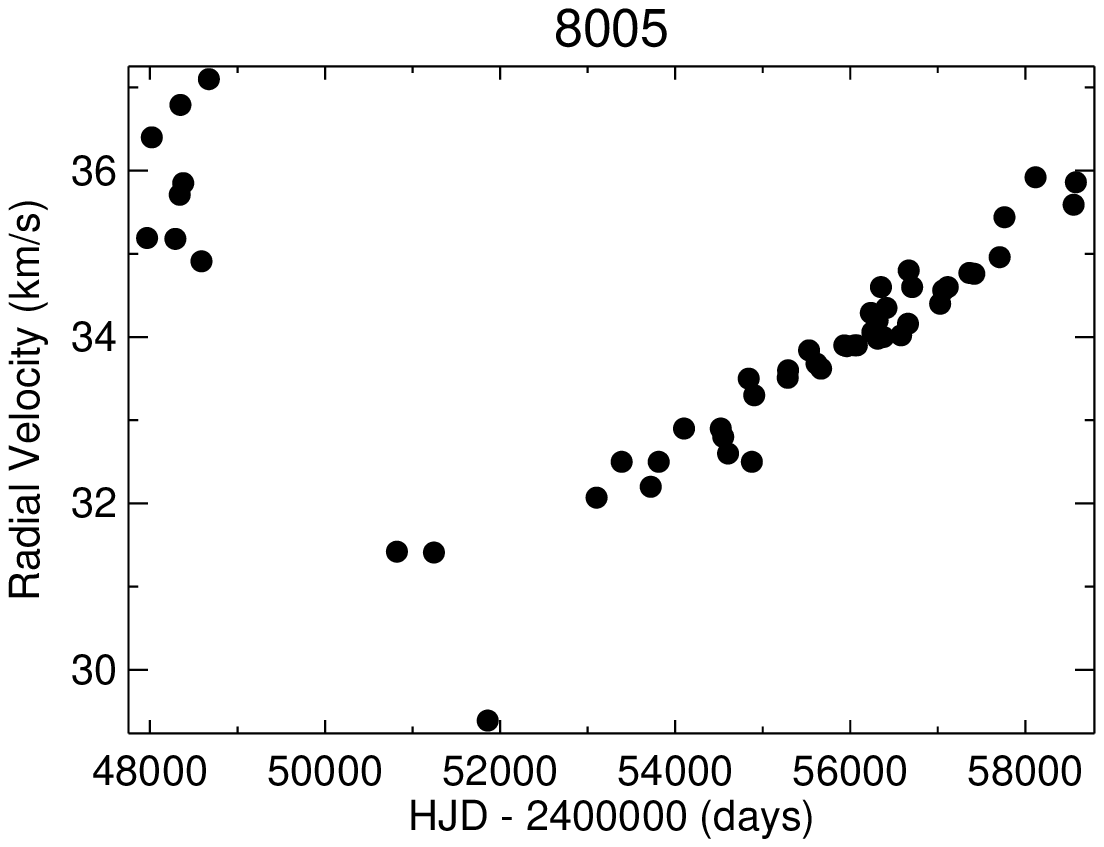}{0.3\textwidth}{}
    \fig{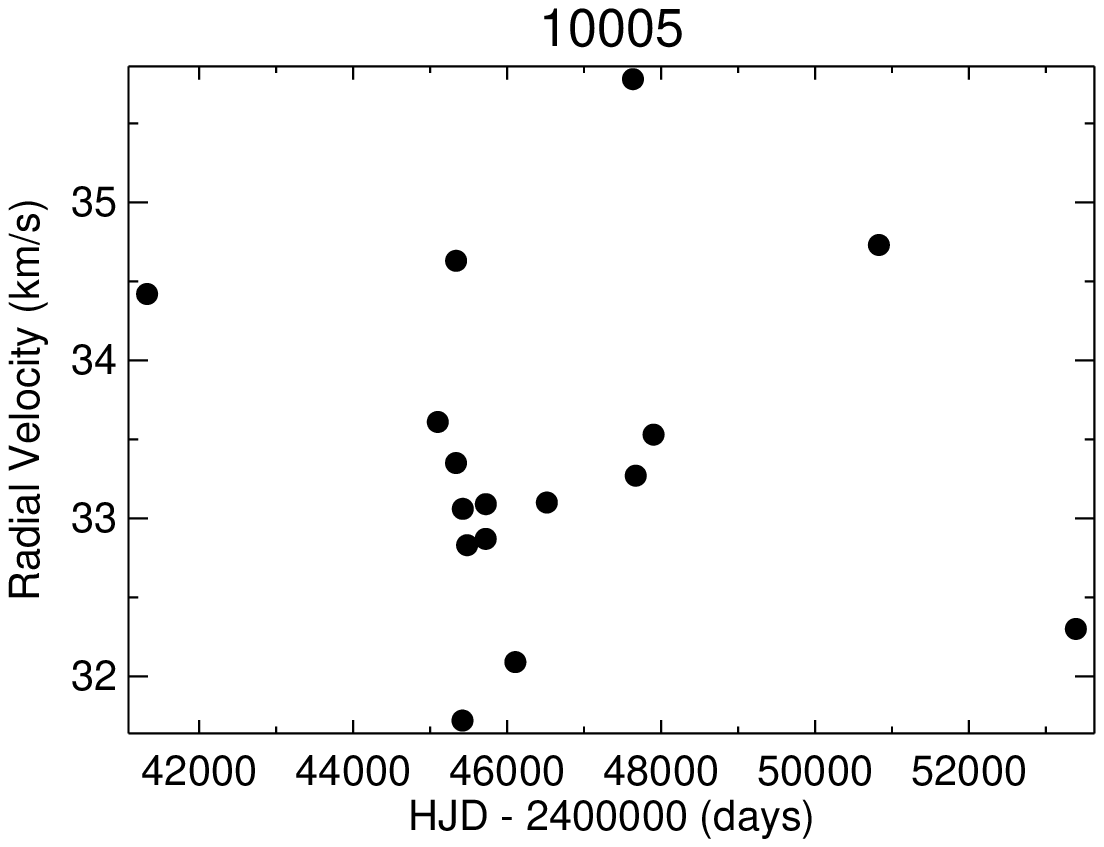}{0.3\textwidth}{}
    \fig{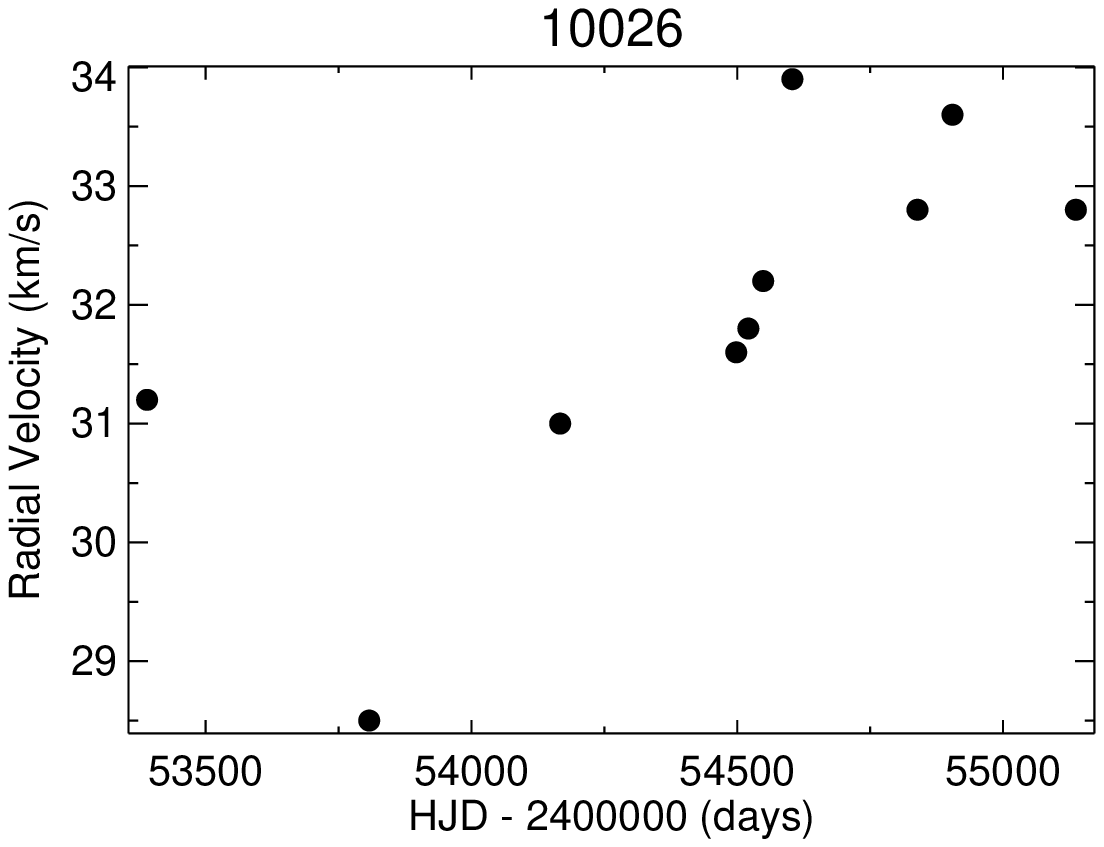}{0.3\textwidth}{}}
\gridline{\fig{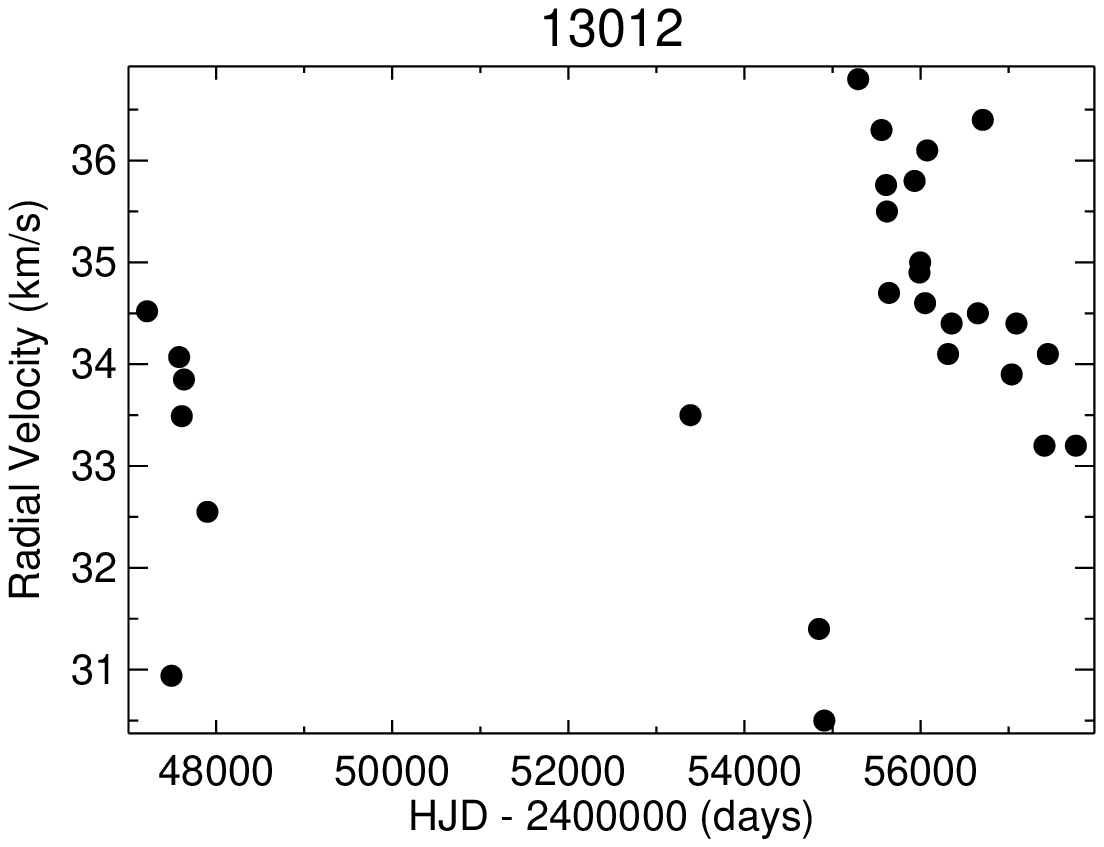}{0.3\textwidth}{}
    \fig{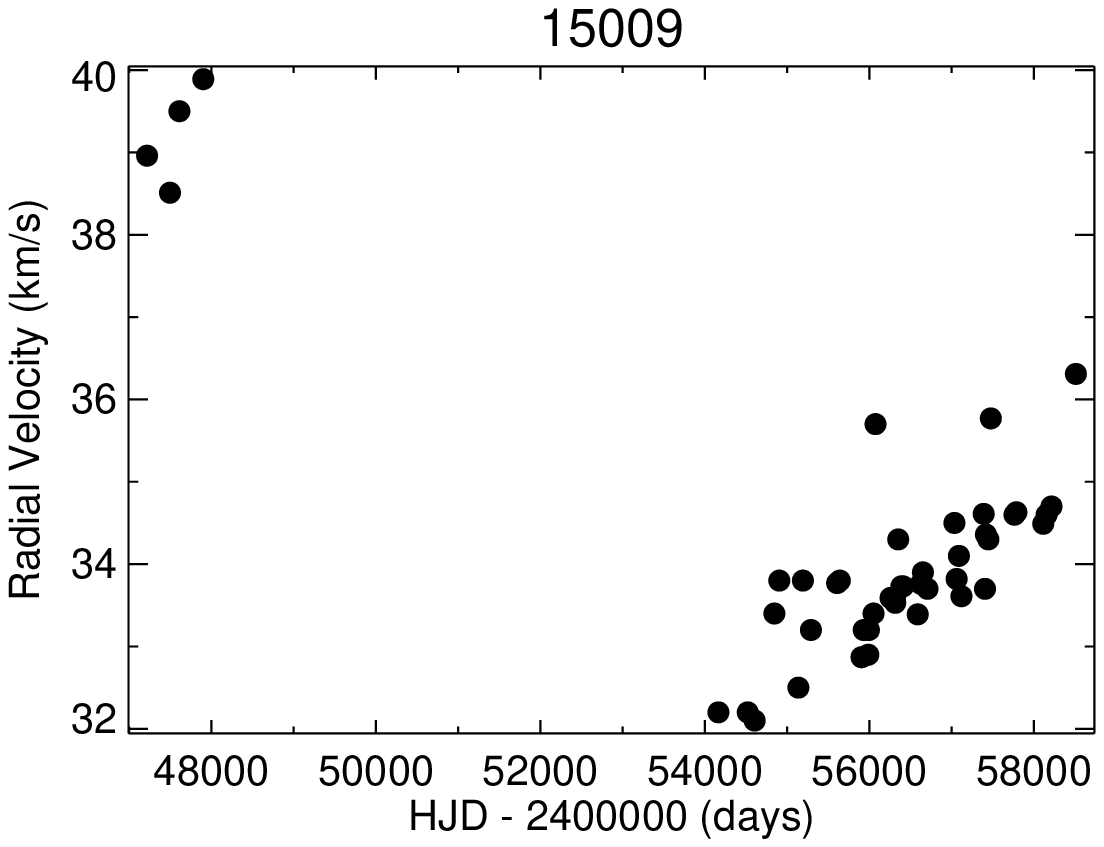}{0.3\textwidth}{}
    \fig{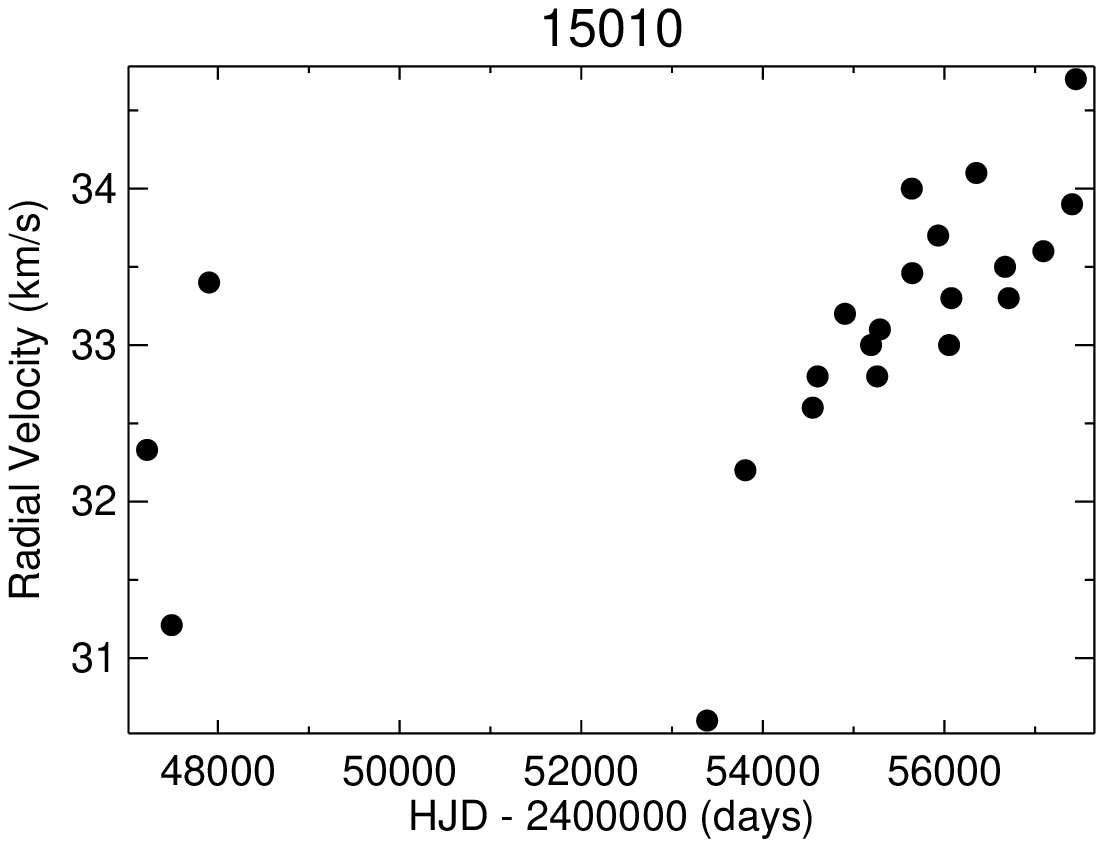}{0.3\textwidth}{}}
\gridline{\fig{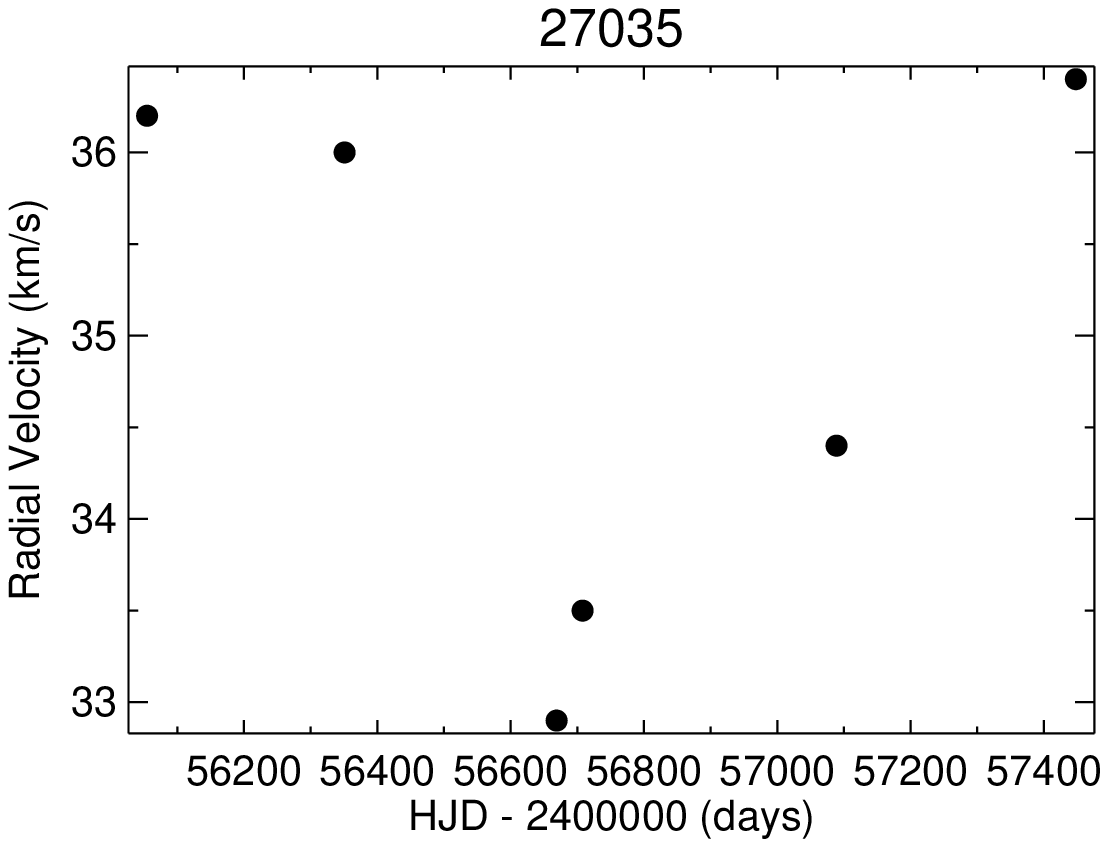}{0.3\textwidth}{}
}
\caption{(Continued.)}
\end{figure*}

\begin{longrotatetable}
\begin{deluxetable*}{llrcrrrrccrcc}
\tabletypesize{\footnotesize}
\tablecaption{Orbital Parameters for M67 Single-Lined Binaries\label{tab:sb1pars}}
\tablehead{\colhead{ID} & \colhead{XID} & \colhead{$P$} & \colhead{Orbital} & \colhead{$\gamma$} & \colhead{$K$} & \colhead{$e$} & \colhead{$\omega$} & \colhead{$T_\circ$} & \colhead{$a\sin i$} & \colhead{$f(m)$} & \colhead{$\sigma$} & \colhead{$N$} \\
\colhead{} & \colhead{} & \colhead{(days)} & \colhead{Cycles} & \colhead{(\kms)} & \colhead{(\kms)} & \colhead{} & \colhead{(deg)} & \colhead{(HJD-2400000 d)} & \colhead{(10$^6$ km)} & \colhead{(\Msolar)} & \colhead{(\kms)} & \colhead{}}
\startdata
    1005 & S1250 &           4422 &    2.8 &           33.64 &            2.59 &           0.529 &              81 &           47587 &             133 &          4.8$\times10^{-3}$ &  0.44 &  111 \\
         &       &       $\pm$ 17 &        &      $\pm$ 0.05 &      $\pm$ 0.08 &     $\pm$ 0.023 &         $\pm$ 3 &        $\pm$ 20 &         $\pm$ 5 &    $\pm$ 0.5$\times10^{-3}$ &       &      \\
    1015 & S1237 &         698.56 &   23.9 &           33.58 &            5.02 &           0.087 &             260 &           46574 &            48.0 &          9.1$\times10^{-3}$ &  0.44 &   75 \\
         &       &     $\pm$ 0.24 &        &      $\pm$ 0.06 &      $\pm$ 0.07 &     $\pm$ 0.015 &        $\pm$ 10 &        $\pm$ 19 &       $\pm$ 0.7 &    $\pm$ 0.4$\times10^{-3}$ &       &      \\
    1033 &  S721 &           6350 &    2.5 &           34.53 &            0.98 &            0.54 &             266 &           46860 &              71 &          3.6$\times10^{-4}$ &  0.41 &   89 \\
         &       &      $\pm$ 220 &        &      $\pm$ 0.06 &      $\pm$ 0.09 &      $\pm$ 0.07 &        $\pm$ 10 &       $\pm$ 220 &         $\pm$ 8 &    $\pm$ 1.1$\times10^{-4}$ &       &      \\
    2002 & S1040 &        42.8251 &  382.4 &           33.00 &            8.47 &           0.016 &              80 &           45539 &            4.99 &         2.70$\times10^{-3}$ &  0.54 &   60 \\
         &       &   $\pm$ 0.0010 &        &      $\pm$ 0.07 &      $\pm$ 0.10 &     $\pm$ 0.012 &        $\pm$ 50 &         $\pm$ 5 &      $\pm$ 0.06 &   $\pm$ 0.10$\times10^{-3}$ &       &      \\
\enddata
\end{deluxetable*}
\end{longrotatetable}

\begin{figure*}
\gridline{\fig{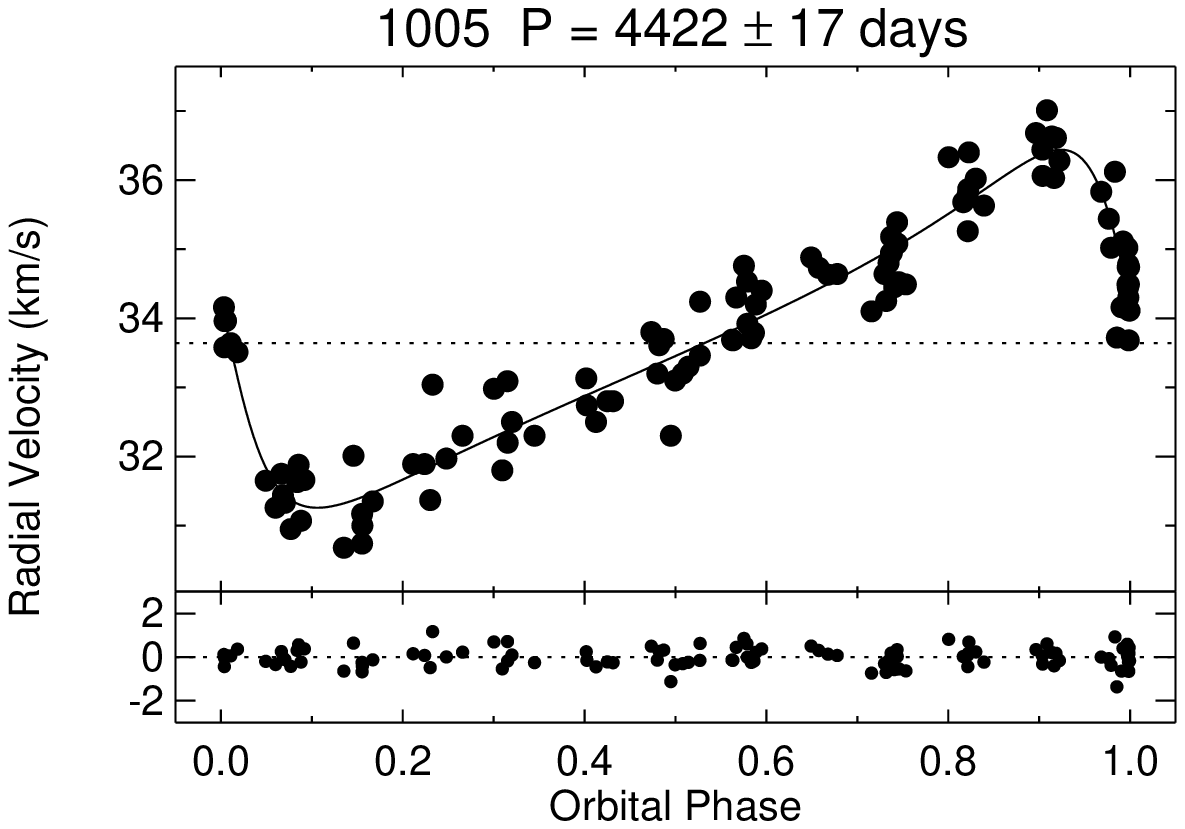}{0.3\textwidth}{}
	\fig{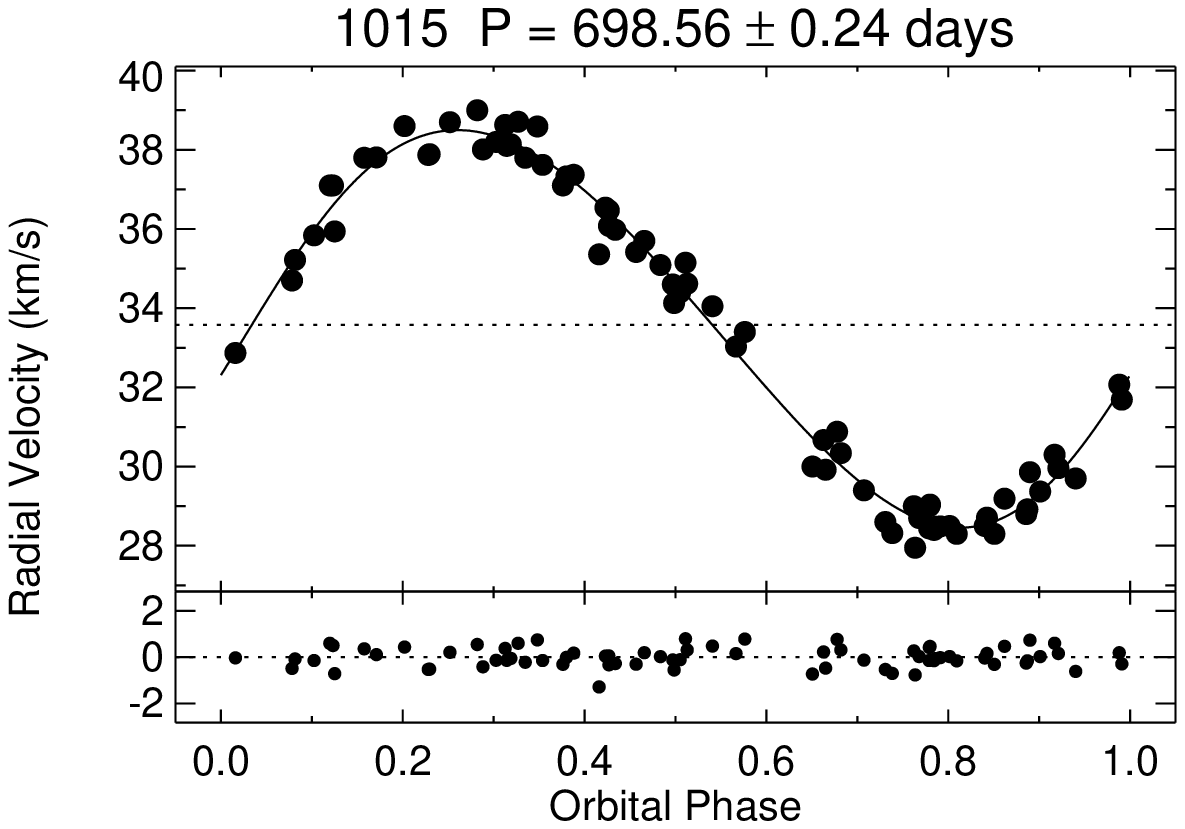}{0.3\textwidth}{}
	\fig{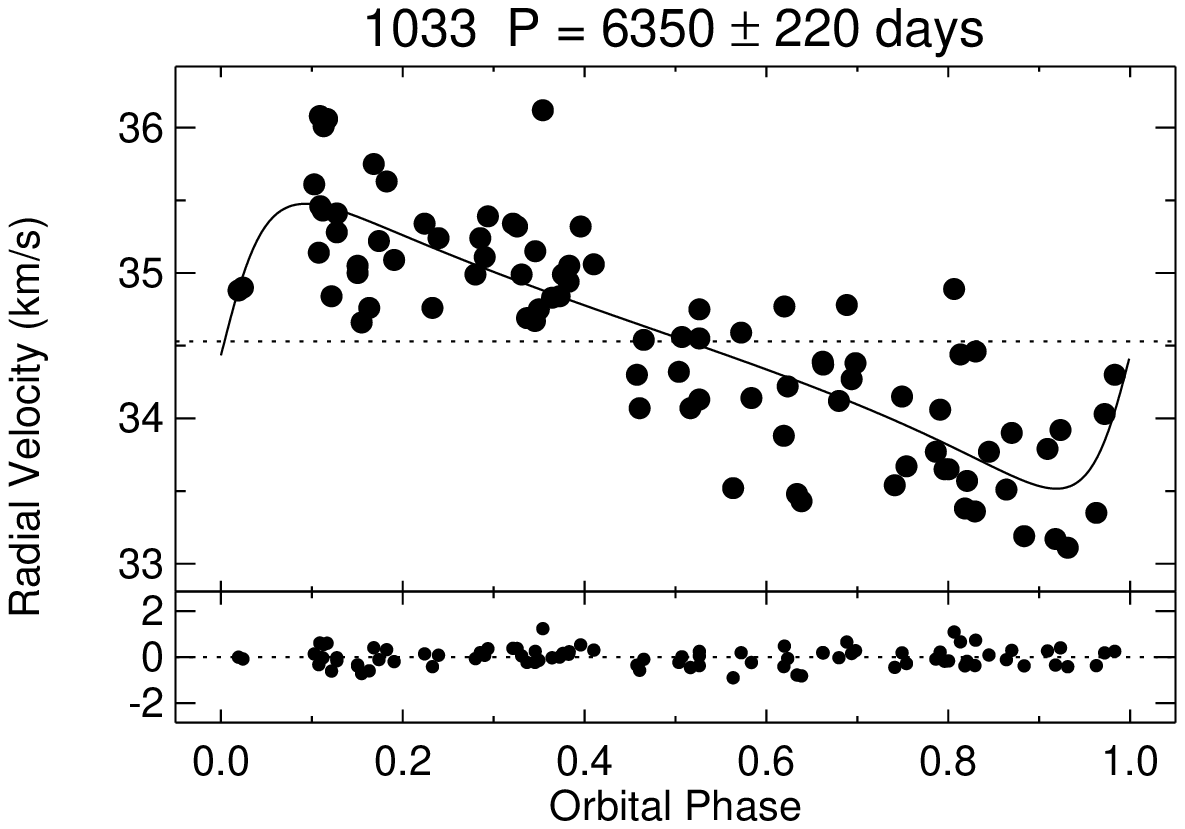}{0.3\textwidth}{}}
\gridline{\fig{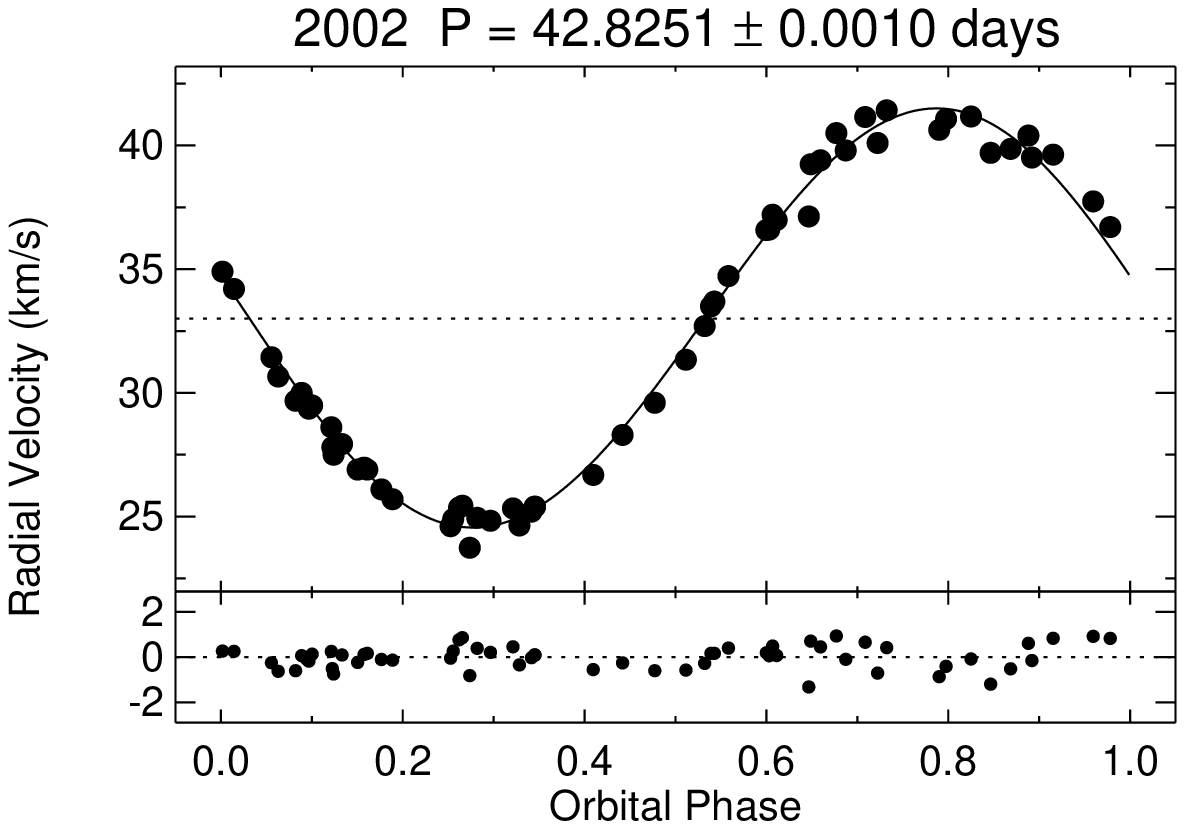}{0.3\textwidth}{}
	\fig{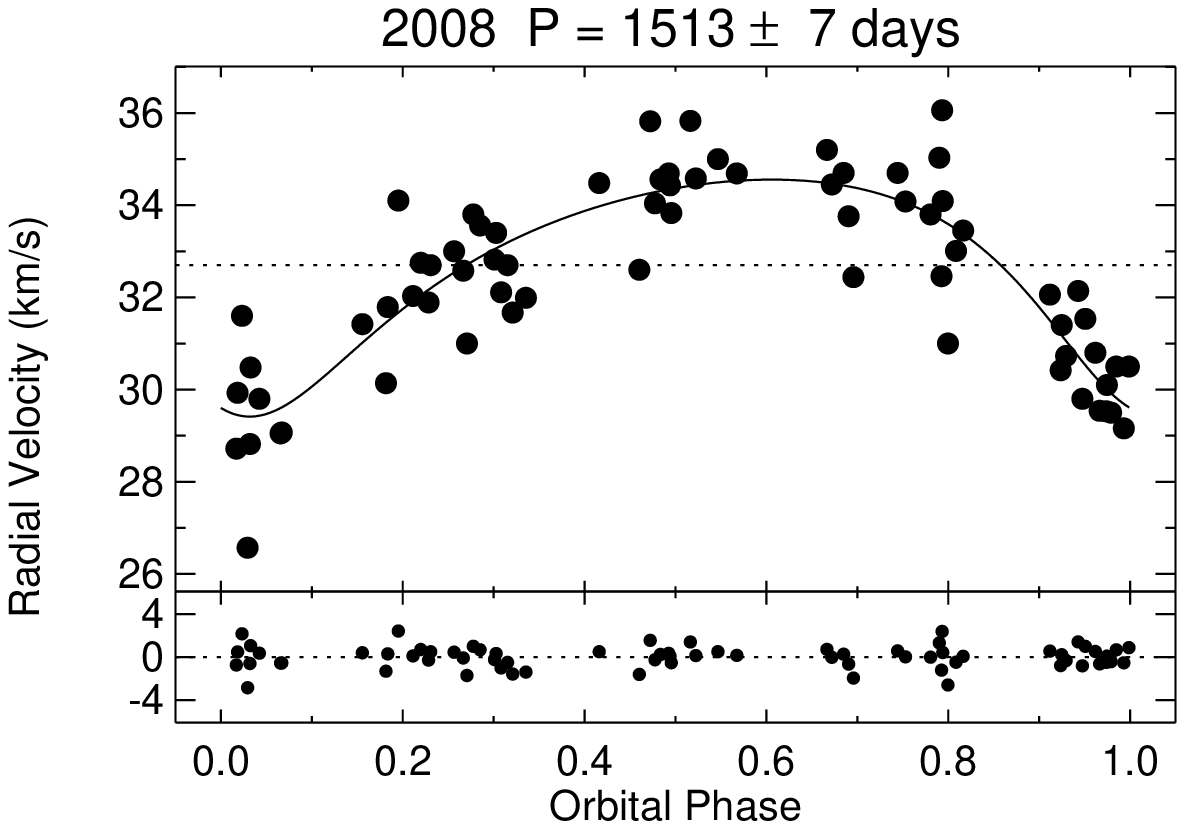}{0.3\textwidth}{}
	\fig{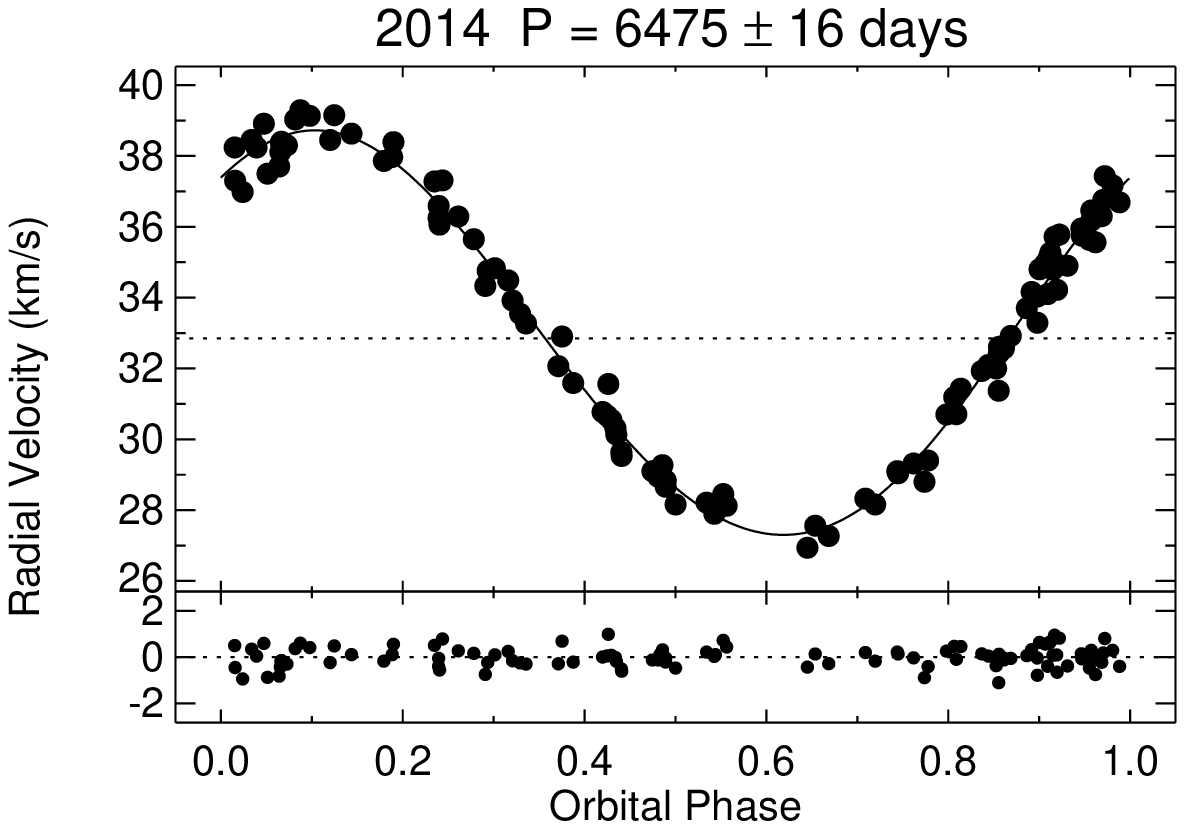}{0.3\textwidth}{}}
\gridline{\fig{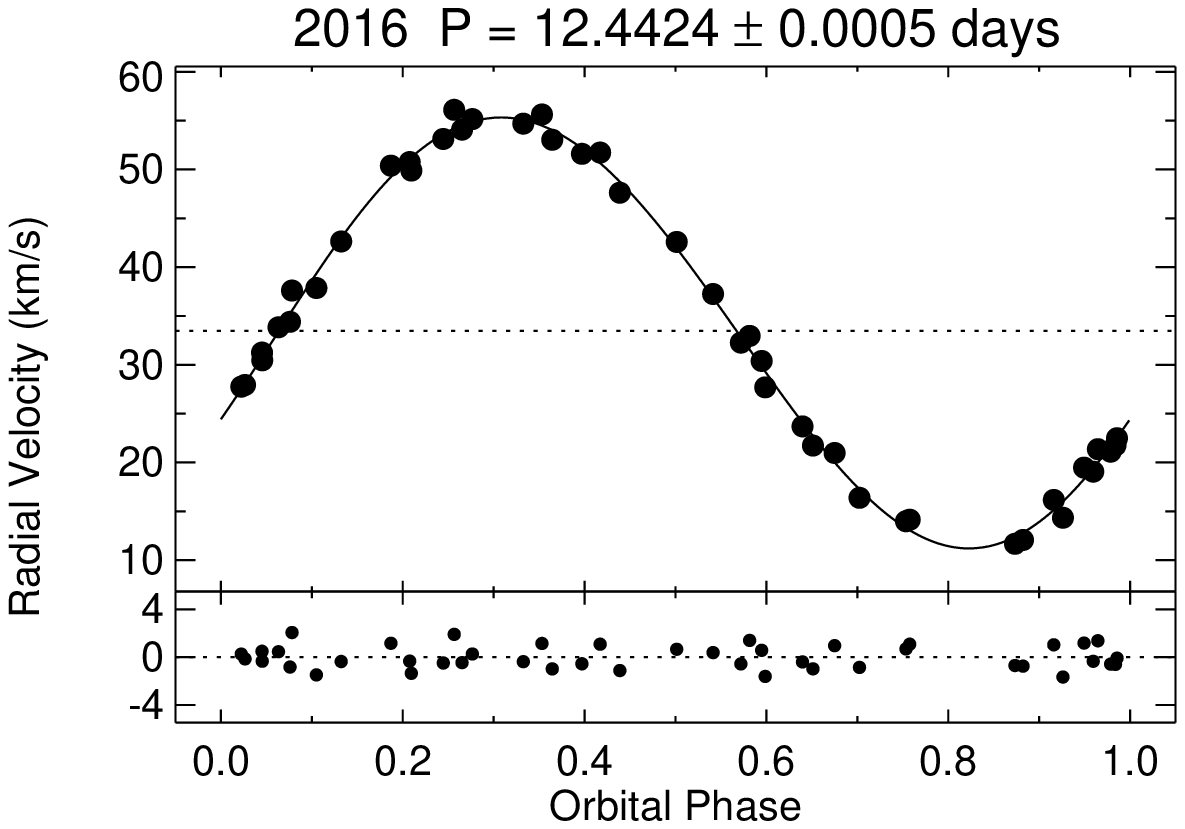}{0.3\textwidth}{}
	\fig{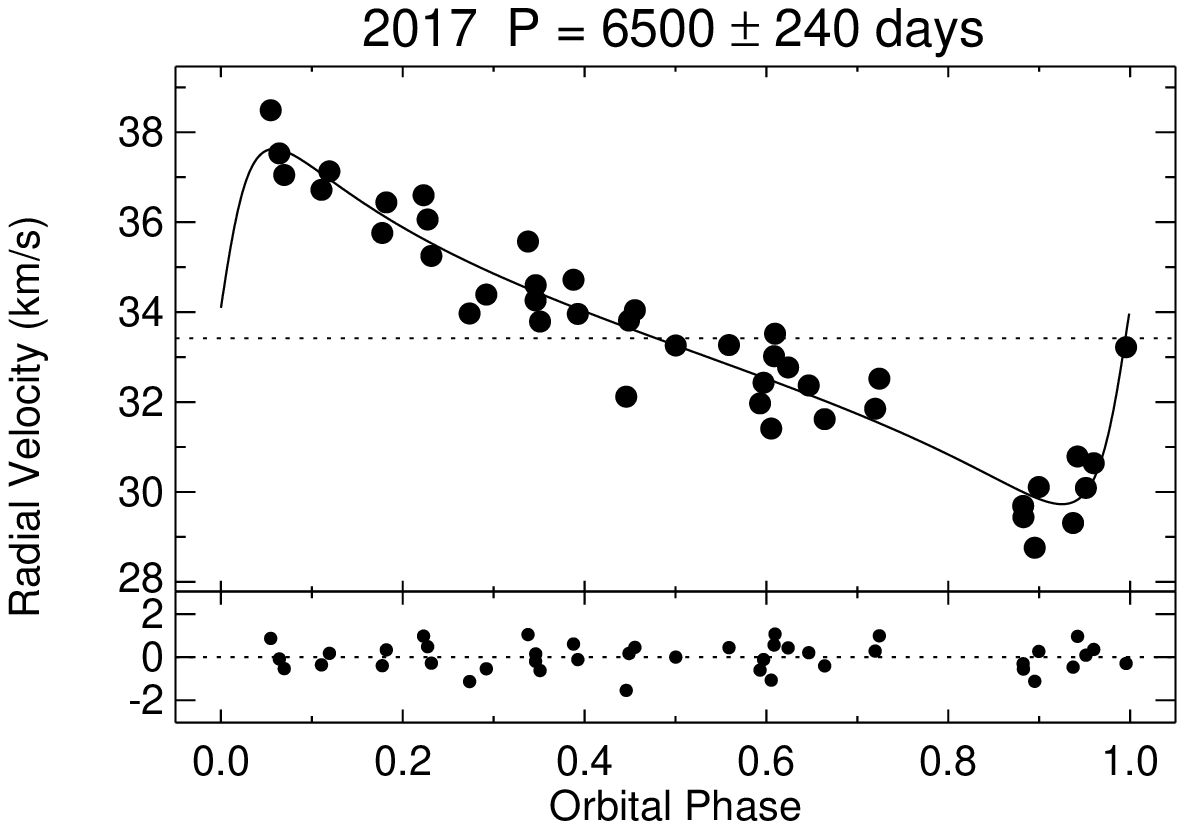}{0.3\textwidth}{}
	\fig{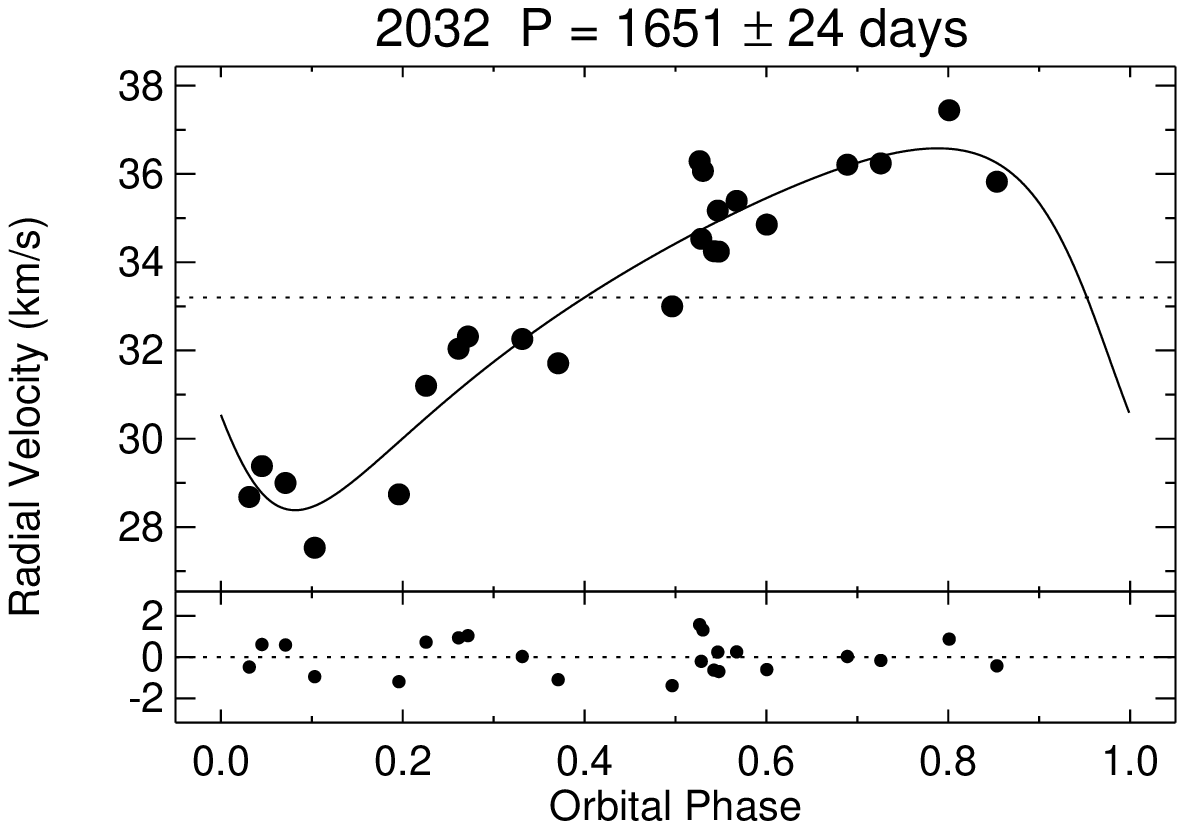}{0.3\textwidth}{}}
\gridline{\fig{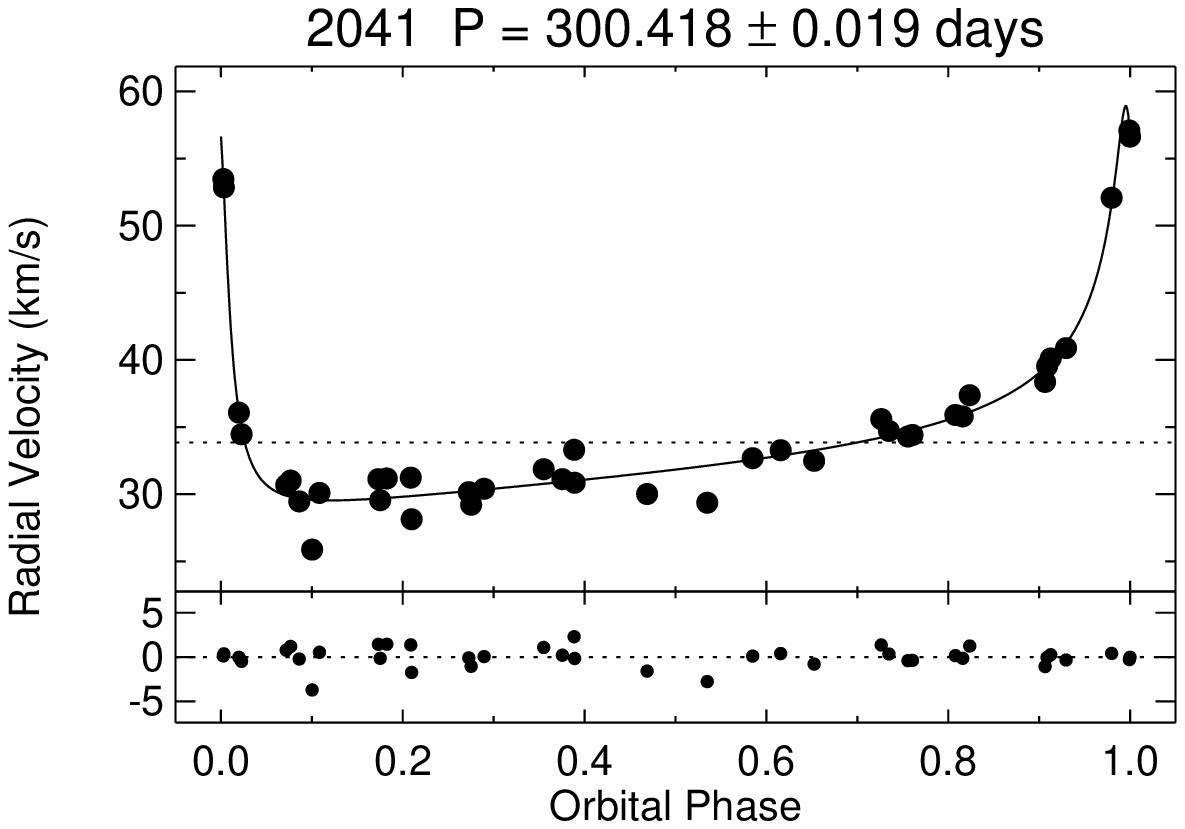}{0.3\textwidth}{}
	\fig{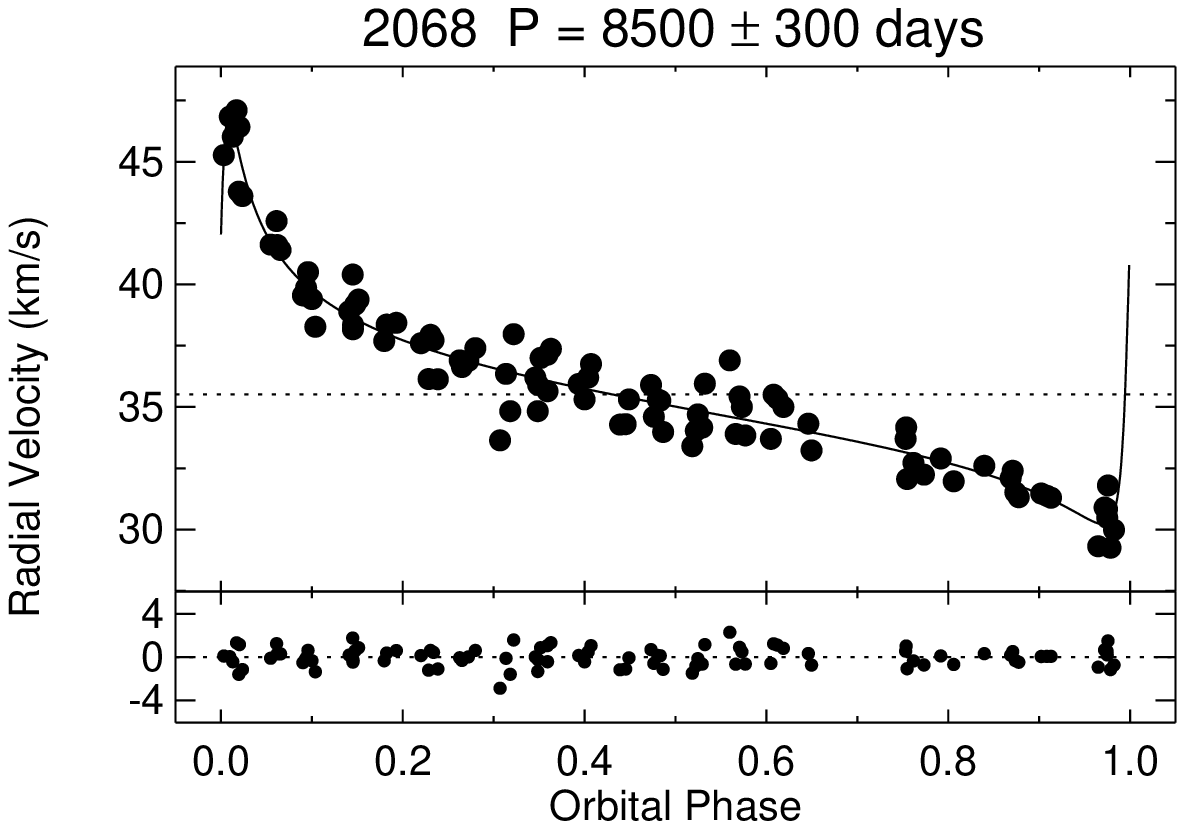}{0.3\textwidth}{}
	\fig{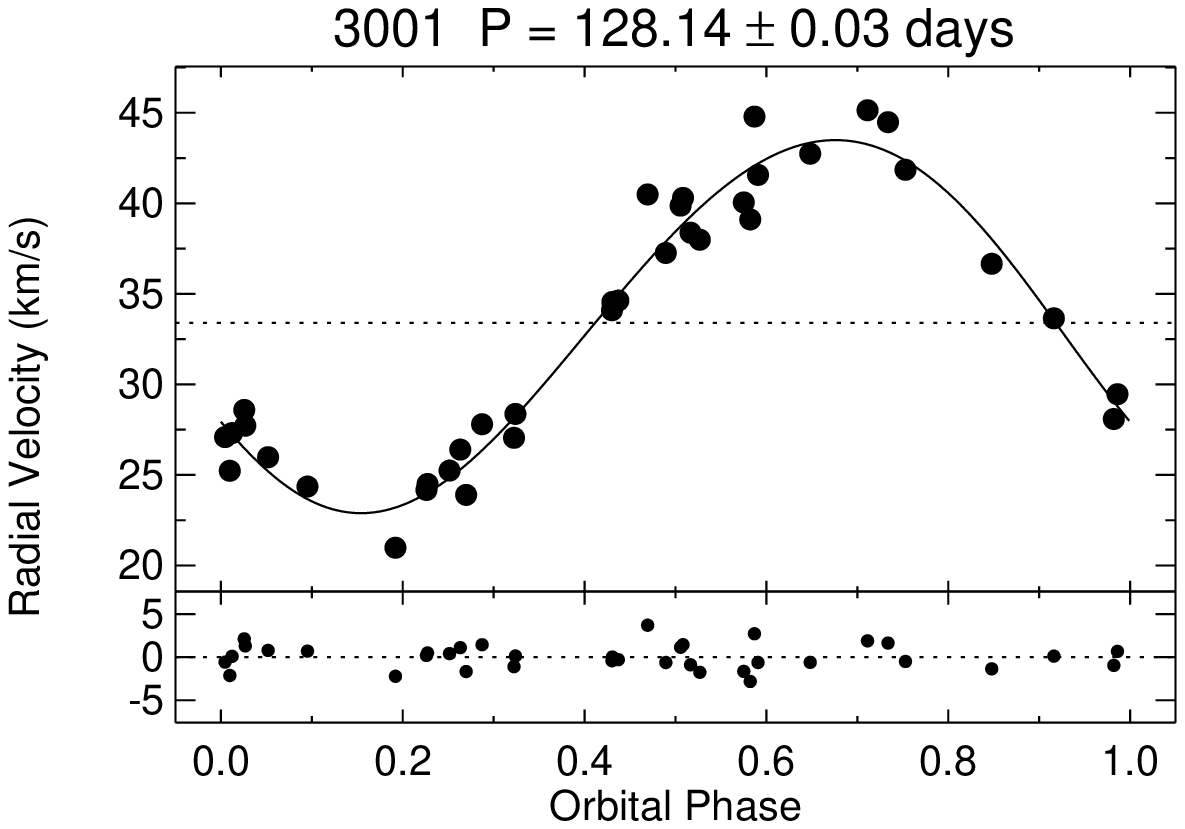}{0.3\textwidth}{}}
\caption{Single-lined M67 member orbit plots of radial velocity (RV) versus phase. RV data points are shown with filled circles, and the orbital solution is plotted as a solid black line. The dotted line marks the $\gamma$ velocity of the binary. Below each plot, the residuals ($O-C$) are given. Above each plot, we list the WOCS ID and orbital period.\label{fig:sb1orbs}}
\end{figure*}

\begin{figure*}
\figurenum{11}
\gridline{\fig{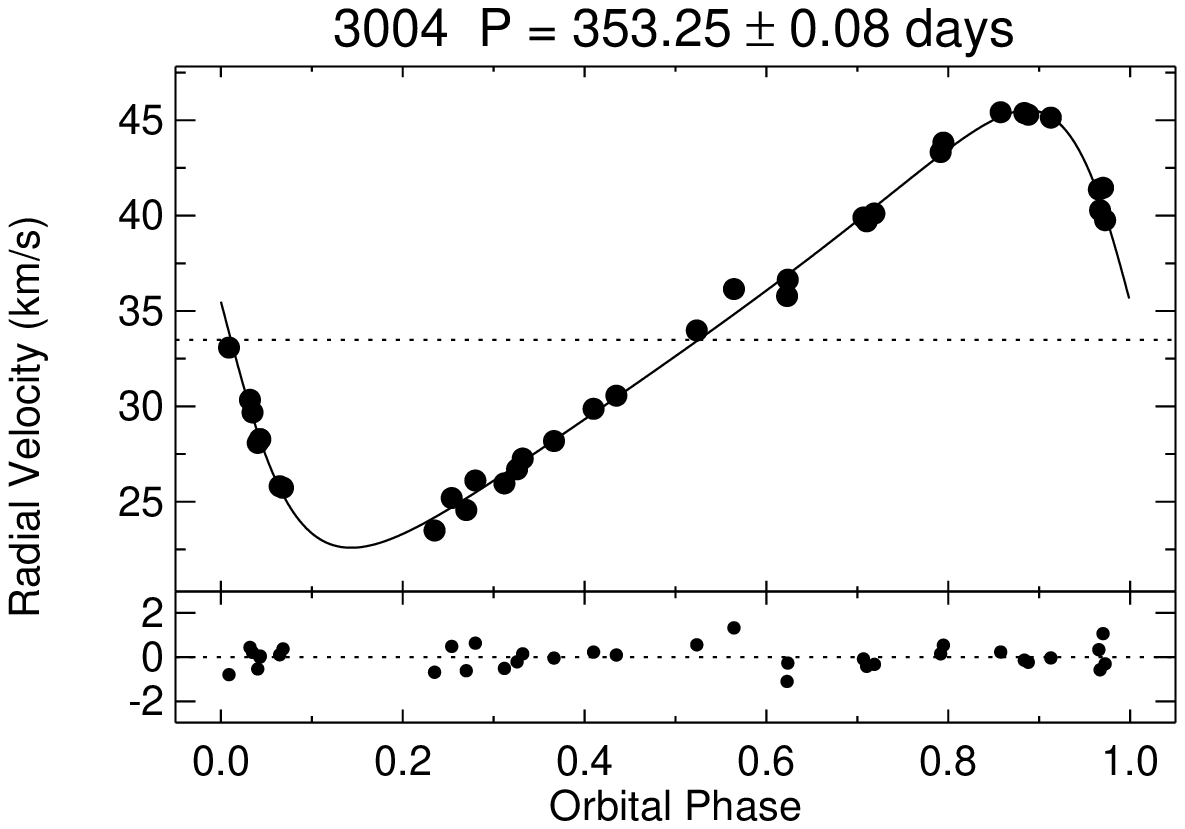}{0.3\textwidth}{}
	\fig{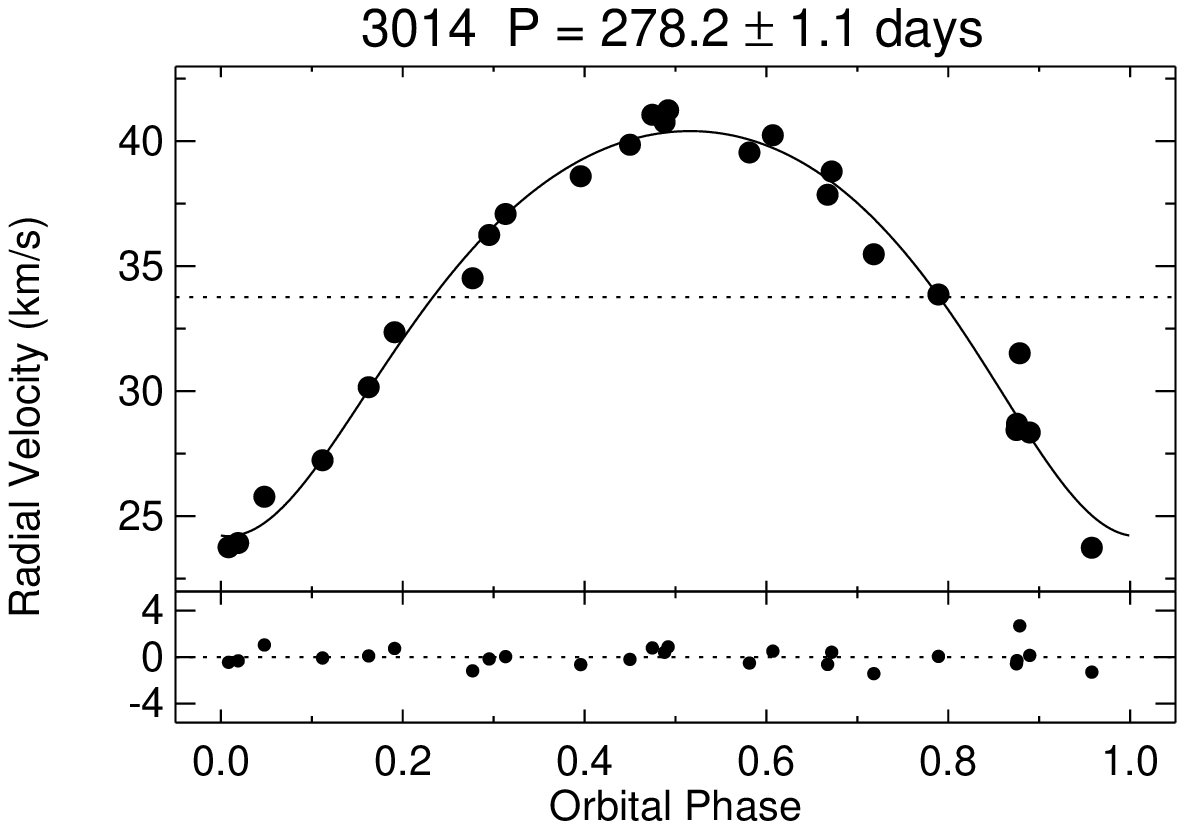}{0.3\textwidth}{}
	\fig{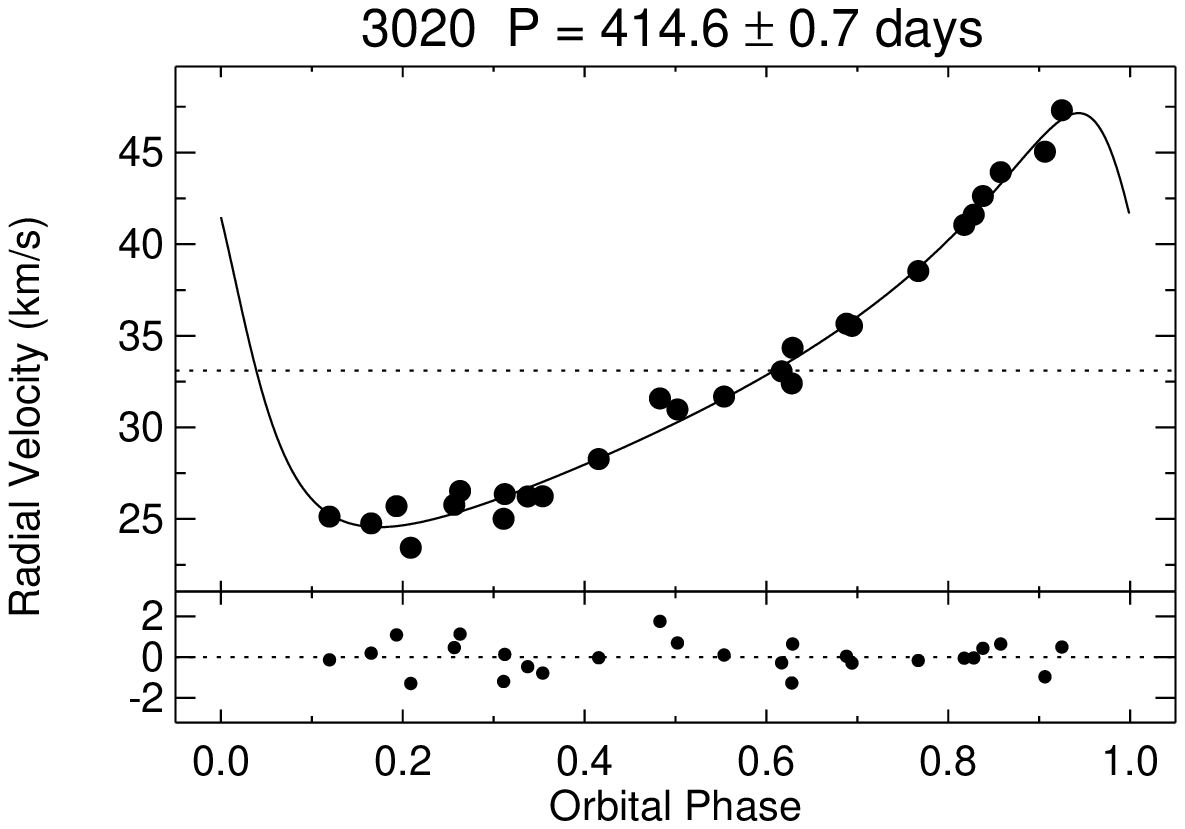}{0.3\textwidth}{}}
\gridline{\fig{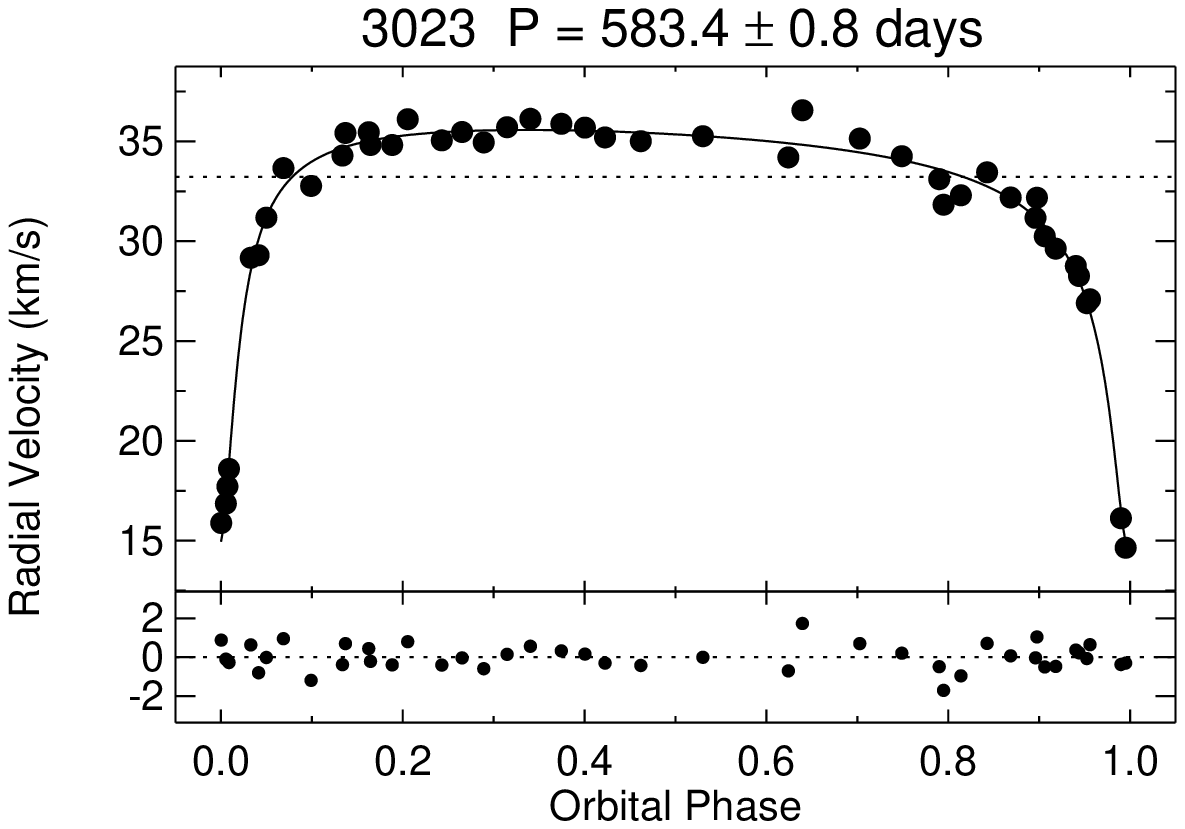}{0.3\textwidth}{}
	\fig{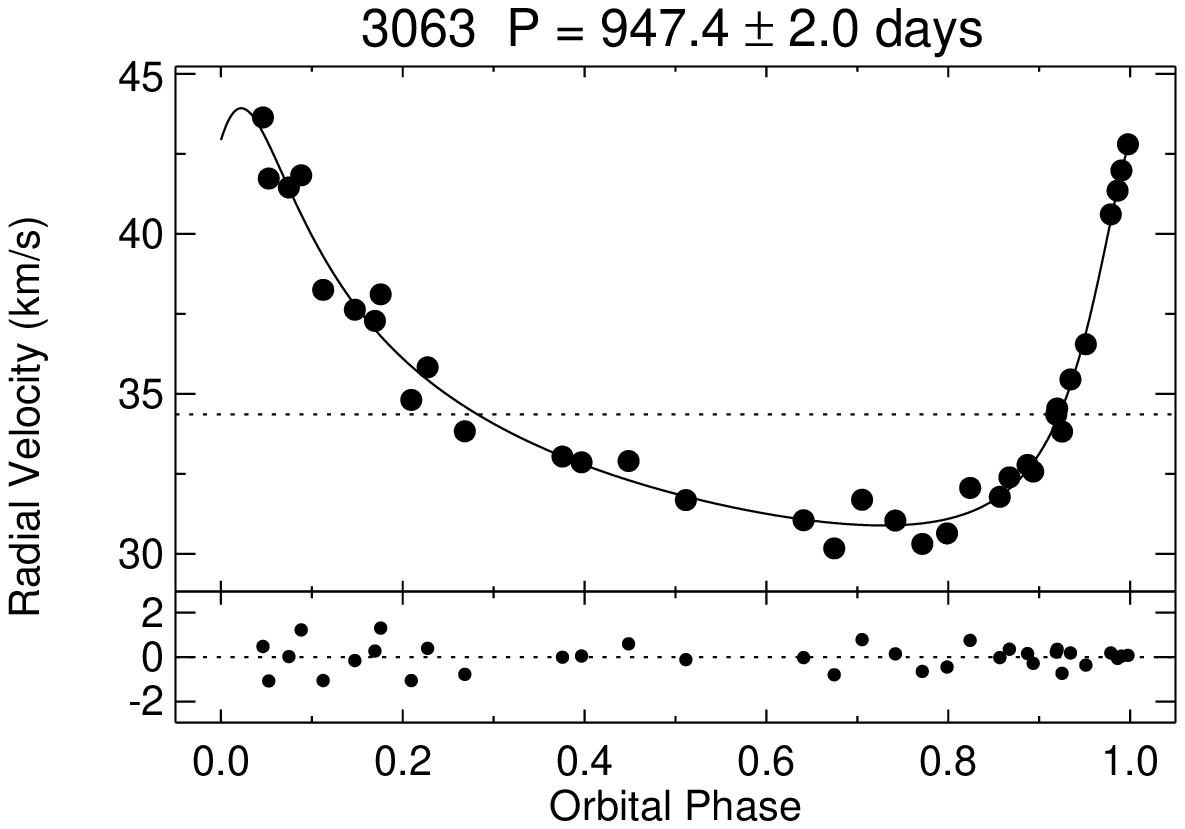}{0.3\textwidth}{}
	\fig{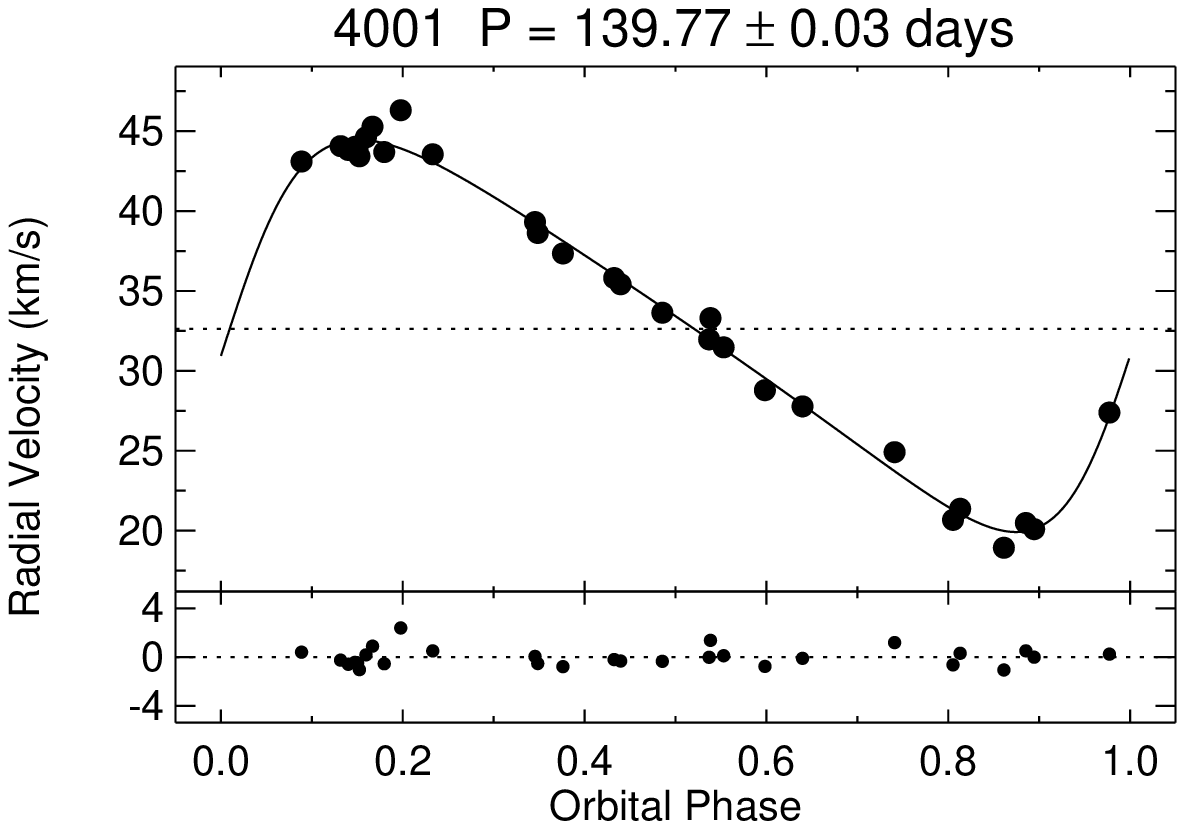}{0.3\textwidth}{}}
\gridline{\fig{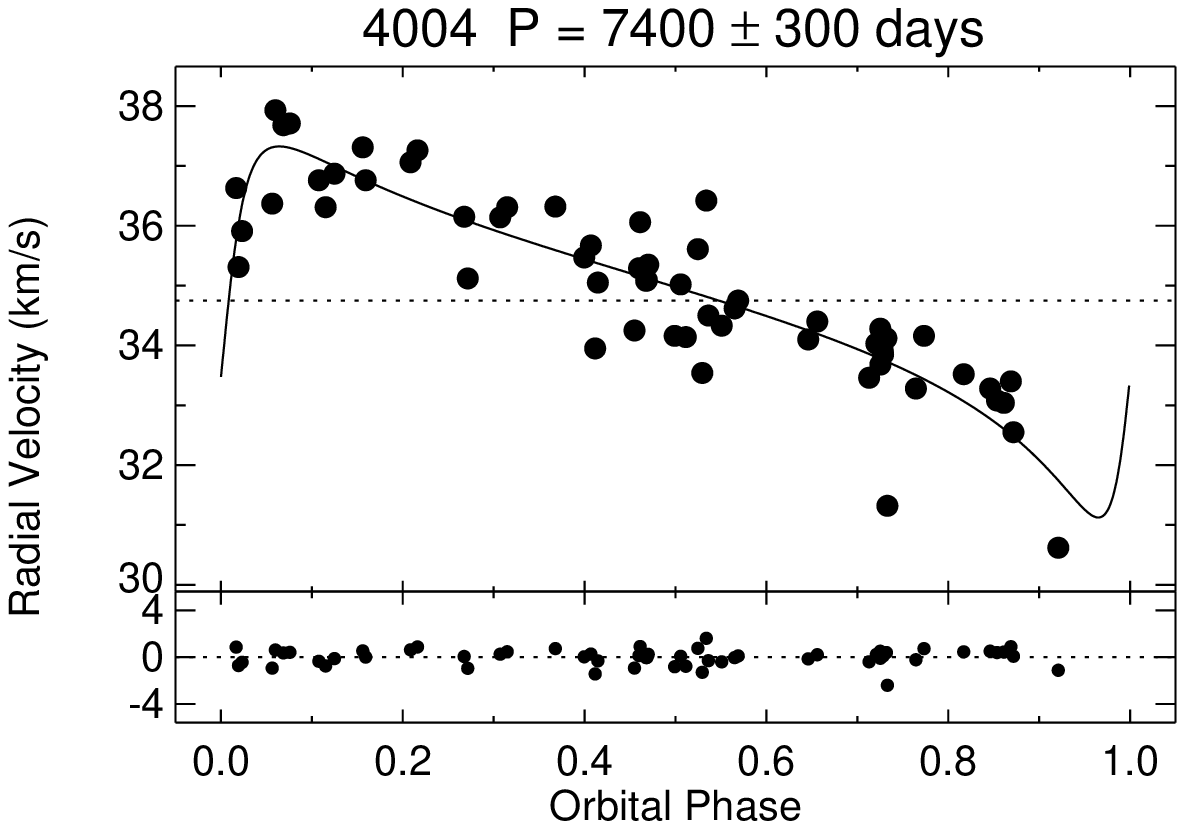}{0.3\textwidth}{}
	\fig{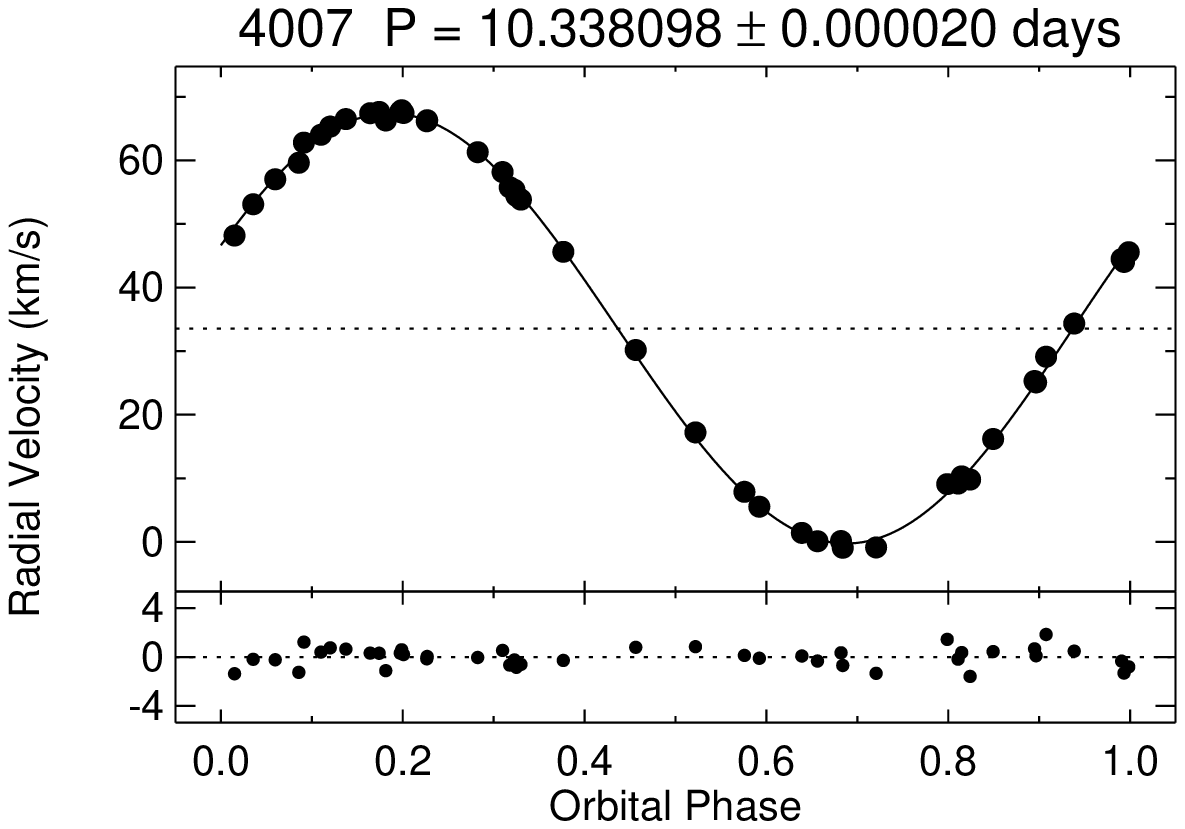}{0.3\textwidth}{}
	\fig{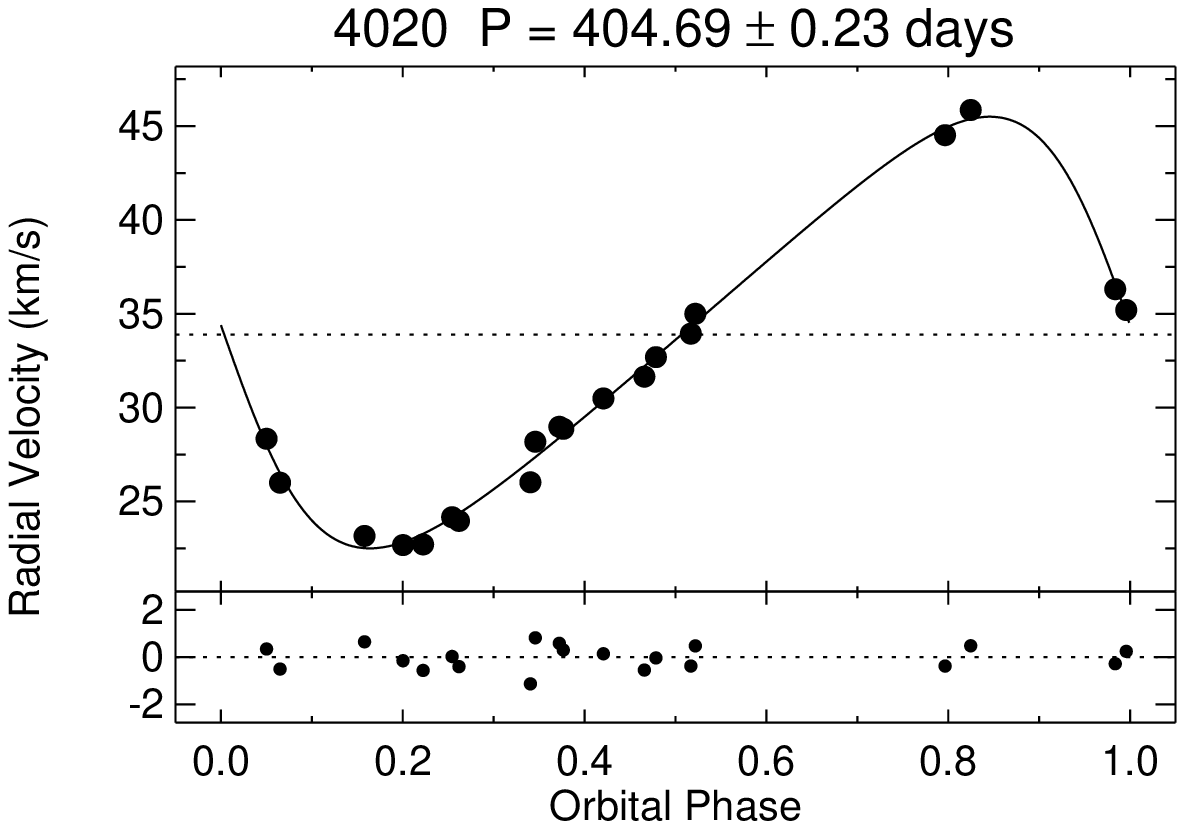}{0.3\textwidth}{}}
\gridline{\fig{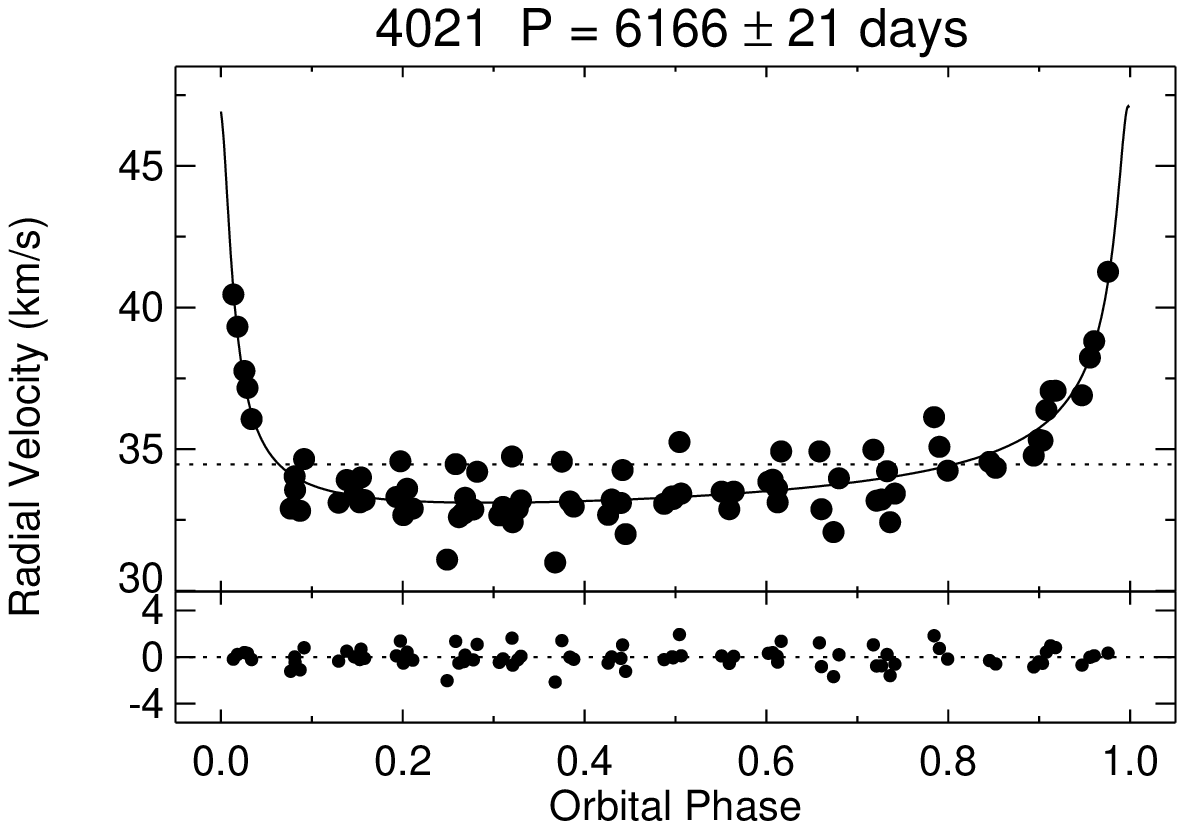}{0.3\textwidth}{}
	\fig{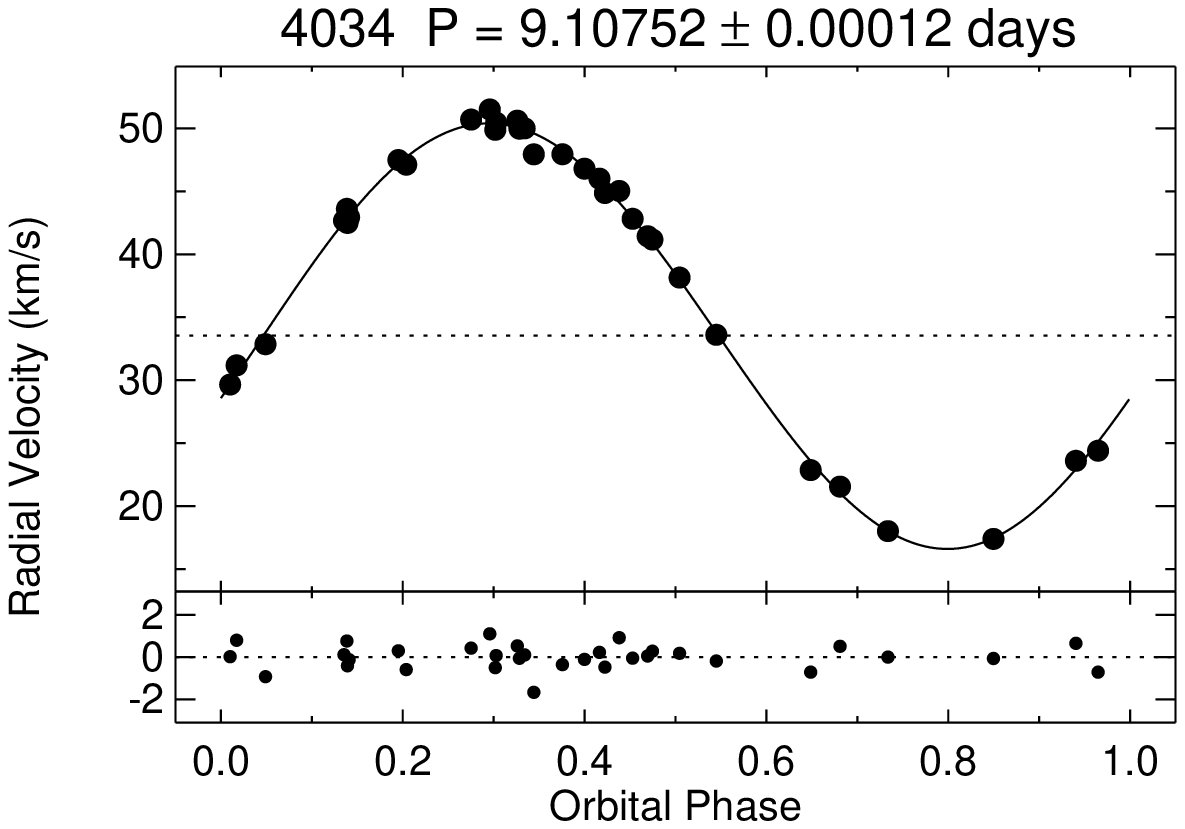}{0.3\textwidth}{}
	\fig{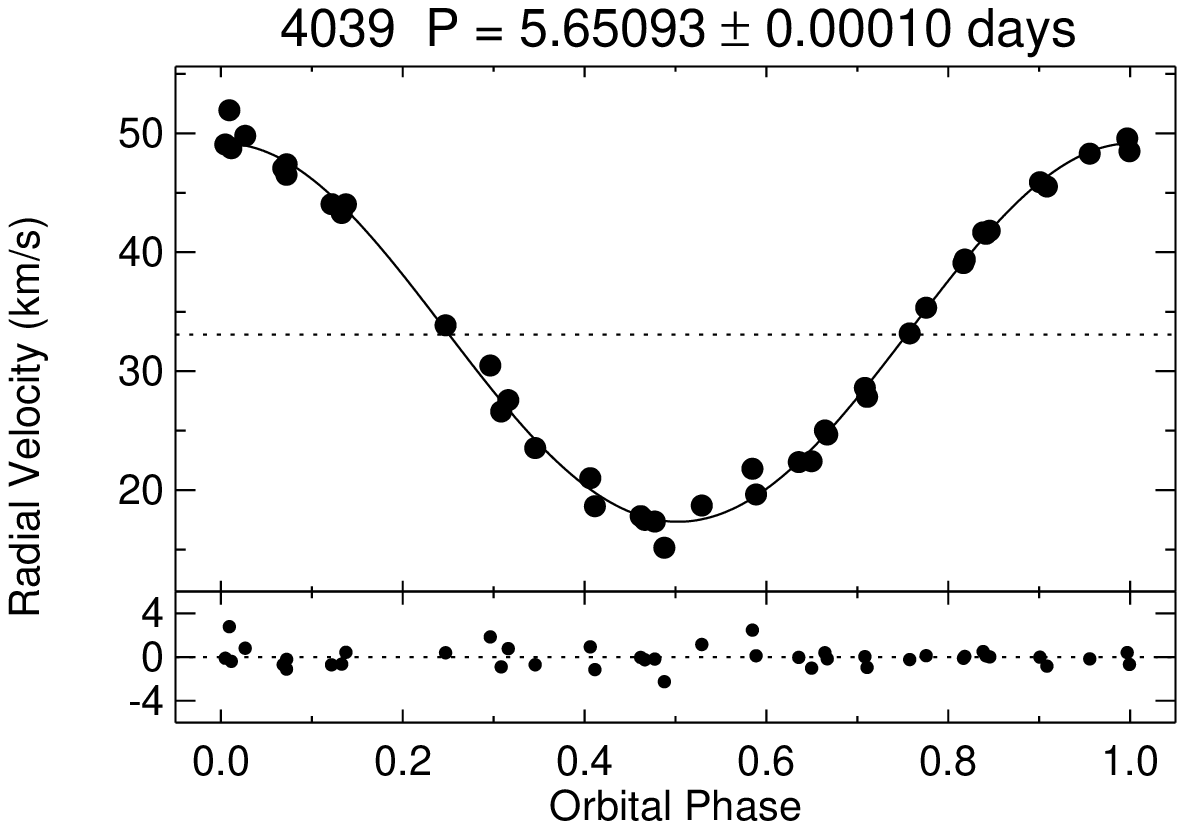}{0.3\textwidth}{}}
\caption{(Continued.)}
\end{figure*}

\begin{figure*}
\figurenum{11}
\gridline{\fig{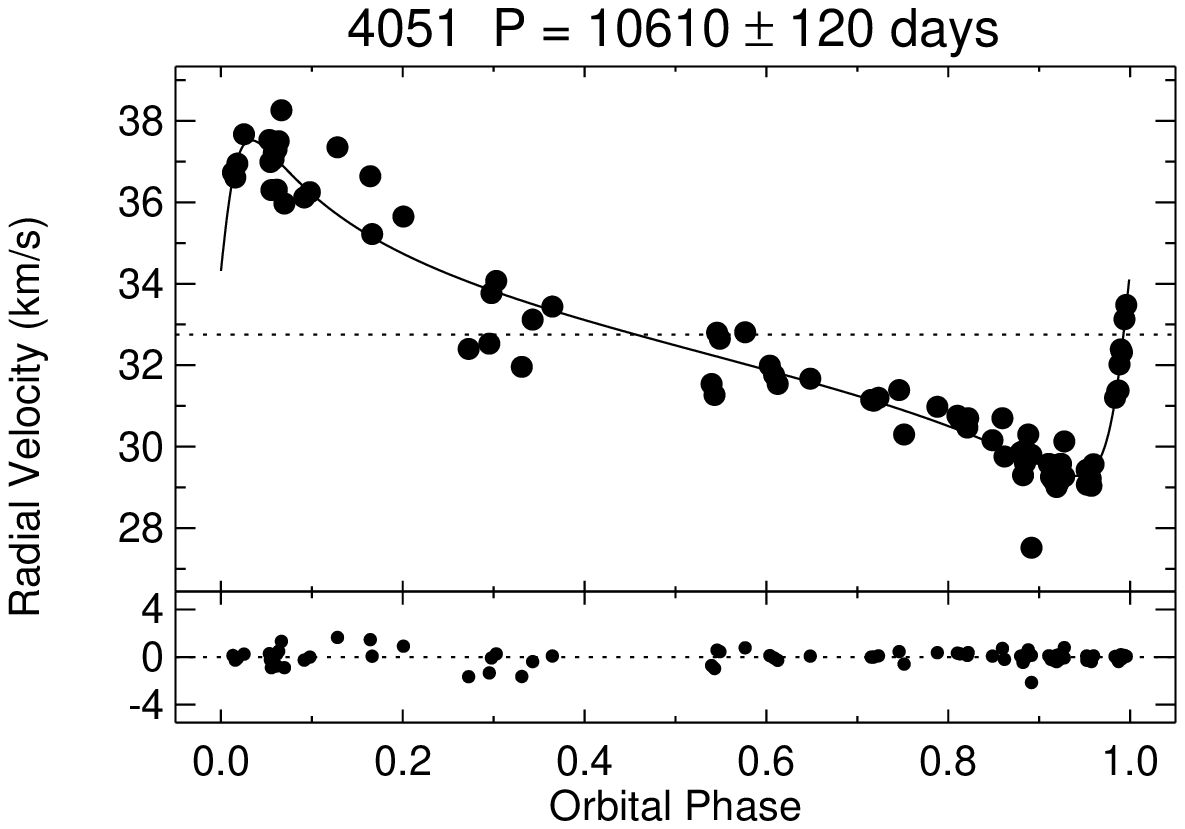}{0.3\textwidth}{}
	\fig{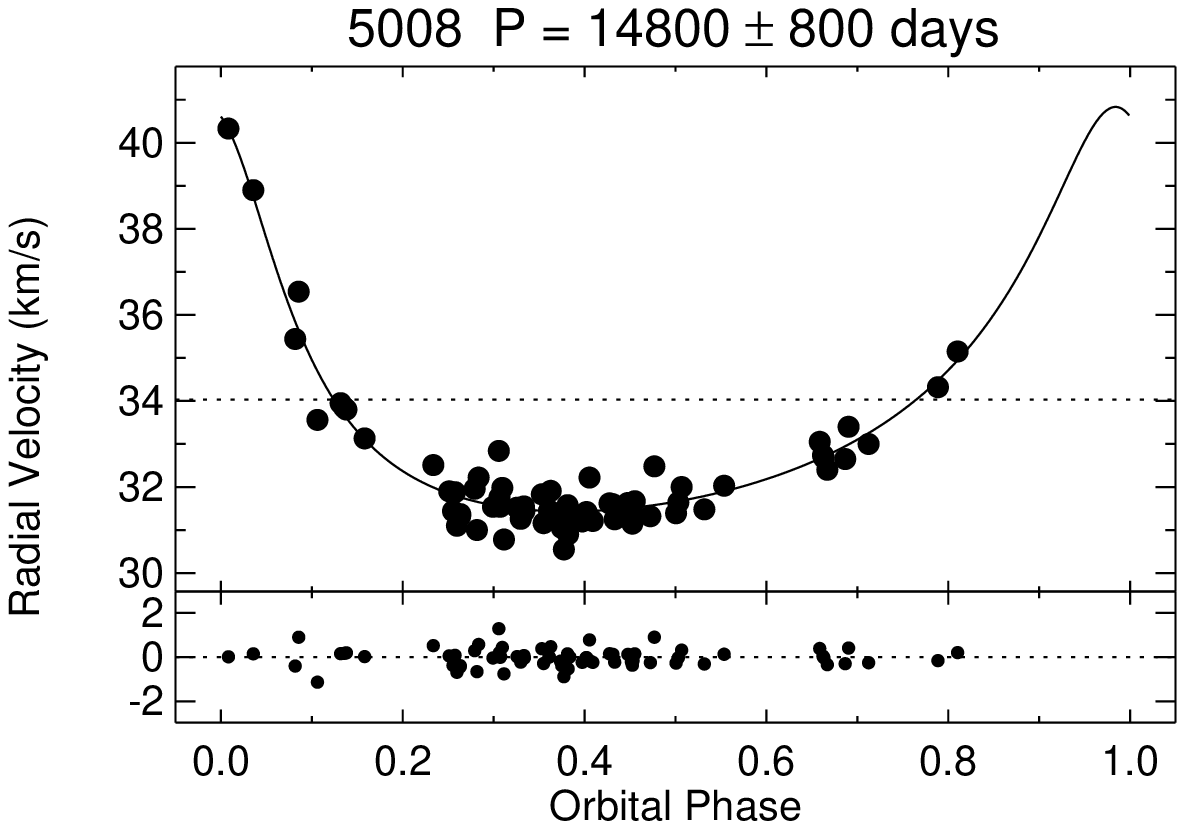}{0.3\textwidth}{}
	\fig{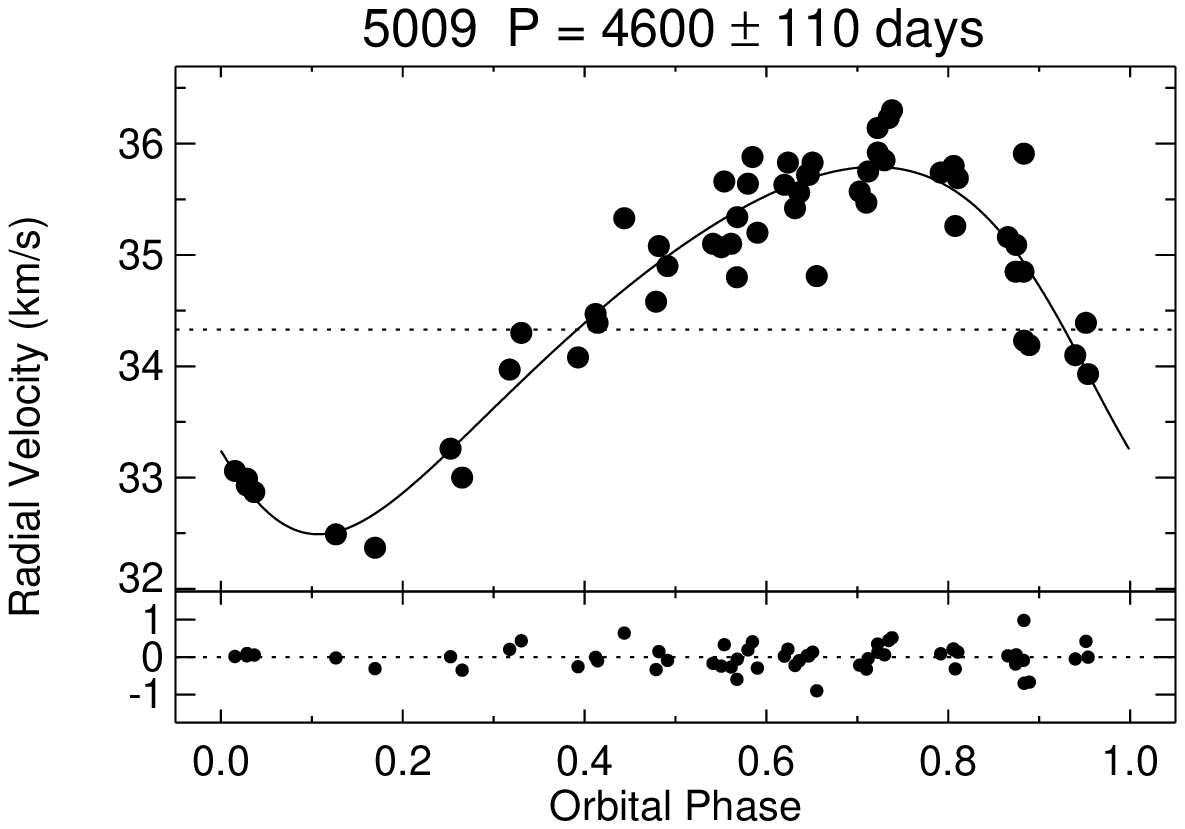}{0.3\textwidth}{}}
\gridline{\fig{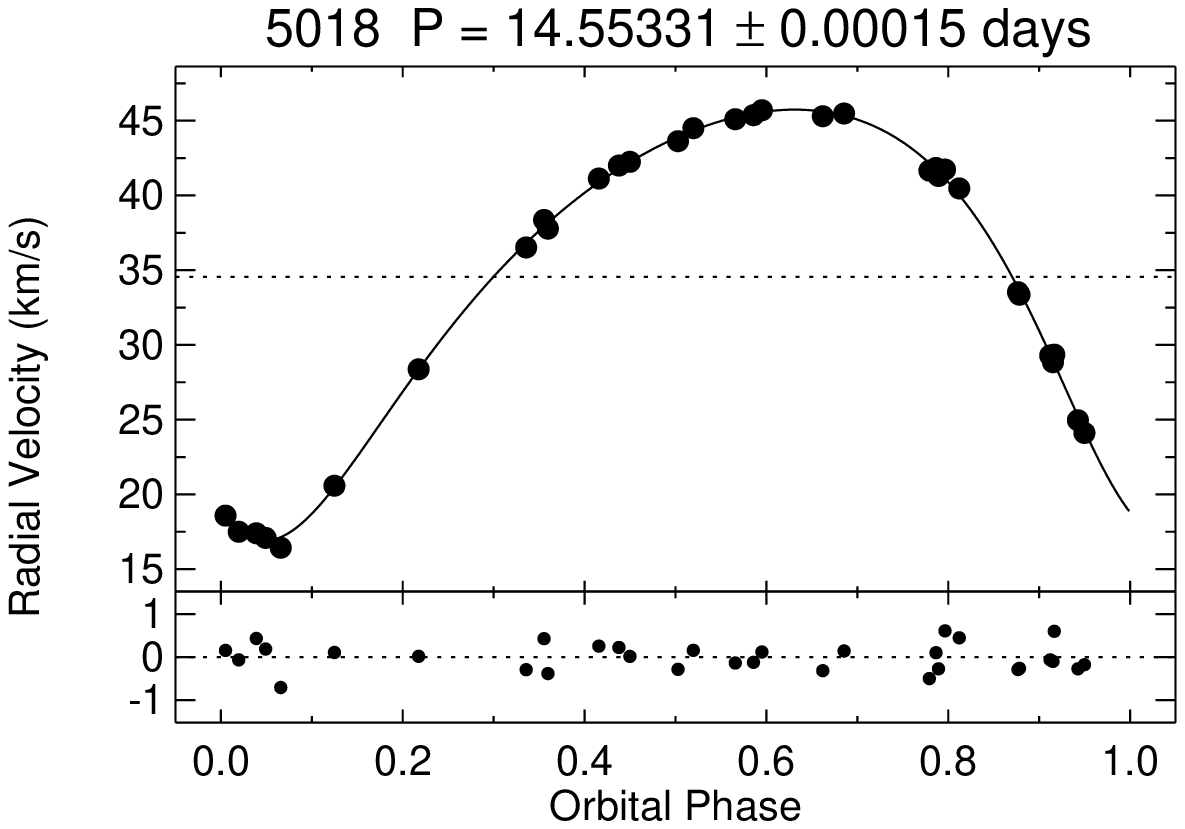}{0.3\textwidth}{}
	\fig{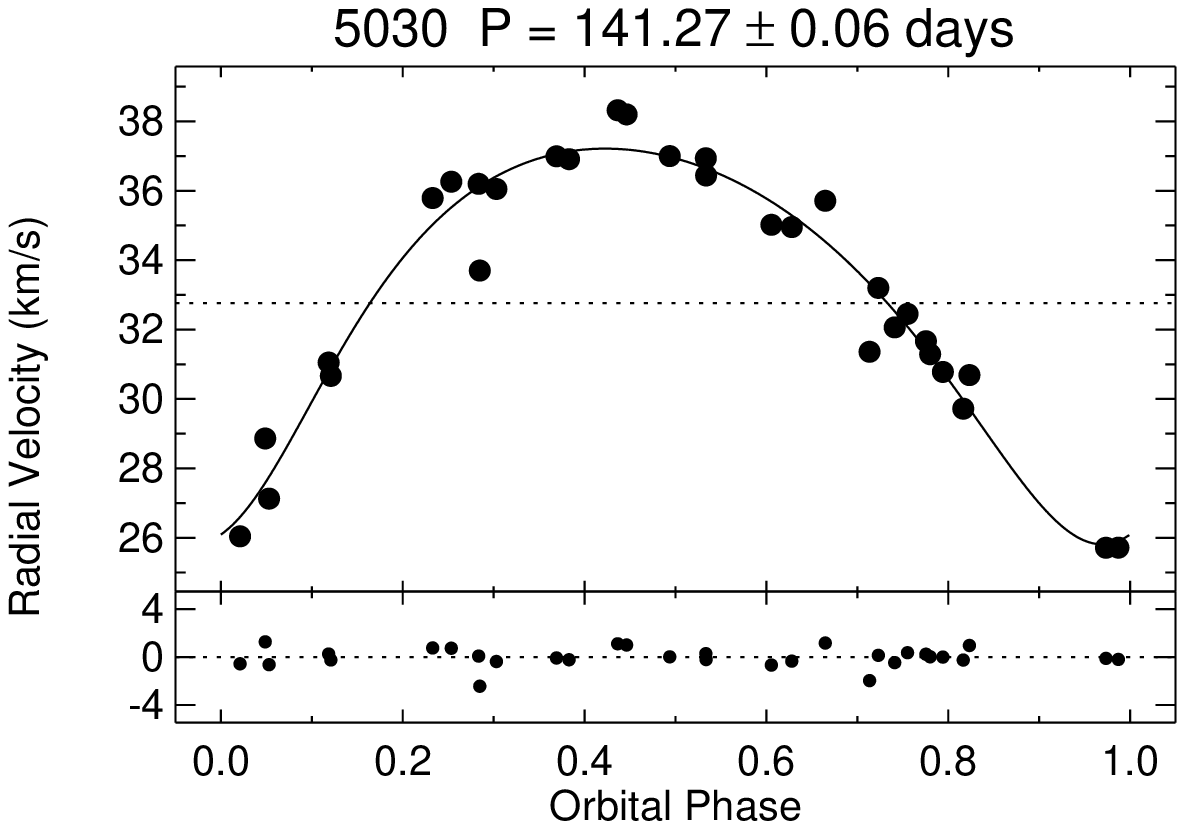}{0.3\textwidth}{}
	\fig{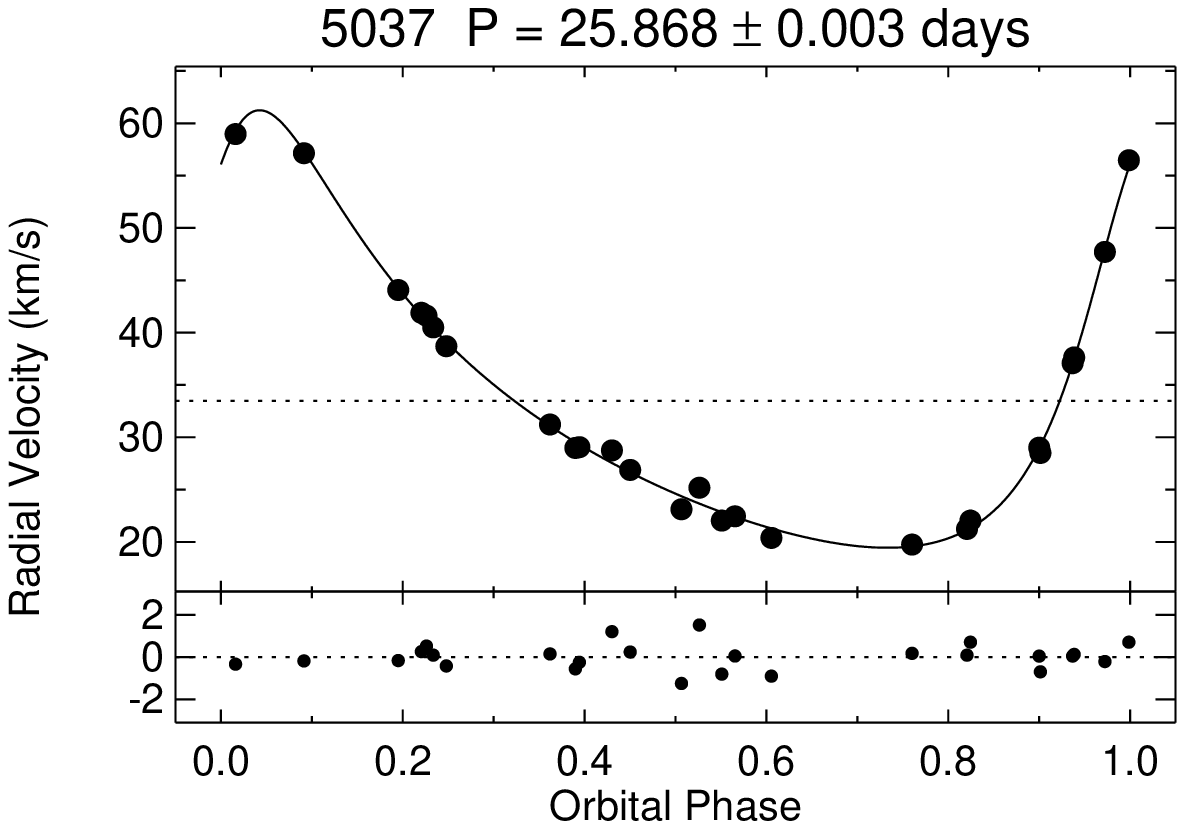}{0.3\textwidth}{}}
\gridline{\fig{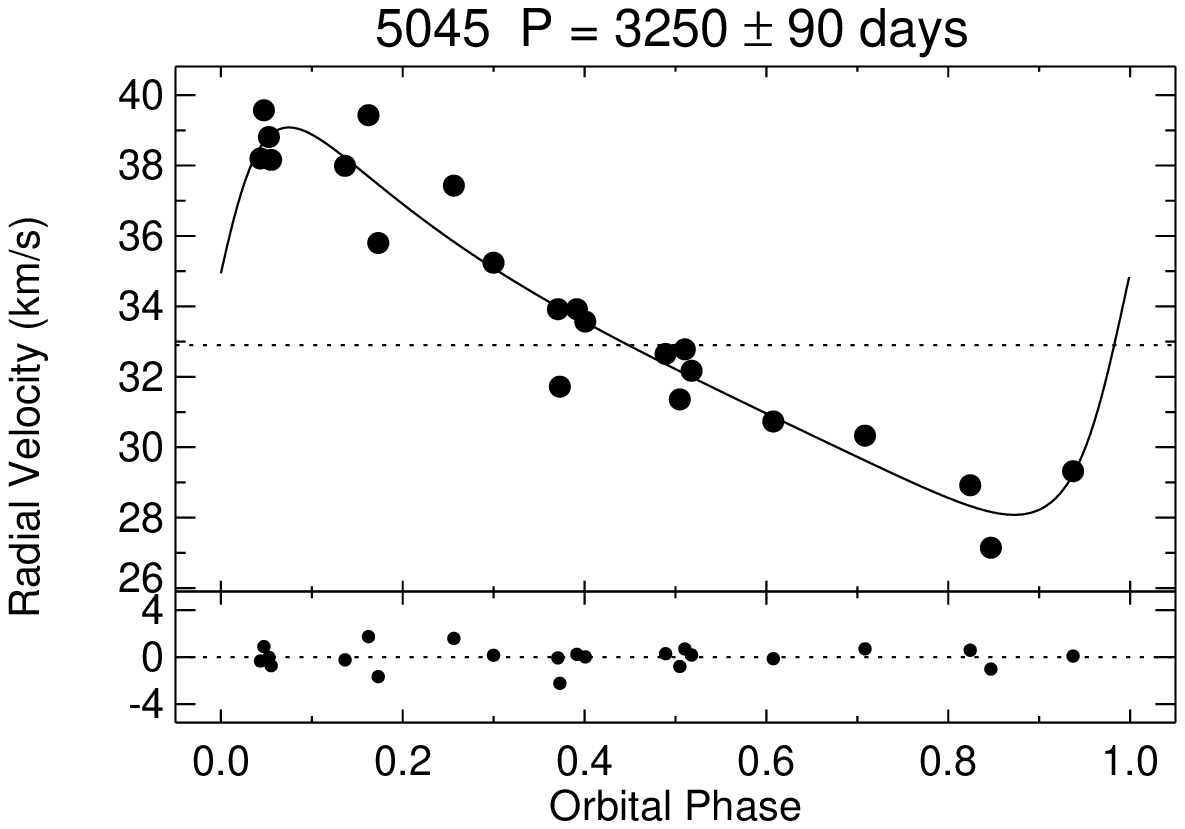}{0.3\textwidth}{}
	\fig{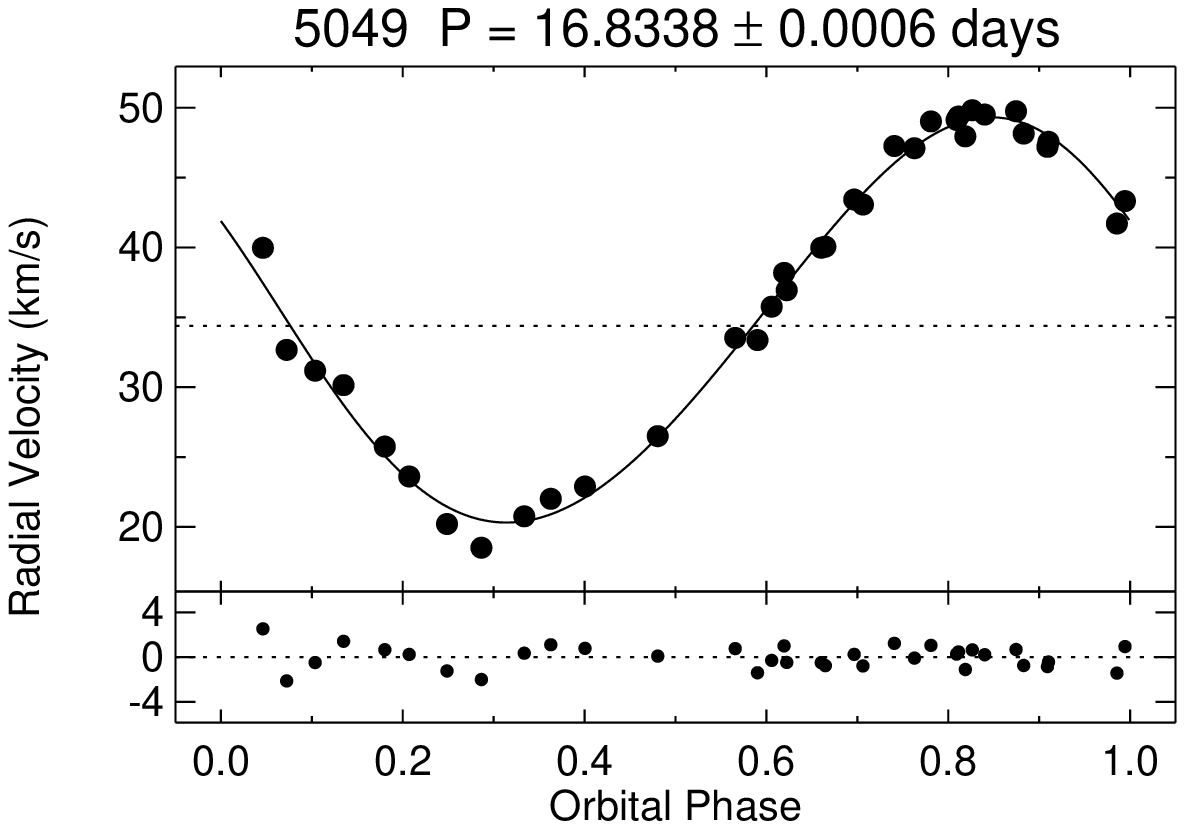}{0.3\textwidth}{}
	\fig{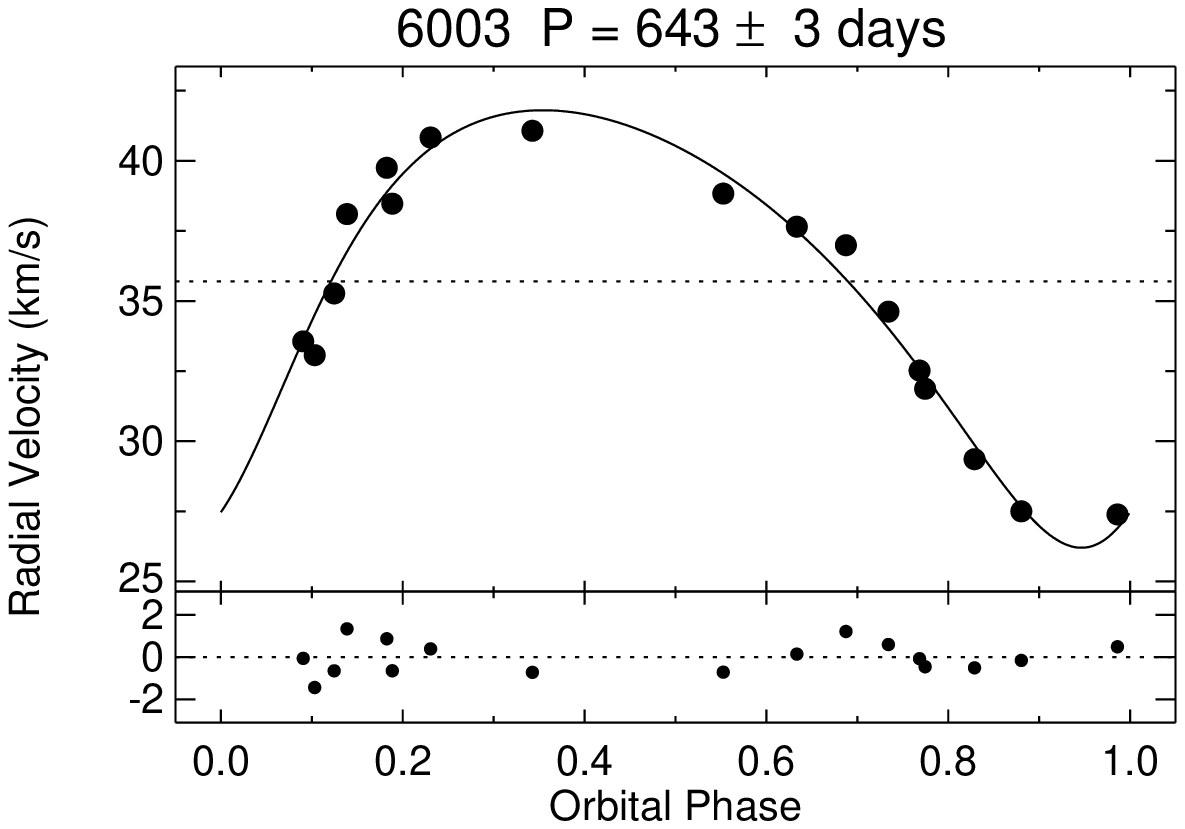}{0.3\textwidth}{}}
\gridline{\fig{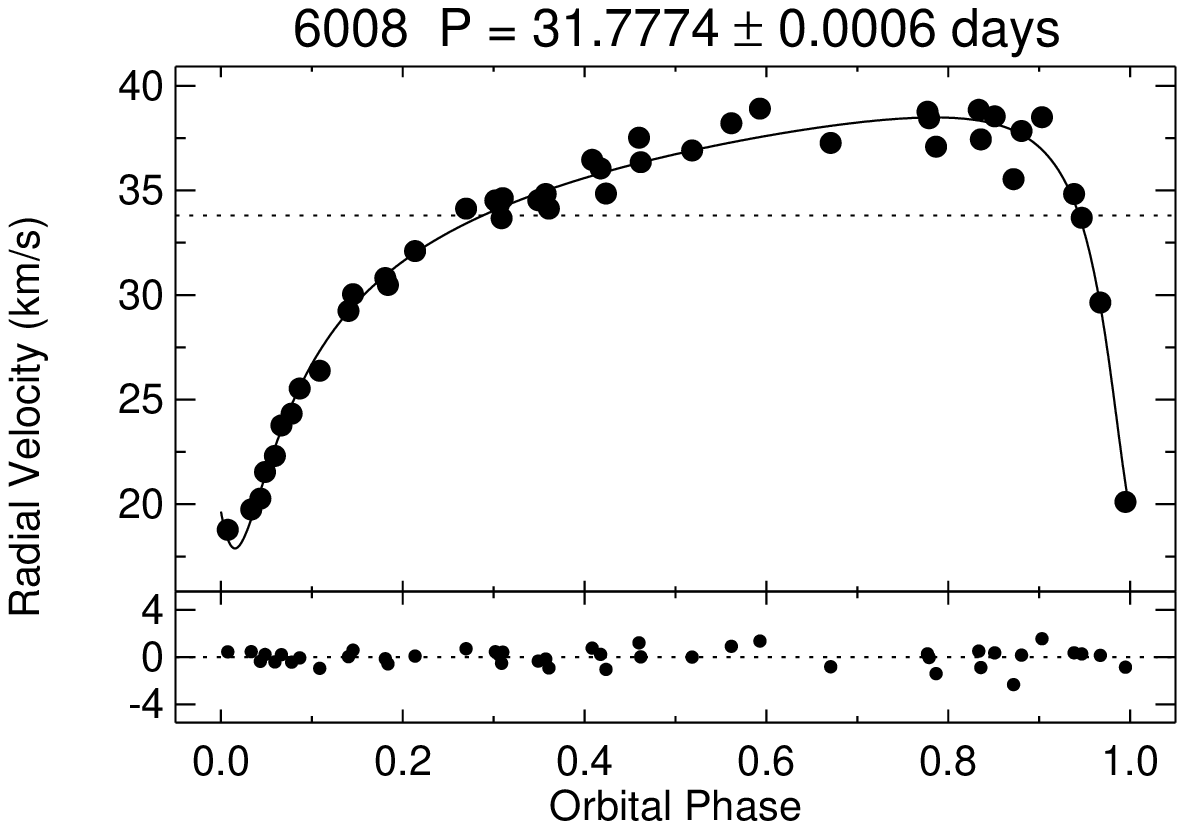}{0.3\textwidth}{}
	\fig{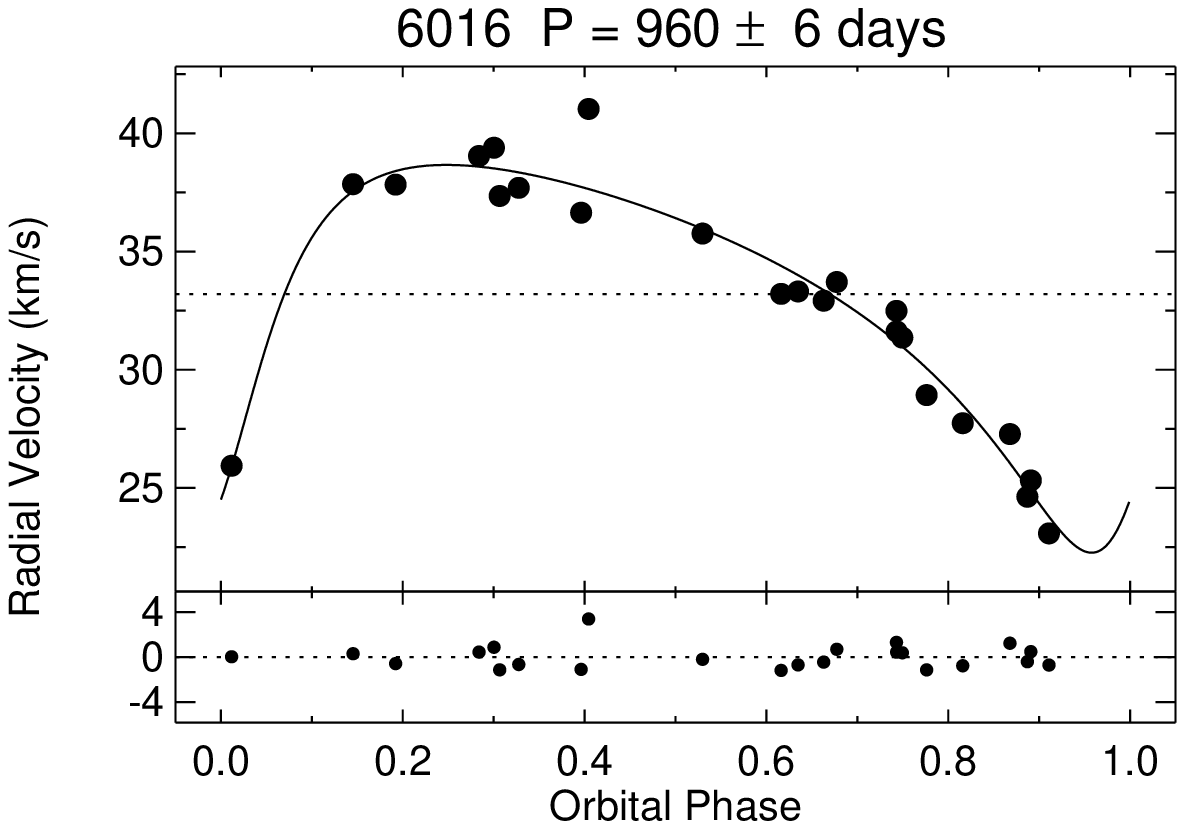}{0.3\textwidth}{}
	\fig{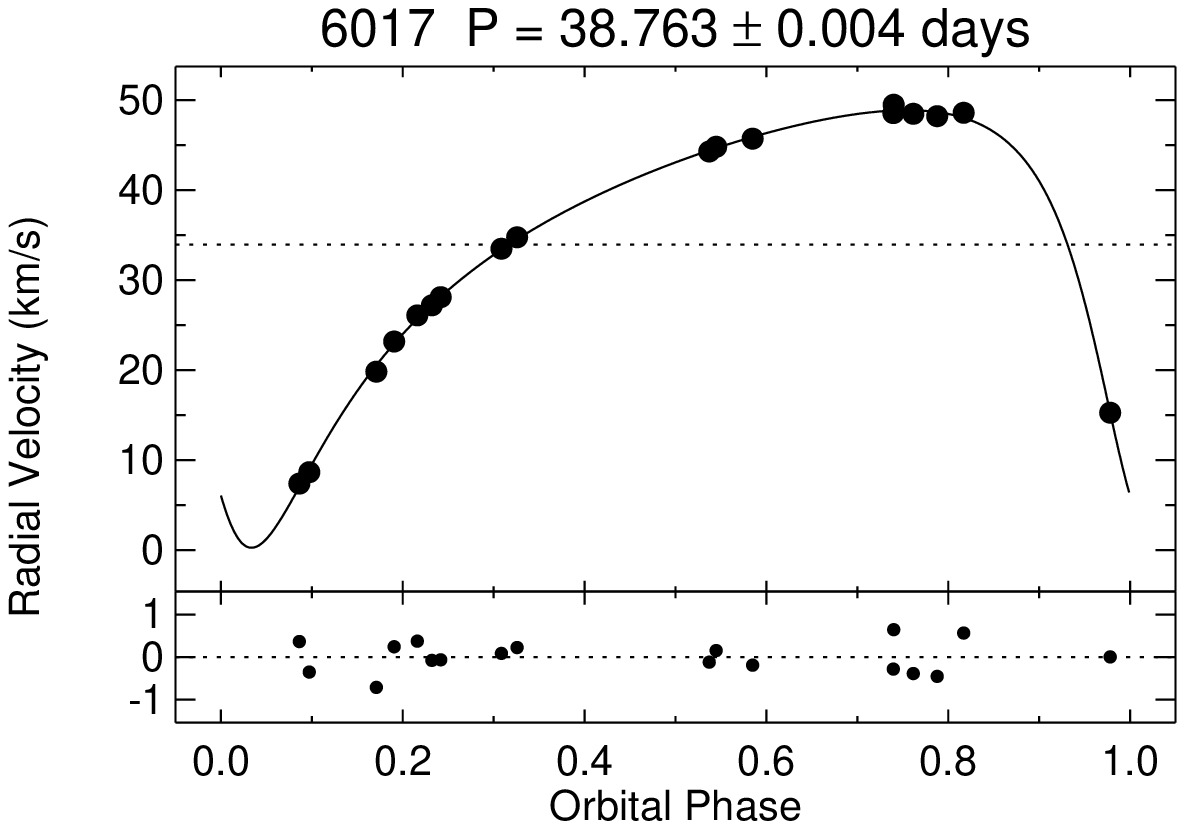}{0.3\textwidth}{}}
\caption{(Continued.)}
\end{figure*}

\begin{figure*}
\figurenum{11}
\gridline{\fig{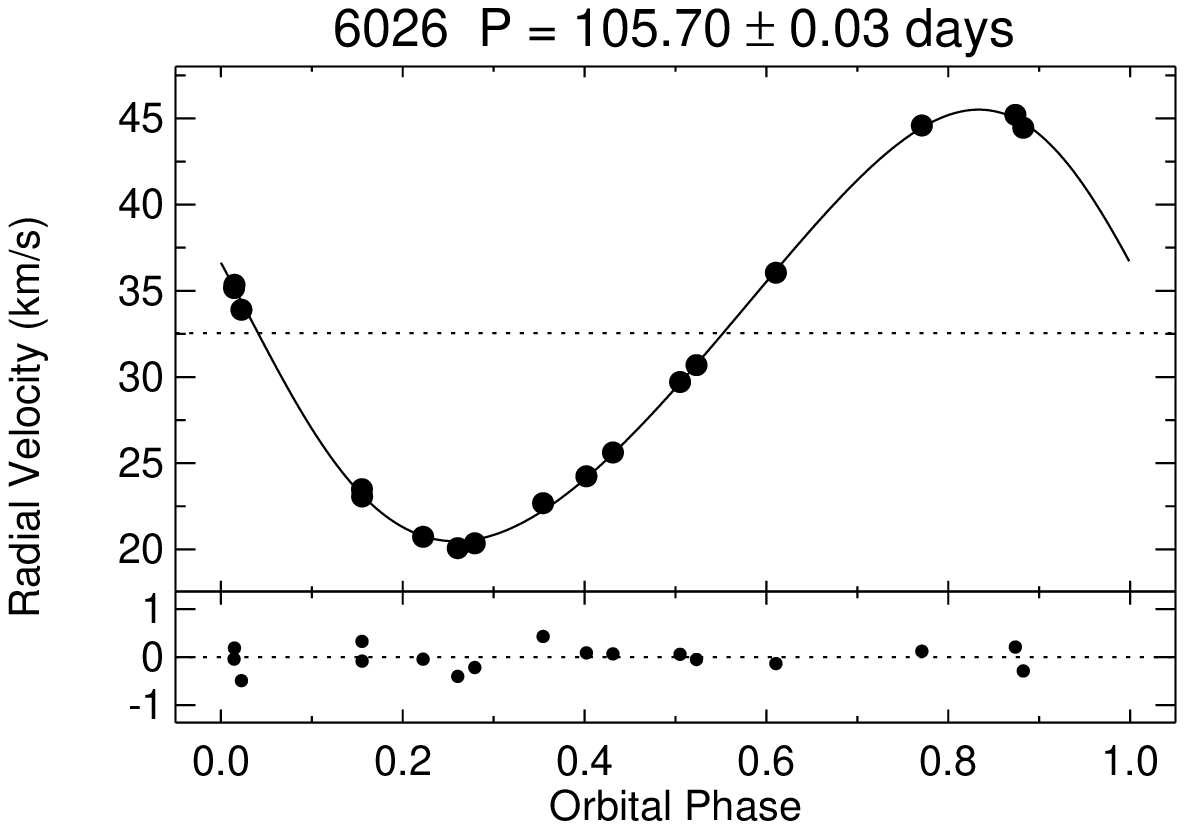}{0.3\textwidth}{}
	\fig{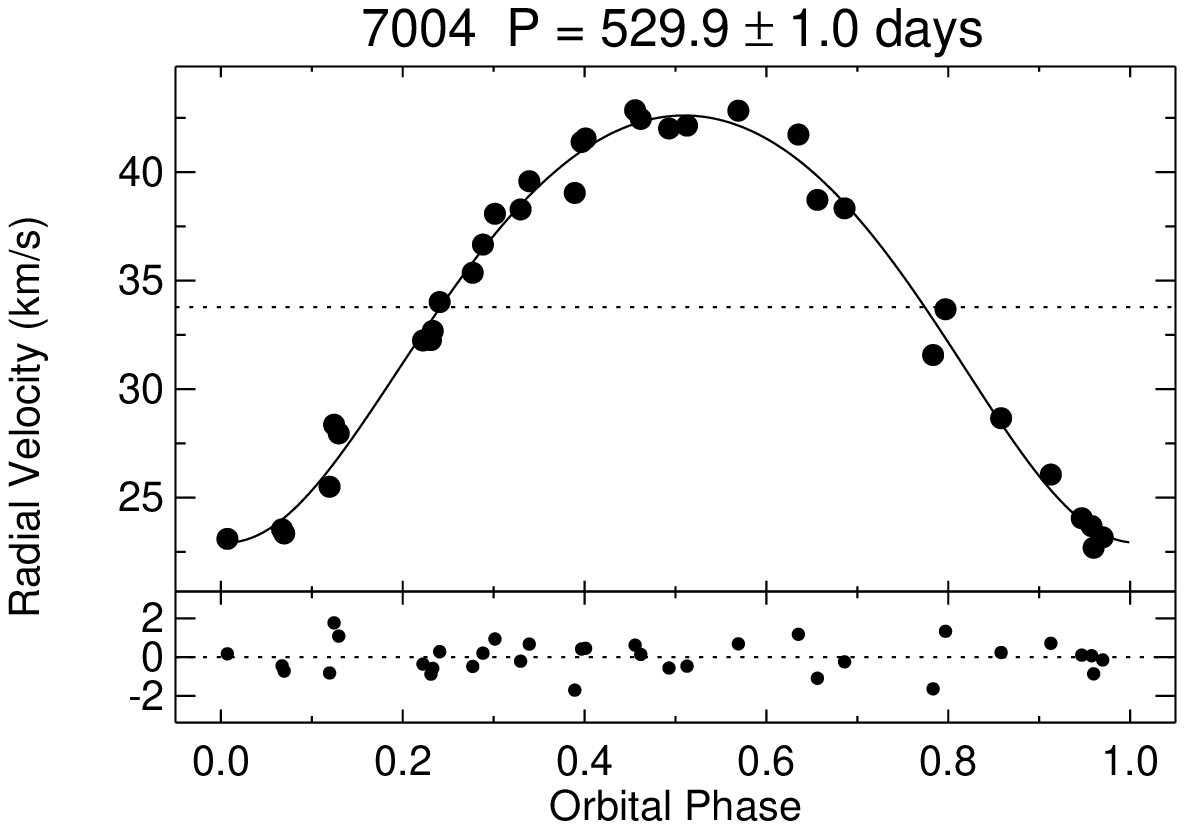}{0.3\textwidth}{}
	\fig{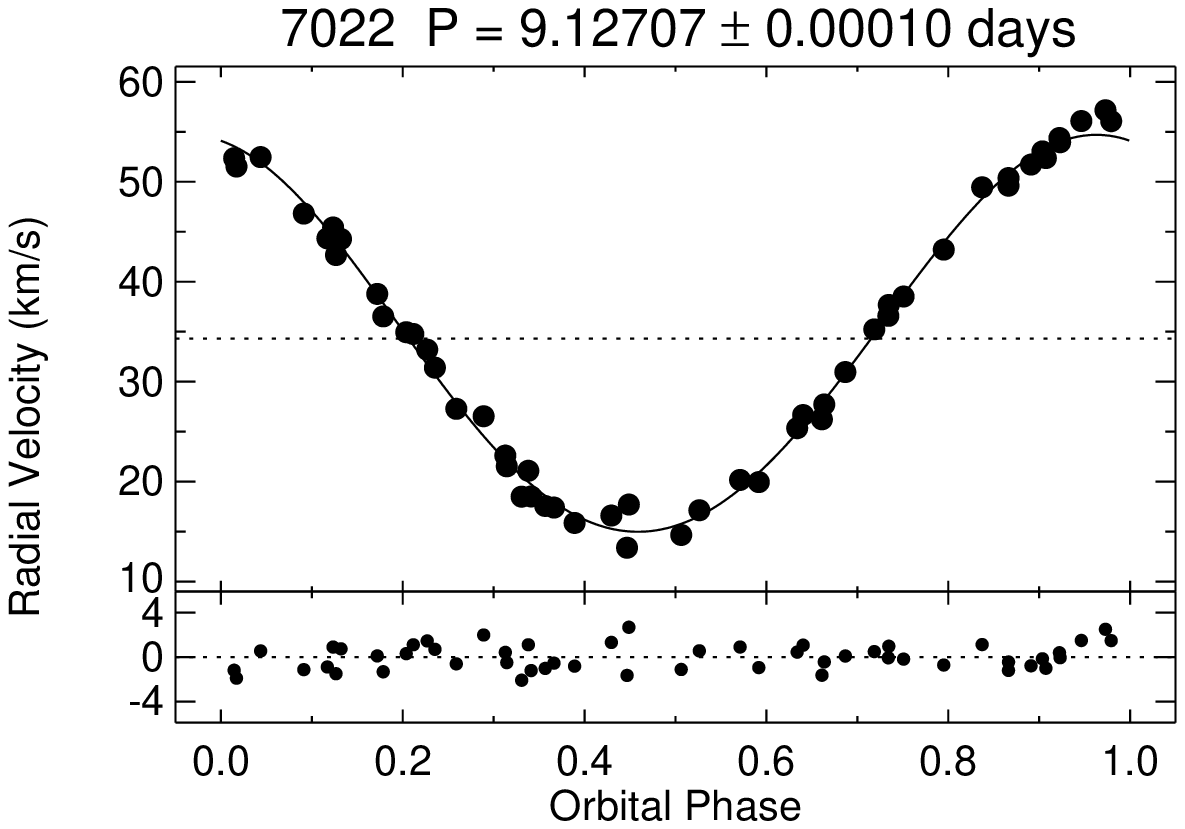}{0.3\textwidth}{}}
\gridline{\fig{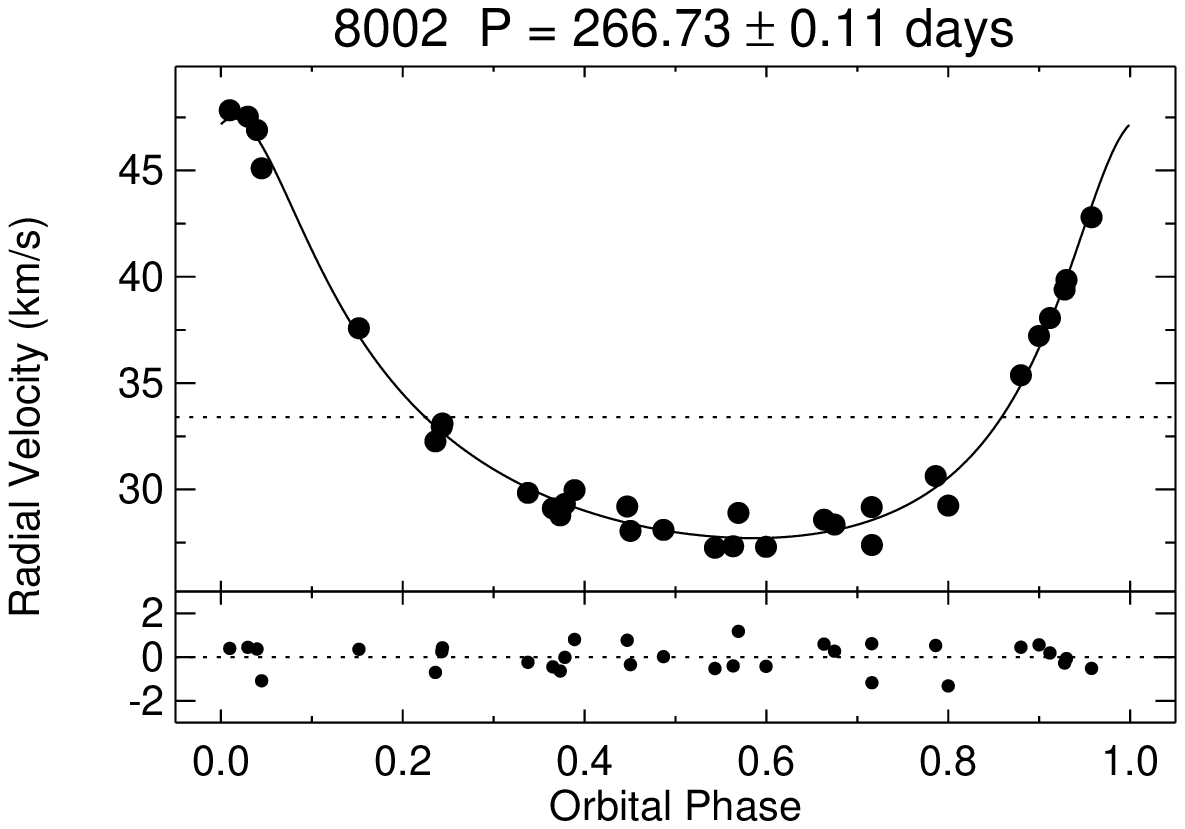}{0.3\textwidth}{}
	\fig{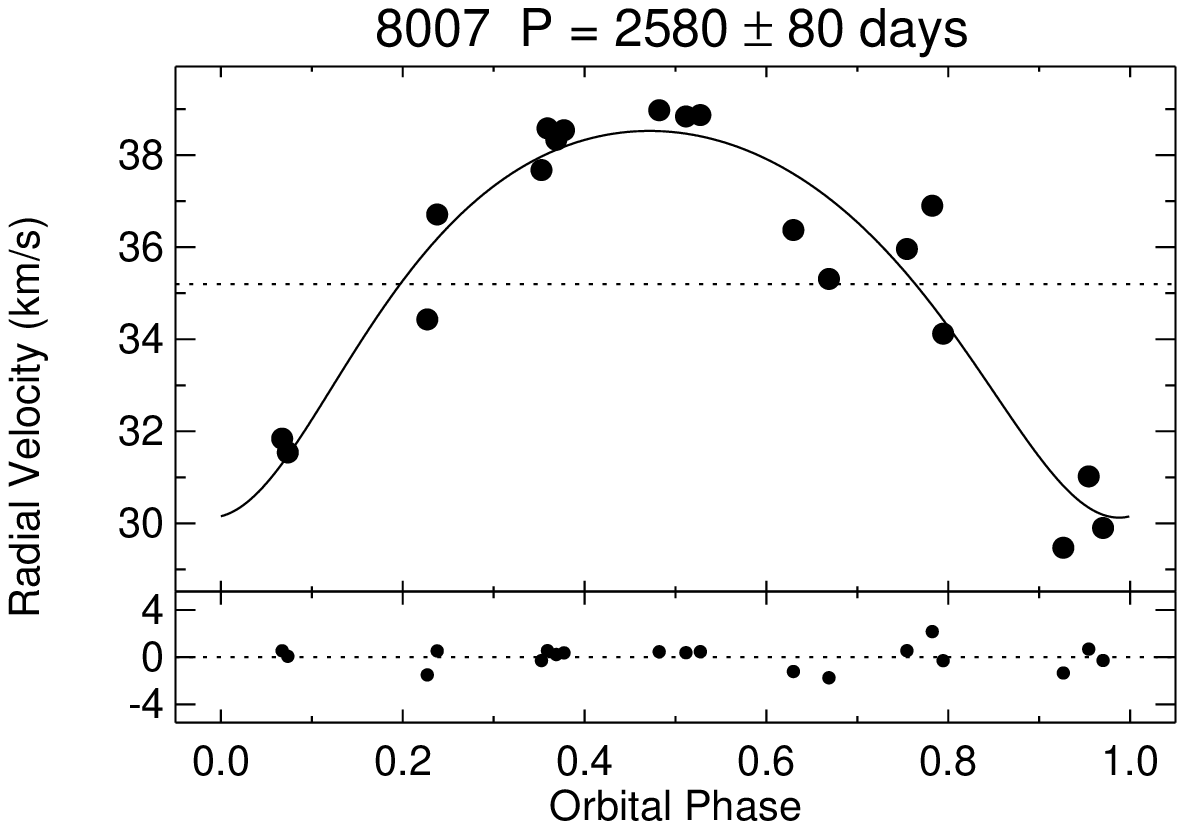}{0.3\textwidth}{}
	\fig{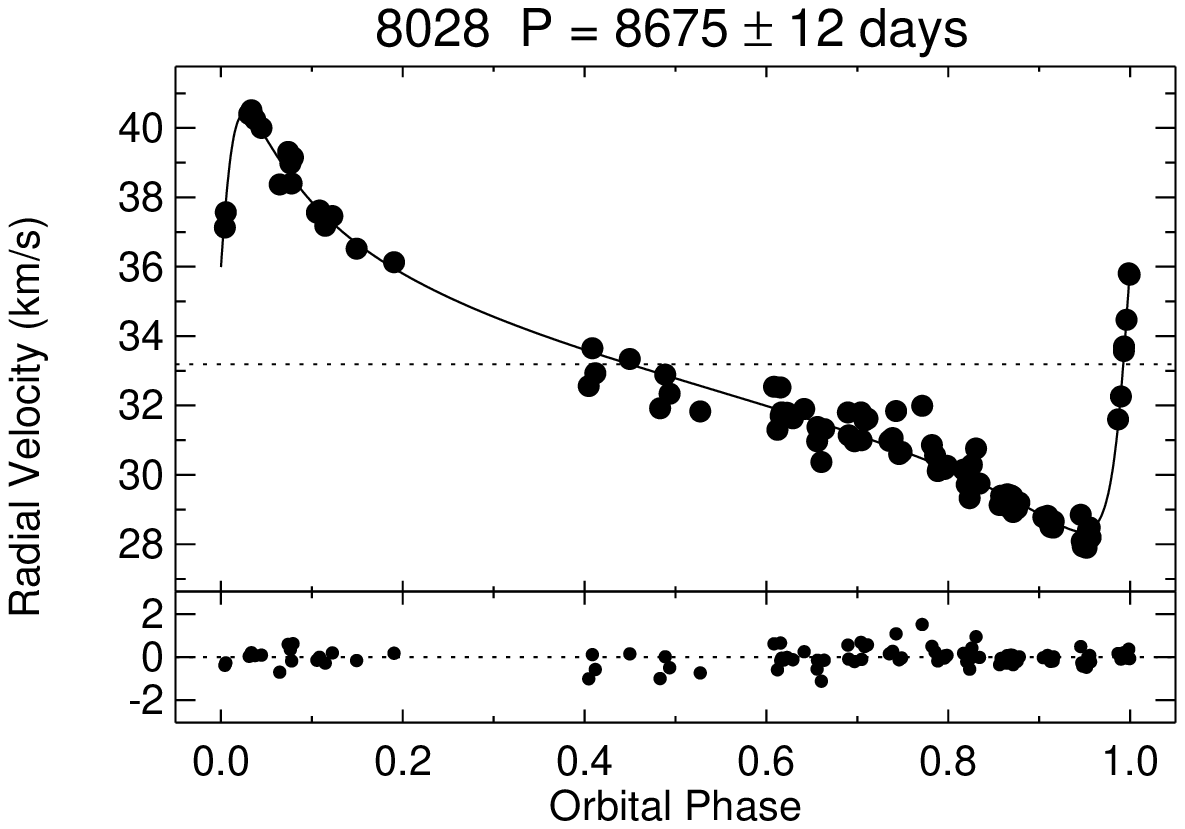}{0.3\textwidth}{}}
\gridline{\fig{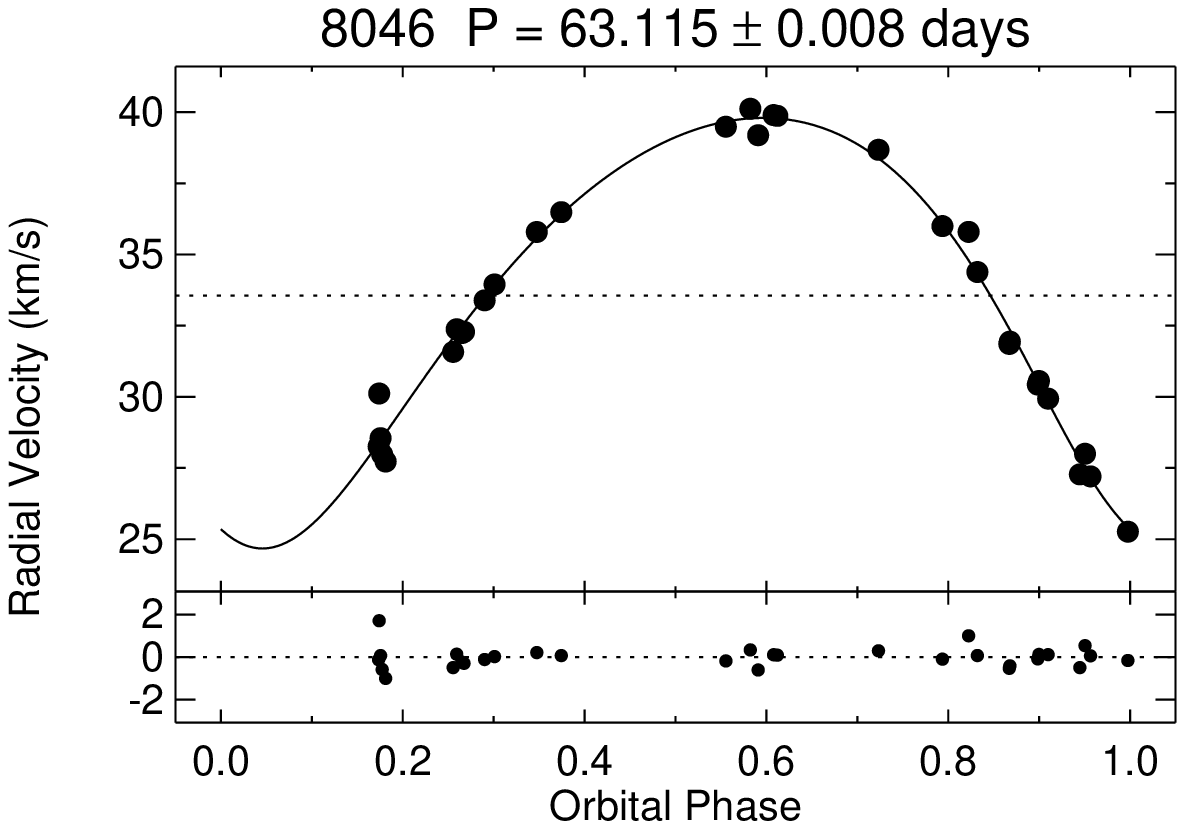}{0.3\textwidth}{}
	\fig{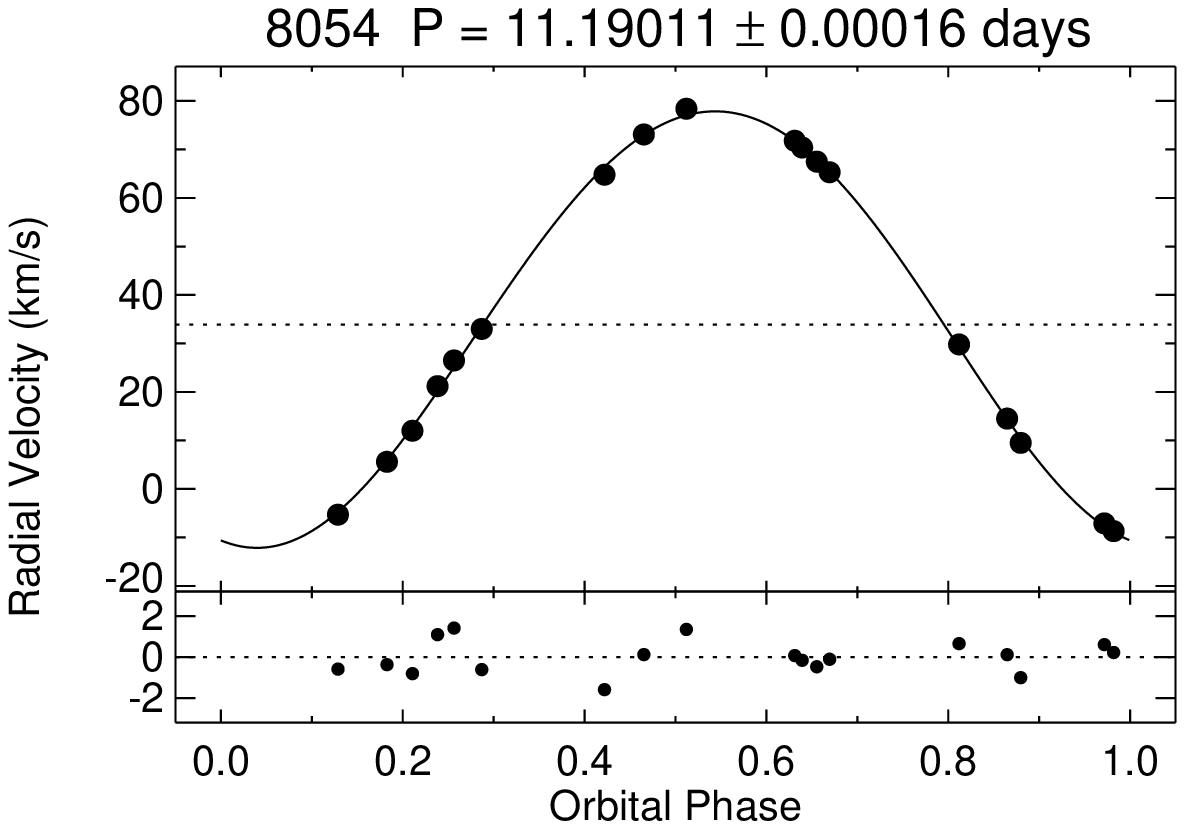}{0.3\textwidth}{}
	\fig{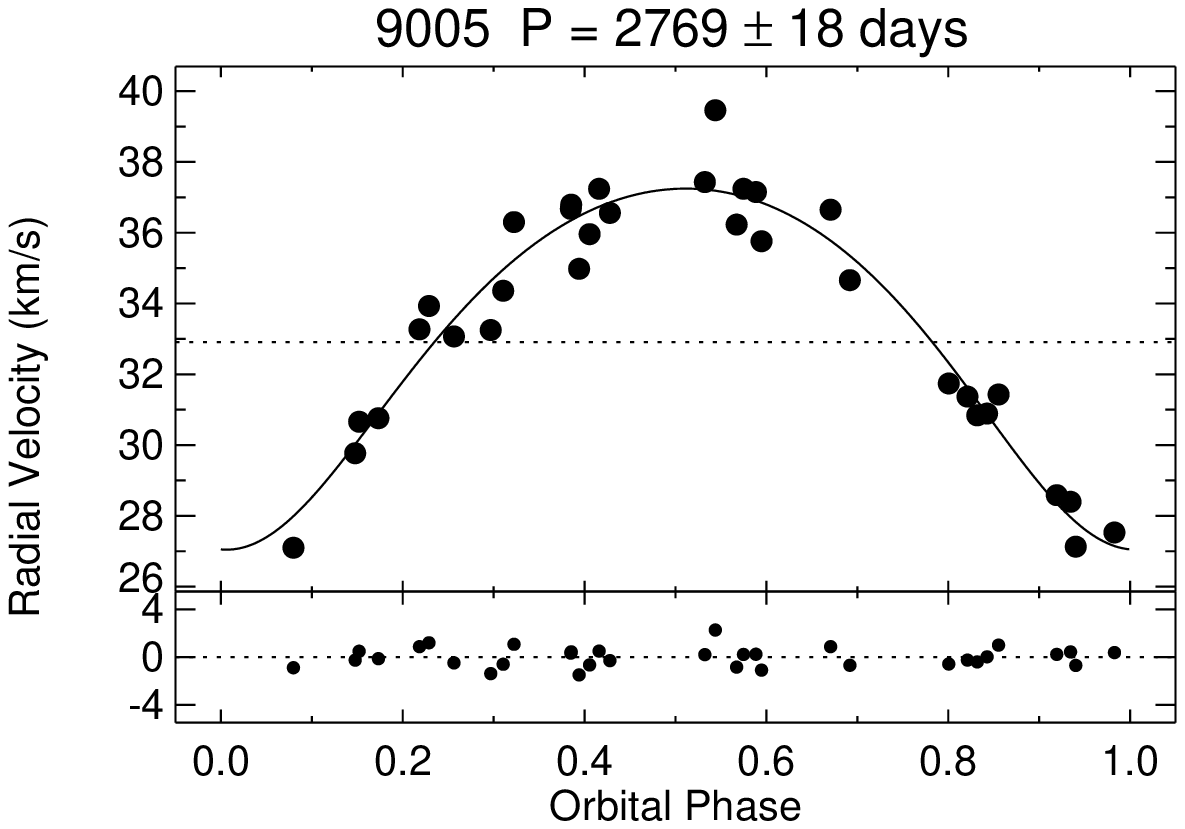}{0.3\textwidth}{}}
\gridline{\fig{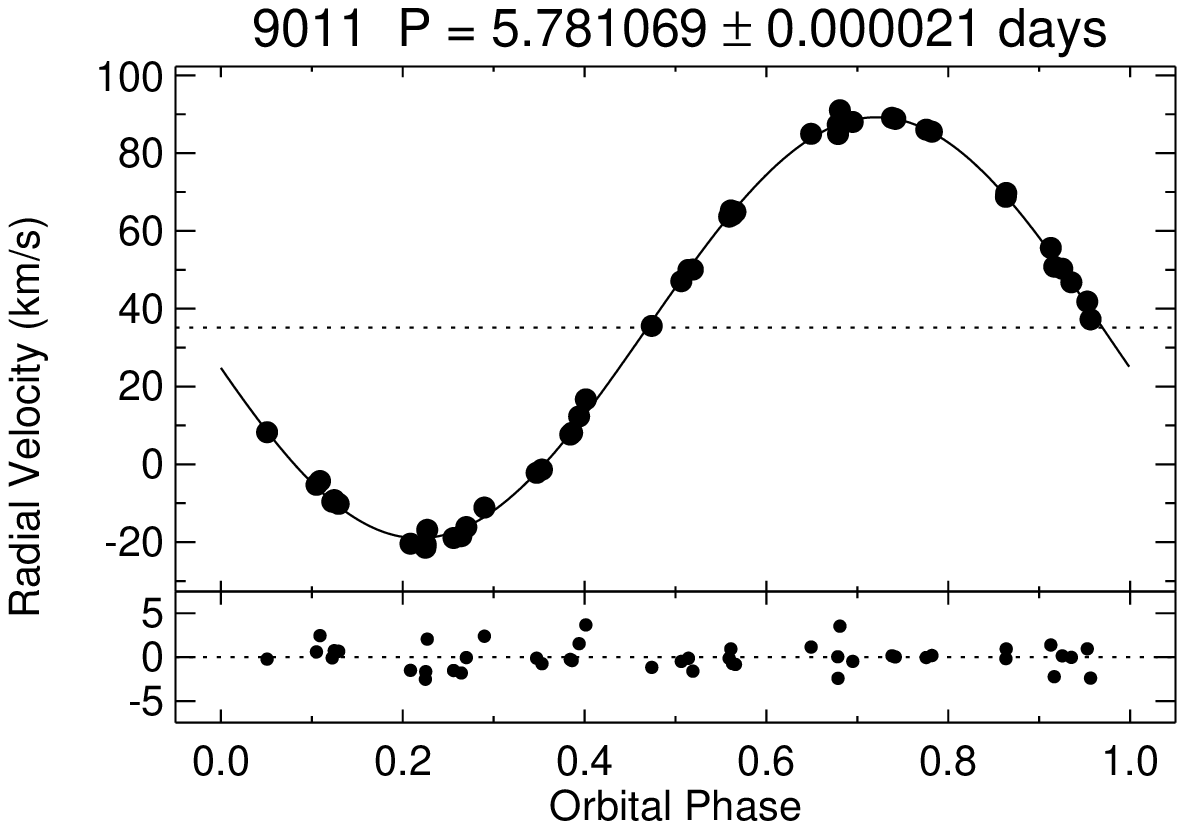}{0.3\textwidth}{}
	\fig{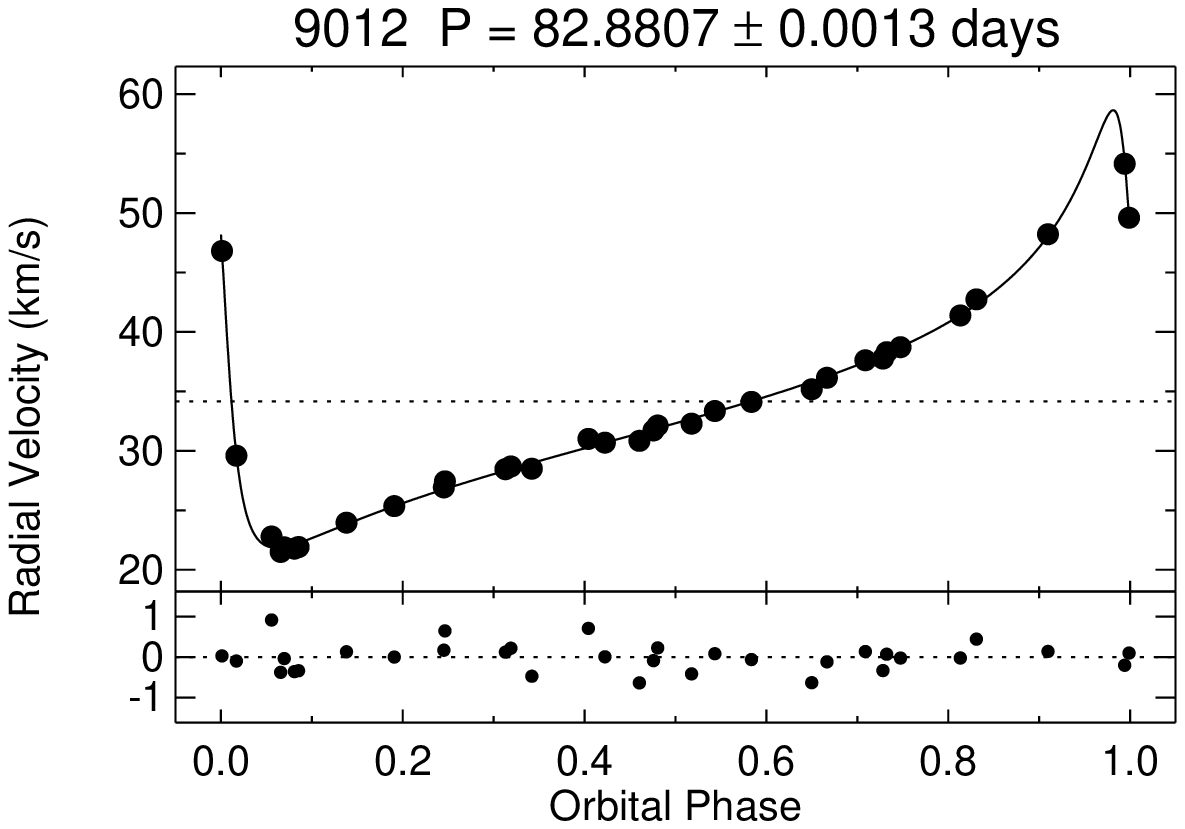}{0.3\textwidth}{}
	\fig{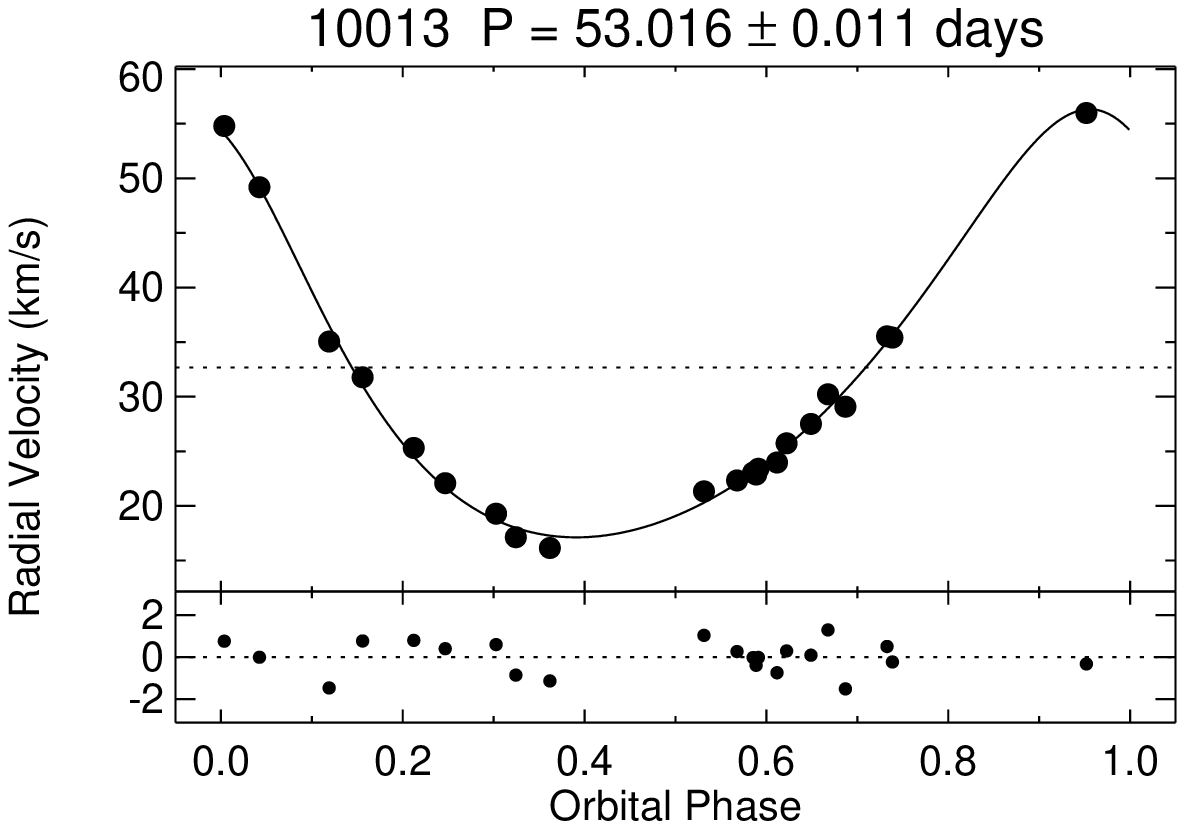}{0.3\textwidth}{}}
\caption{(Continued.)}
\end{figure*}

\begin{figure*}
\figurenum{11}
\gridline{\fig{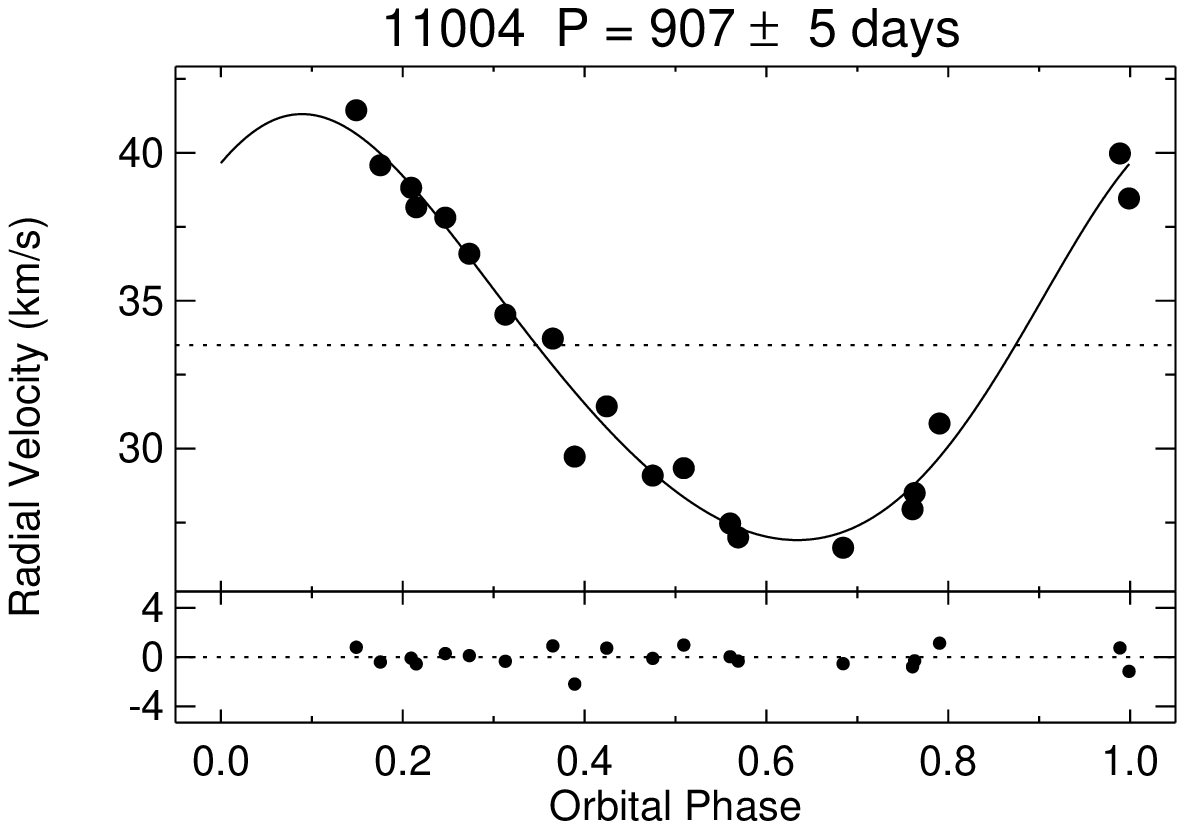}{0.3\textwidth}{}
	\fig{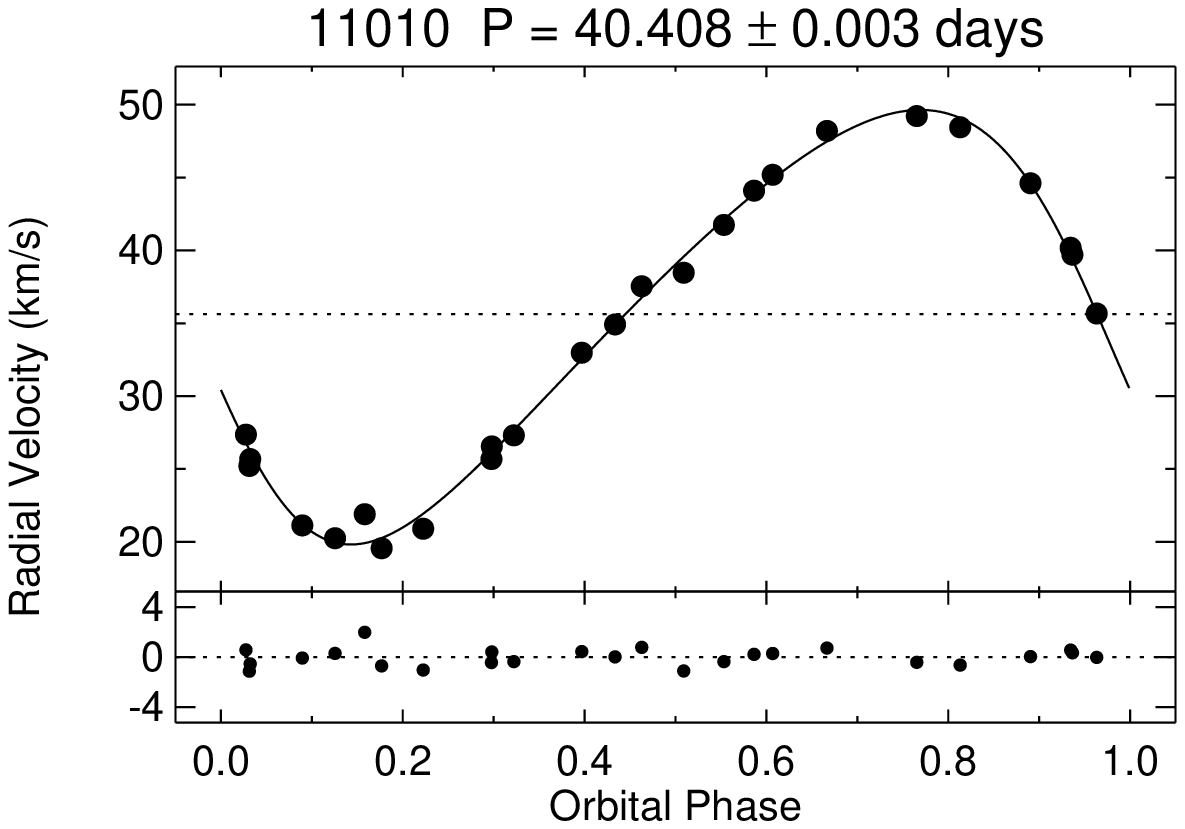}{0.3\textwidth}{}
	\fig{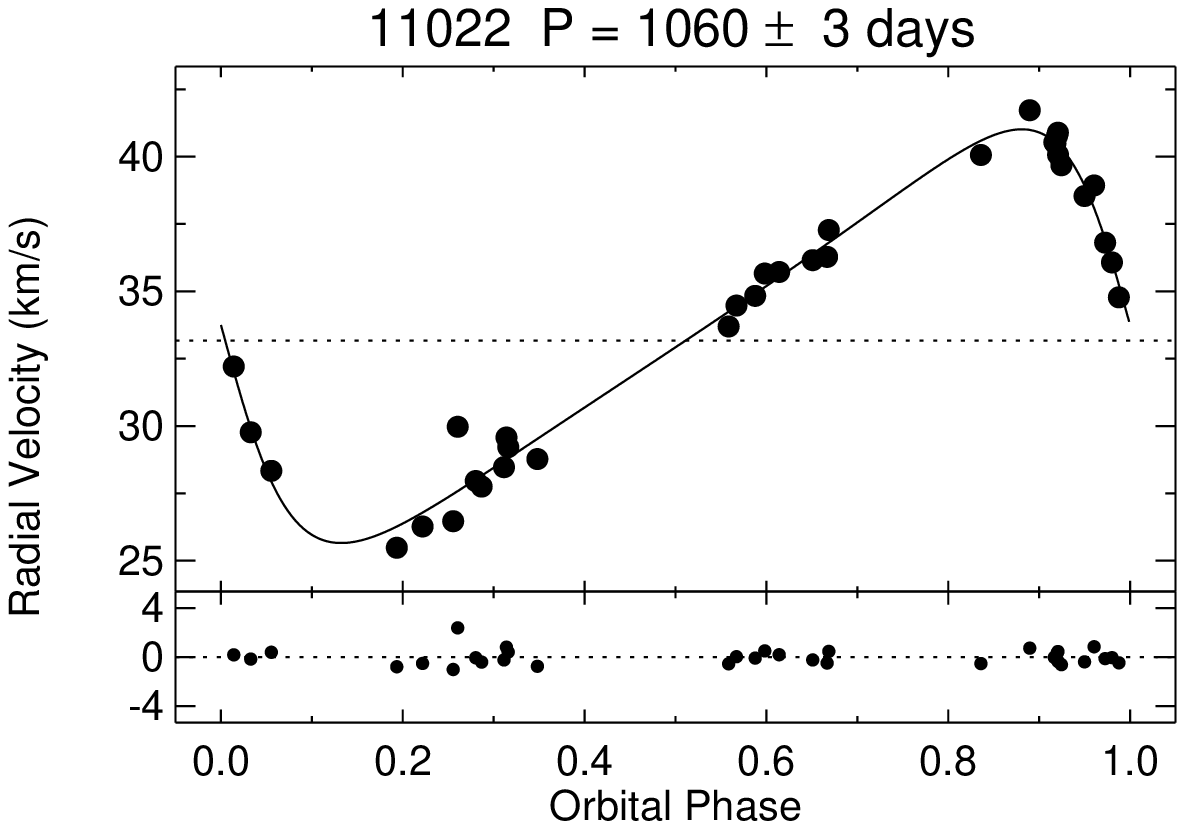}{0.3\textwidth}{}}
\gridline{\fig{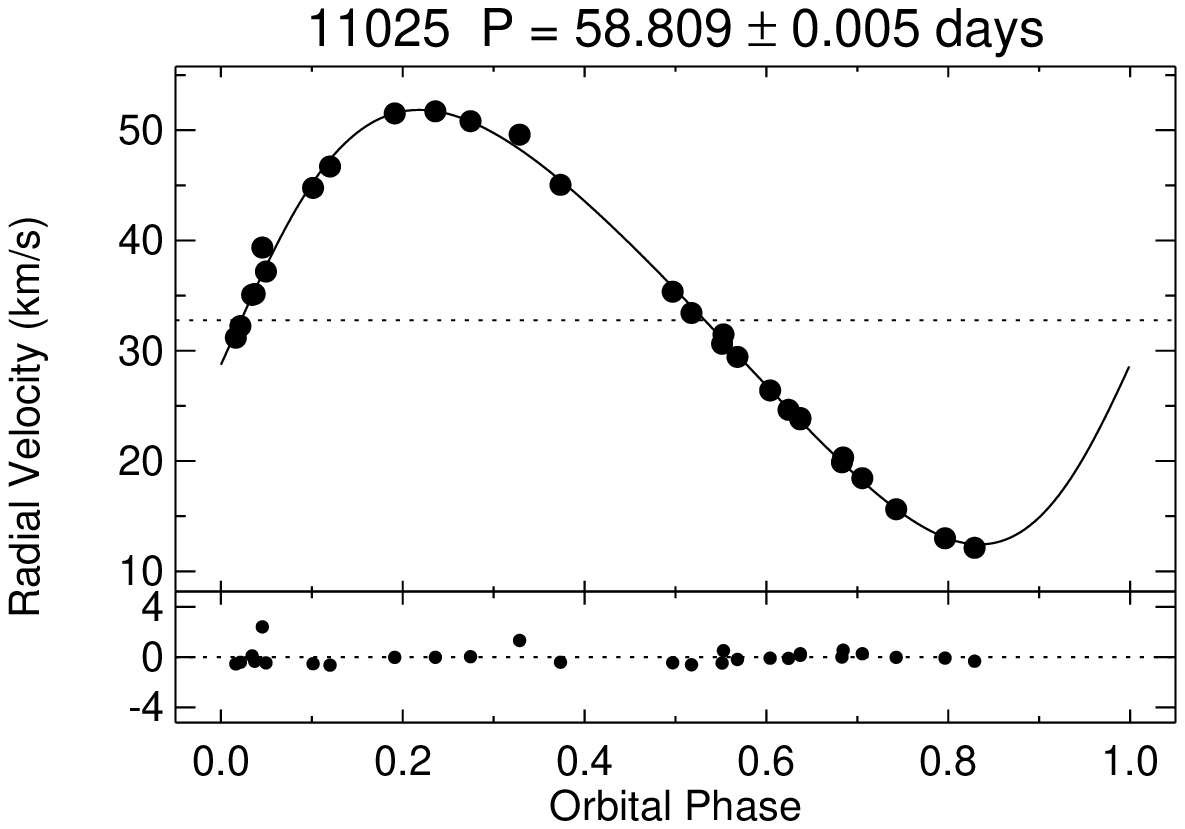}{0.3\textwidth}{}
	\fig{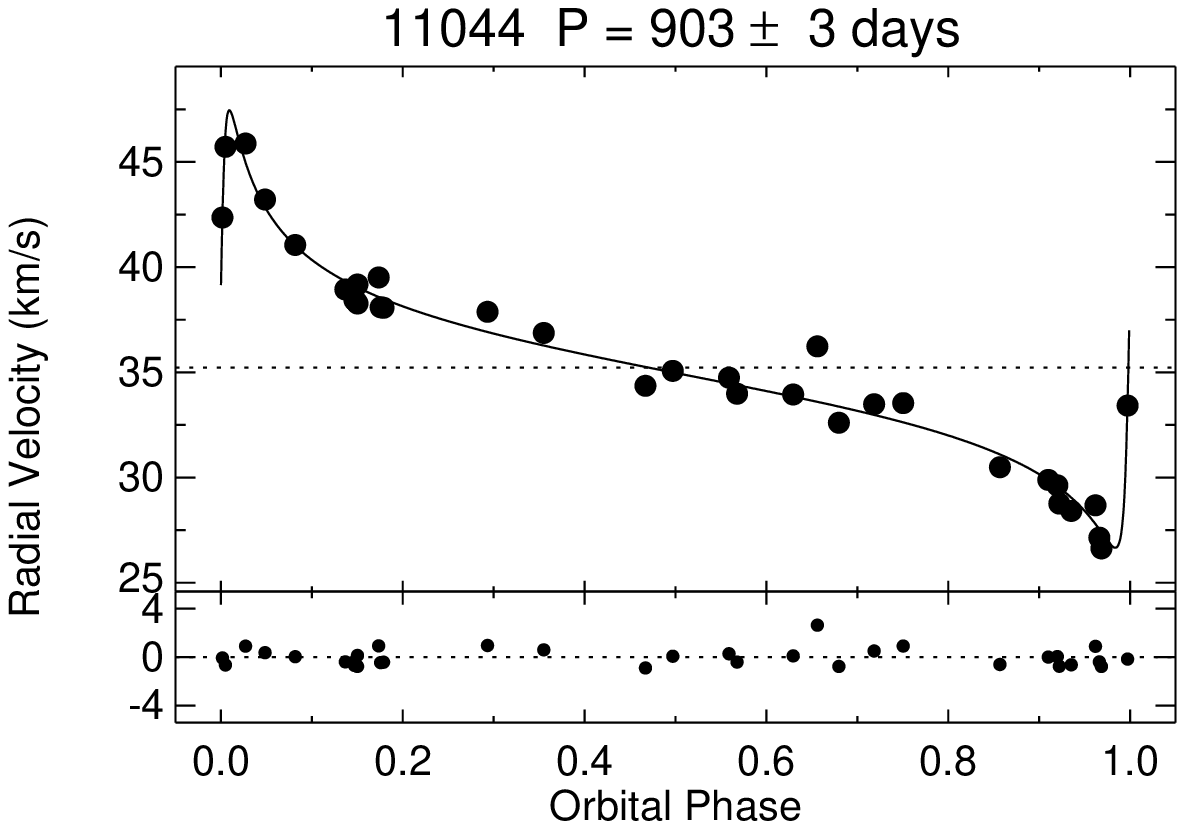}{0.3\textwidth}{}
	\fig{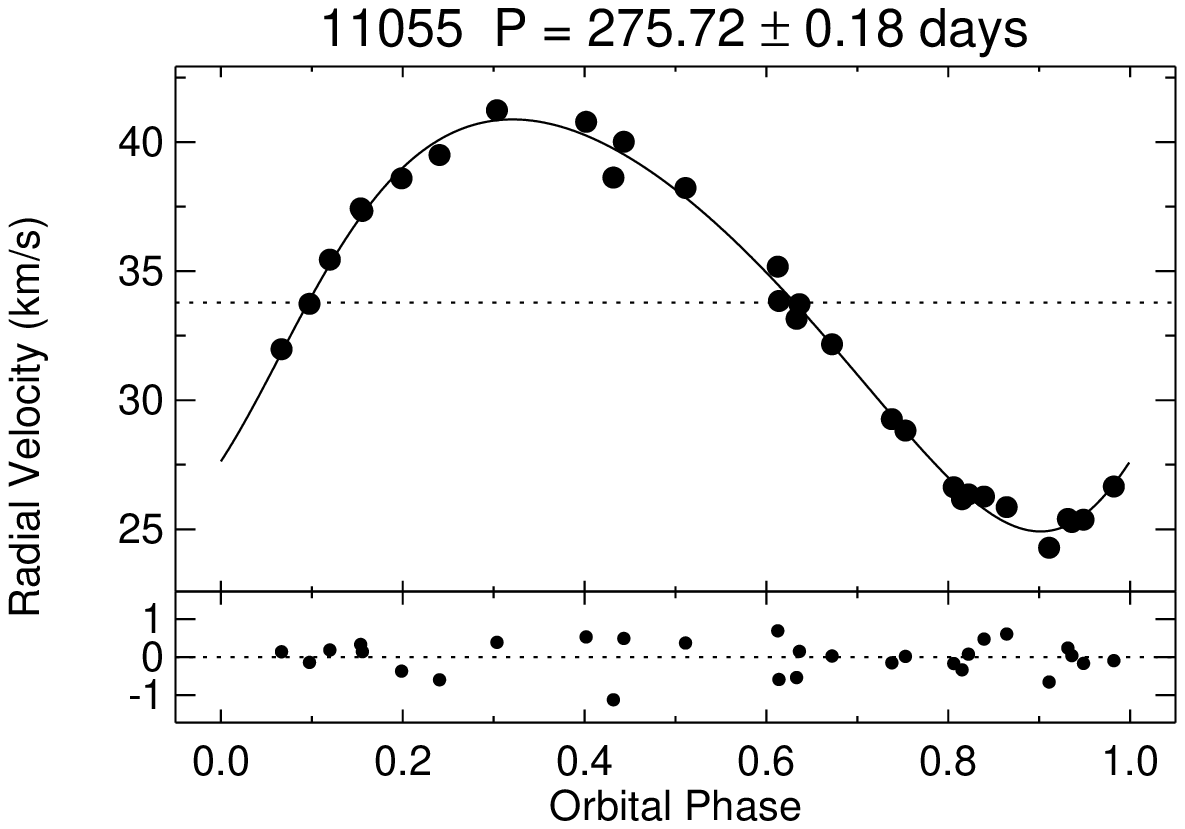}{0.3\textwidth}{}}
\gridline{\fig{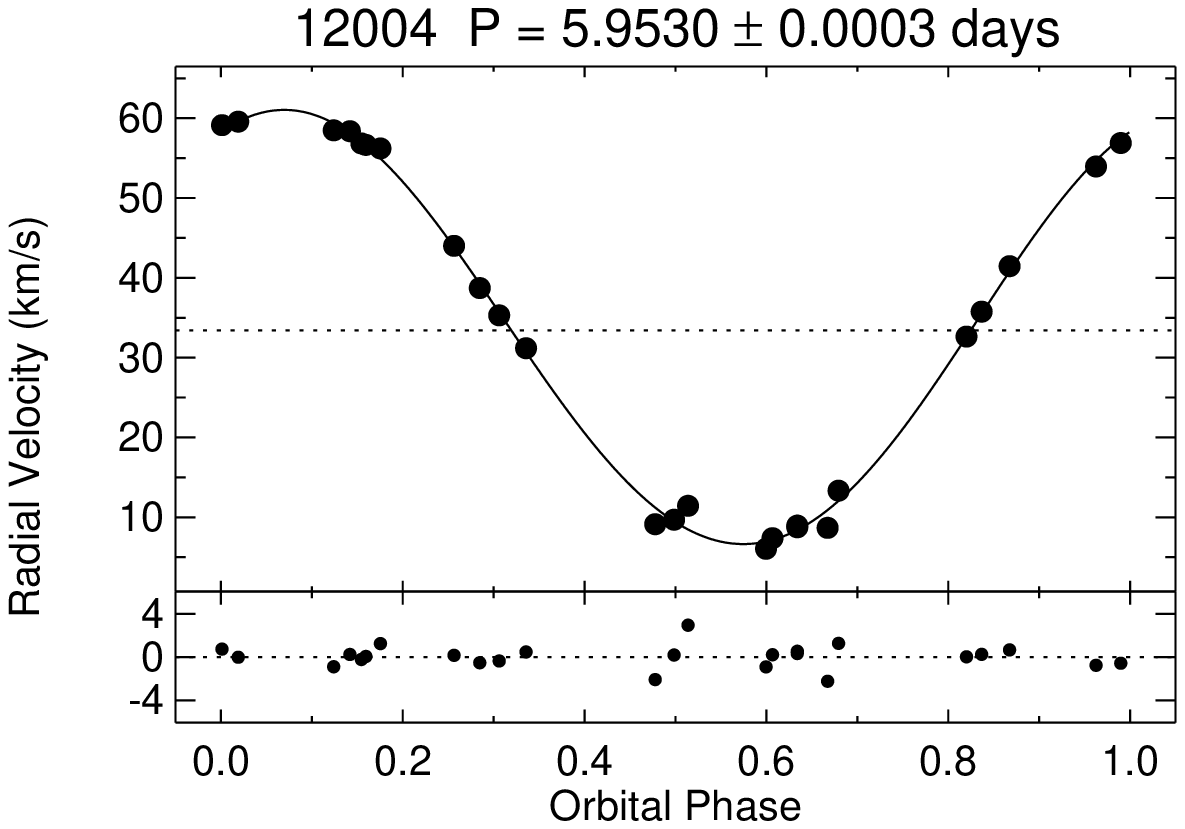}{0.3\textwidth}{}
	\fig{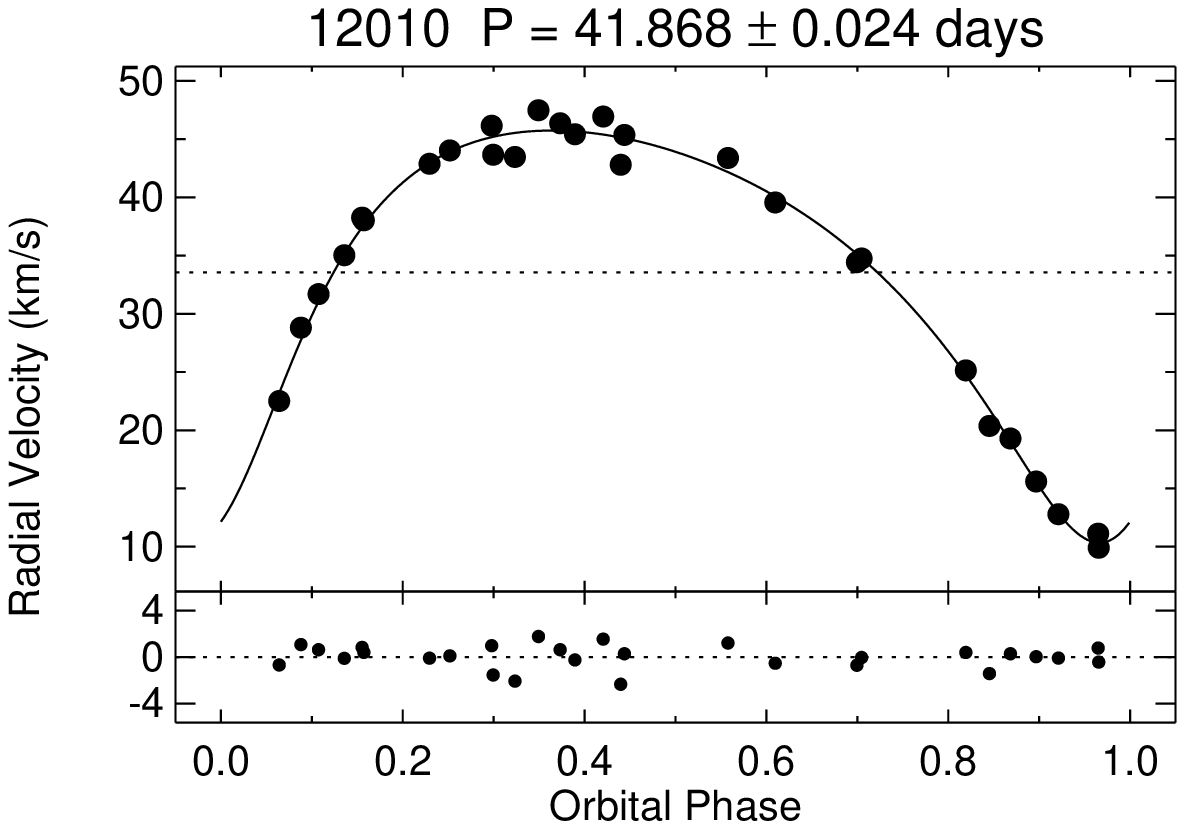}{0.3\textwidth}{}
	\fig{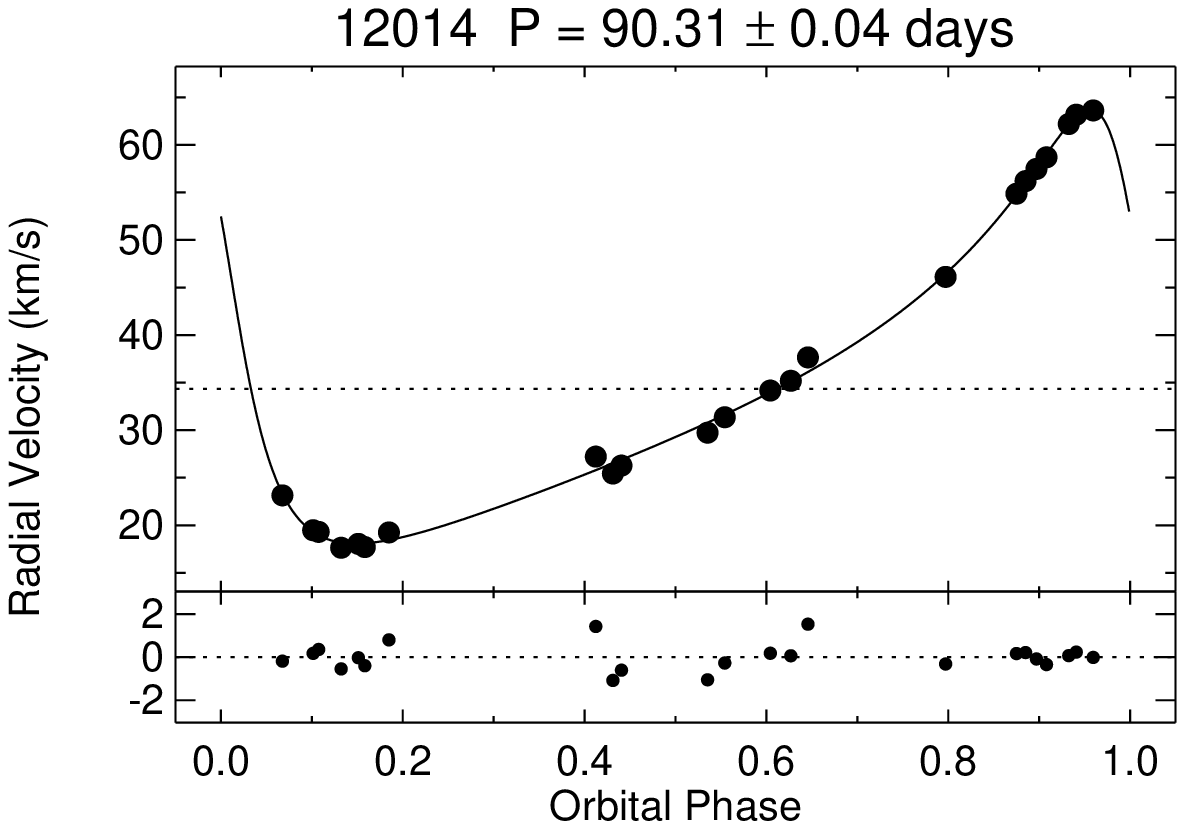}{0.3\textwidth}{}}
\gridline{\fig{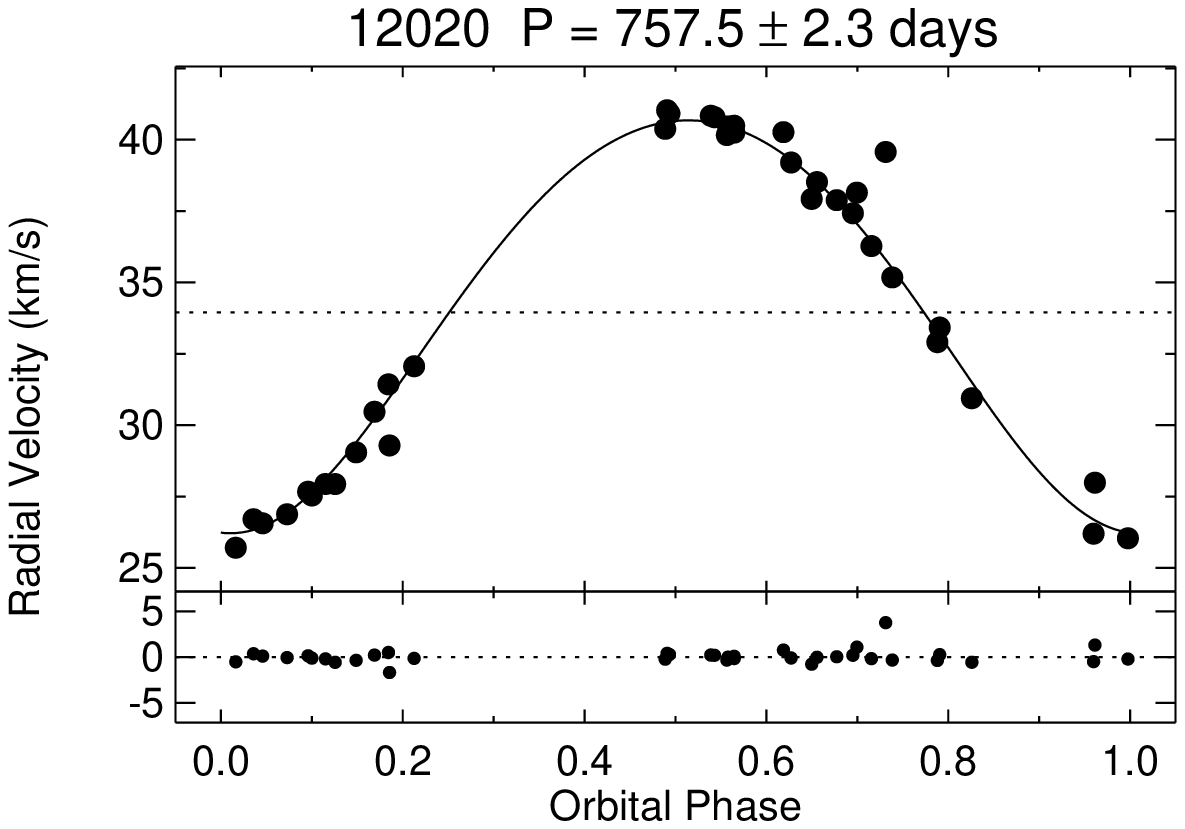}{0.3\textwidth}{}
	\fig{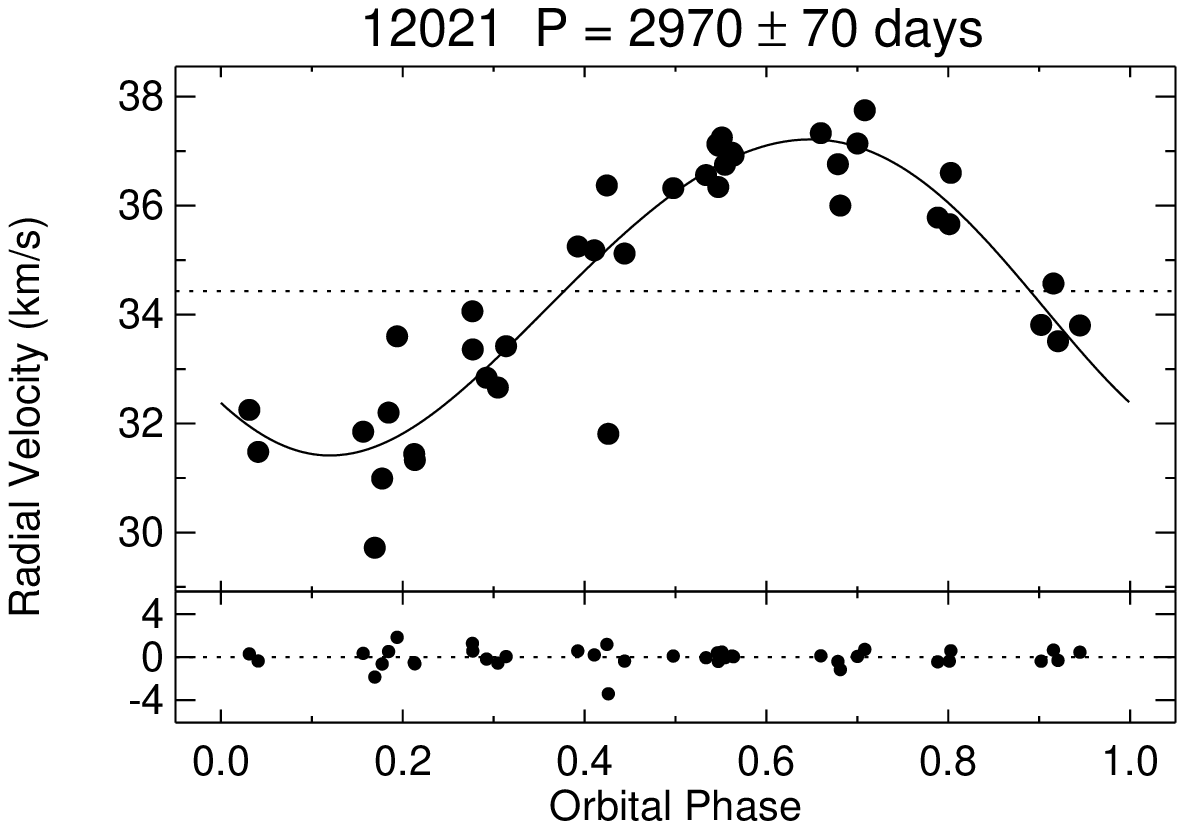}{0.3\textwidth}{}
	\fig{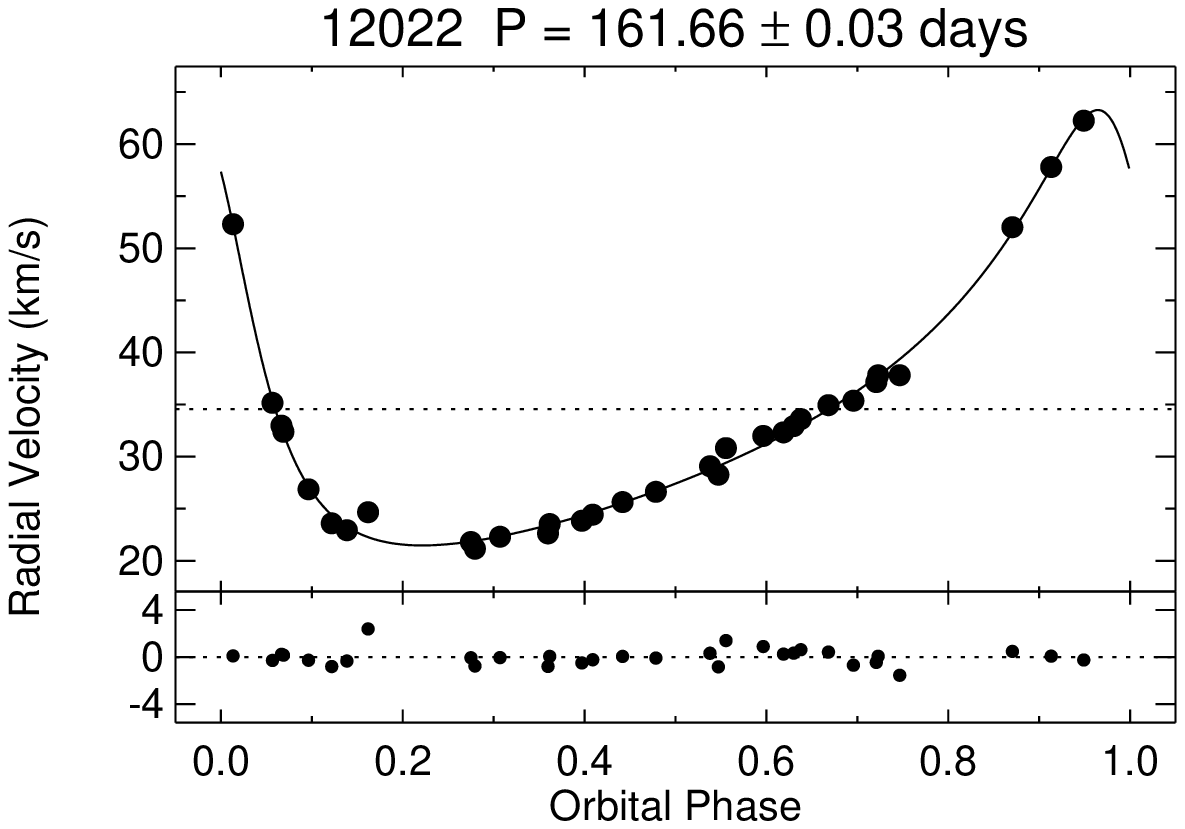}{0.3\textwidth}{}}
\caption{(Continued.)}
\end{figure*}

\begin{figure*}
\figurenum{11}
\gridline{\fig{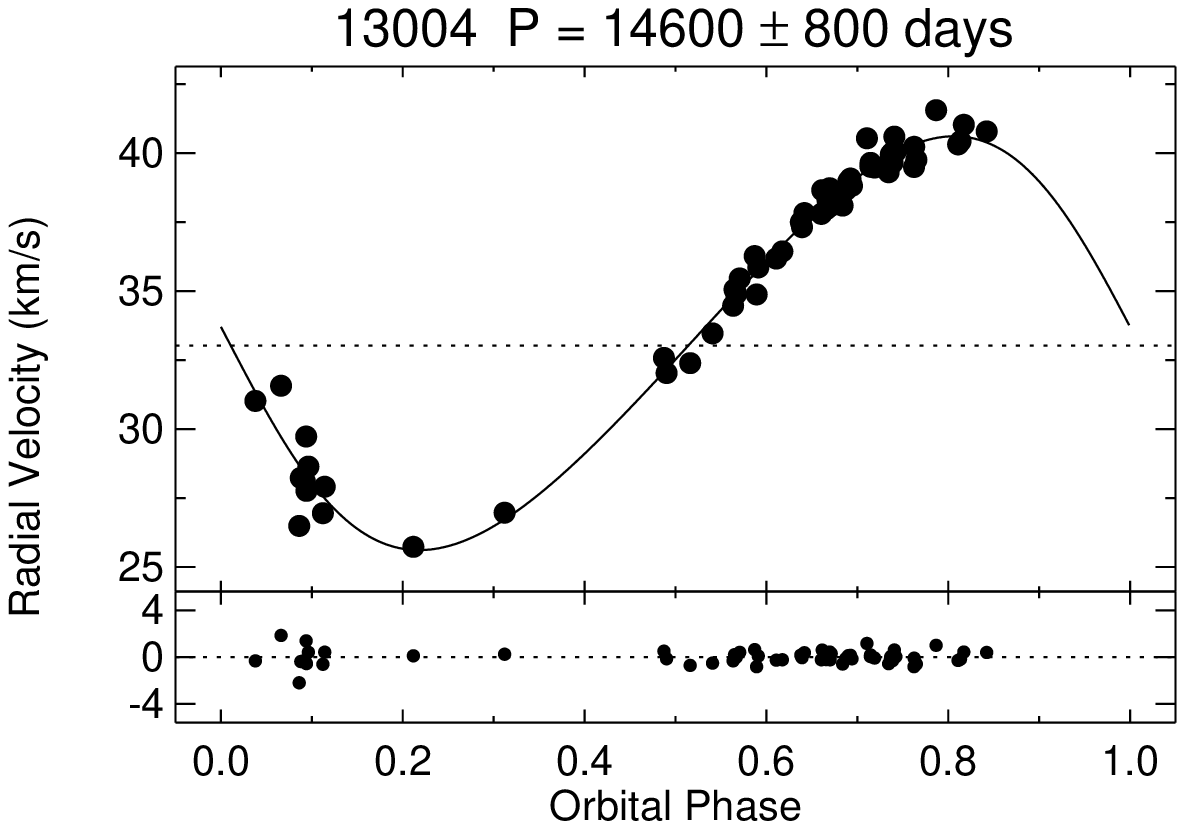}{0.3\textwidth}{}
	\fig{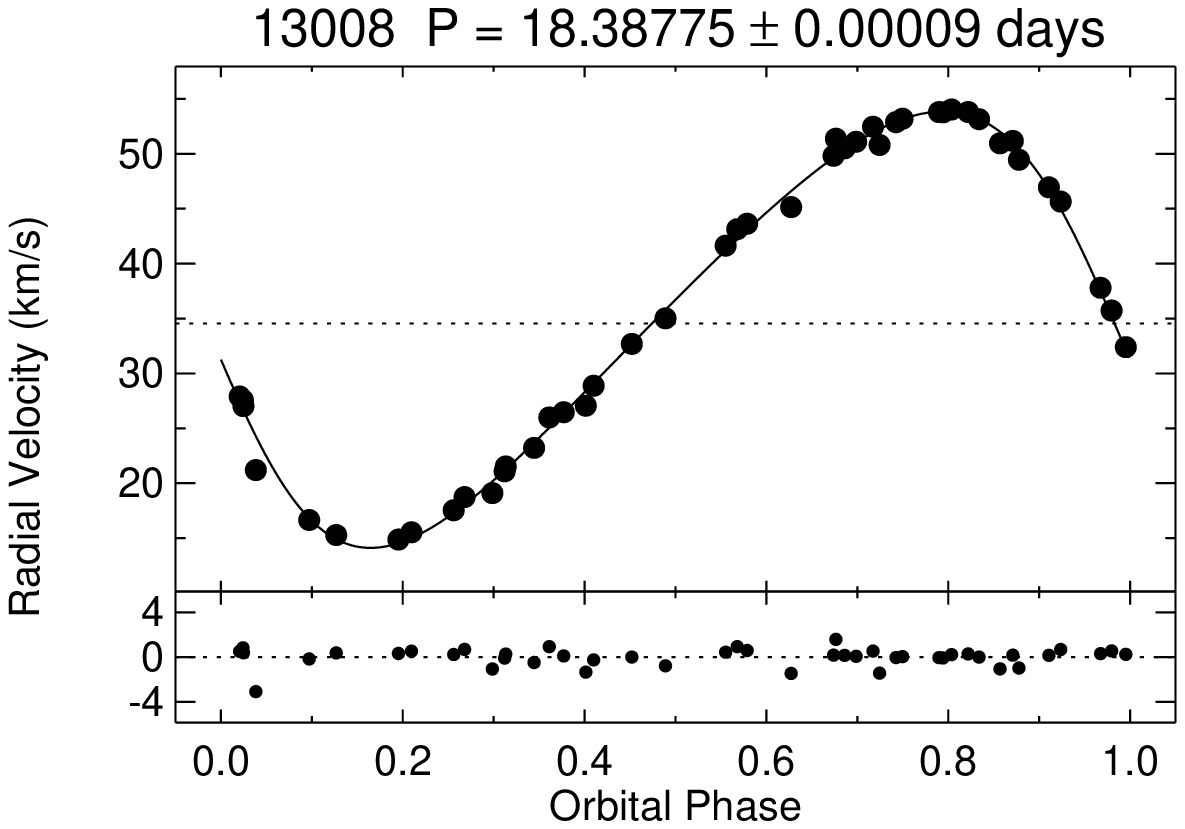}{0.3\textwidth}{}
	\fig{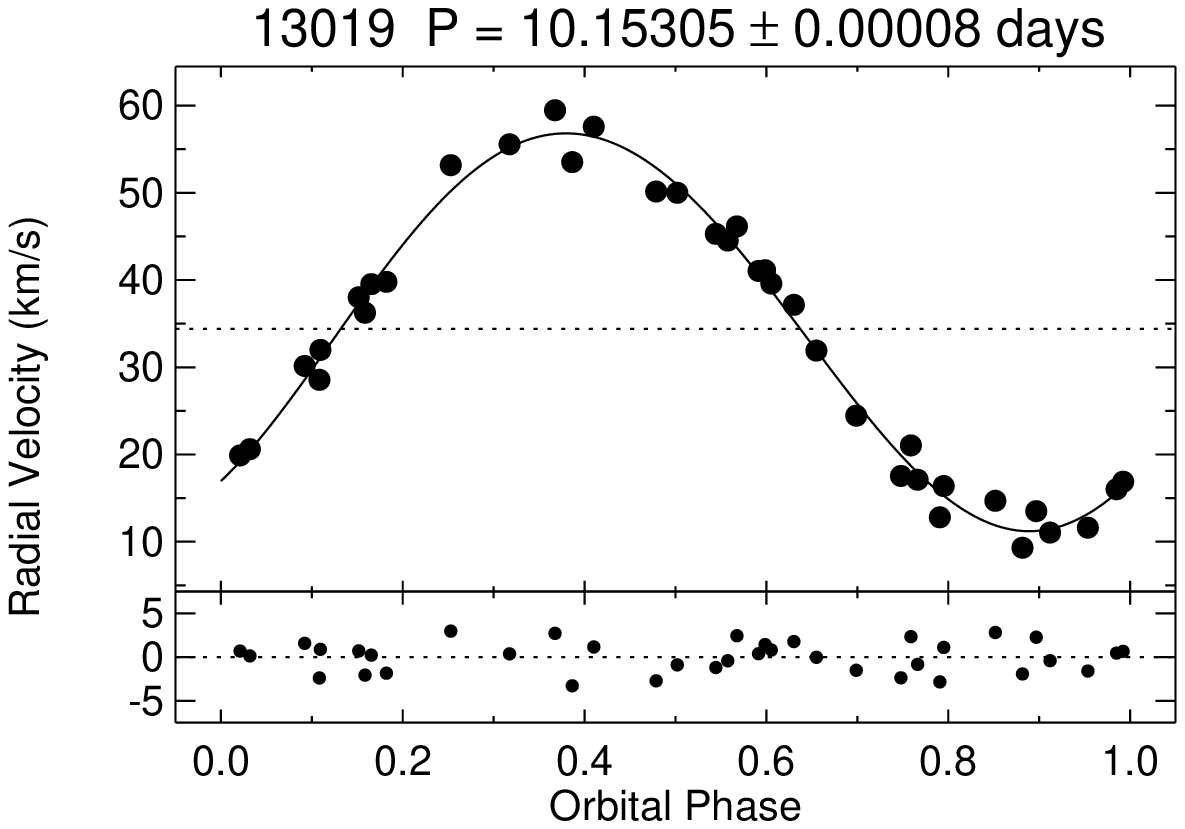}{0.3\textwidth}{}}
\gridline{\fig{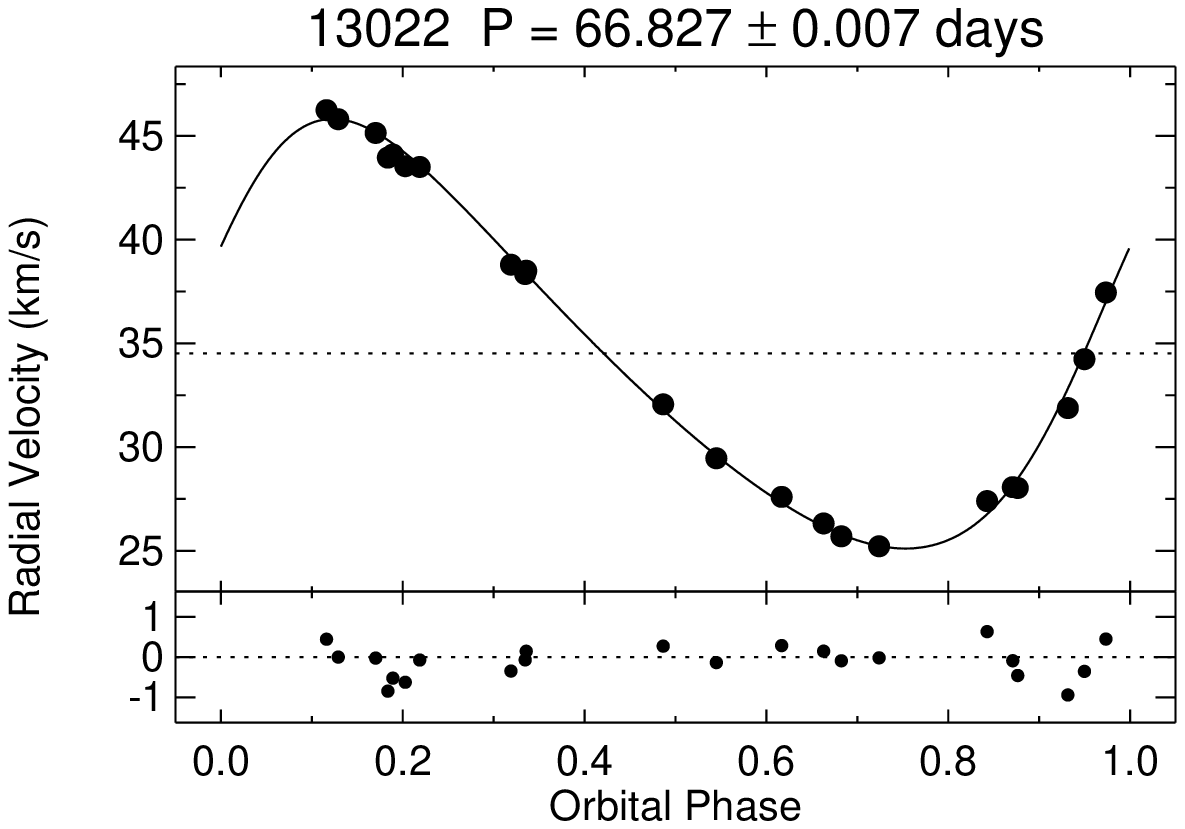}{0.3\textwidth}{}
	\fig{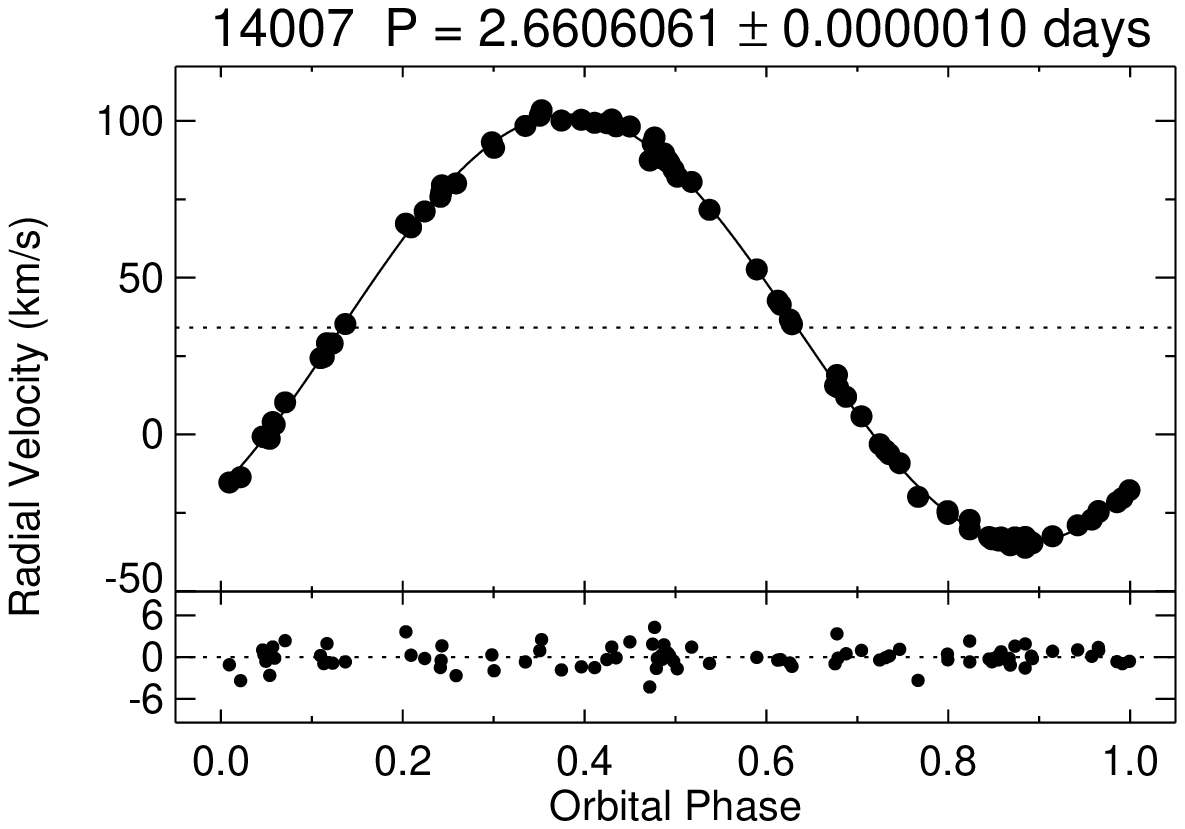}{0.3\textwidth}{}
	\fig{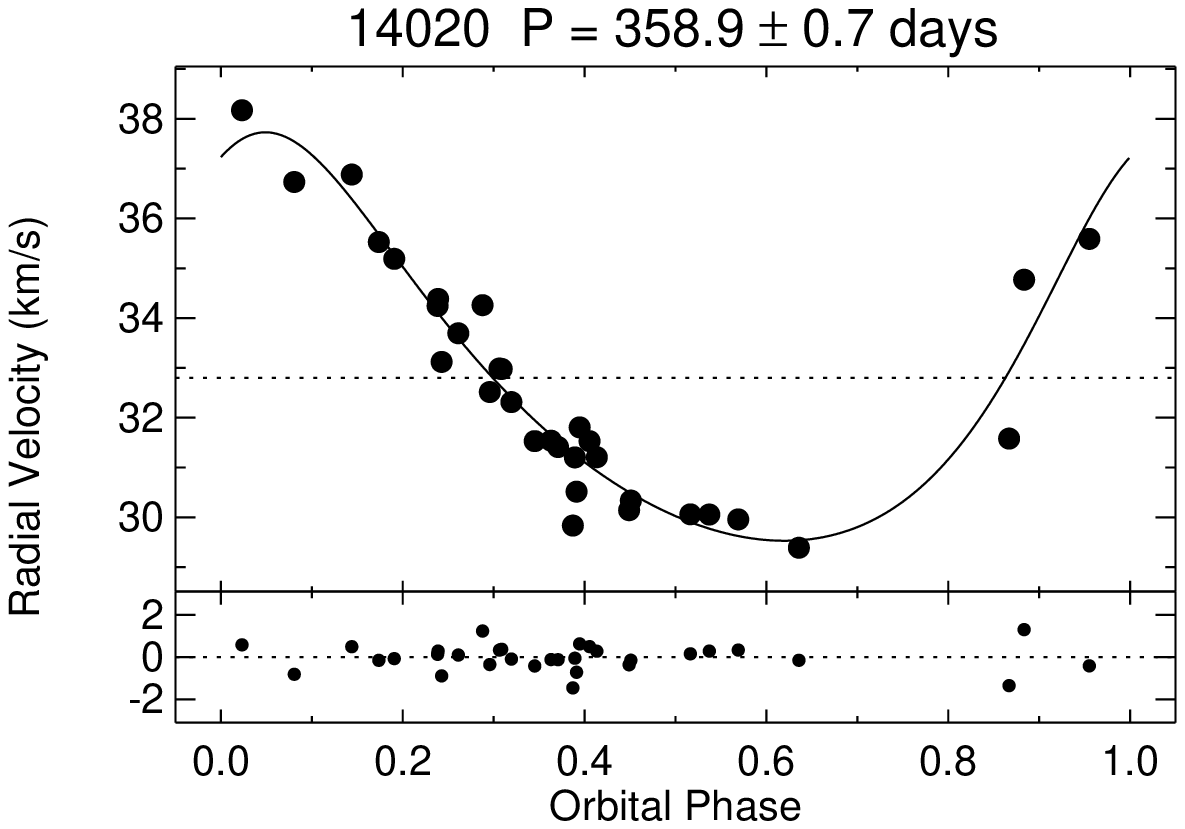}{0.3\textwidth}{}}
\gridline{\fig{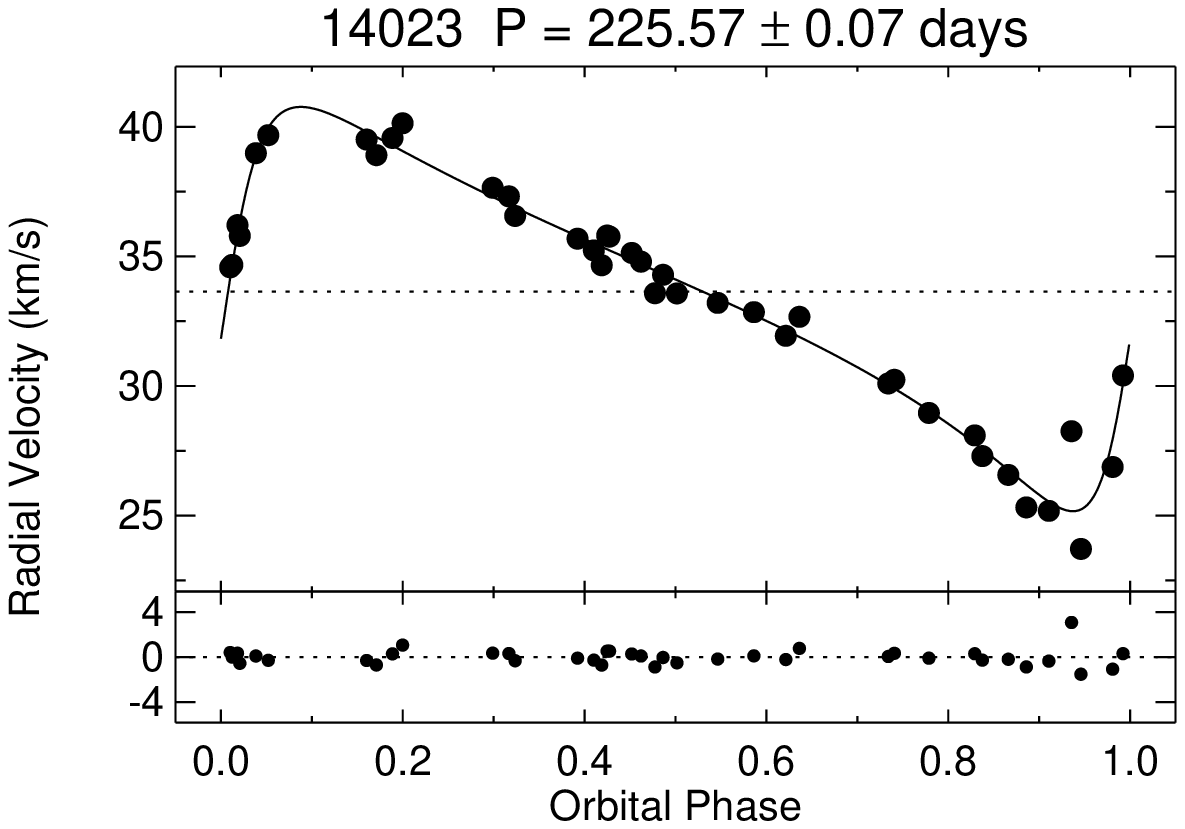}{0.3\textwidth}{}
	\fig{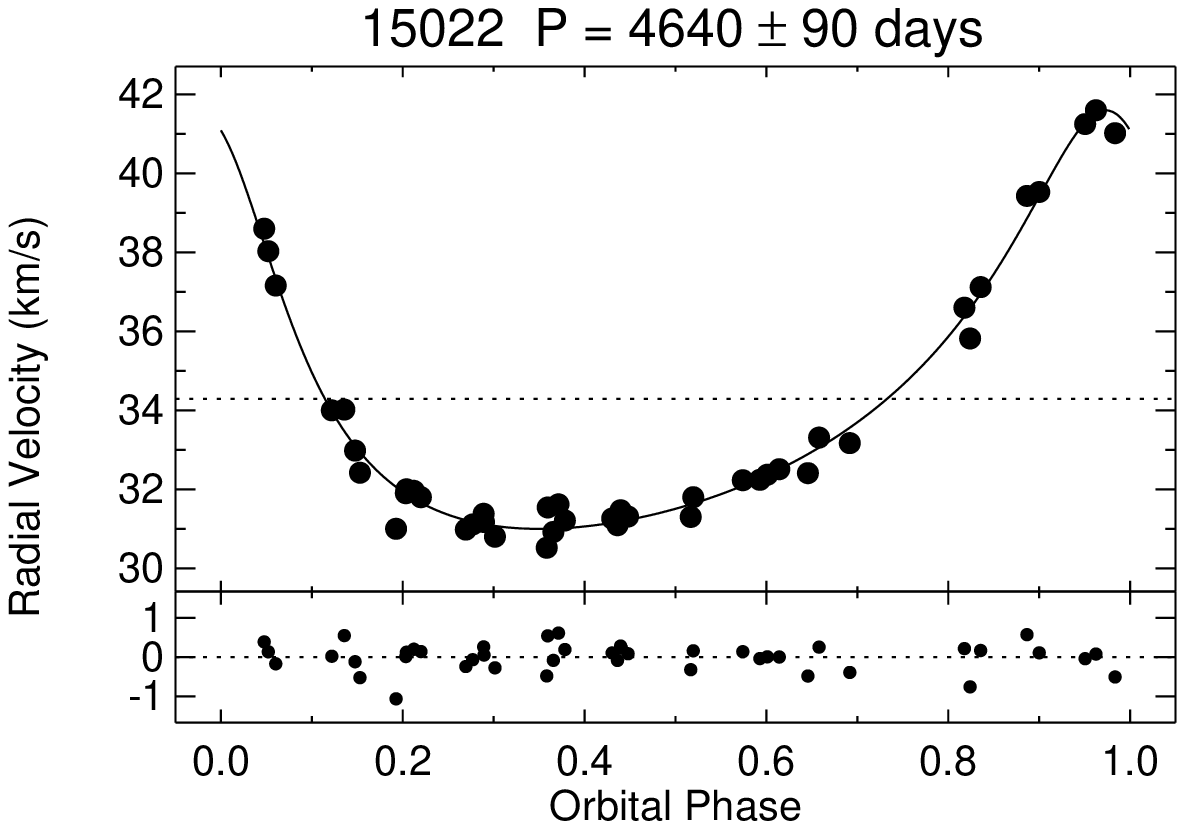}{0.3\textwidth}{}
	\fig{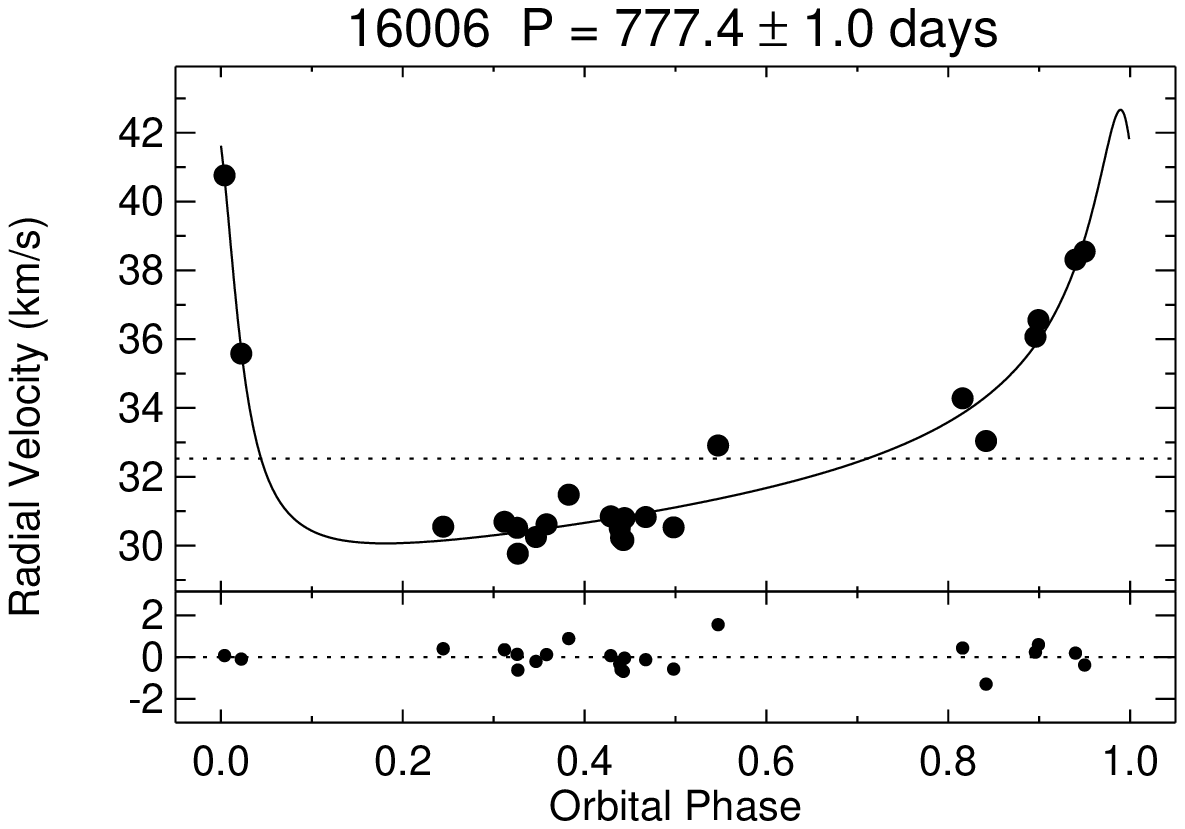}{0.3\textwidth}{}}
\gridline{\fig{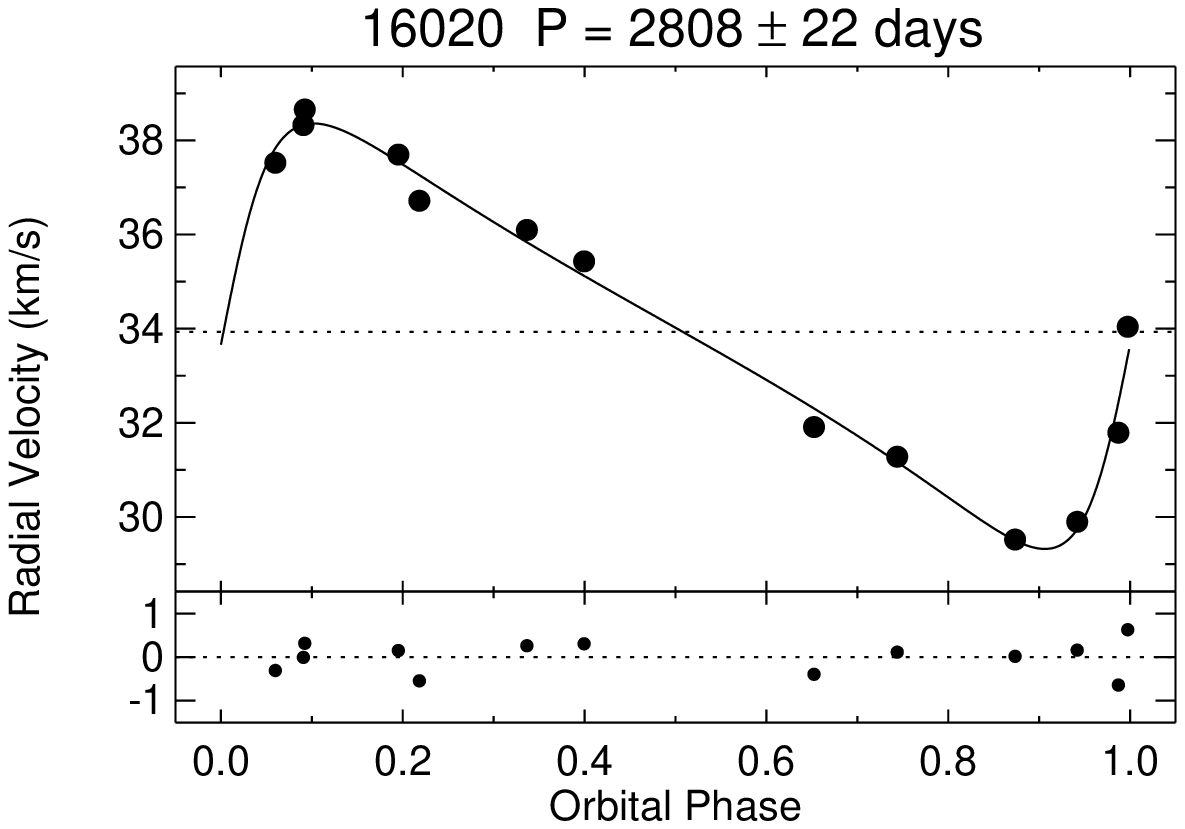}{0.3\textwidth}{}
	\fig{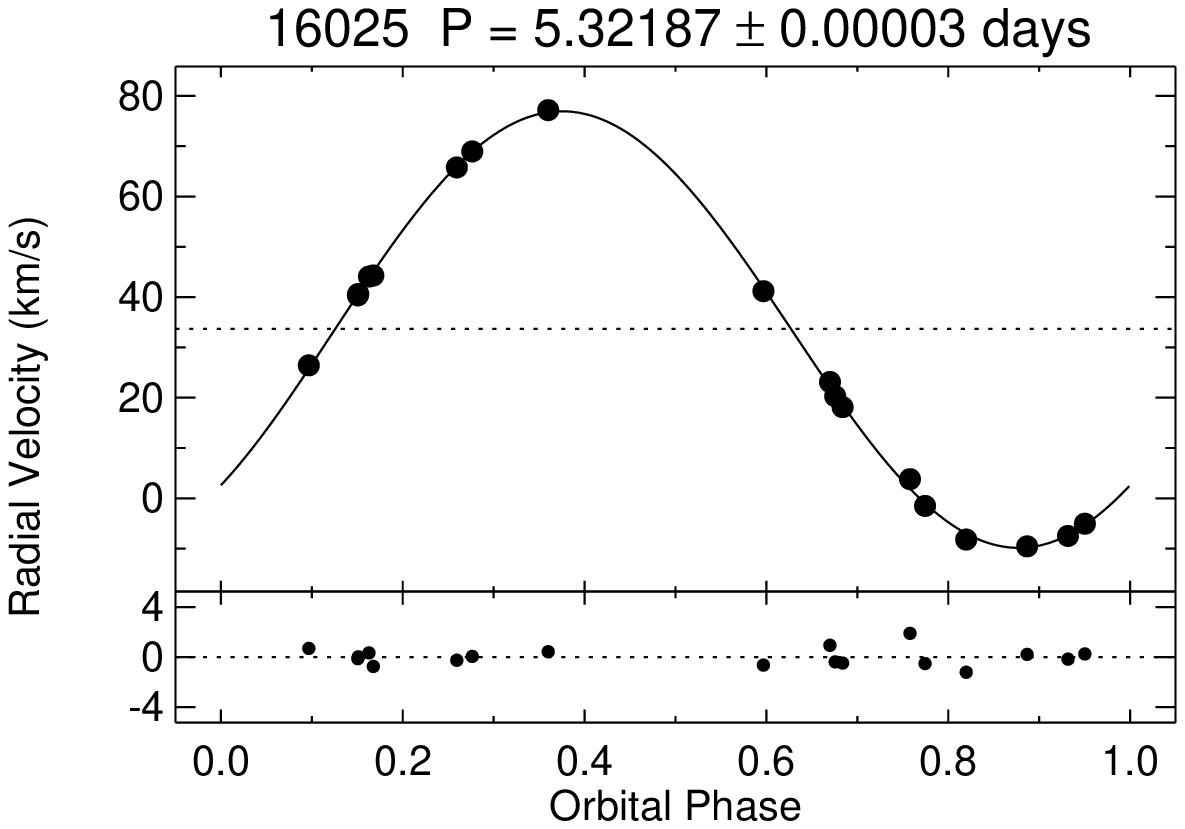}{0.3\textwidth}{}
	\fig{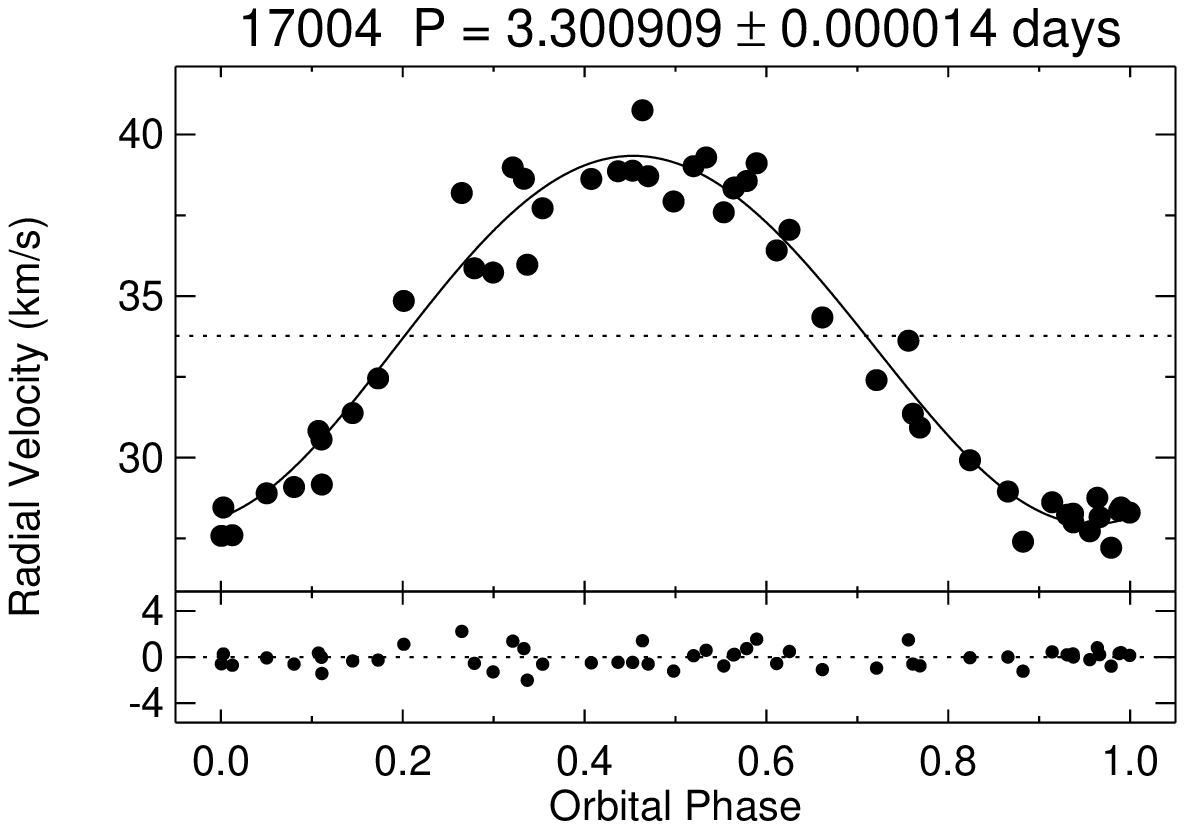}{0.3\textwidth}{}}
\caption{(Continued.)}
\end{figure*}

\begin{figure*}
\figurenum{11}
\gridline{\fig{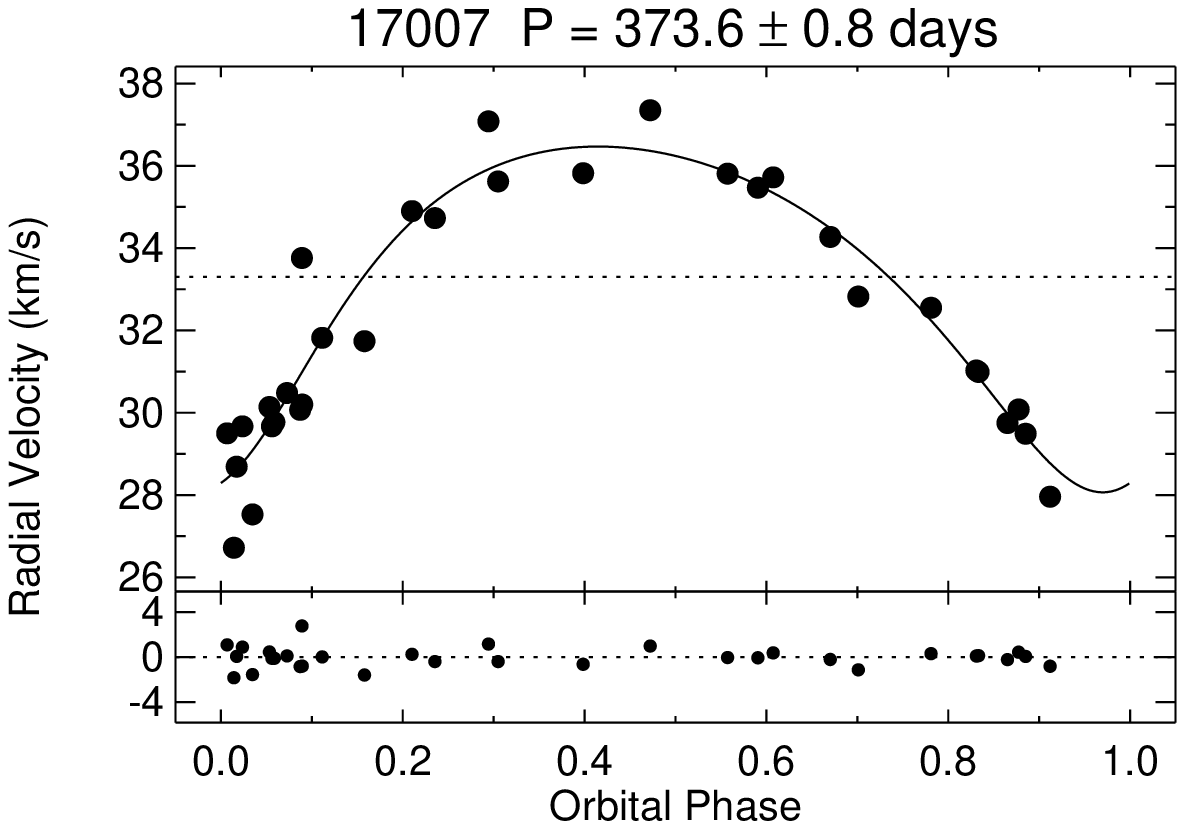}{0.3\textwidth}{}
	\fig{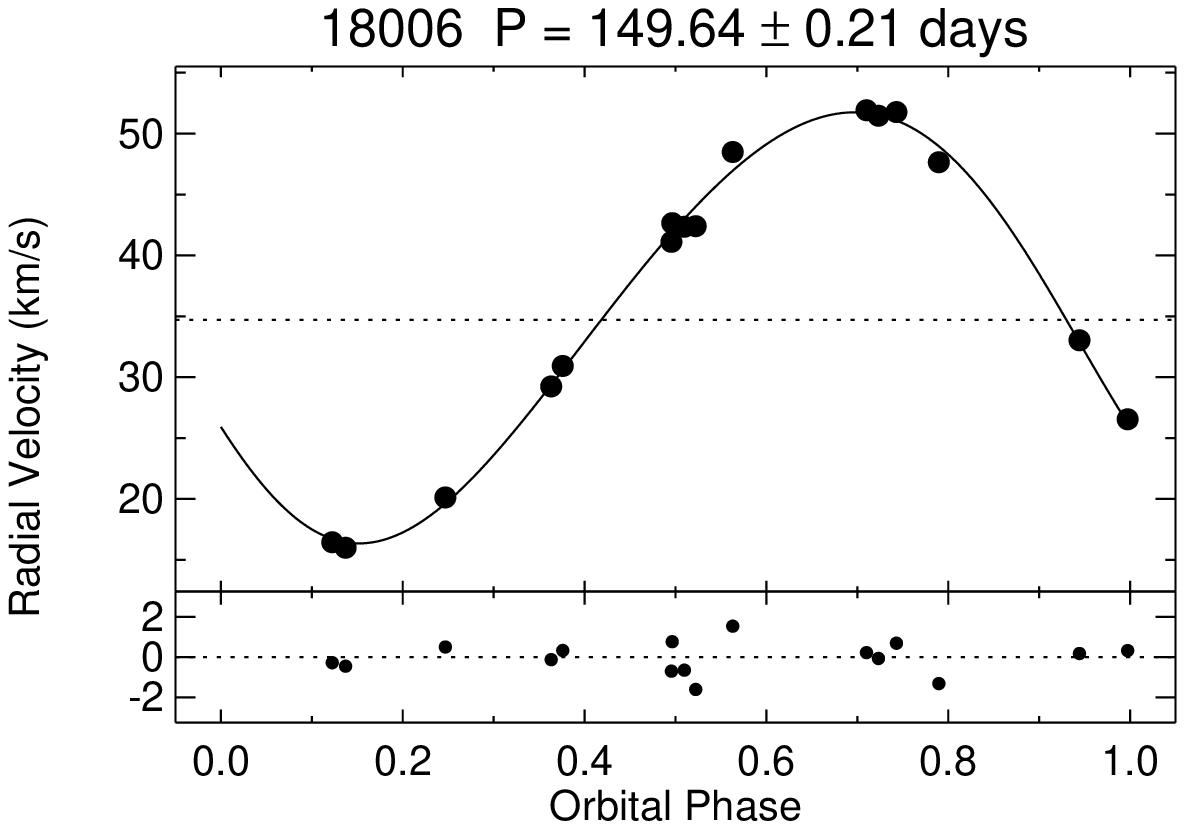}{0.3\textwidth}{}
	\fig{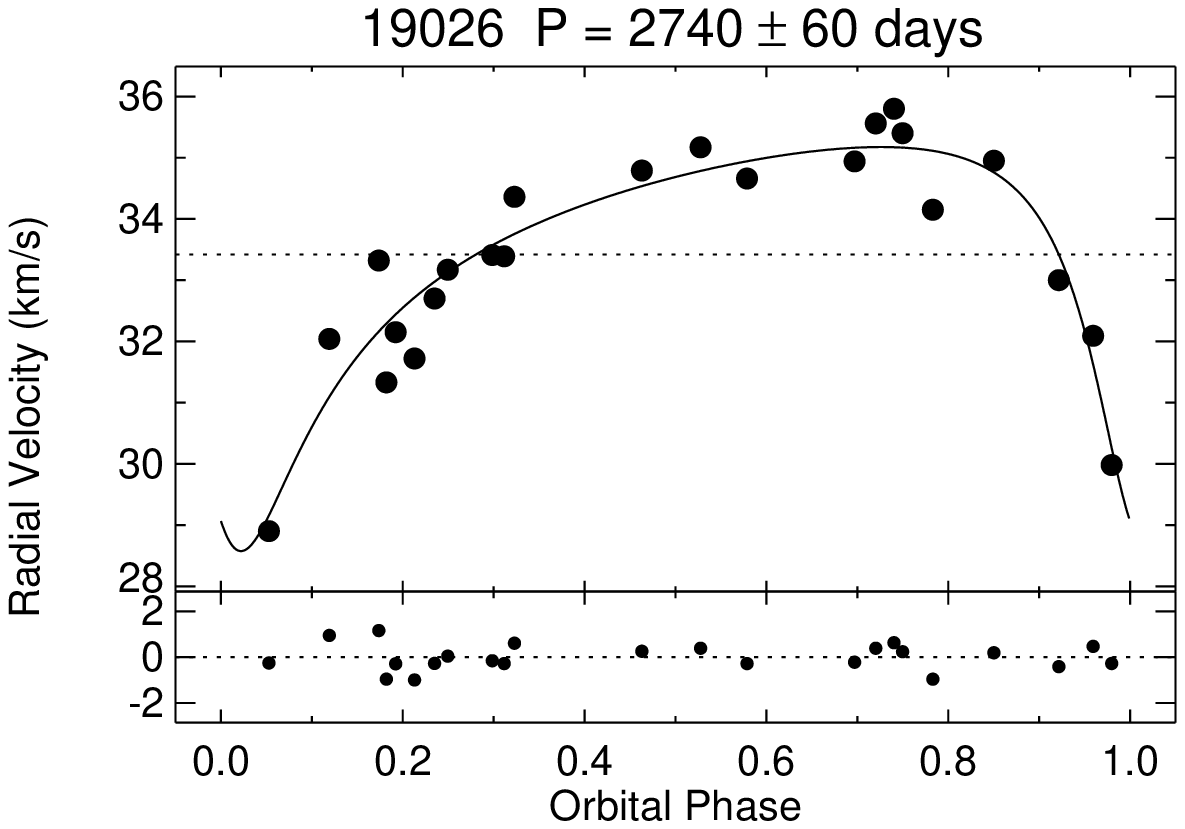}{0.3\textwidth}{}}
\gridline{\fig{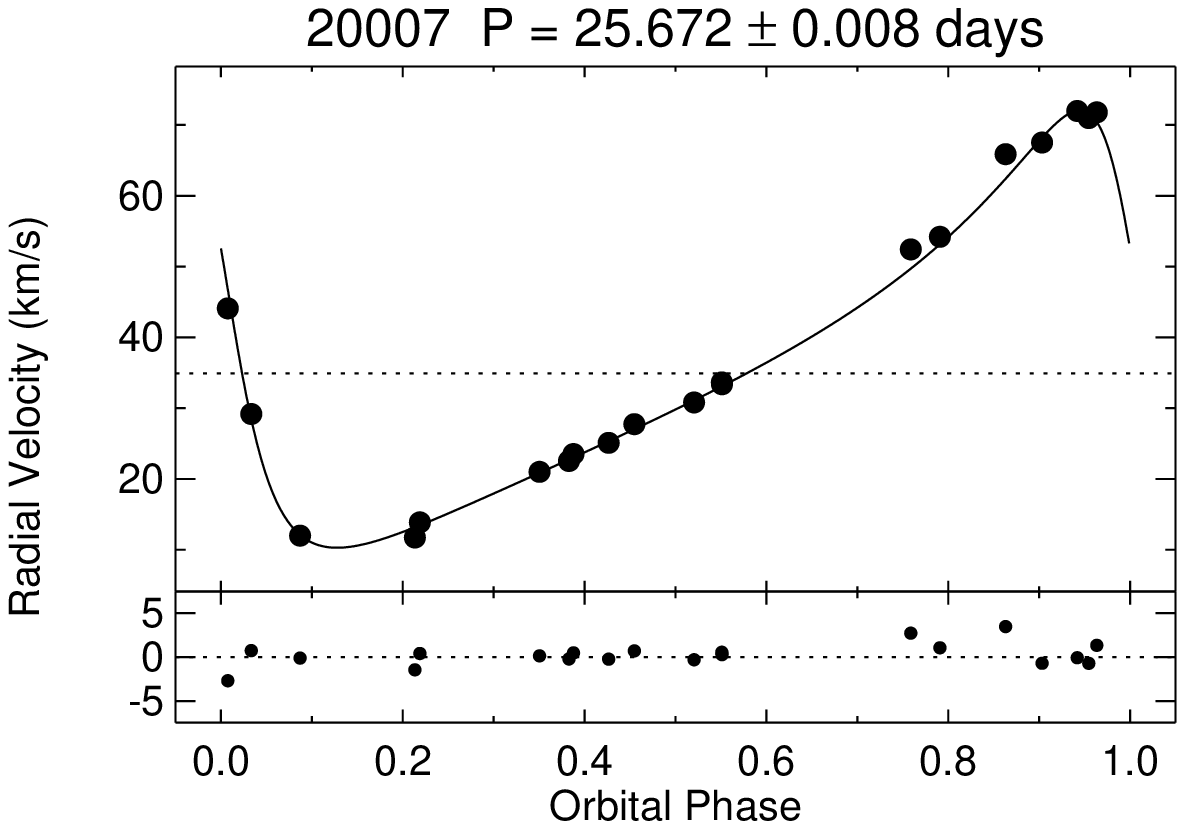}{0.3\textwidth}{}
	\fig{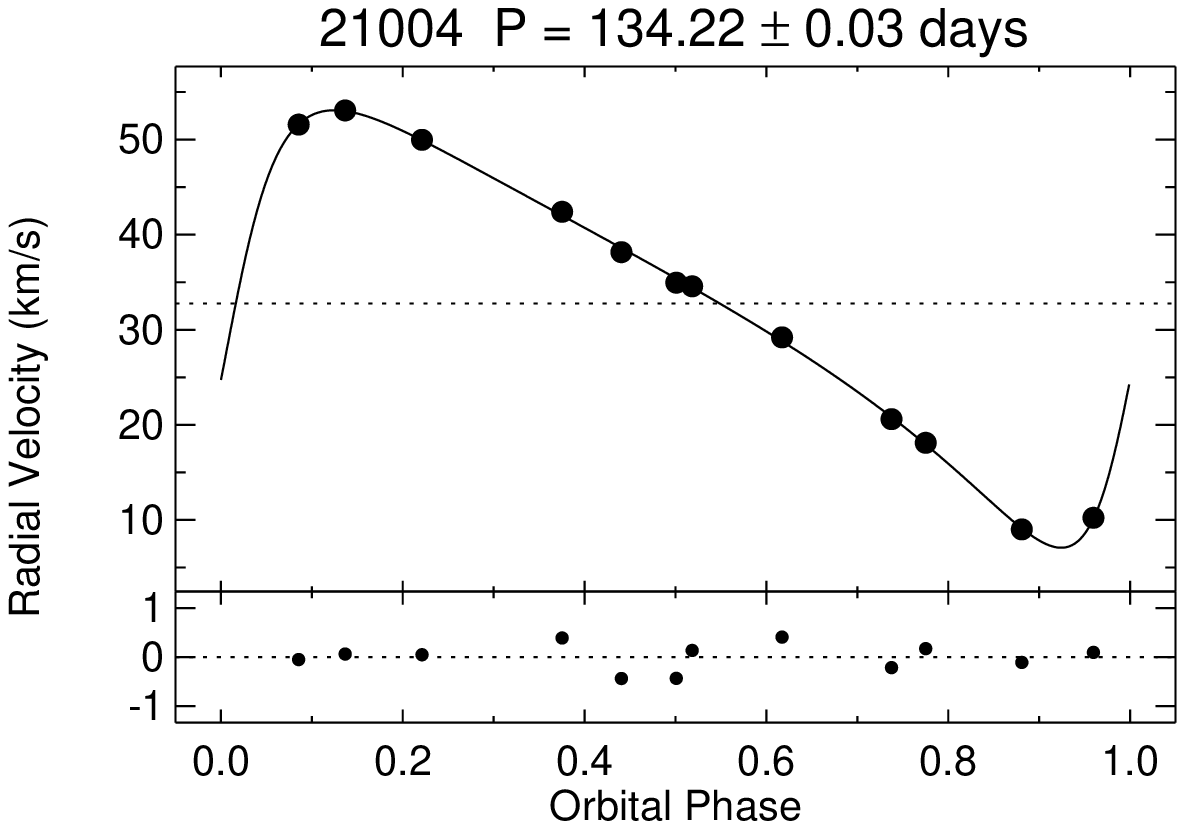}{0.3\textwidth}{}
	\fig{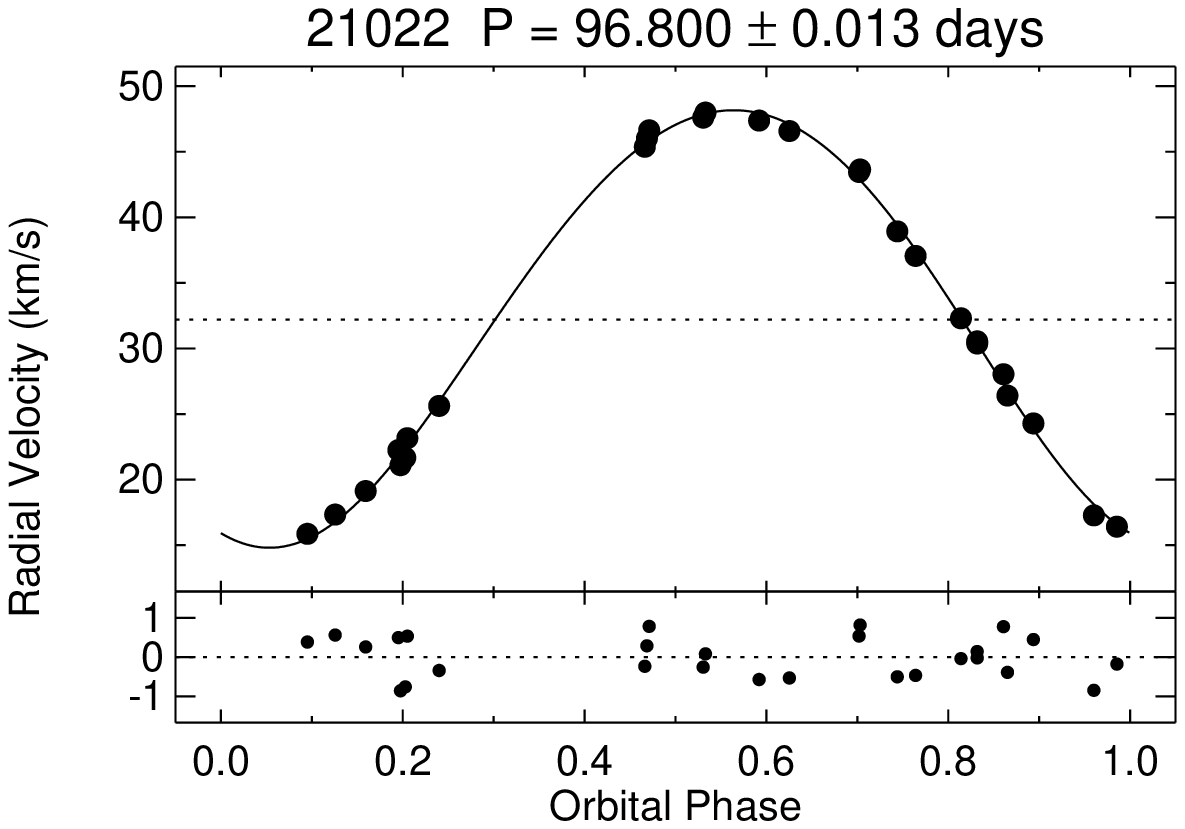}{0.3\textwidth}{}}
\gridline{\fig{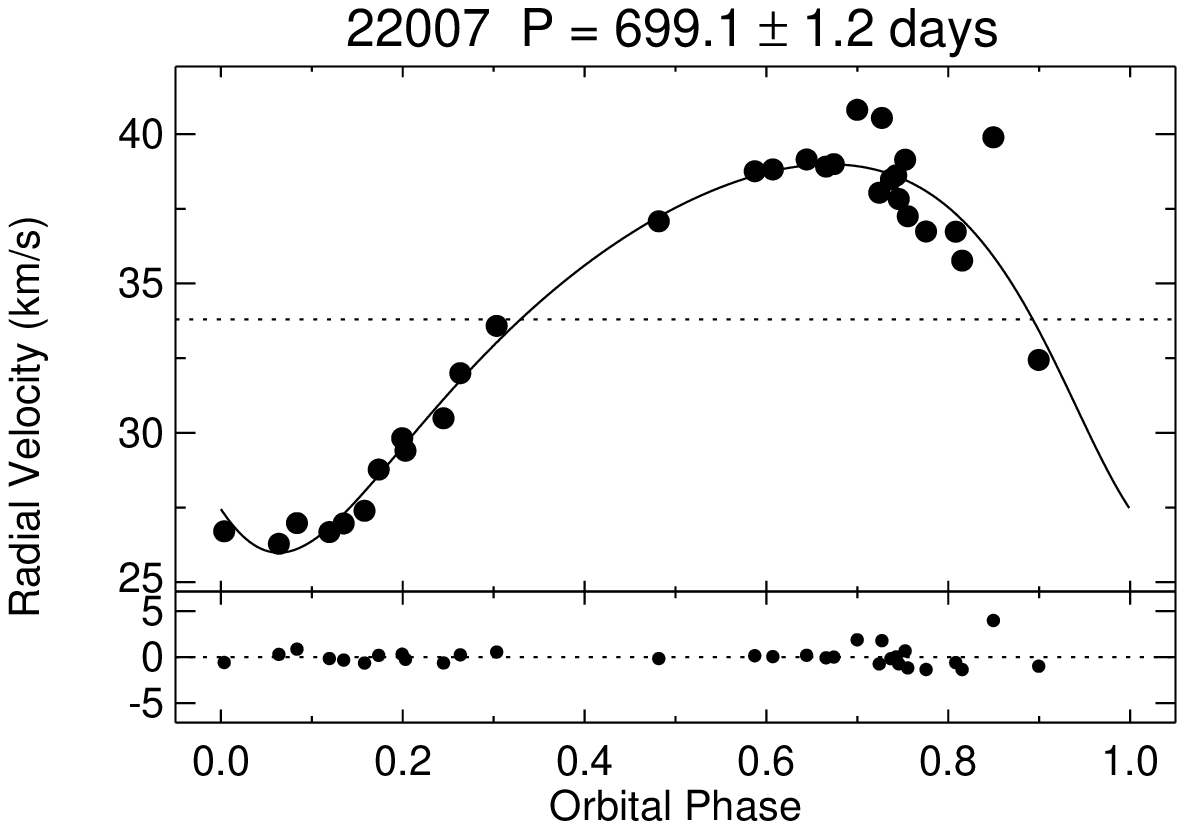}{0.3\textwidth}{}
	\fig{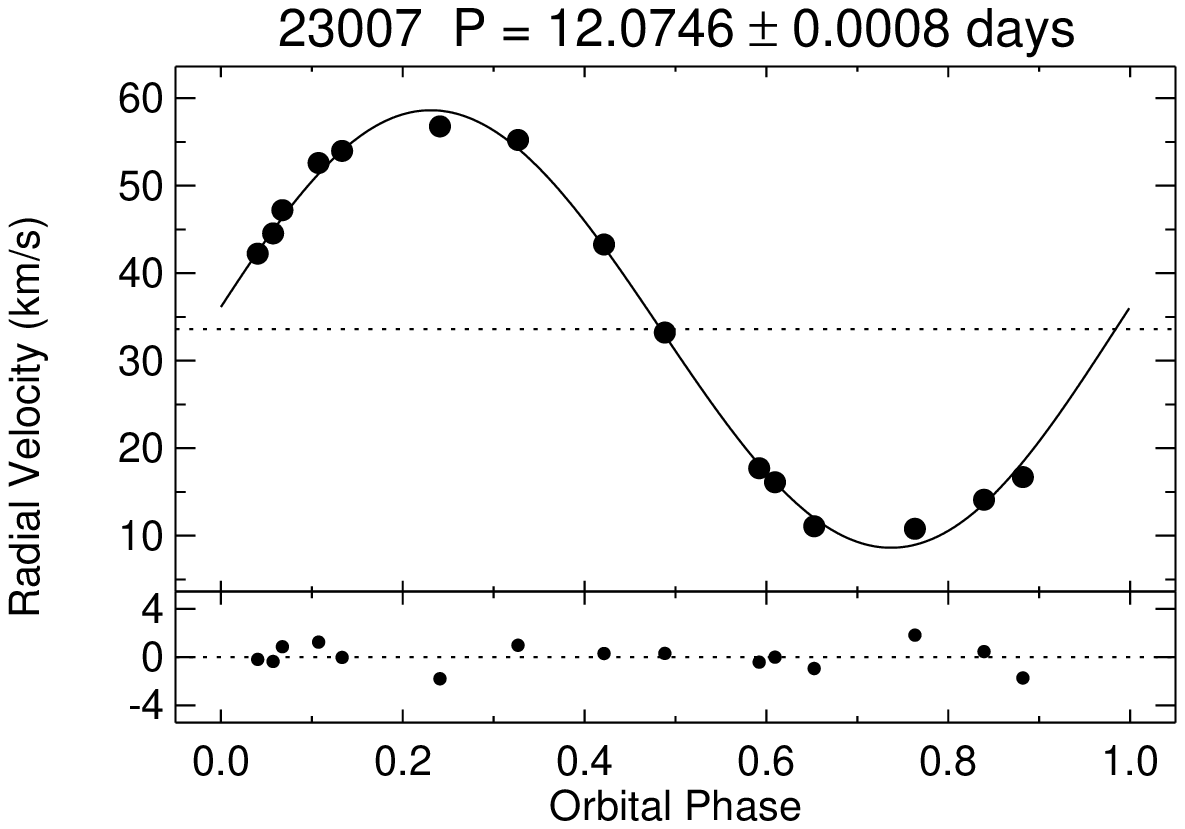}{0.3\textwidth}{}
	\fig{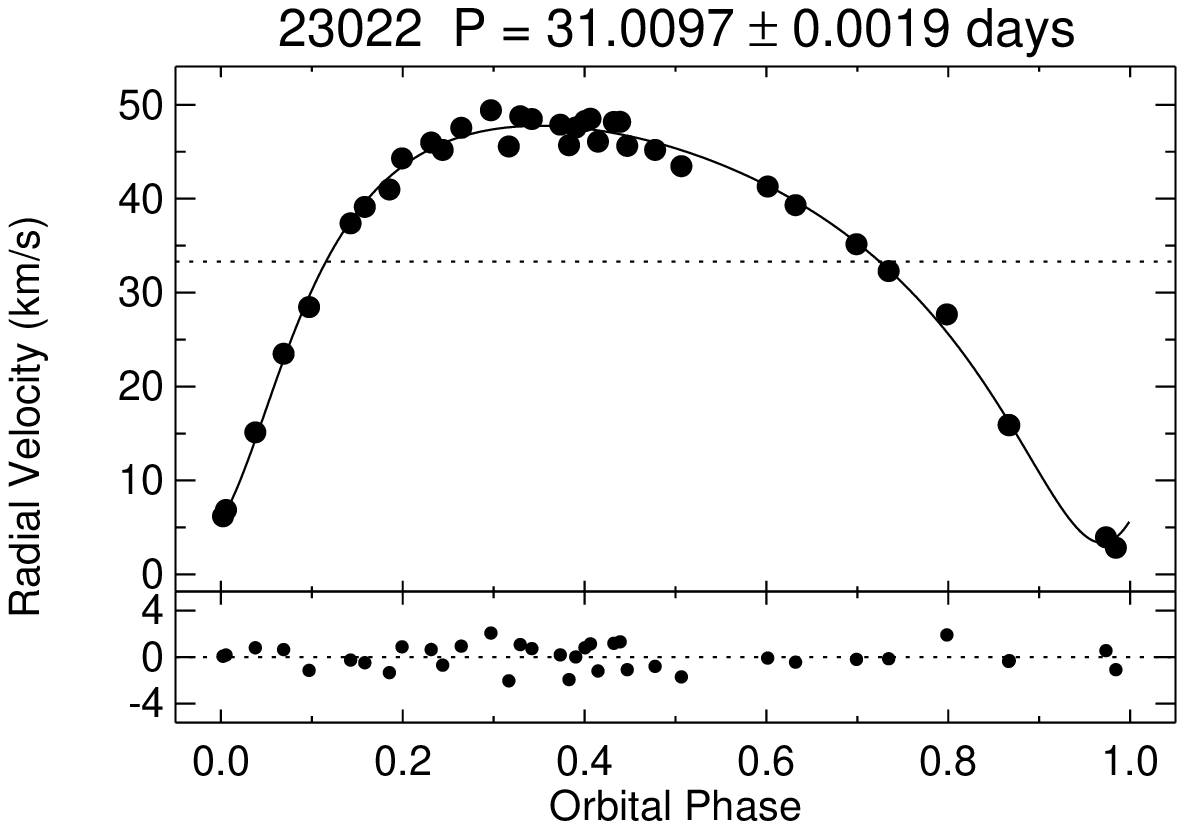}{0.3\textwidth}{}}
\gridline{\fig{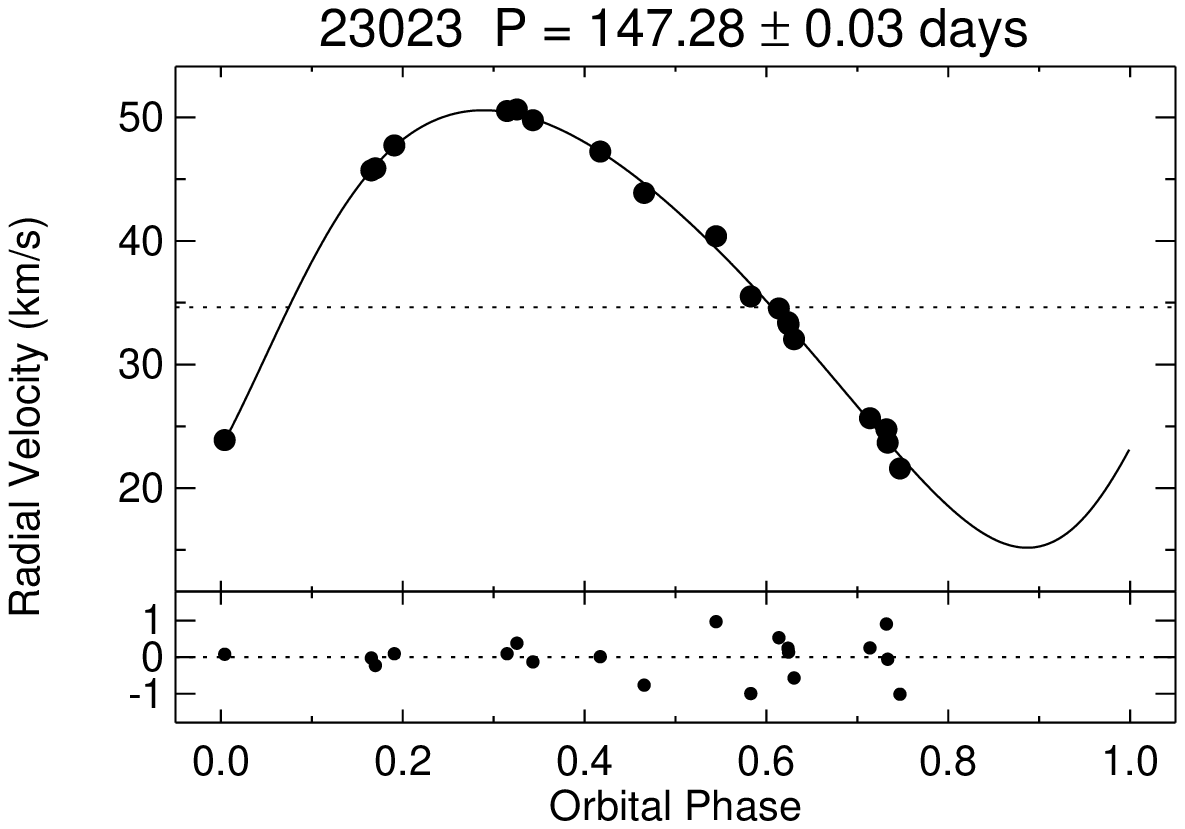}{0.3\textwidth}{}
	\fig{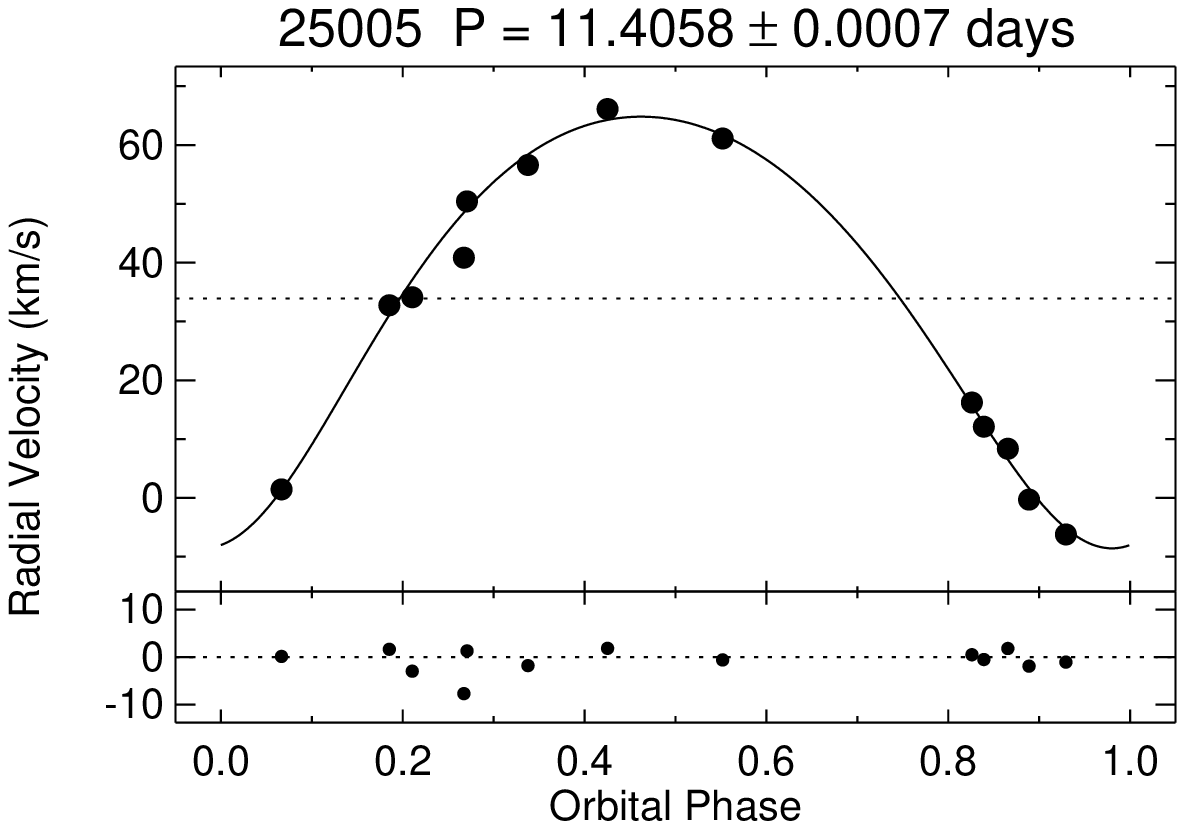}{0.3\textwidth}{}}
\caption{(Continued.)}
\end{figure*}

\begin{longrotatetable}
\begin{deluxetable*}{llrcrrrrcccrcc}
\tabletypesize{\footnotesize}
\tablecaption{Orbital Parameters for M67 Double-Lined Binaries\label{tab:sb2pars}}
\tablehead{\colhead{ID} & \colhead{XID} & \colhead{$P$} & \colhead{Orbital} & \colhead{$\gamma$} & \colhead{$K$} & \colhead{$e$} & \colhead{$\omega$} & \colhead{$T_\circ$} & \colhead{$a\sin i$} & \colhead{$m\sin^3 i$} & \colhead{$q$} & \colhead{$\sigma$} & \colhead{$N$} \\
\colhead{} & \colhead{} & \colhead{(days)} & \colhead{Cycles} & \colhead{(\kms)} & \colhead{(\kms)} & \colhead{} & \colhead{(deg)} & \colhead{(HJD-2400000 d)} & \colhead{(10$^6$ km)} & \colhead{(\Msolar)} & \colhead{} & \colhead{(\kms)} & \colhead{}}
\startdata
    1001 & S1024 &         7.15957 & 242.3 &           33.33 &            68.0 &           0.005 &              90 &         46593.4 &            6.70 &           0.950 &           0.992 &  1.41 &   34 \\
         &       &   $\pm$ 0.00007 &       &      $\pm$ 0.17 &       $\pm$ 0.3 &     $\pm$ 0.004 &        $\pm$ 40 &       $\pm$ 0.8 &      $\pm$ 0.03 &     $\pm$ 0.011 &     $\pm$ 0.007 &       &      \\
         &       &                 &       &                 &            68.6 &                 &                 &                 &            6.75 &           0.942 &                 &  1.46 &   34 \\
         &       &                 &       &                 &       $\pm$ 0.3 &                 &                 &                 &      $\pm$ 0.04 &     $\pm$ 0.011 &                 &       &      \\
    2003 & S1045 &         7.64528 & 243.8 &           34.13 &            63.4 &           0.008 &             340 &         46322.8 &            6.66 &           0.818 &           0.993 &  1.76 &   37 \\
         &       &   $\pm$ 0.00008 &       &      $\pm$ 0.20 &       $\pm$ 0.4 &     $\pm$ 0.004 &        $\pm$ 30 &       $\pm$ 0.7 &      $\pm$ 0.04 &     $\pm$ 0.012 &     $\pm$ 0.009 &       &      \\
         &       &                 &       &                 &            63.8 &                 &                 &                 &            6.71 &           0.812 &                 &  1.69 &   37 \\
         &       &                 &       &                 &       $\pm$ 0.4 &                 &                 &                 &      $\pm$ 0.04 &     $\pm$ 0.012 &                 &       &      \\
\enddata
\end{deluxetable*}
\end{longrotatetable}

\begin{figure*}
\gridline{\fig{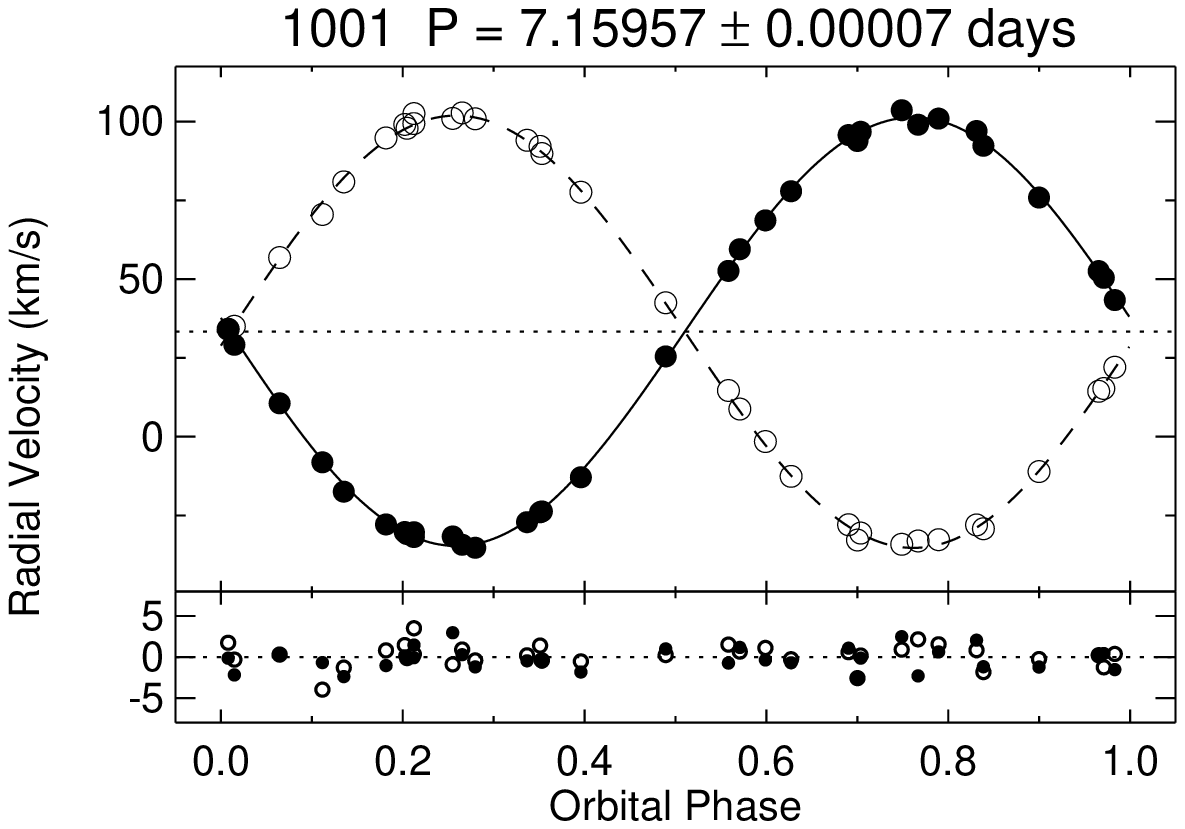}{0.3\textwidth}{}
	\fig{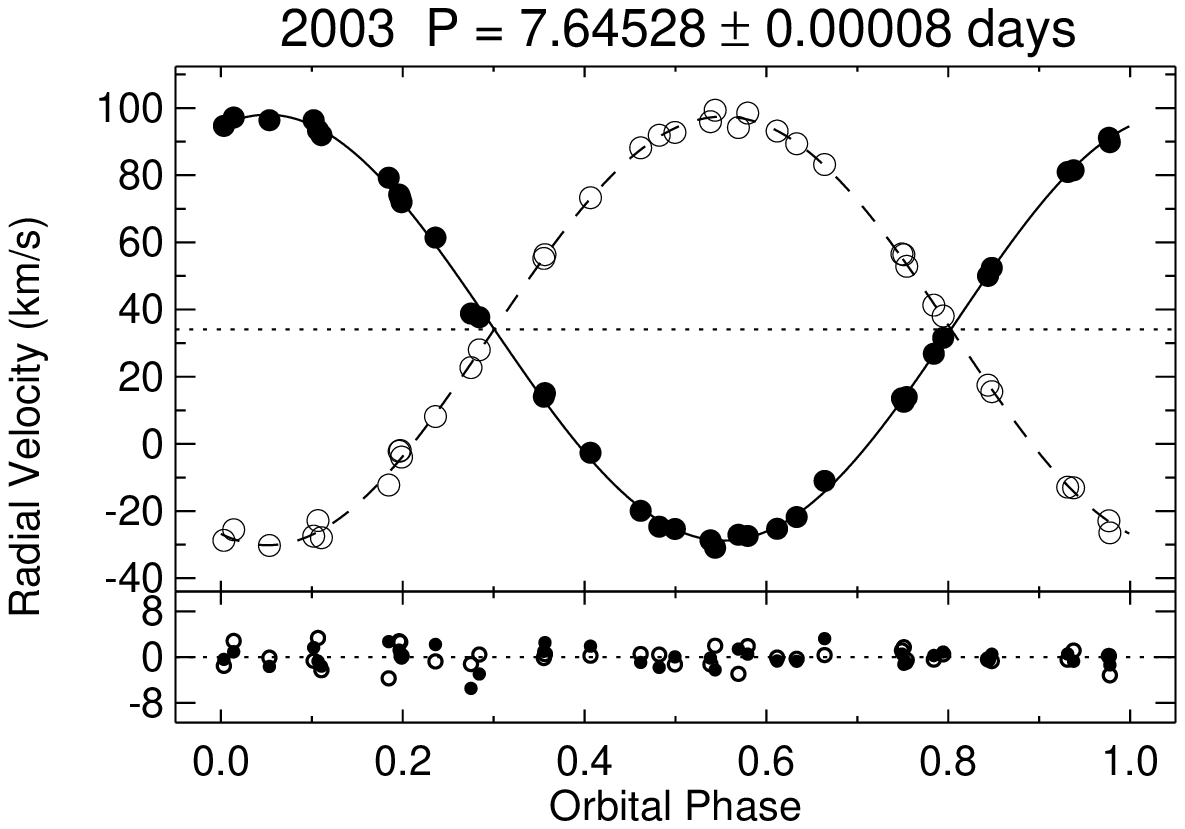}{0.3\textwidth}{}
	\fig{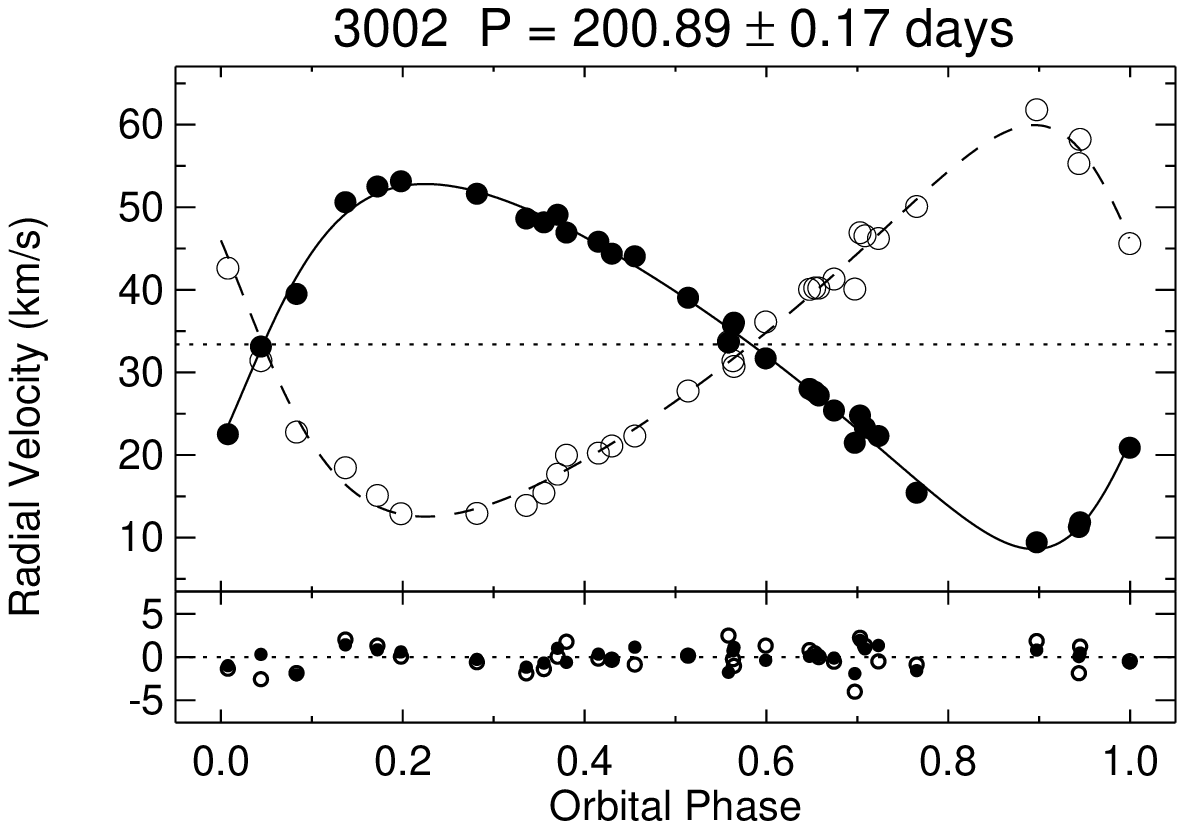}{0.3\textwidth}{}}
\gridline{\fig{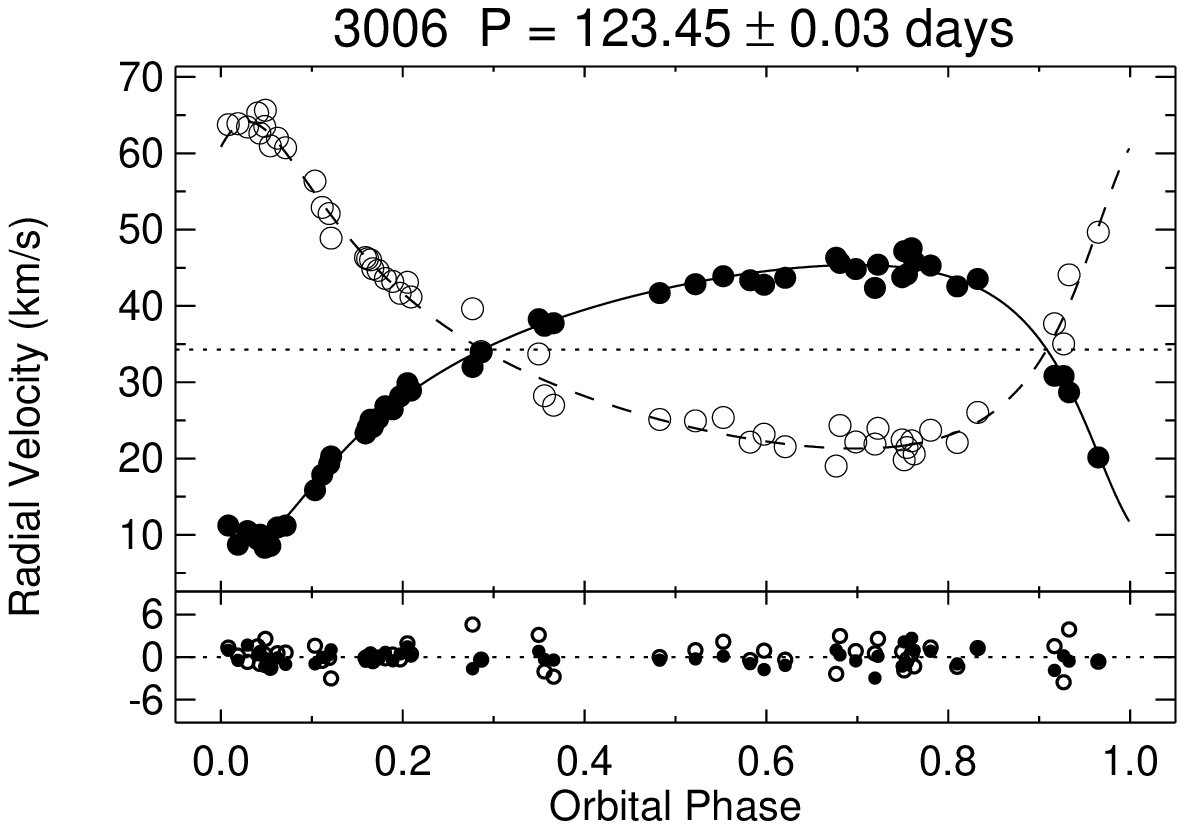}{0.3\textwidth}{}
	\fig{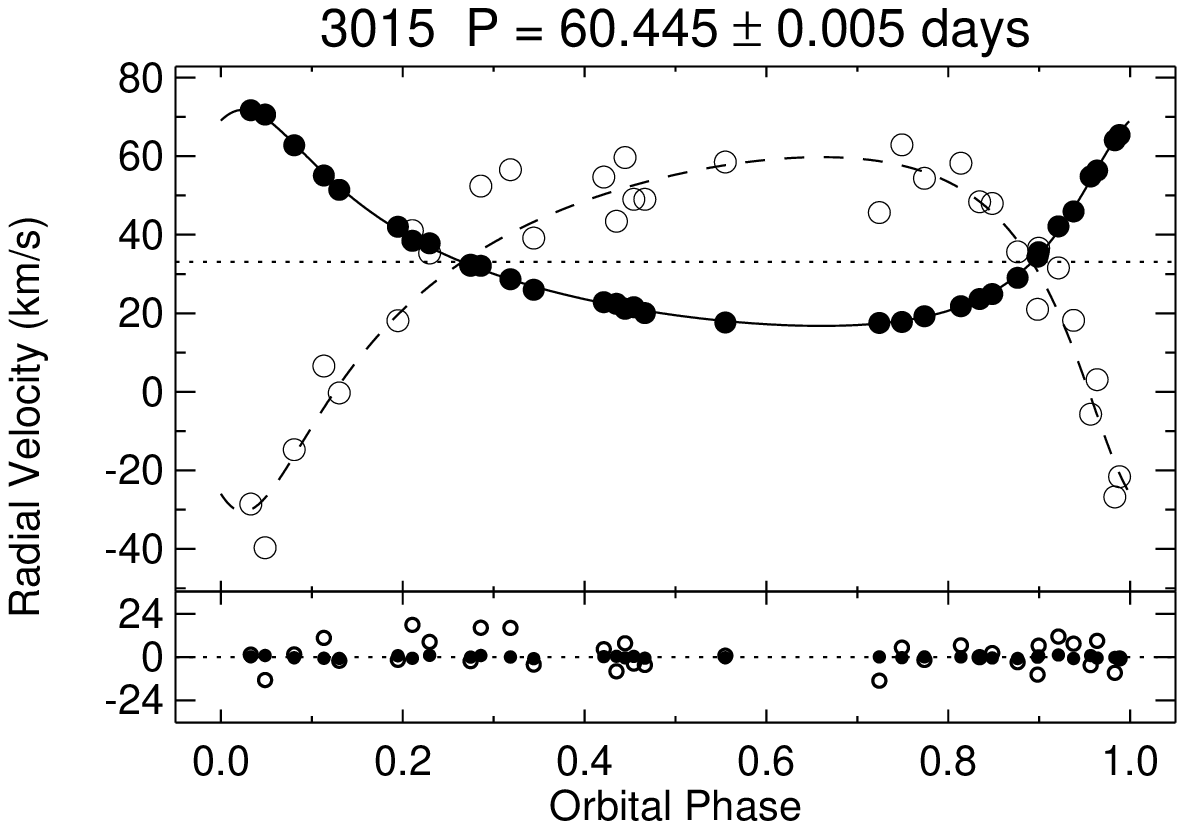}{0.3\textwidth}{}
	\fig{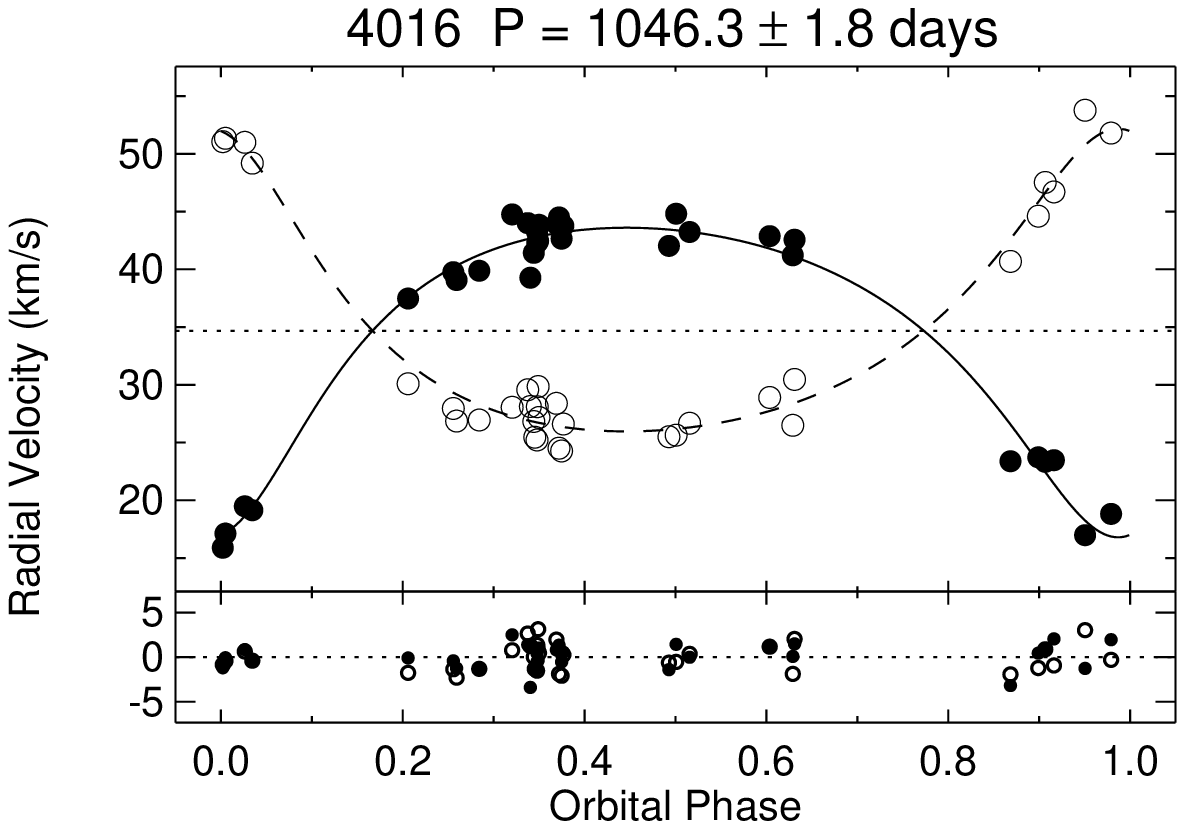}{0.3\textwidth}{}}
\gridline{\fig{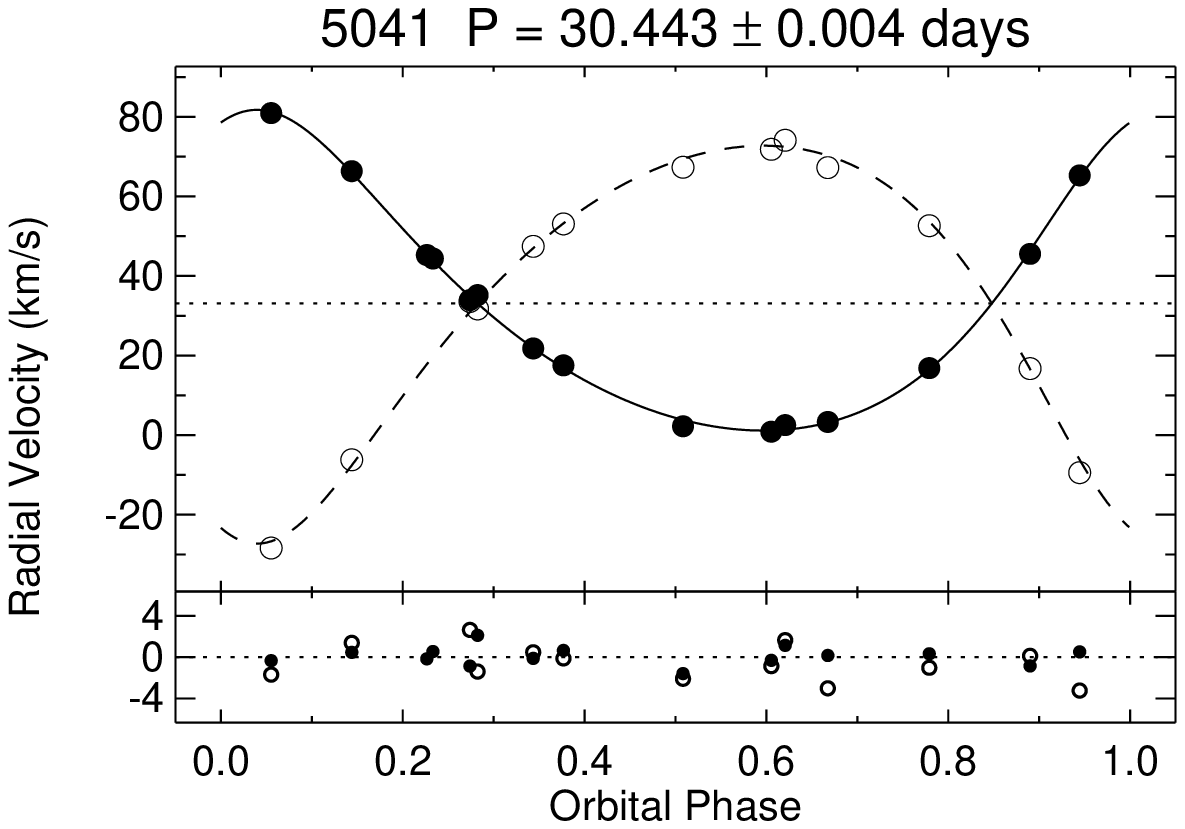}{0.3\textwidth}{}
	\fig{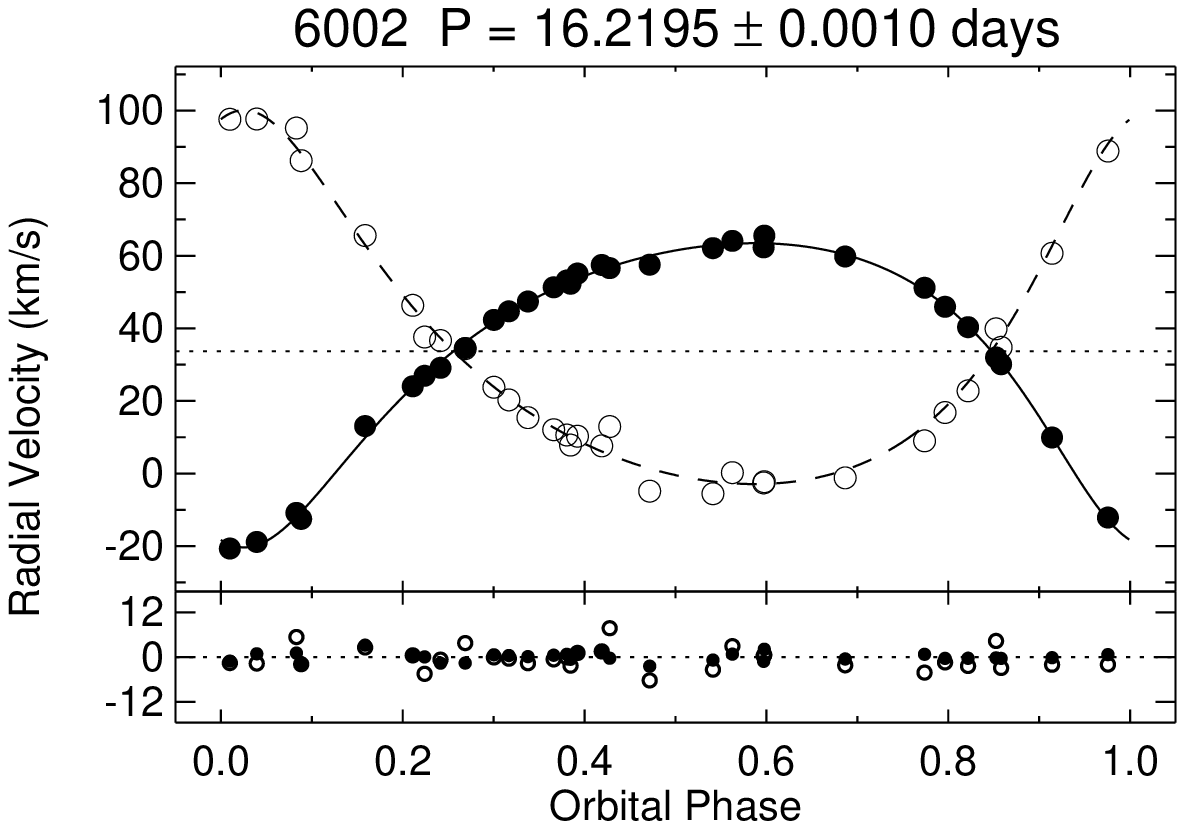}{0.3\textwidth}{}
	\fig{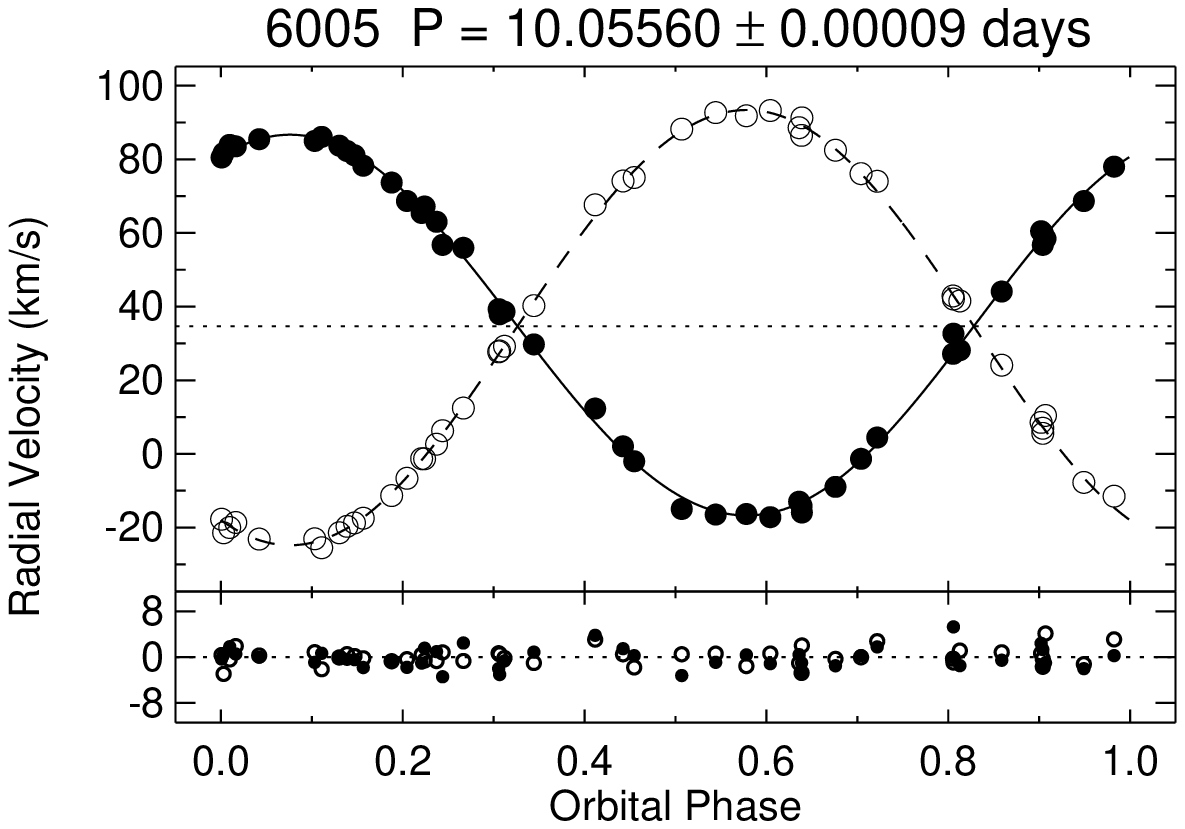}{0.3\textwidth}{}}
\gridline{\fig{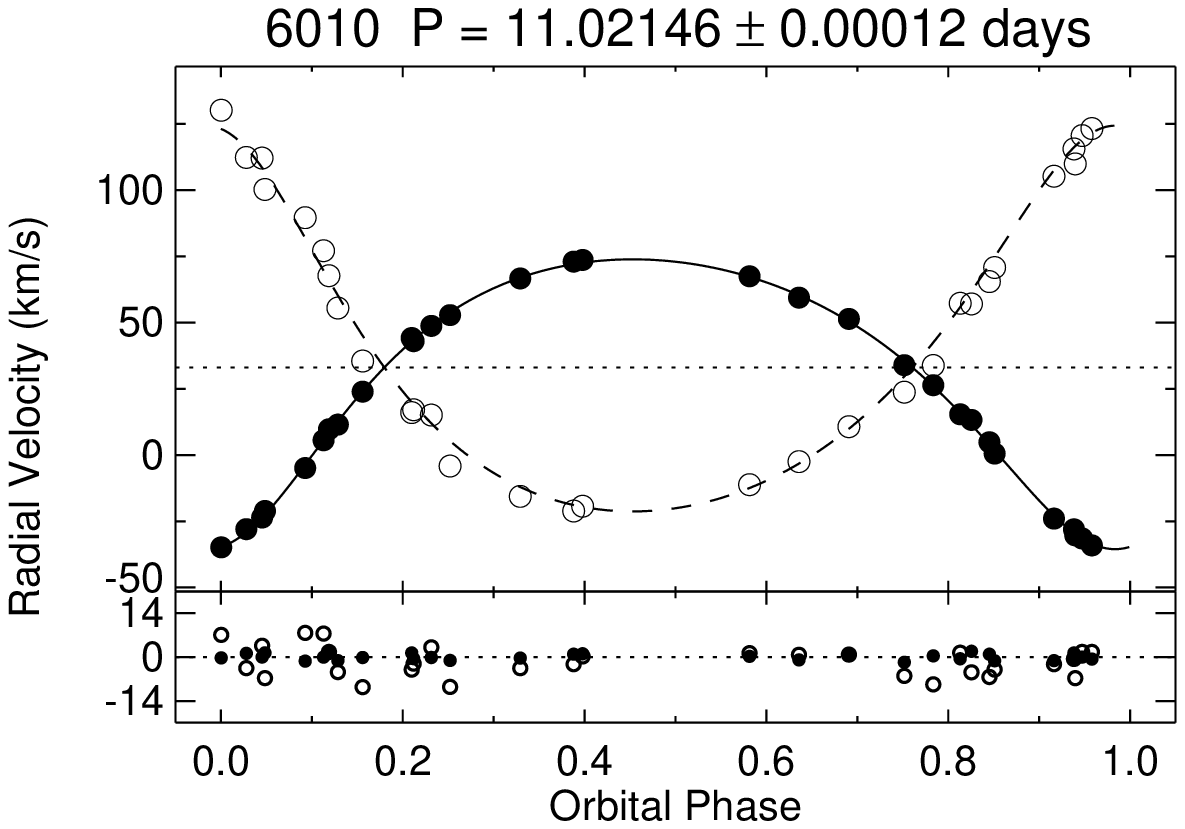}{0.3\textwidth}{}
	\fig{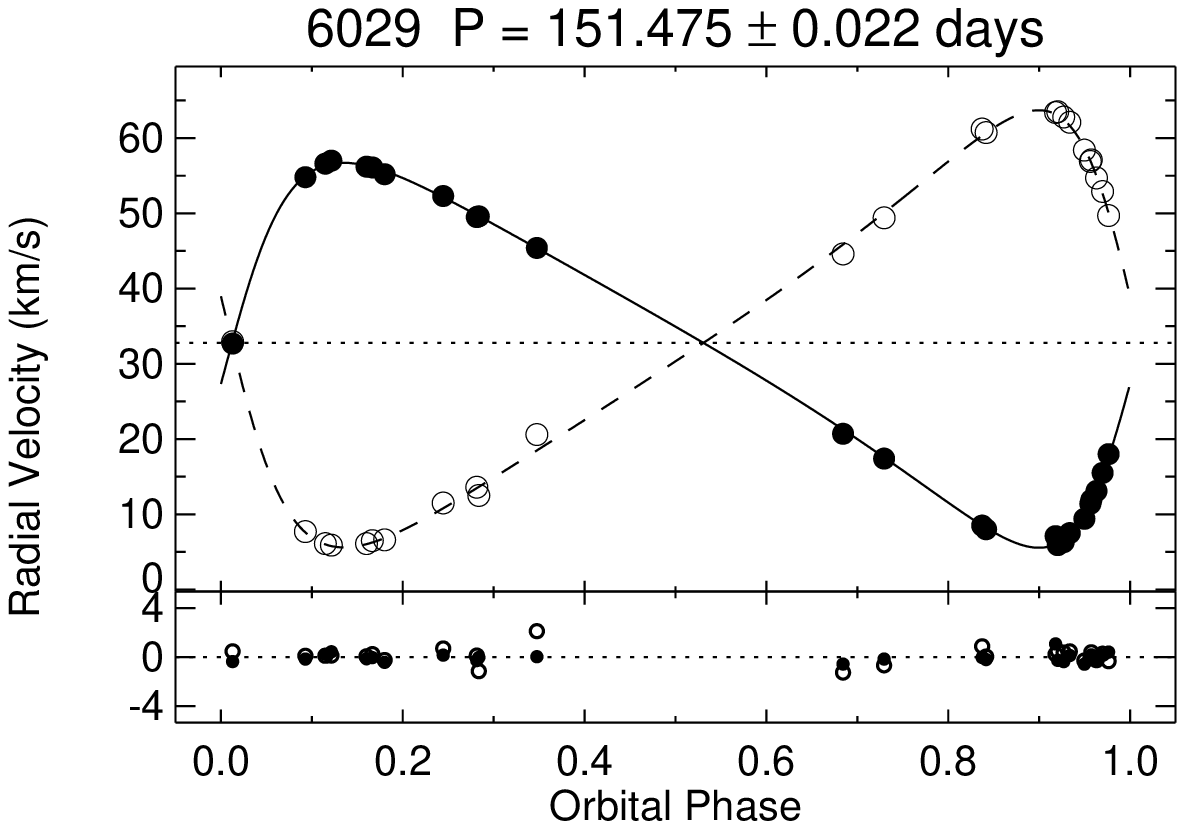}{0.3\textwidth}{}
	\fig{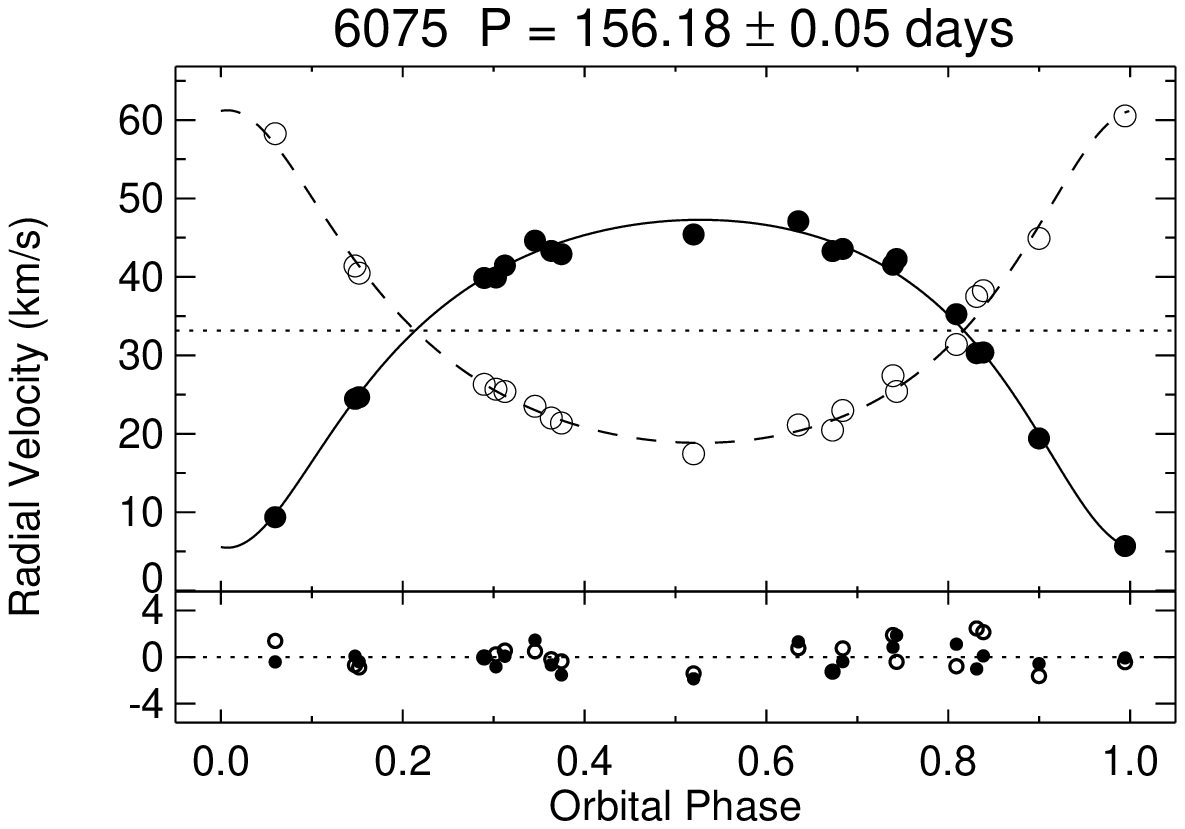}{0.3\textwidth}{}}
\caption{Double-lined M67 member orbit plots. Radial velocity (RV) is plotted versus phase, with primary RVs shown as filled circles, and secondary RVs in open circles. The primary orbital solution is shown as a solid black line, the secondary solution as a dashed black line, with the dotted black line marking the $\gamma$ velocity of the binary. Below each plot, the primary and secondary residuals ($O-C$) are given. Above each plot, we list the WOCS ID and orbital period.\label{fig:sb2orbs}}
\end{figure*}

\begin{figure*}
\figurenum{12}
\gridline{\fig{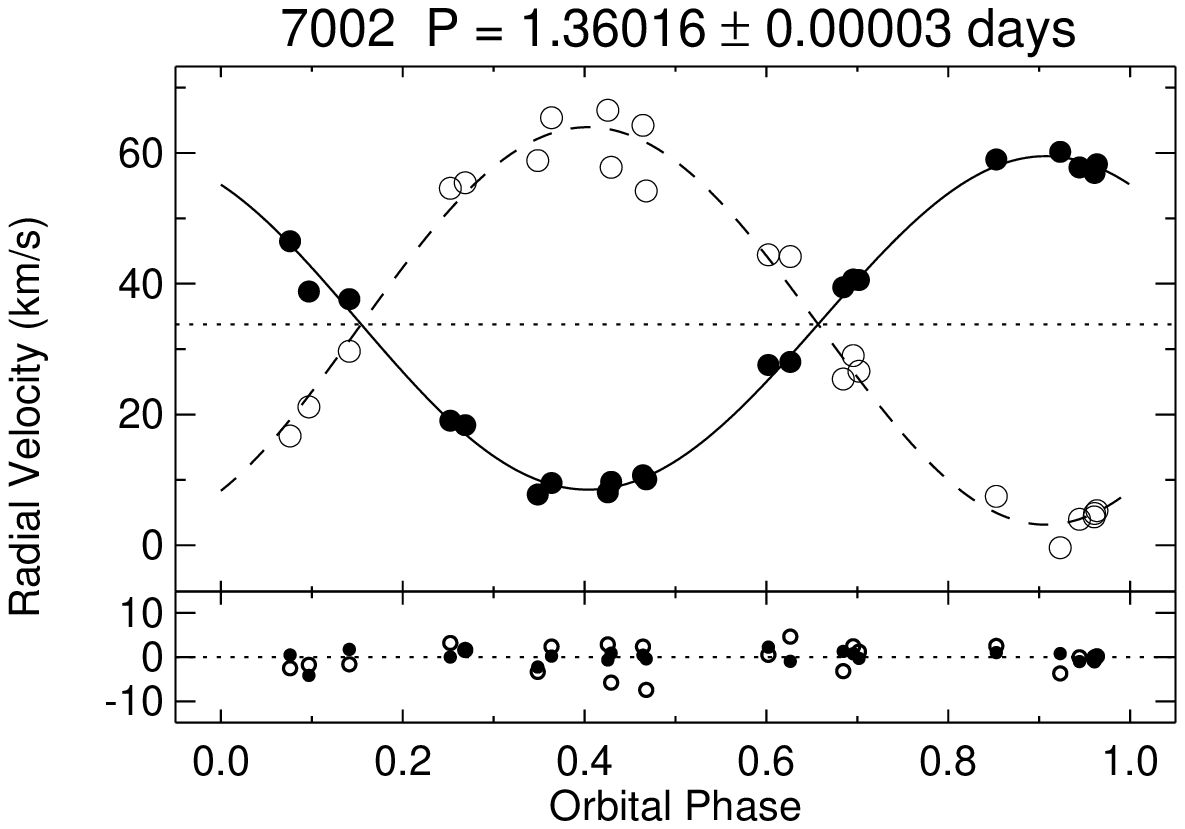}{0.3\textwidth}{}
	\fig{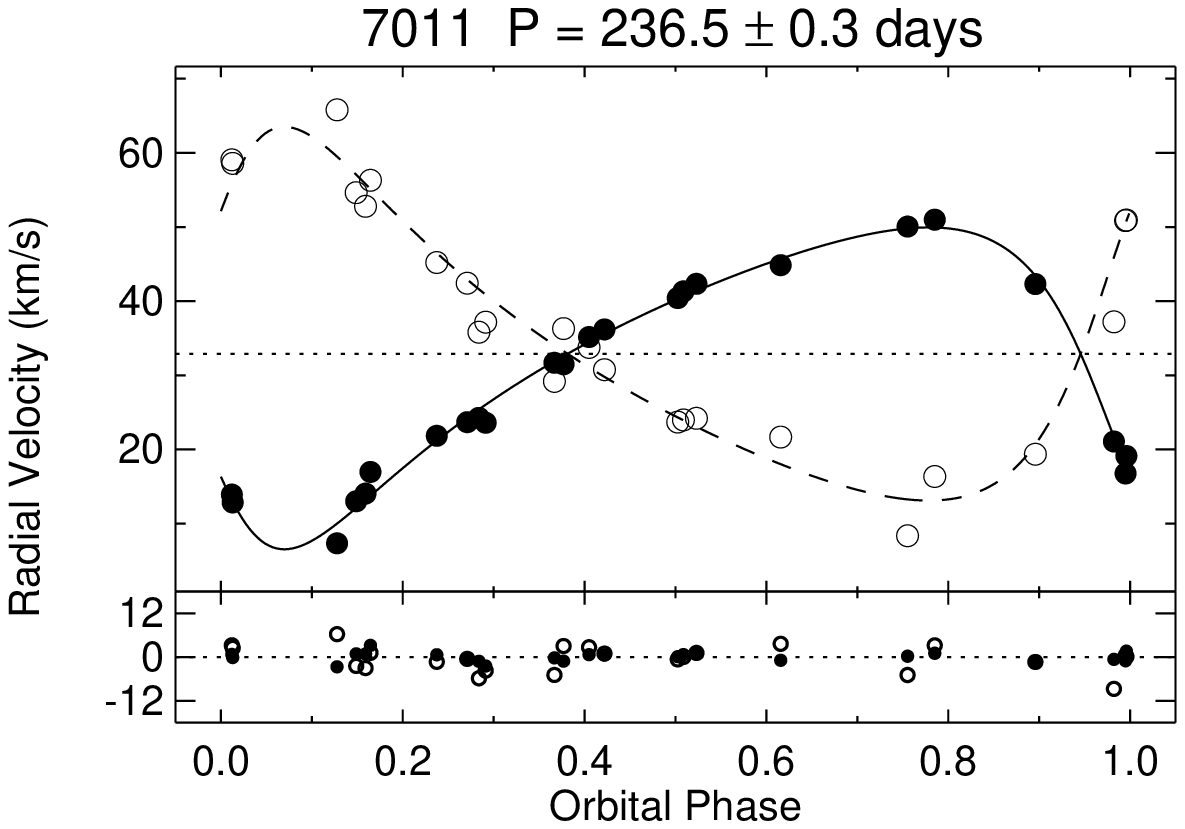}{0.3\textwidth}{}
	\fig{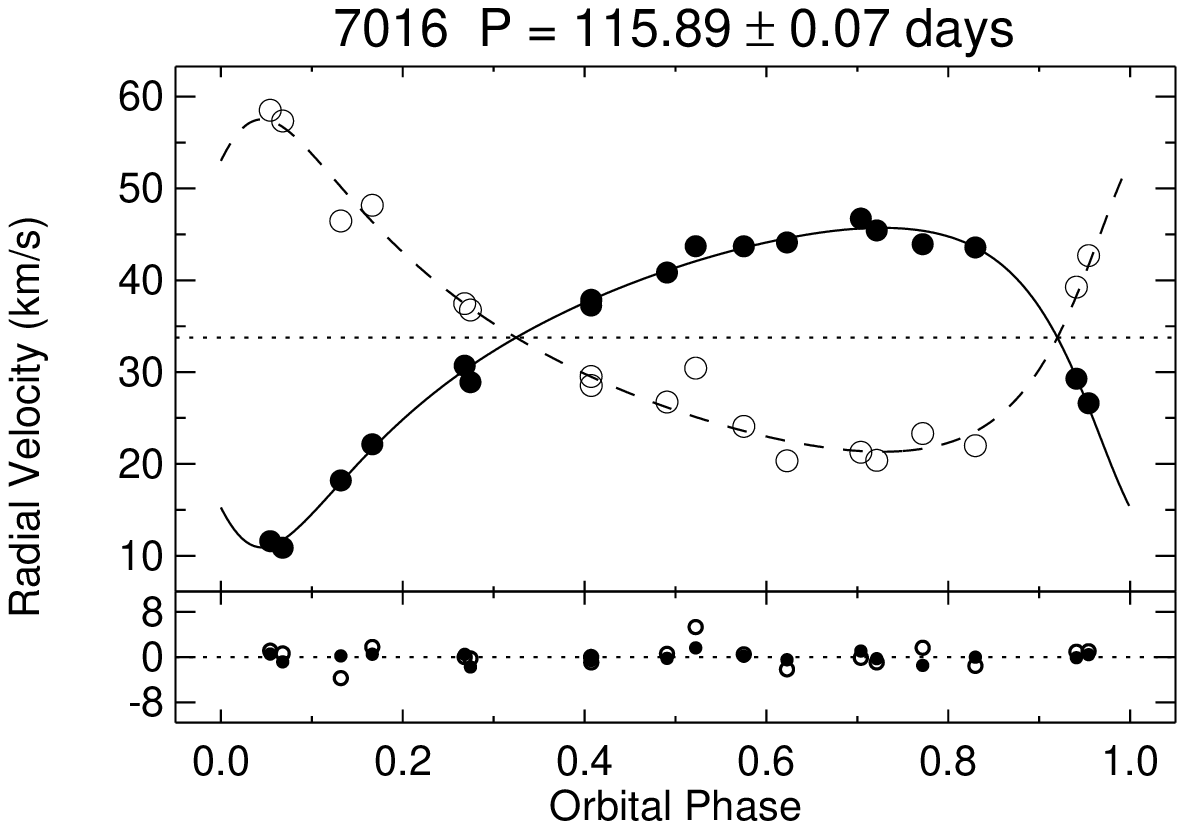}{0.3\textwidth}{}}
\gridline{\fig{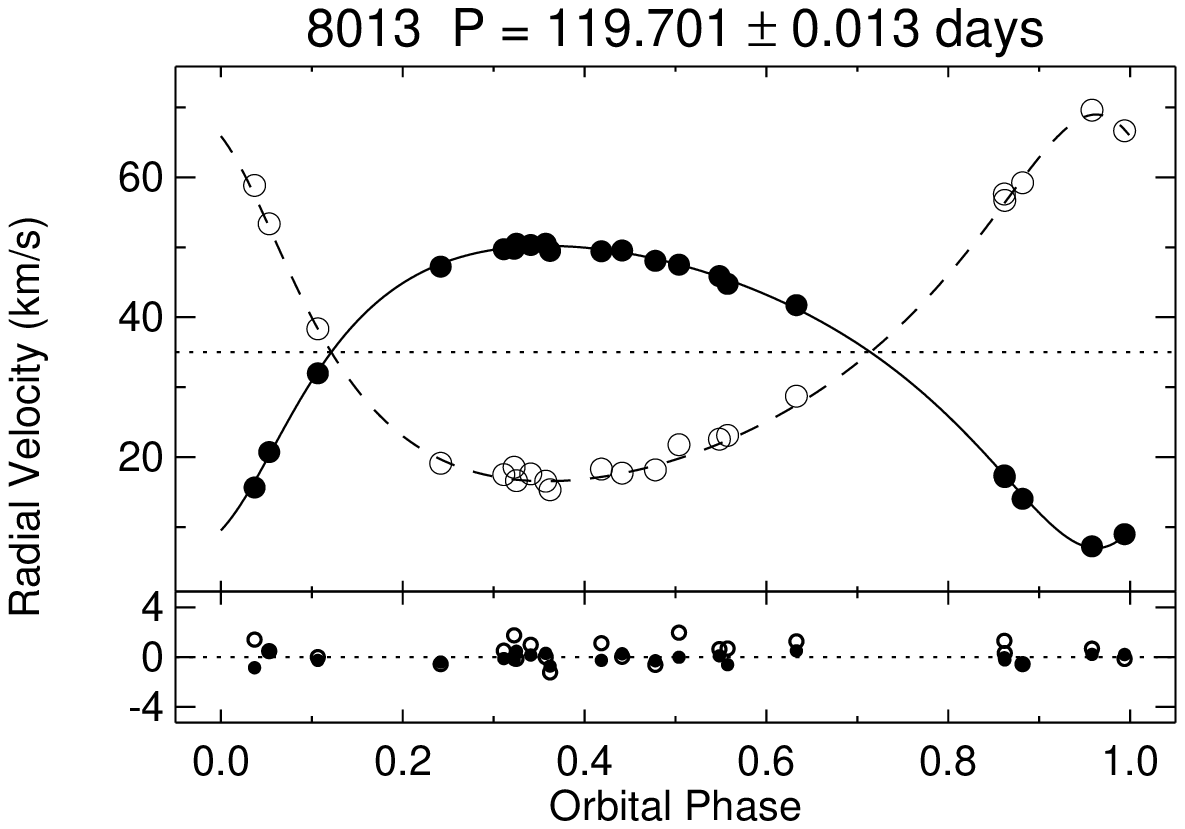}{0.3\textwidth}{}
	\fig{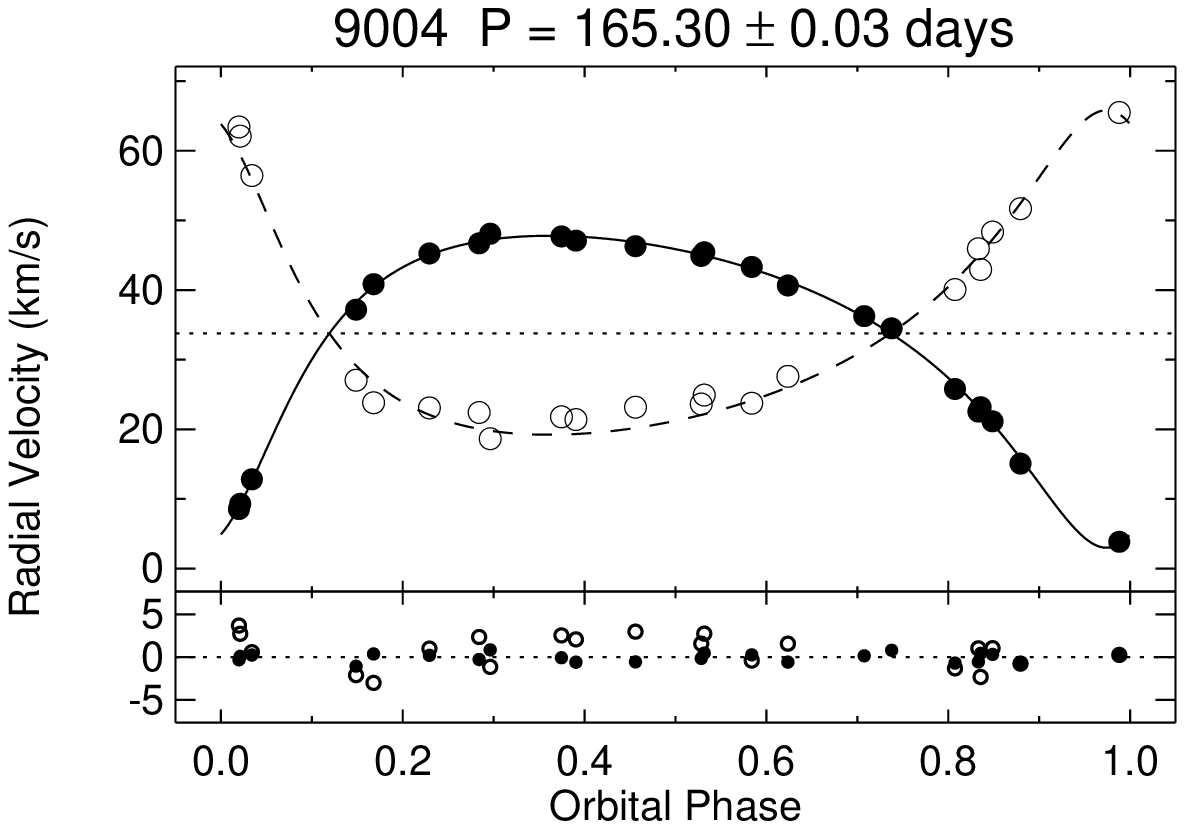}{0.3\textwidth}{}
	\fig{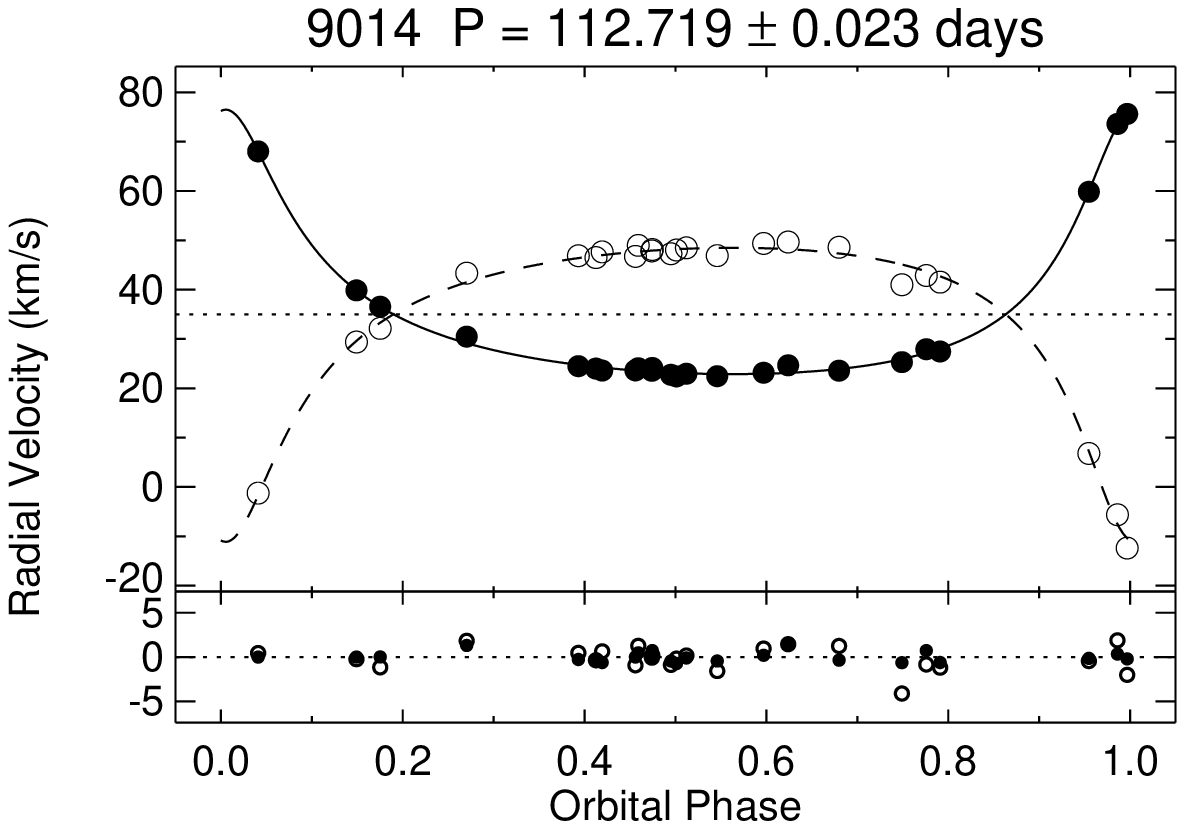}{0.3\textwidth}{}}
\gridline{\fig{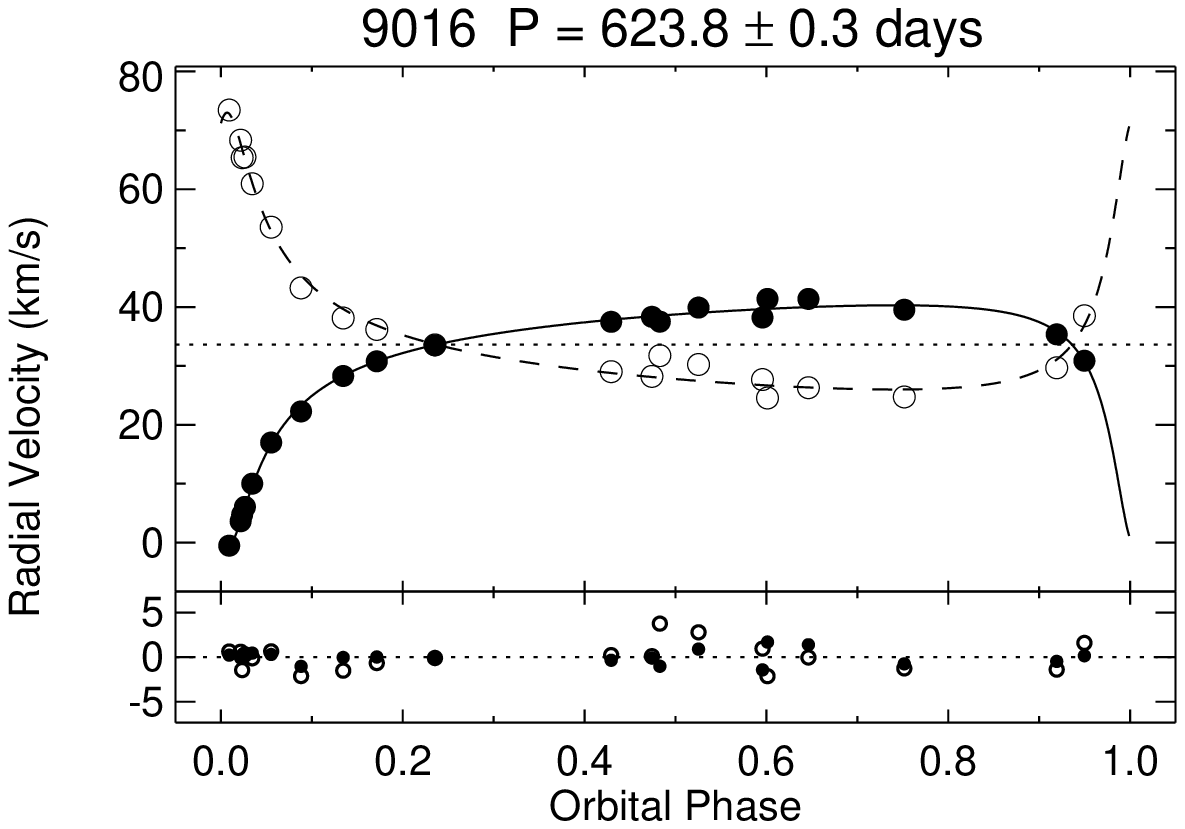}{0.3\textwidth}{}
	\fig{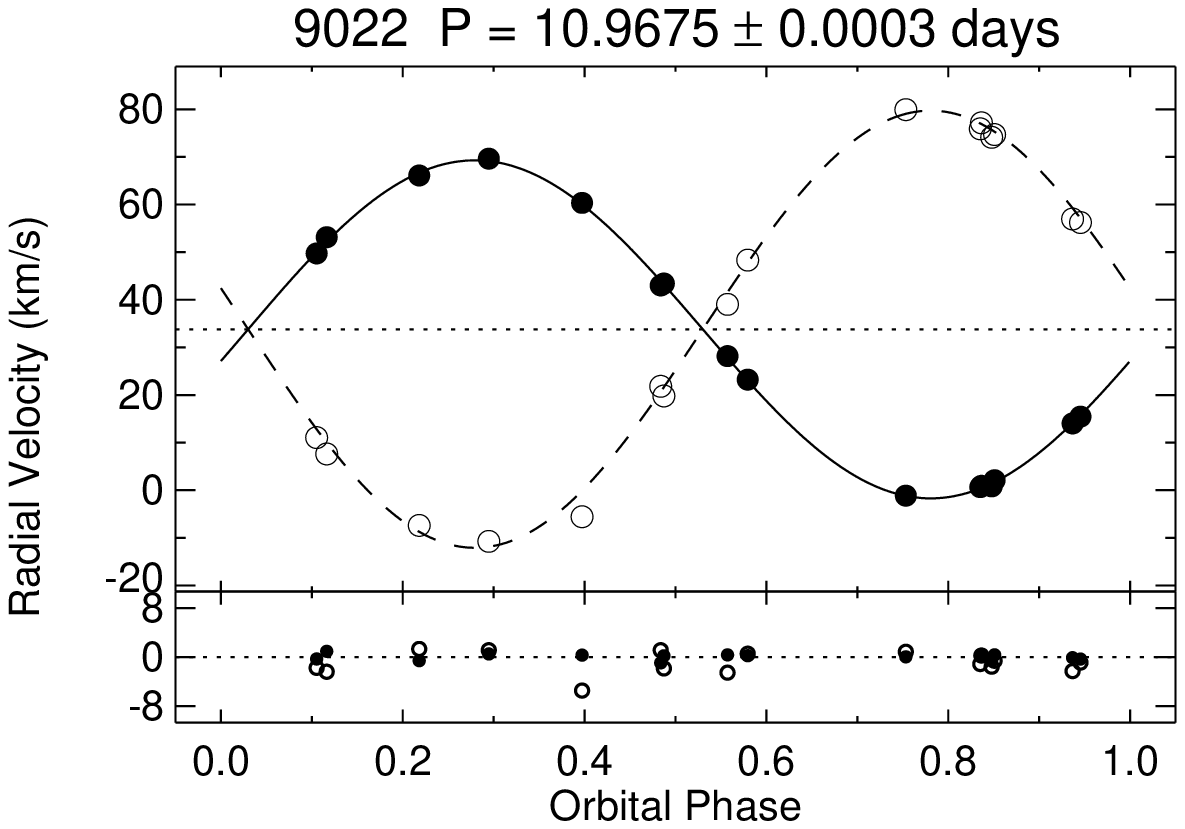}{0.3\textwidth}{}
	\fig{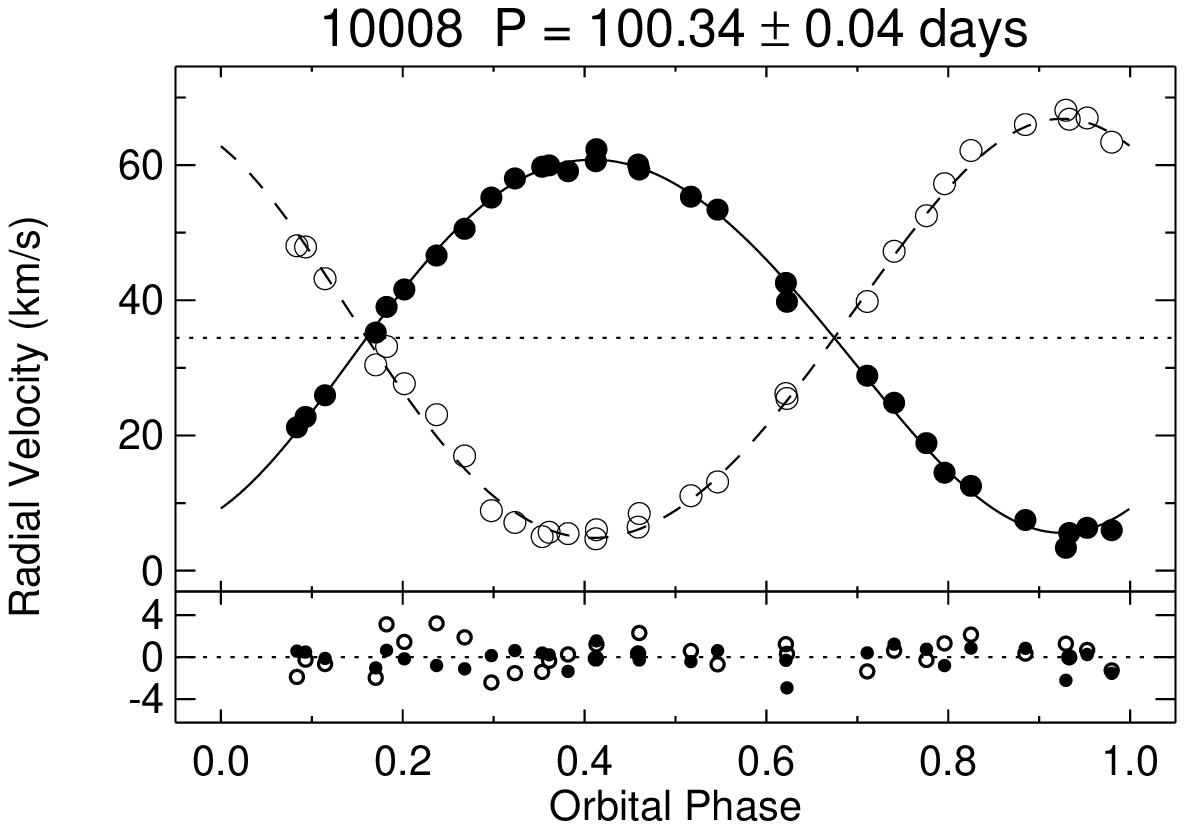}{0.3\textwidth}{}}
\gridline{\fig{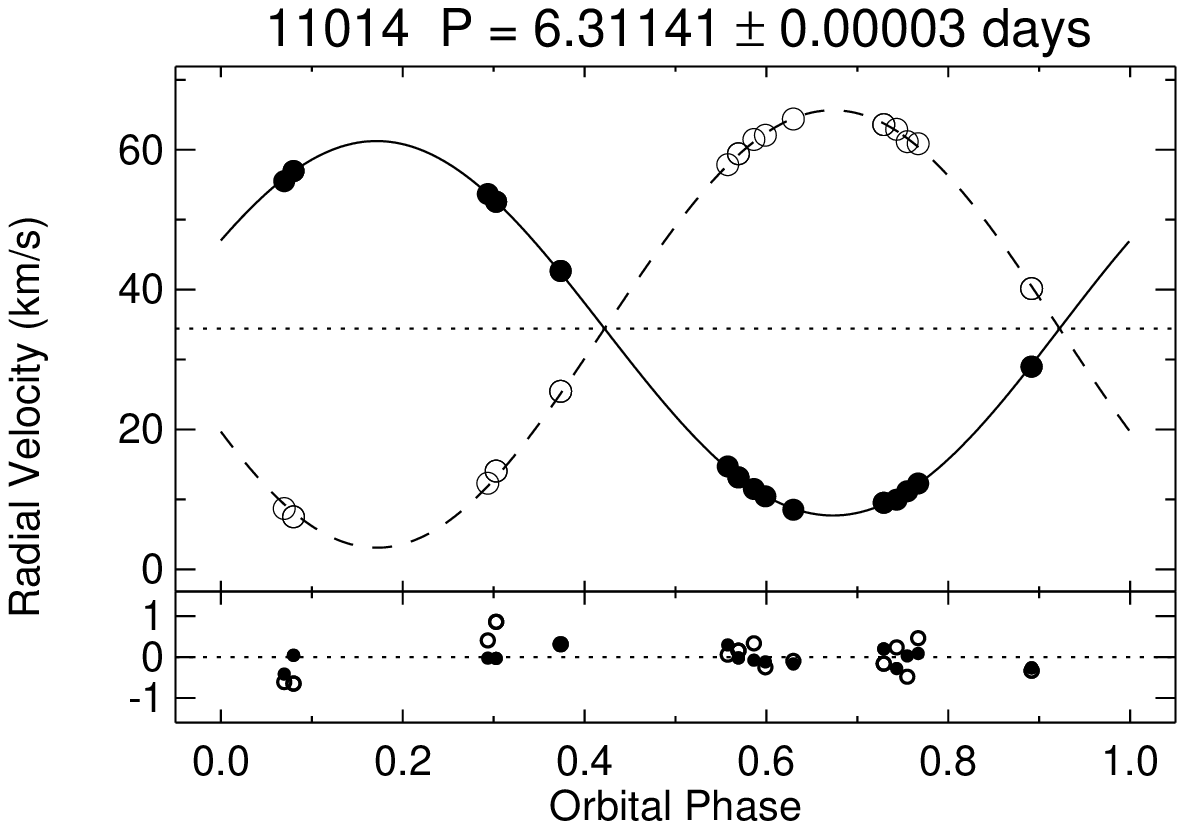}{0.3\textwidth}{}
	\fig{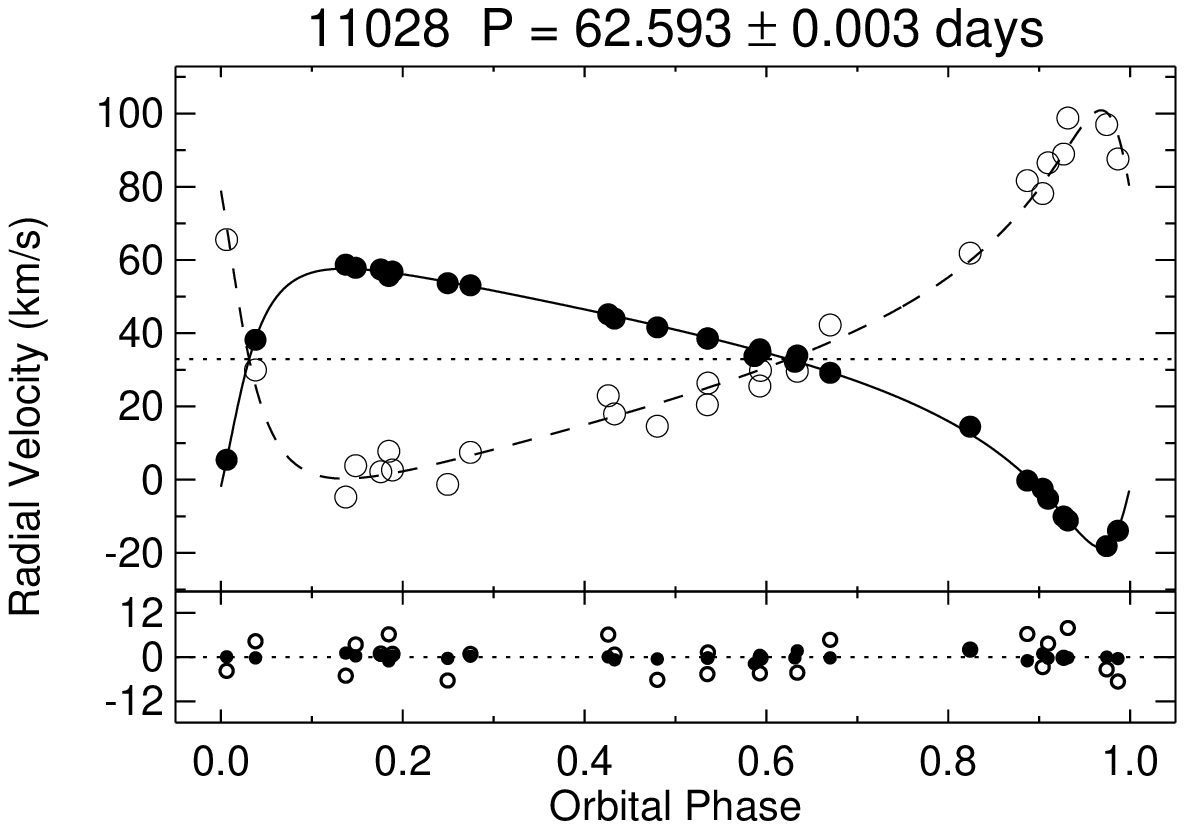}{0.3\textwidth}{}
	\fig{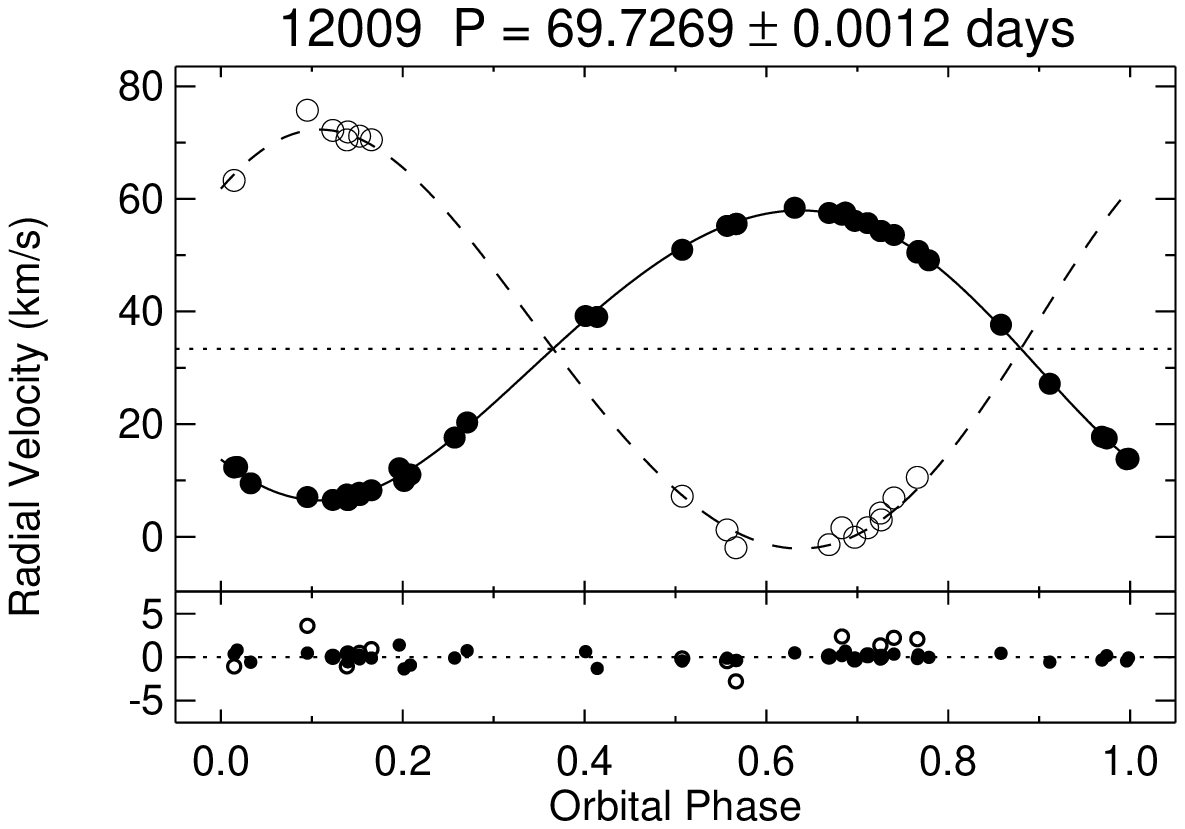}{0.3\textwidth}{}}
\caption{(Continued.)}
\end{figure*}

\begin{figure*}
\figurenum{12}
\gridline{\fig{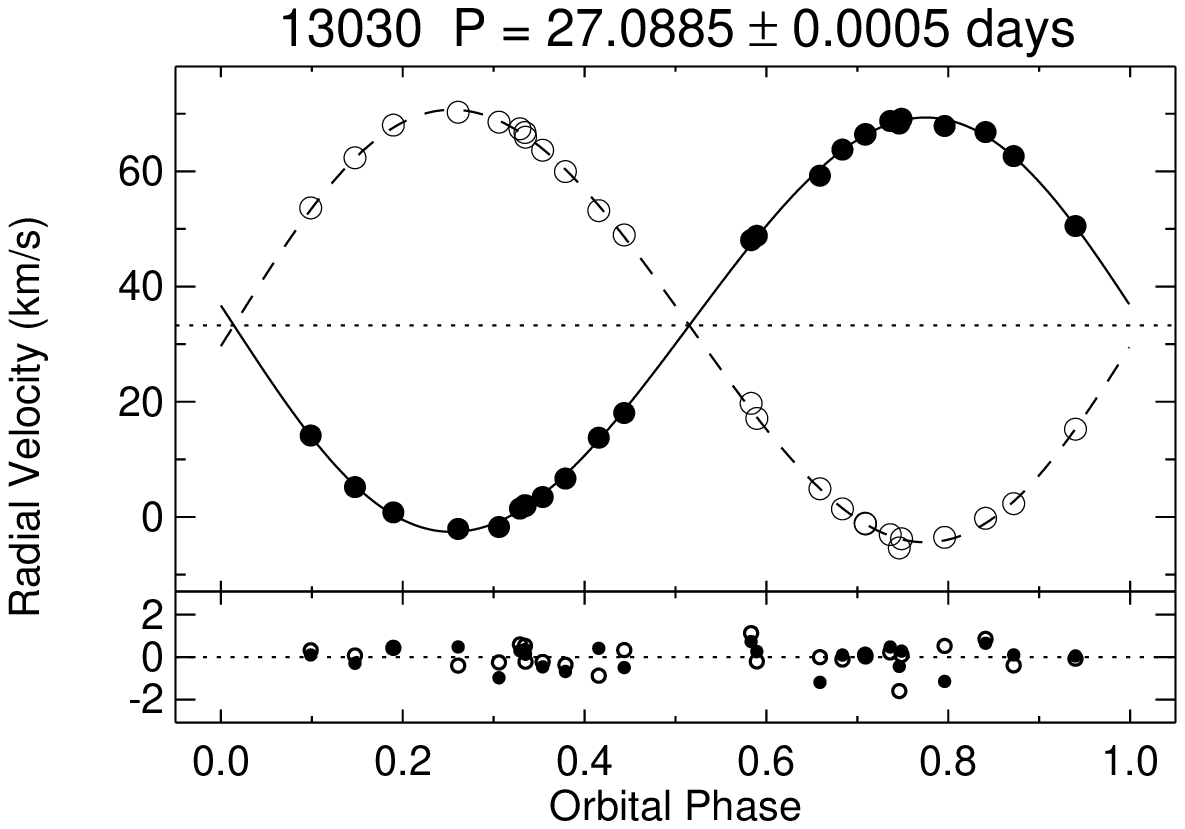}{0.3\textwidth}{}
	\fig{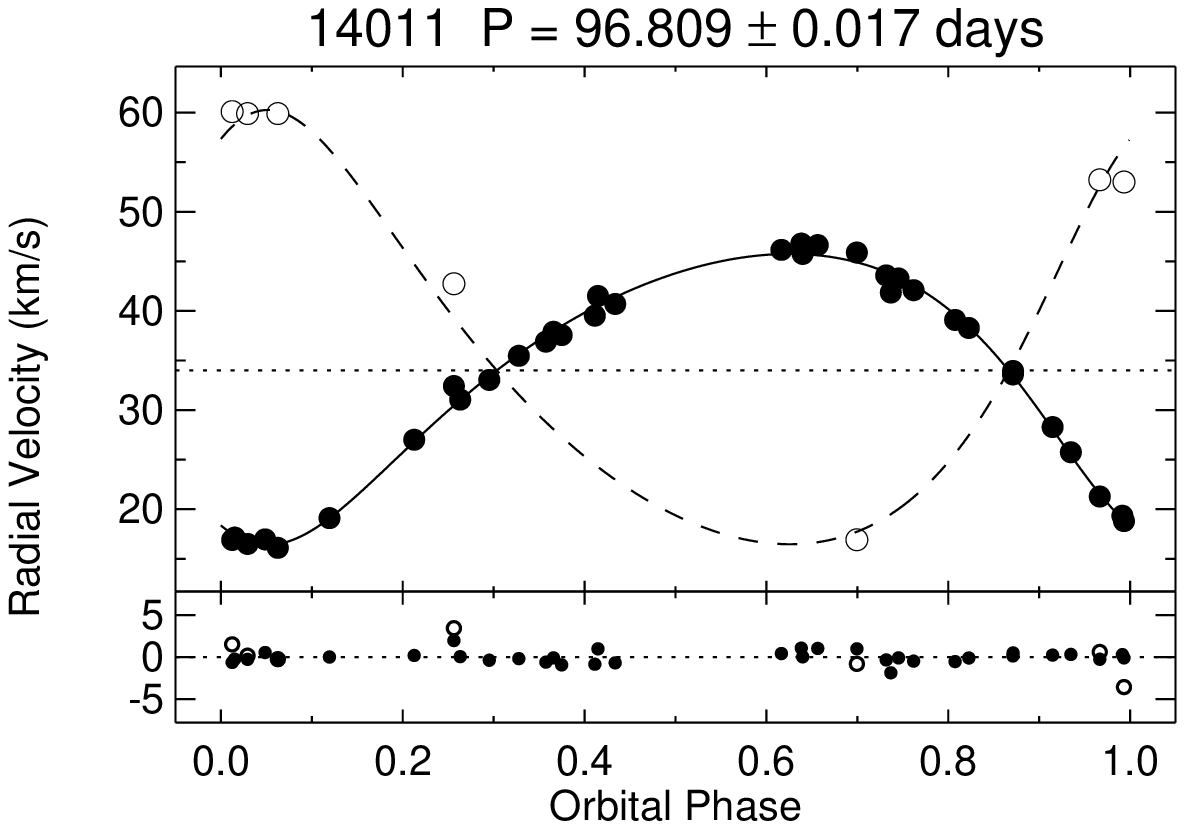}{0.3\textwidth}{}
	\fig{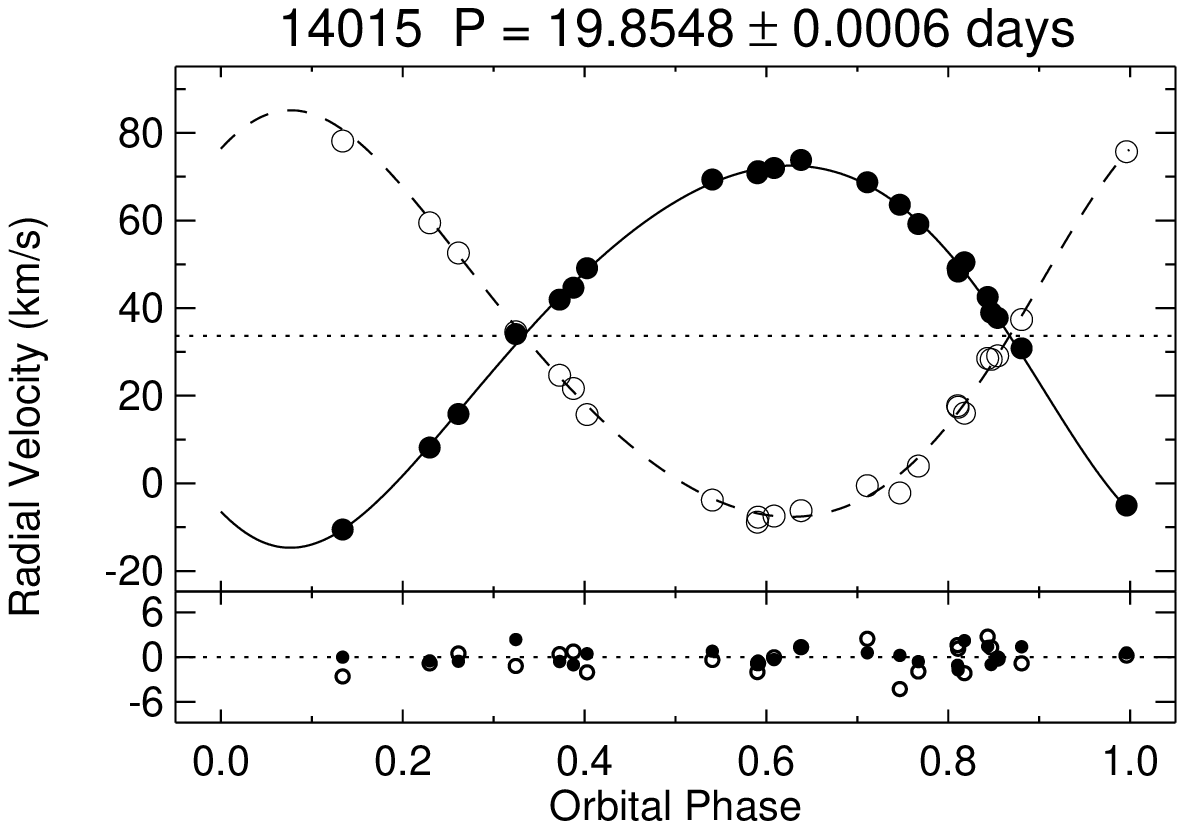}{0.3\textwidth}{}}
\gridline{\fig{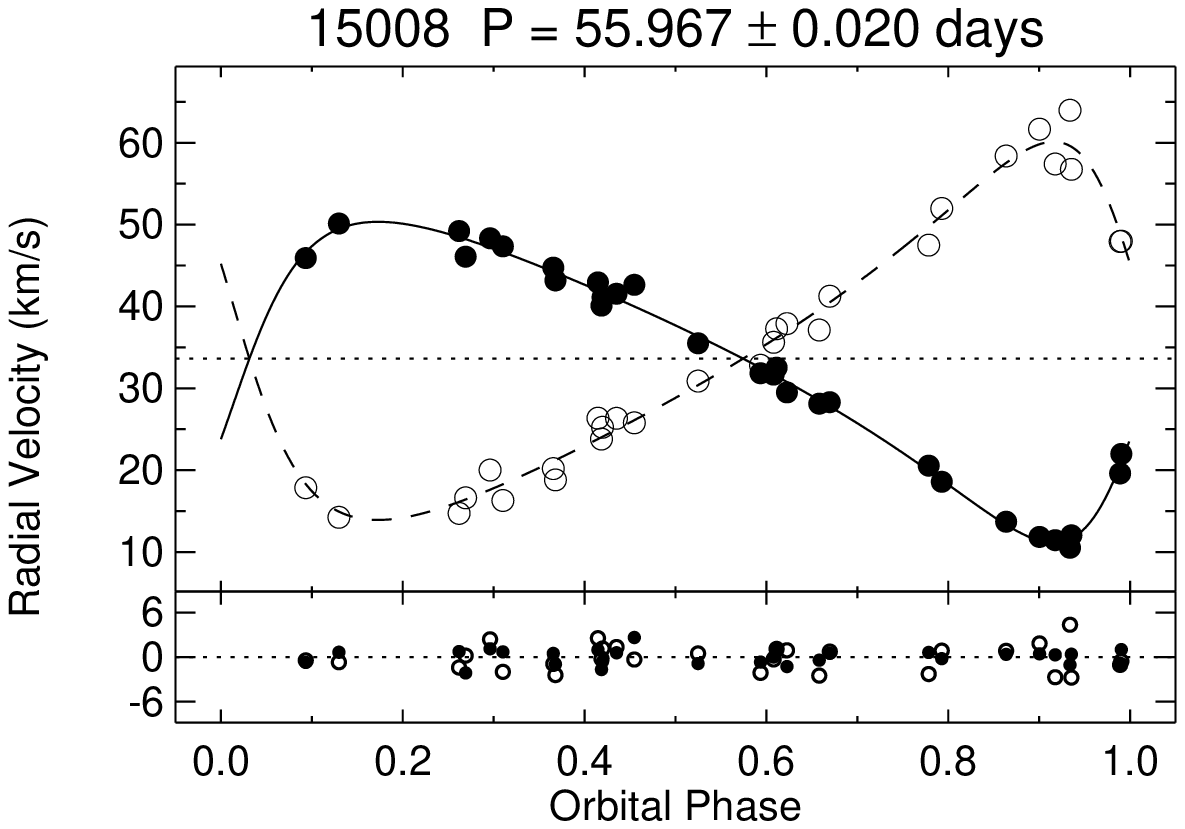}{0.3\textwidth}{}
	\fig{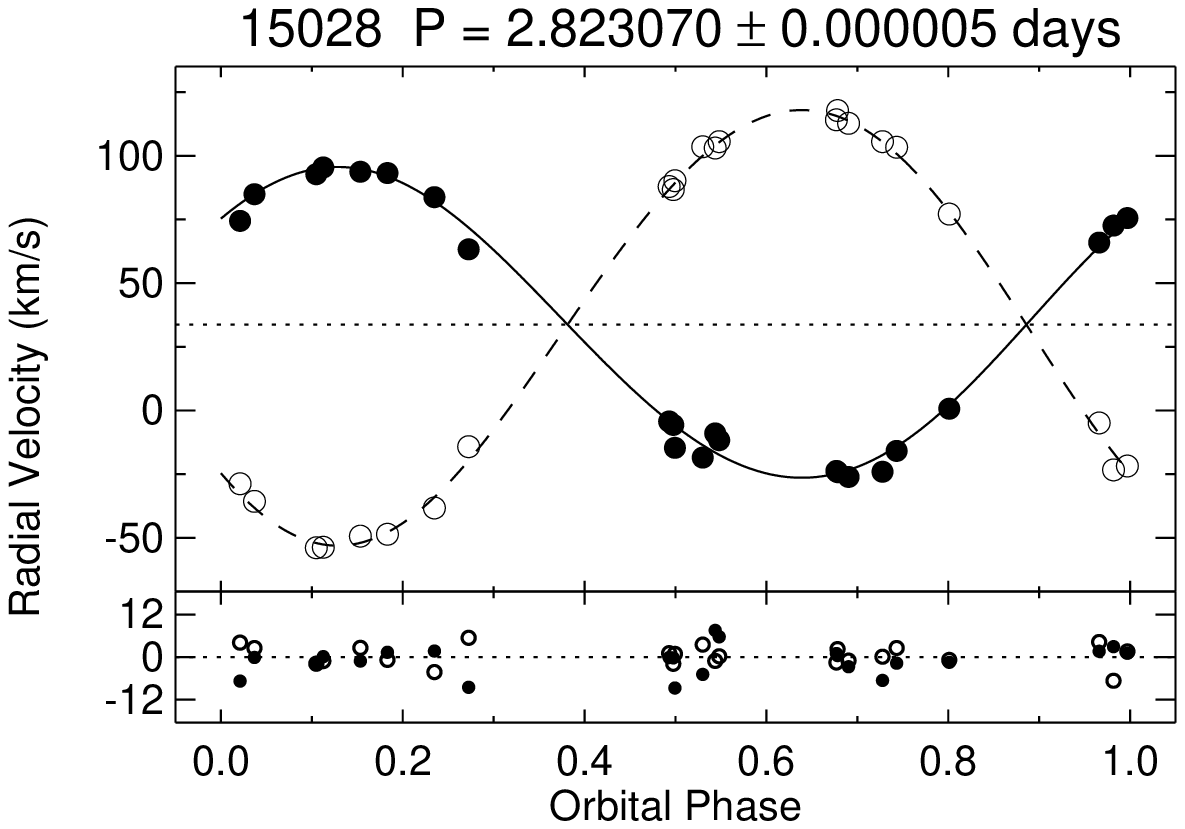}{0.3\textwidth}{}
	\fig{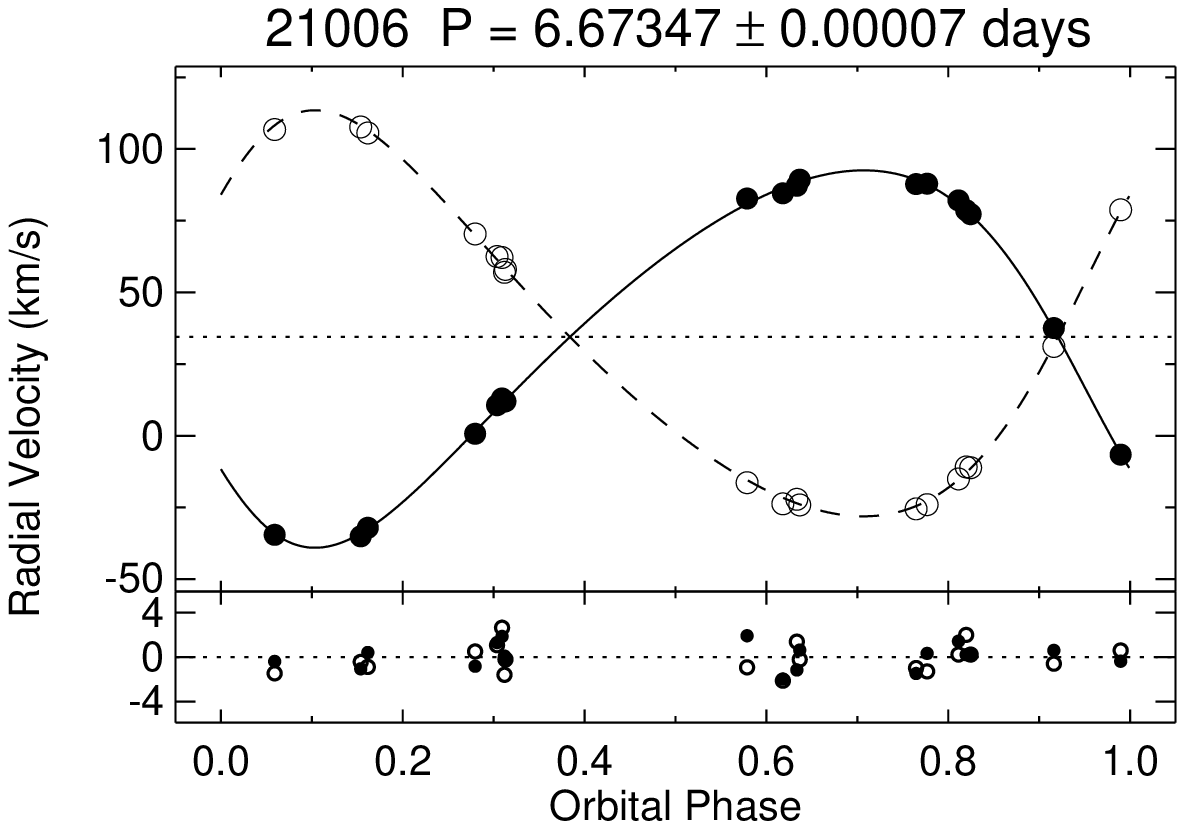}{0.3\textwidth}{}}
\gridline{\fig{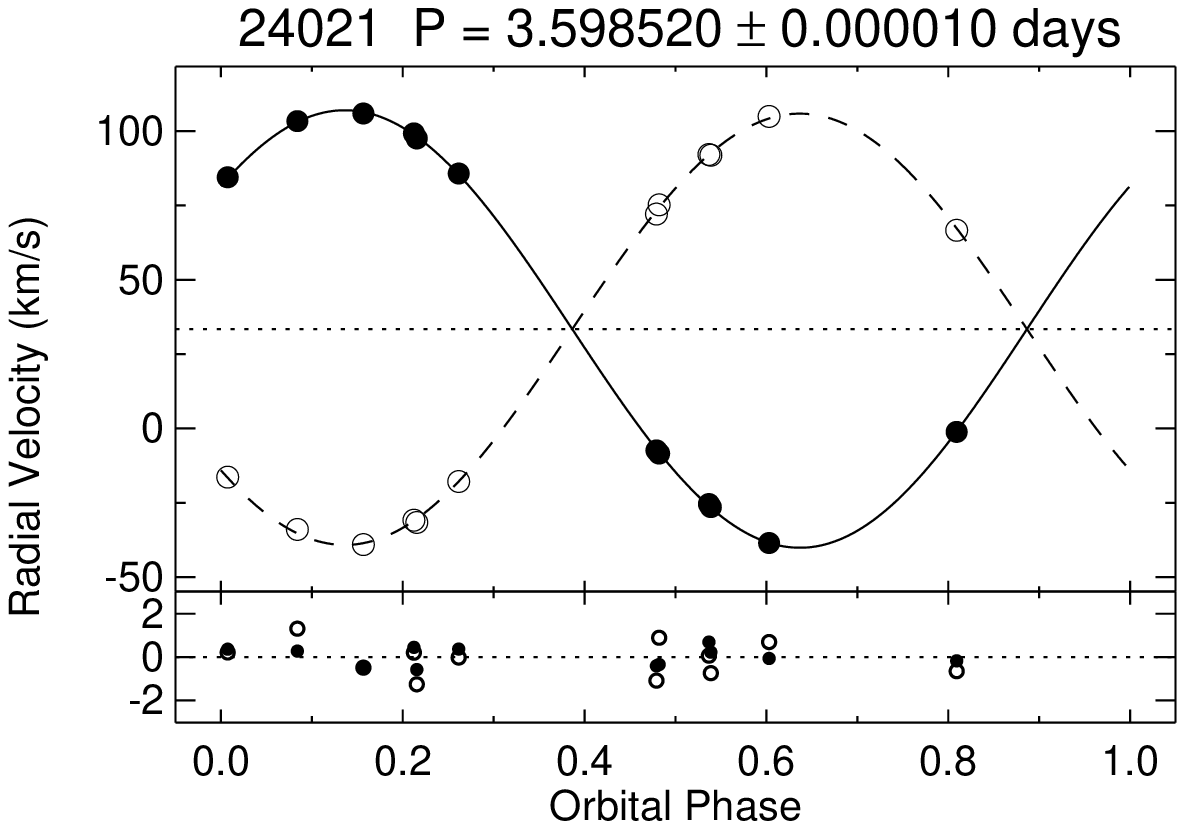}{0.3\textwidth}{}
}
\caption{(Continued.)}
\end{figure*}

\begin{deluxetable*}{llrcrrrrccrcrcc}
\tabletypesize{\footnotesize}
\tablecaption{Orbital Parameters for M67 Triples\label{tab:triplepars}}
\tablehead{\colhead{ID} & \colhead{XID} & \colhead{$P$} & \colhead{Orbital} & \colhead{$\gamma$} & \colhead{$K$} & \colhead{$e$} & \colhead{$\omega$} & \colhead{$T_\circ$} & \colhead{$a\sin i$} & \colhead{$f(m)$} & \colhead{$m\sin^3 i$} & \colhead{$q$} & \colhead{$\sigma$} & \colhead{$N$} \\
\colhead{} & \colhead{} & \colhead{(days)} & \colhead{Cycles} & \colhead{(\kms)} & \colhead{(\kms)} & \colhead{} & \colhead{(deg)} & \colhead{(HJD-2400000 d)} & \colhead{(10$^6$ km)} & \colhead{(\Msolar)} & \colhead{(\Msolar)} & \colhead{} & \colhead{(\kms)} & \colhead{}}
\rotate
\startdata
\arrayrulecolor{lightgray}
   3012  & S1077 &&&&&&&&&&&&&\\
   \multicolumn{2}{c}{inner} &       1.3587758 & 8679.8 &         33.45 &            48.3 &           0.015 &             330 &        52100.29 &           0.902 &                             &       0.112 &           0.76 &  3.64 &   99 \\
         &                   & $\pm$ 0.0000007 &      &      $\pm$ 0.21 &       $\pm$ 0.5 &     $\pm$ 0.011 &        $\pm$ 40 &      $\pm$ 0.16 &     $\pm$ 0.010 &                             & $\pm$ 0.014 &     $\pm$ 0.05 &       &      \\
         &                   &                &       &                 &              64 &                 &                 &                 &            1.19 &                             &       0.085 &                & 11.60 &   16 \\
         &                   &                &       &                 &         $\pm$ 4 &                 &                 &                 &      $\pm$ 0.07 &                             & $\pm$ 0.006 &                &       &      \\
   \multicolumn{2}{c}{outer} &           3613 &   3.3 &           33.45 &             6.0 &            0.33 &             233 &           47150 &             280 &                             &       1.30 &            0.49 &  3.64 &   99 \\
         &                   &       $\pm$ 13 &       &      $\pm$ 0.21 &       $\pm$ 0.6 &      $\pm$ 0.03 &         $\pm$ 5 &        $\pm$ 57 &        $\pm$ 26 &                             & $\pm$ 0.14 &      $\pm$ 0.04 &       &      \\
         &                   &                &       &                 &            12.3 &                 &                 &                 &             577 &                             &       0.64 &                 &  2.13 &   99 \\
         &                   &                &       &                 &       $\pm$ 0.5 &                 &                 &                 &        $\pm$ 20 &                             & $\pm$ 0.11 &                 &       &      \\
\hline
  4008   & S2206 &&&&&&&&&&&&&\\
  \multicolumn{2}{c}{inner} &        18.3768 &  449.8 &            31.3 &            12.3 &            0.12 &             100 &         47955.9 &            3.08 &          3.5$\times10^{-3}$ &            &                 &  3.91 &   91 \\
         &                  &   $\pm$ 0.0012 &        &       $\pm$ 0.4 &       $\pm$ 0.6 &      $\pm$ 0.05 &        $\pm$ 23 &       $\pm$ 1.1 &      $\pm$ 0.16 &    $\pm$ 0.5$\times10^{-3}$ &            &                 &       &      \\
\hline
  4030   & S1416 &&&&&&&&&&&&&\\
  \multicolumn{2}{c}{inner} &         8.9918 &  233.1 &            33.4 &            39.4 &           0.333 &           162.5 &        49770.39 &            4.59 &         4.77$\times10^{-2}$ &            &                 &  0.92 &   21 \\
         &                  &   $\pm$ 0.0003 &        &       $\pm$ 0.3 &       $\pm$ 0.7 &     $\pm$ 0.014 &       $\pm$ 1.6 &      $\pm$ 0.03 &      $\pm$ 0.07 &   $\pm$ 0.22$\times10^{-2}$ &            &                 &       &      \\
  \multicolumn{2}{c}{outer} &            673 &    3.1 &            33.4 &             7.0 &            0.42 &              75 &           49837 &              58 &          1.8$\times10^{-2}$ &            &                 &  0.92 &   21 \\   
         &                  &       $\pm$ 14 &        &       $\pm$ 0.3 &       $\pm$ 0.4 &      $\pm$ 0.05 &        $\pm$ 10 &        $\pm$ 12 &         $\pm$ 4 &    $\pm$ 0.3$\times10^{-2}$ &            &                 &       &      \\
\hline
  7008   & S1234 &&&&&&&&&&&&&\\
  \multicolumn{2}{c}{inner} &       4.355797 & 3503.0 &           33.26 &           23.34 &           0.028 &             314 &        50613.62 &           1.397 &         5.73$\times10^{-3}$ &            &                 &  1.22 &   70 \\
         &                  & $\pm$ 0.000007 &        &      $\pm$ 0.13 &      $\pm$ 0.19 &     $\pm$ 0.010 &        $\pm$ 18 &      $\pm$ 0.22 &     $\pm$ 0.012 &   $\pm$ 0.14$\times10^{-3}$ &            &                 &       &      \\
  \multicolumn{2}{c}{outer} &          10360 &    1.1 &           33.26 &             3.4 &            0.65 &             148 &           44560 &             360 &                             &       0.29 &            0.53 &  1.22 &   70 \\
         &                  &       $\pm$ 90 &        &      $\pm$ 0.13 &       $\pm$ 0.4 &      $\pm$ 0.04 &         $\pm$ 5 &       $\pm$ 140 &        $\pm$ 40 &                             & $\pm$ 0.05 &      $\pm$ 0.06 &       &      \\
         &                  &                &        &                 &             6.4 &                 &                 &                 &             690 &                             &       0.15 &                 &  1.07 &   77 \\
         &                  &                &        &                 &       $\pm$ 0.6 &                 &                 &                 &        $\pm$ 50 &                             & $\pm$ 0.03 &                 &       &      \\
\hline
  10012  & S796 &&&&&&&&&&&&&\\
  \multicolumn{2}{c}{inner} &           1910 &    2.5 &           33.88 &            1.28 &            0.45 &             208 &           55490 &              30 &          3.0$\times10^{-4}$ &            &                 &  0.33 &   28 \\
         &                  &      $\pm$  31 &        &      $\pm$ 0.07 &      $\pm$ 0.20 &      $\pm$ 0.08 &        $\pm$ 13 &        $\pm$ 45 &         $\pm$ 5 &    $\pm$ 1.4$\times10^{-4}$ &            &                 &       &      \\
\hline
  21005  & S1278 &&&&&&&&&&&&&\\
  \multicolumn{2}{c}{inner} &          284.21 &  34.9 &           33.66 &            10.5 &           0.209 &             324 &           51979 &            40.0 &                             &      0.150 &            0.92 &  1.10 &   32 \\
         &                  &      $\pm$ 0.08 &       &      $\pm$ 0.18 &       $\pm$ 0.3 &     $\pm$ 0.025 &         $\pm$ 7 &         $\pm$ 5 &       $\pm$ 1.0 &                             & $\pm$ 0.016 &     $\pm$ 0.05 &       &      \\
         &                  &                 &       &                 &            11.4 &                 &                 &                 &            43.6 &                             &      0.138 &                 &  2.29 &   32 \\
         &                  &                 &       &                 &       $\pm$ 0.6 &                 &                 &                 &       $\pm$ 2.1 &                             & $\pm$ 0.010 &                &       &      \\
\arrayrulecolor{black}
\enddata
\end{deluxetable*}

\begin{figure*}
\plottwo{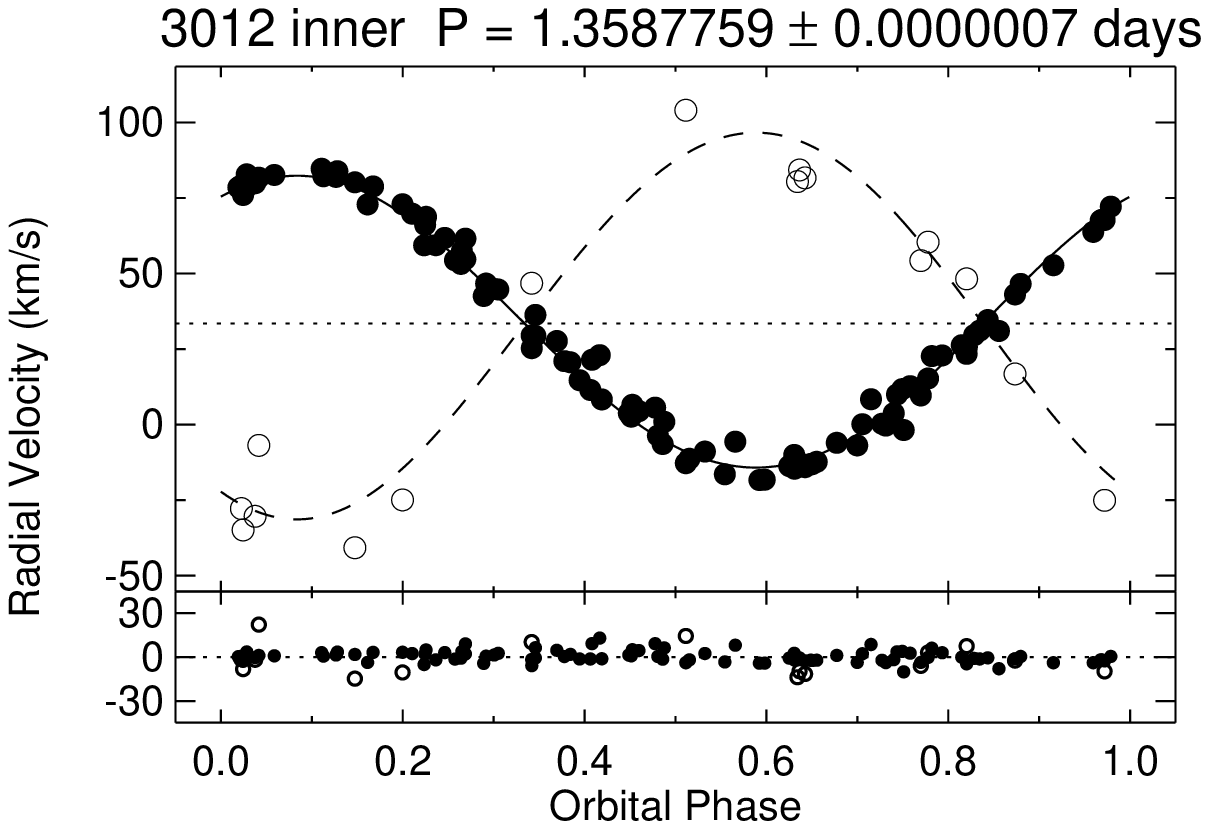}{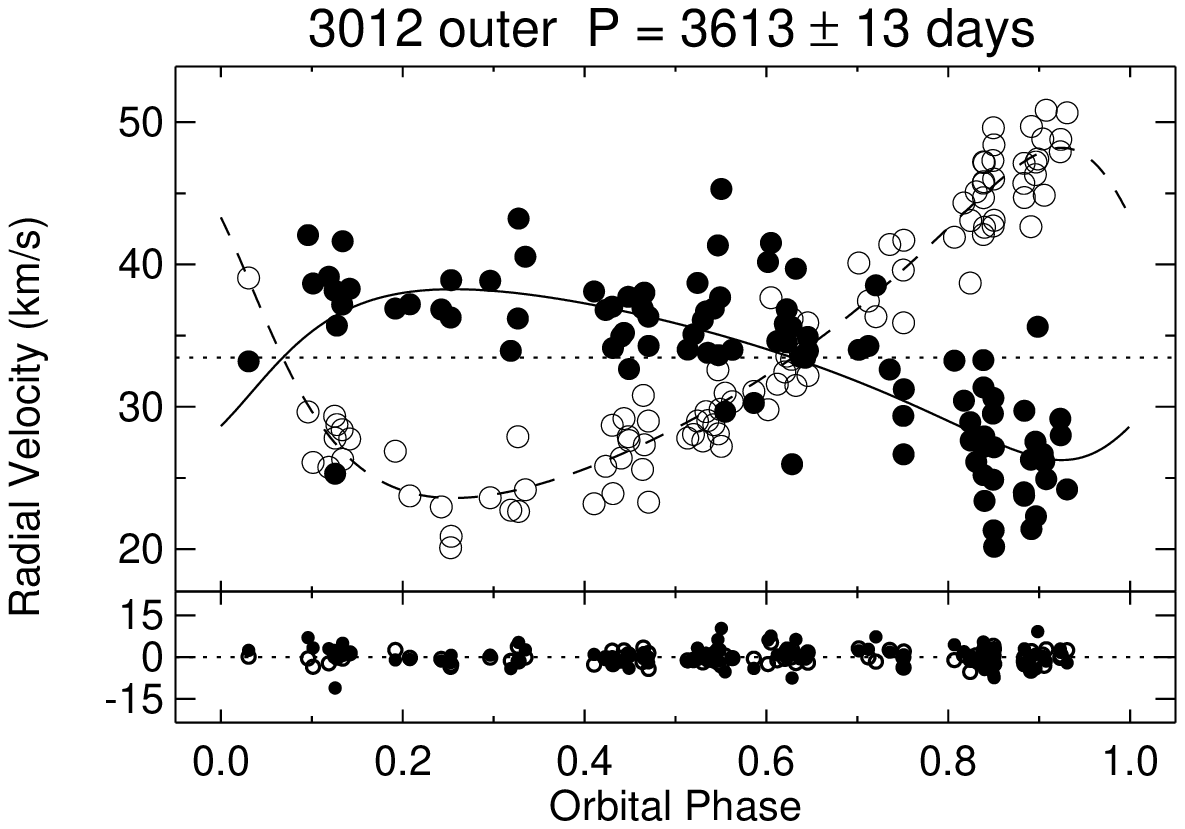}
\plottwo{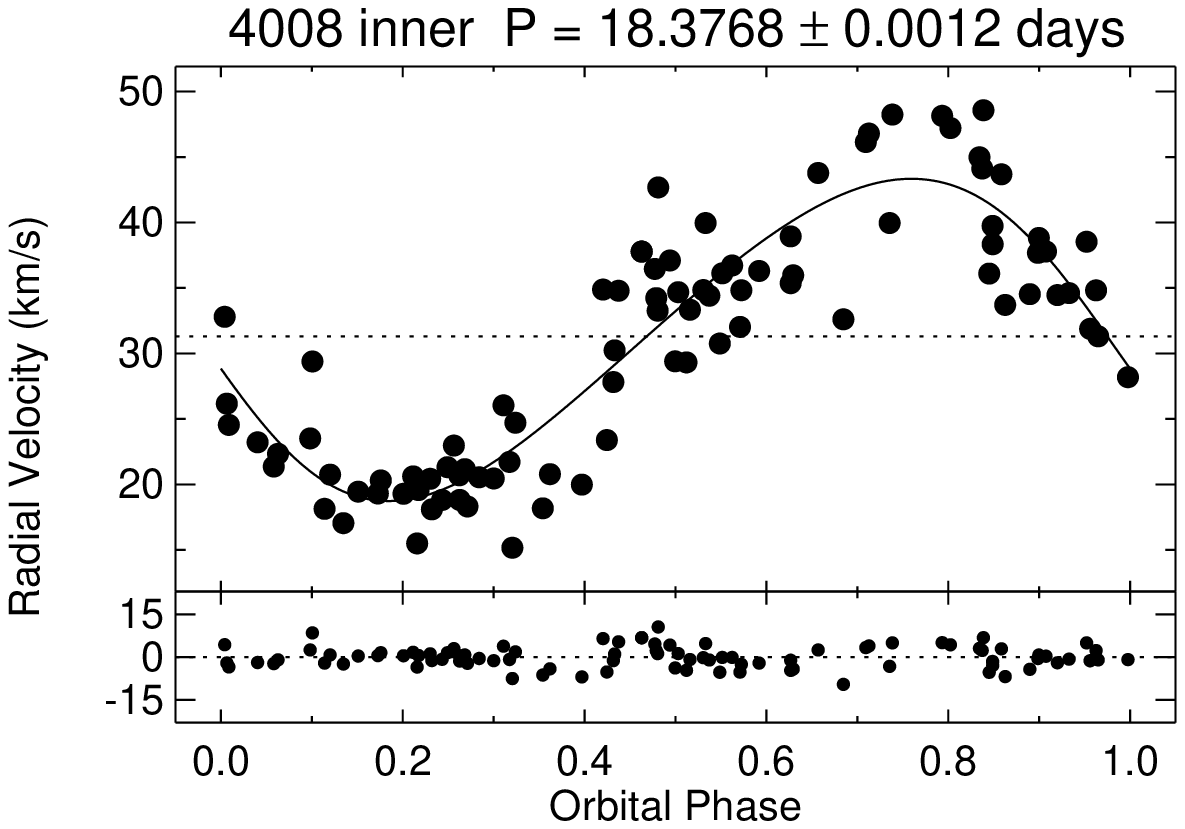}{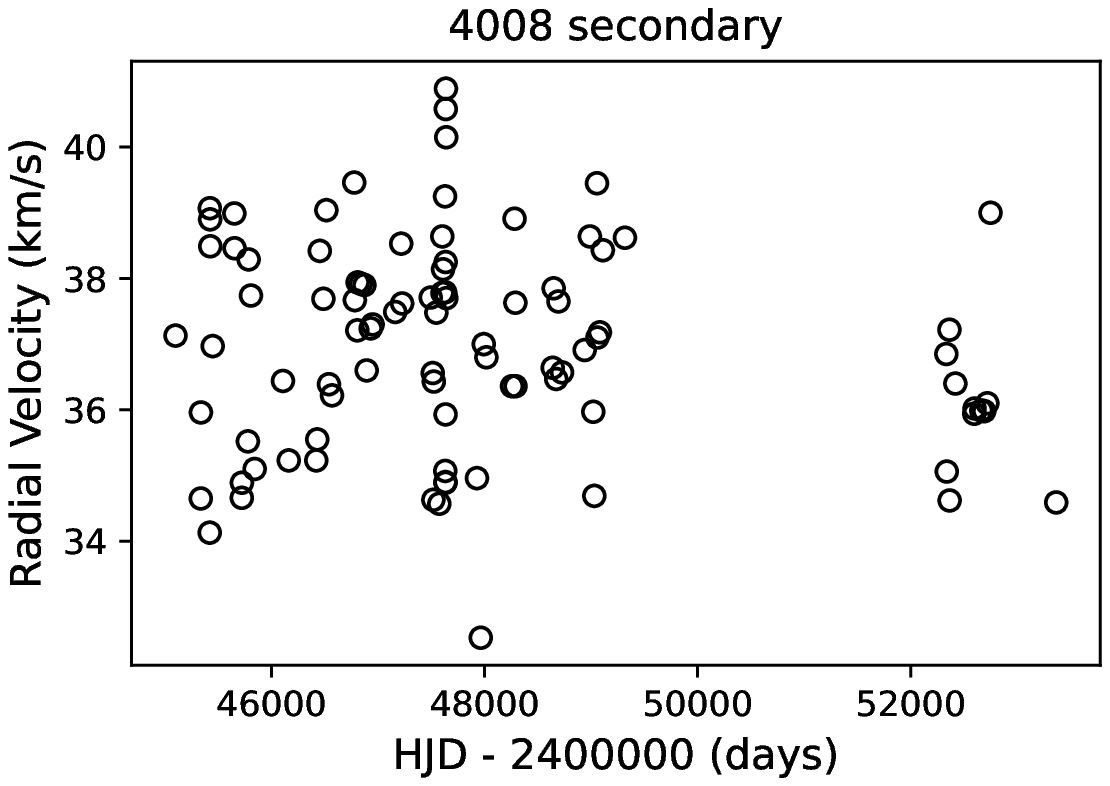}
\plottwo{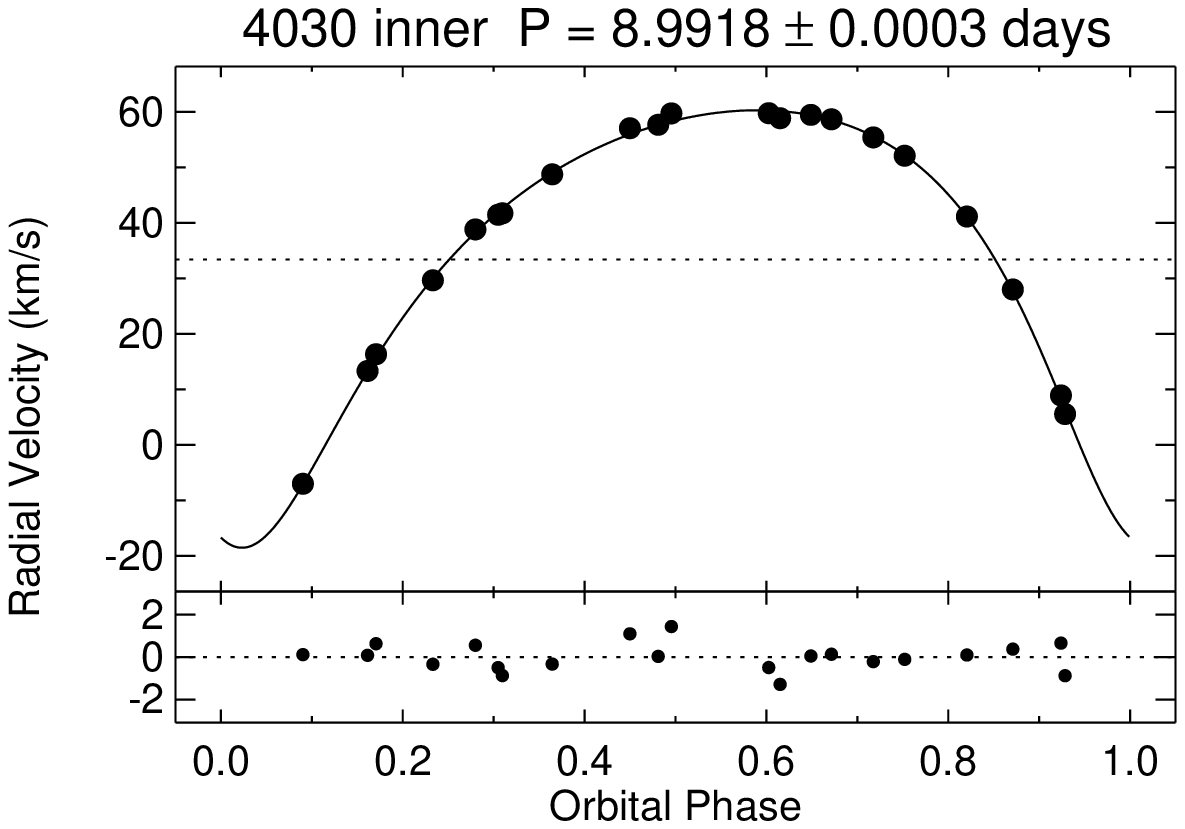}{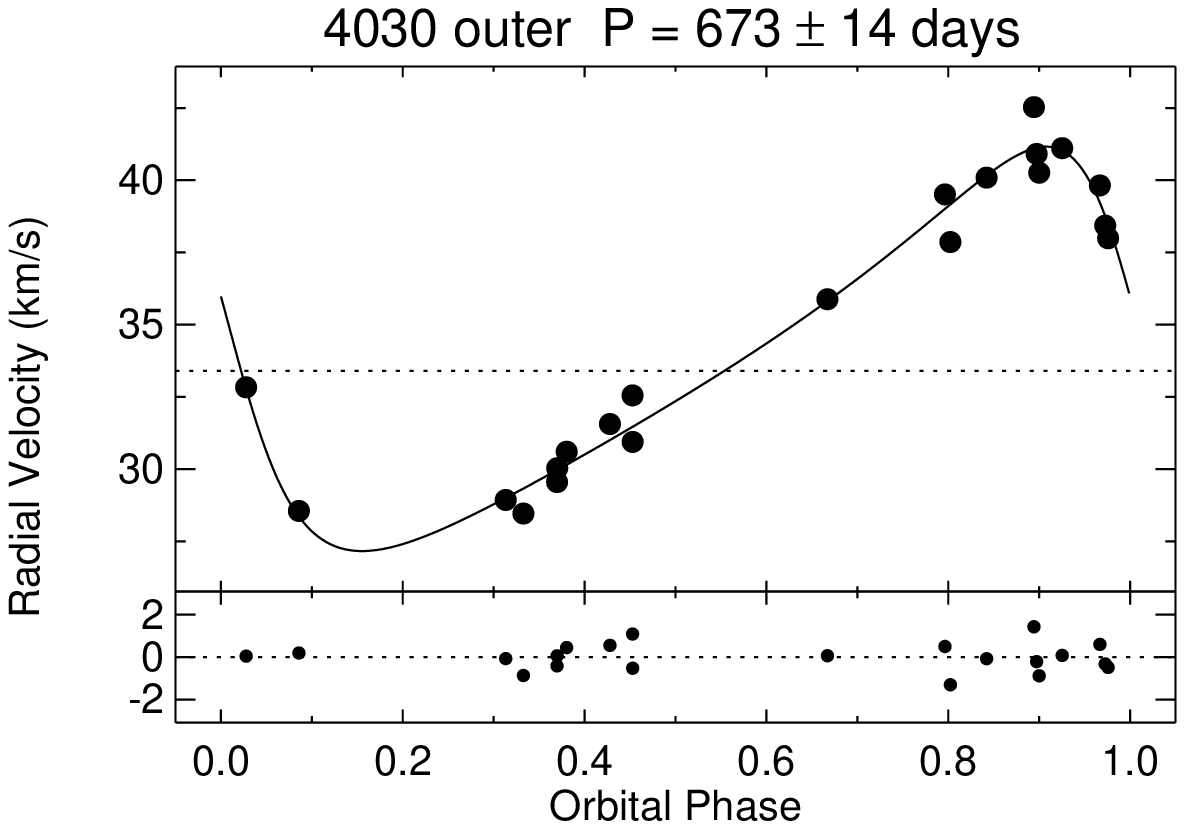}
\caption{M67 triple member plots. Each row in the figure corresponds to one triple system.  Orbit plots follow a similar format to Figures~\ref{fig:sb1orbs}~and~\ref{fig:sb2orbs}; radial velocity (RV) is plotted versus phase, with primary RVs shown as filled circles, and secondary RVs in open circles (where relevant). For two triples (WOCS 3012 and WOCS 7008), outer orbits are fit using center-of-mass RVs of the inner binary and secondary RVs of the outer triple companion---these center-of-mass RVs are plotted with filled circles. The primary orbital solution is shown as a solid black line, the secondary solution as a dashed black line (where relevant), with the dotted black line marking the $\gamma$ velocity of the binary. Below each orbit plot, the primary and secondary residuals ($O-C$) are given. Above each plot, we list the WOCS ID, the component of the triple and orbital period (if relevant).  For WOCS 4008, on the right we include a plot of the secondary RVs as a function of time.  For WOCS 10012, on the right we include a plot of the observed RVs as a function of time;  the top panel shows the observed RVs (black points) and a linear trend fit to the RVs (gray dashed line), while in the bottom panel the linear trend is subtracted from the observed RVs, and the resulting de-trended RVs are plotted. Finally for WOCS 21005, on the right we plot the ($O-C$) residuals for the primary (filled circles) and secondary (open circles) stars as a function of time.  \label{fig:tripleorbs}}
\end{figure*}

\begin{figure*}
\figurenum{13}
\plottwo{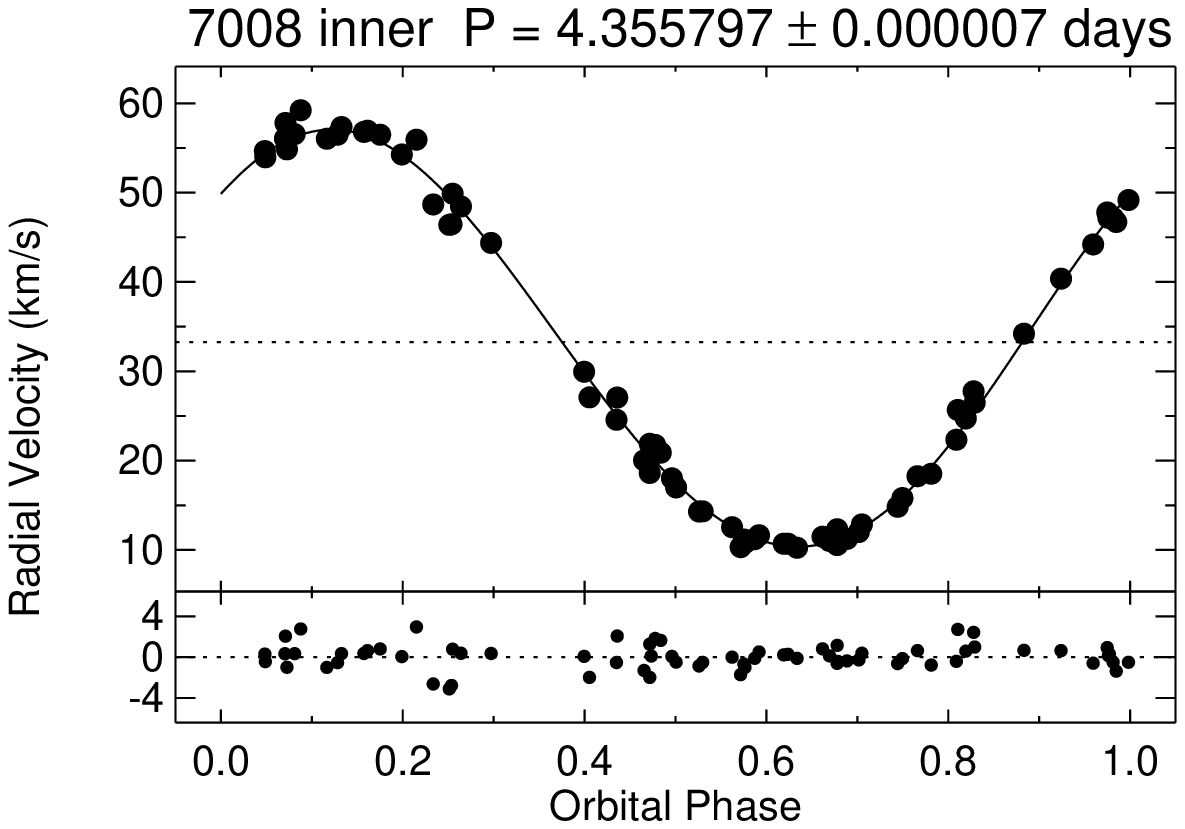}{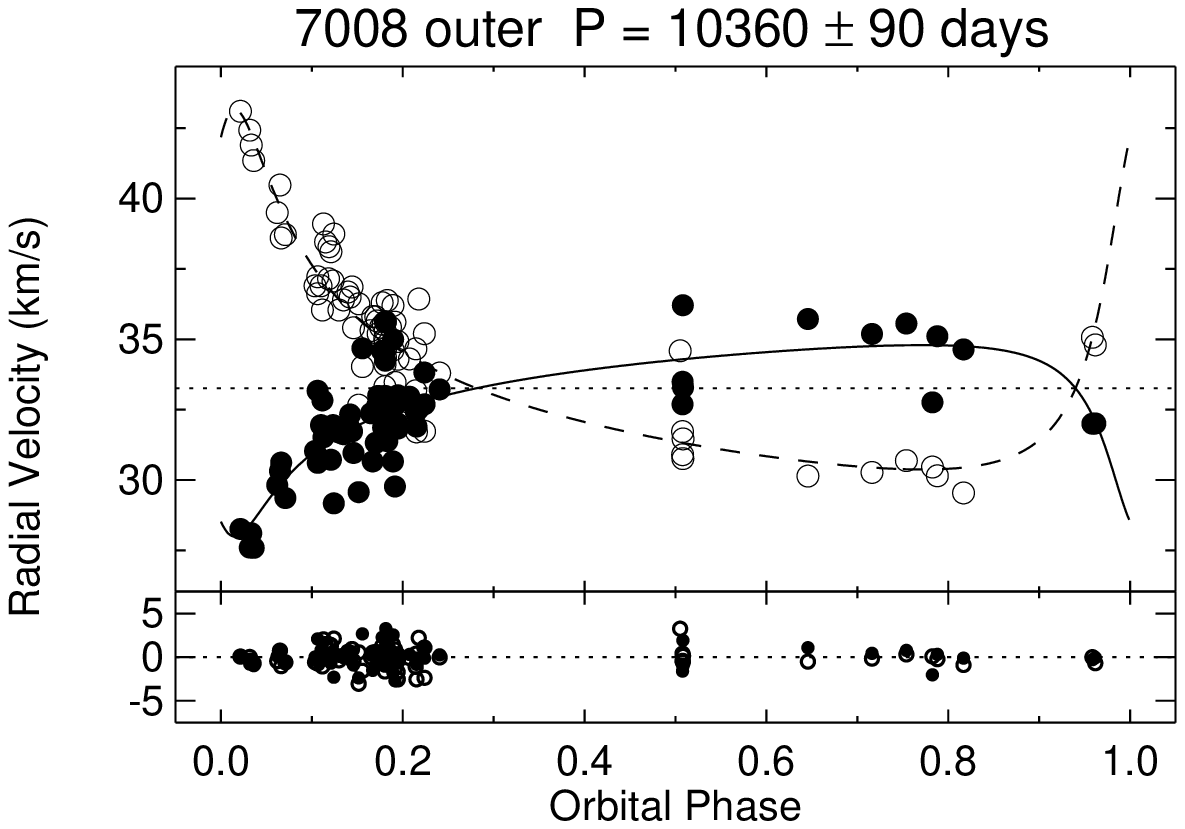}
\end{figure*}
\begin{figure*}[]
    \centering
    \begin{minipage}{0.48\textwidth}
        \centering
        \includegraphics[width=0.9\textwidth]{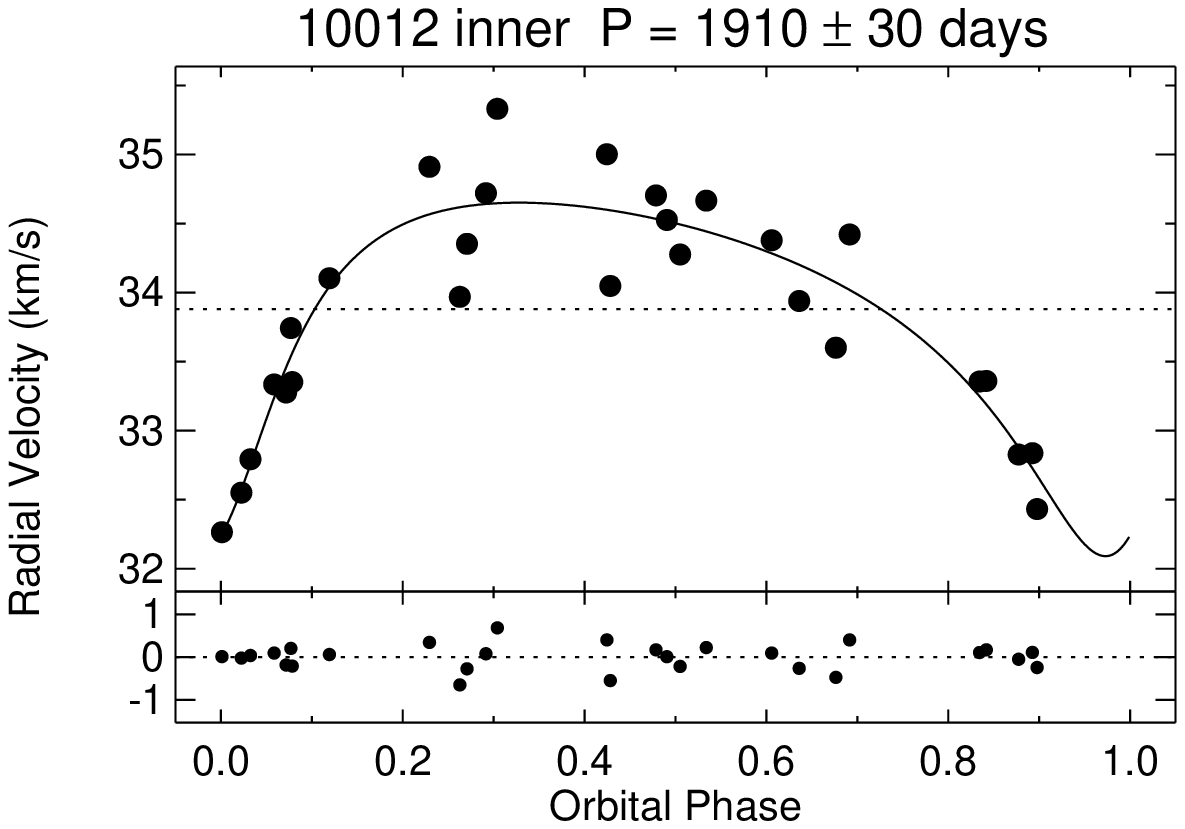}
    \end{minipage}
    \begin{minipage}{0.45\textwidth}
        \centering
        \includegraphics[width=0.95\textwidth]{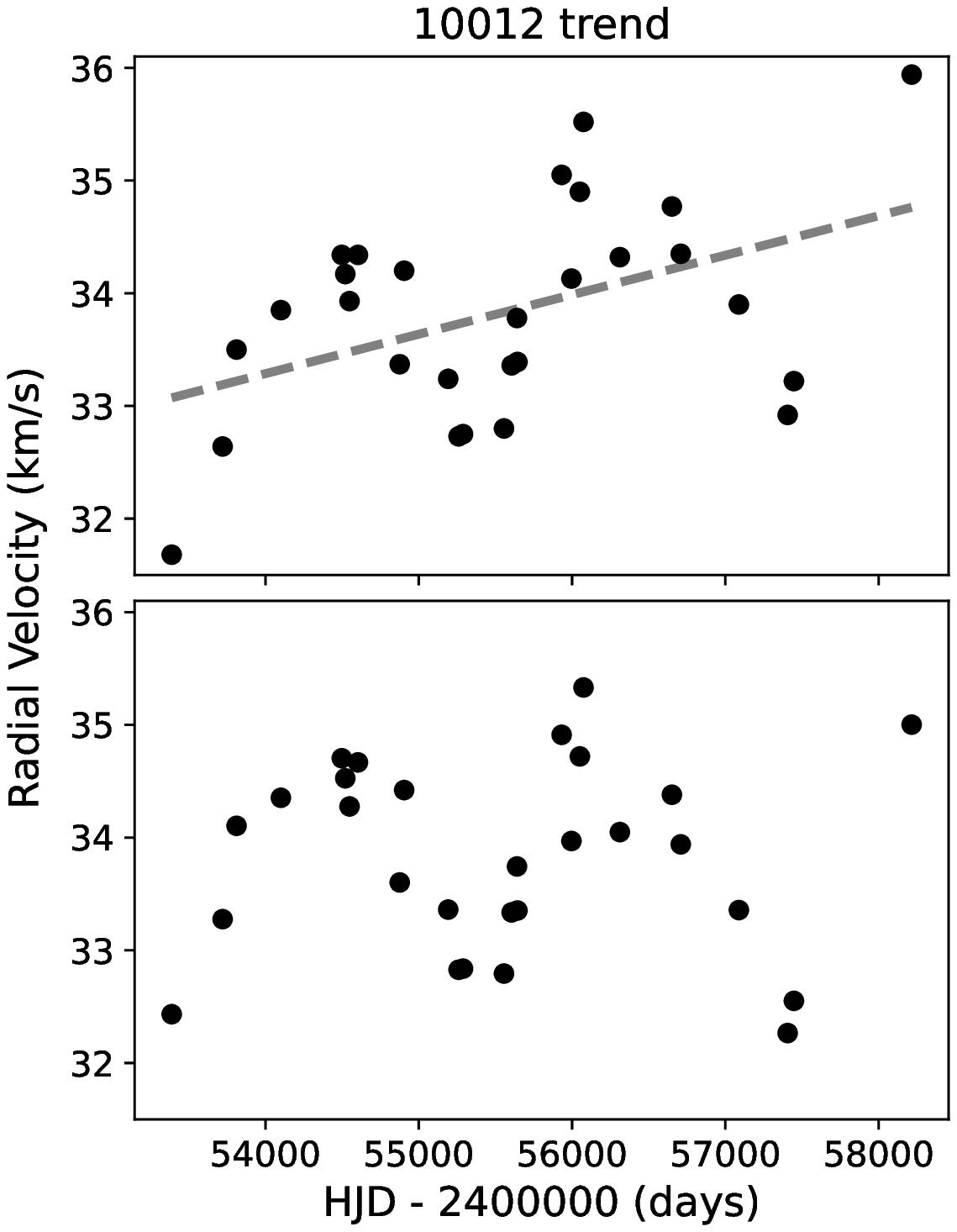}
    \end{minipage}
\end{figure*}
\begin{figure*}
\figurenum{13}
\plottwo{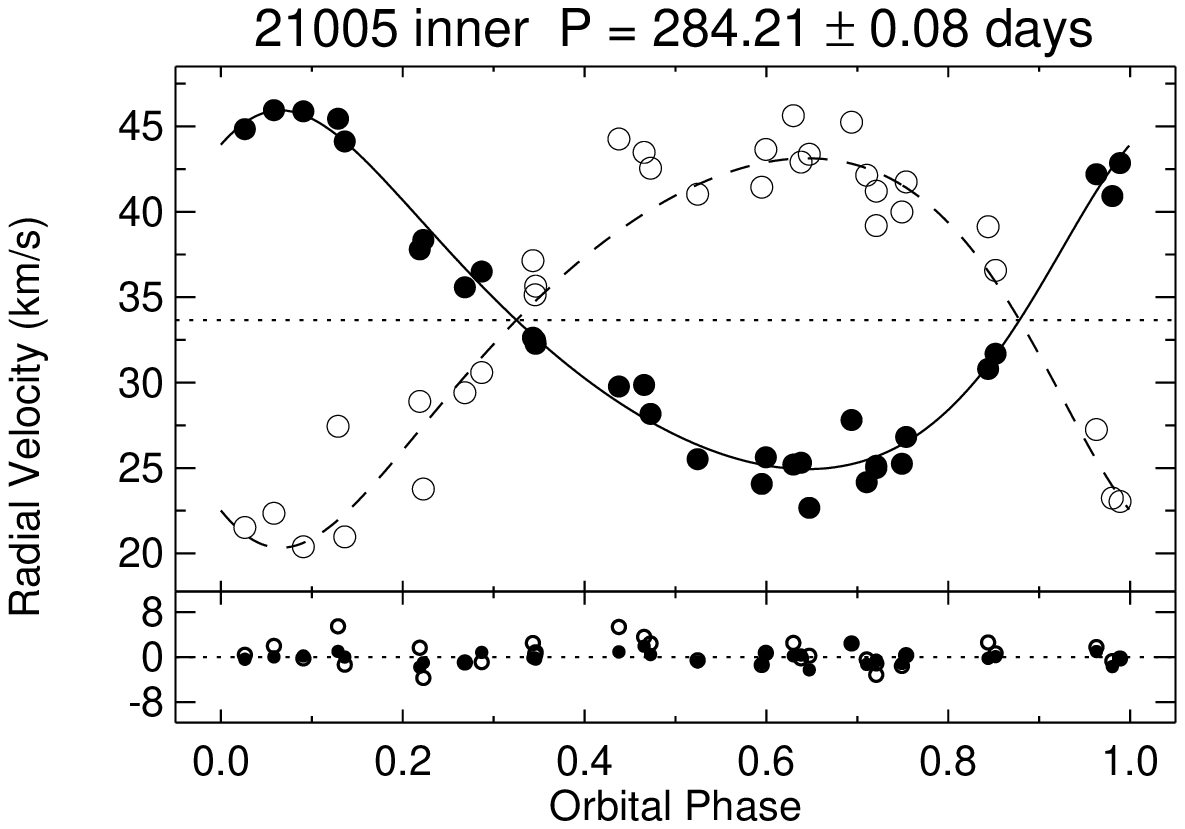}{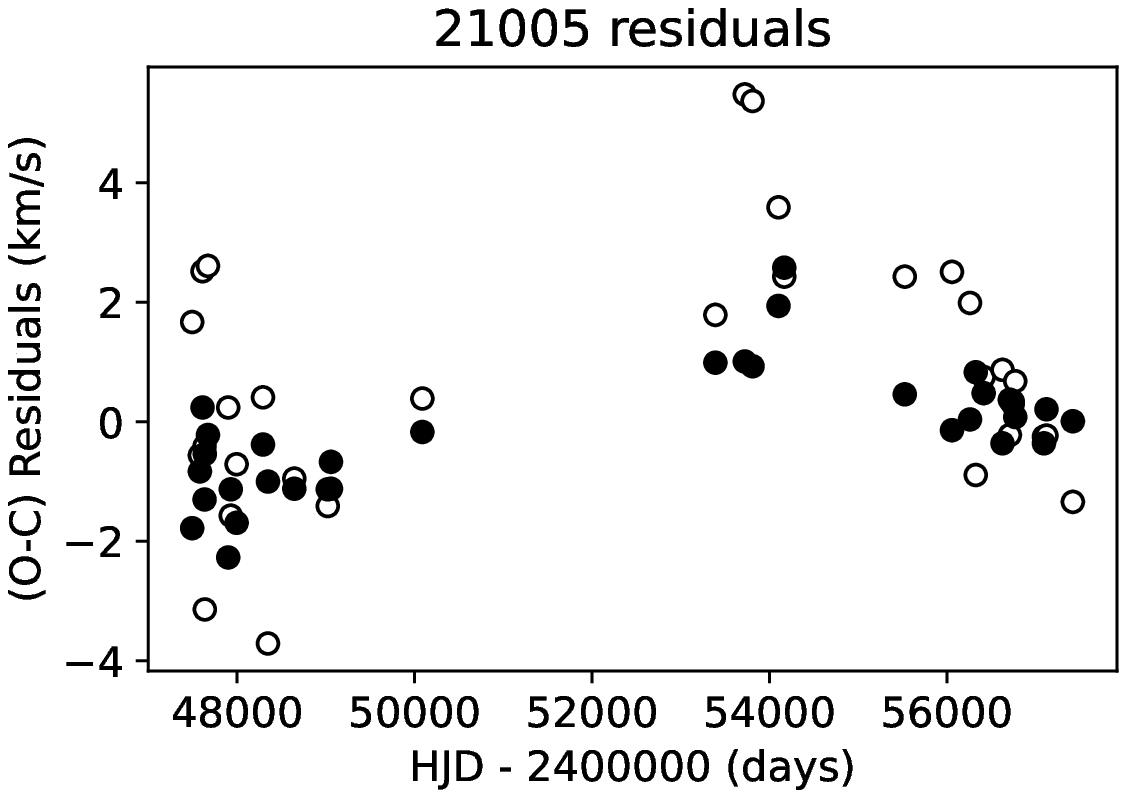}
\caption{(Continued.)}
\end{figure*}

\end{document}